\pdfoutput=1
\documentclass[rmp,aps,twocolumn,showpacs,preprintnumbers,amsmath,amssymb,floatfix,10pt]{revtex4-1}

\usepackage{graphicx}
\usepackage{dcolumn}
\usepackage{bm}
\usepackage{multirow}
\usepackage{hyperref}

\begin{document}


\title{The \texorpdfstring{$^{12}$C$(\alpha,\gamma)^{16}$O}{12C(a,g)16O} reaction and its implications for stellar helium burning} 

\author{R.~J.~deBoer}
\email{Electronic address: rdeboer1@nd.edu}
\affiliation{The Joint Institute for Nuclear Astrophysics}
\affiliation{Department of Physics, University of Notre Dame, Notre Dame, Indiana 46556 USA}

\author{R.E.~Azuma}
\altaffiliation[Deceased]{}
\affiliation{Department of Physics, University of Toronto, Toronto, Ontario M5S 1A7, Canada}
\affiliation{The Joint Institute for Nuclear Astrophysics, Department of Physics, University of Notre Dame, Notre Dame, Indiana 46556 USA}

\author{A.~Best}
\altaffiliation[Present address: ]{Universit\`{a} degli Studi di Napoli ``Federico II'' and INFN, Napoli, Italy}
\affiliation{INFN, Laboratori Nazionali del Gran Sasso, 67100 Assergi, Italy}

\author{C.R.~Brune}
\affiliation{Edwards Accelerator Laboratory, Department of Physics and Astronomy, Ohio University, Athens, Ohio 45701, USA}

\author{C.E.~Fields}
\altaffiliation[Ford Foundation Predoctoral Fellow]{}
\affiliation{The Joint Institute for Nuclear Astrophysics}
\affiliation{Department of Physics and Astronomy, Michigan State University, East Lansing, MI 48824, USA}

\author{J.~G\"{o}rres}
\affiliation{The Joint Institute for Nuclear Astrophysics}
\affiliation{Department of Physics, University of Notre Dame, Notre Dame, Indiana 46556 USA}

\author{S.~Jones}
\affiliation{Heidelberg Institute for Theoretical Studies, Schloss-Wolfsbrunnenweg 35, D-69118 Heidelberg, Germany}
\affiliation{NuGrid Collaboration, \href{http://nugridstars.org}{http://nugridstars.org}}

\author{M.~Pignatari}
\affiliation{E.A. Milne Centre for Astrophysics, Department of Physics \& Mathematics, University of Hull, HU6 7RX, United Kingdom}
\affiliation{Konkoly Observatory, Research Centre for Astronomy and Earth Sciences, Hungarian Academy of Sciences, Konkoly Thege Miklos ut 15-17, H-1121 Budapest, Hungary}
\affiliation{NuGrid Collaboration, \href{http://nugridstars.org}{http://nugridstars.org}}

\author{D.~Sayre}
\affiliation{Lawrence Livermore National Laboratory, Livermore, California 94550, USA}

\author{K.~Smith}
\affiliation{Department of Physics \& Astronomy, University of Tennessee Knoxville, Knoxville, Tennessee 37996 USA}
\altaffiliation[Present address: ]{Los Alamos National Laboratory, Los Alamos, NM 87545, USA}

\author{F.X.~Timmes}
\affiliation{The Joint Institute for Nuclear Astrophysics}
\affiliation{School of Earth and Space Exploration, Arizona State University, Tempe, AZ, USA}

\author{E.~Uberseder}
\affiliation{Cyclotron Institute, Texas A\&M University, College Station, TX 77843 USA}
\altaffiliation[Present address: ]{Nuclear Engineering and Radiological Sciences, University of Michigan, Ann Arbor, MI 48109, USA}

\author{M.~Wiescher}%
\affiliation{The Joint Institute for Nuclear Astrophysics}
\affiliation{Department of Physics, University of Notre Dame, Notre Dame, Indiana 46556 USA}

\date{\today}

\begin{abstract}
The creation of carbon and oxygen in our universe is one of the forefront questions in nuclear astrophysics. The determination of the abundance of these elements is key to both our understanding of the formation of life on earth and to the life cycles of stars. While nearly all models of different nucleosynthesis environments are affected by the production of carbon and oxygen, a key ingredient, the precise determination of the reaction rate of $^{12}$C$(\alpha,\gamma)^{16}$O, has long remained elusive. This is owed to the reaction's inaccessibility, both experimentally and theoretically. Nuclear theory has struggled to calculate this reaction rate because the cross section is produced through different underlying nuclear mechanisms. Isospin selection rules suppress the $E1$ component of the ground state cross section, creating a unique situation where the $E1$ and $E2$ contributions are of nearly equal amplitudes. Experimentally there have also been great challenges. Measurements have been pushed to the limits of state of the art techniques, often developed for just these measurements. The data have been plagued by uncharacterized uncertainties, often the result of the novel measurement techniques, that have made the different results challenging to reconcile. However, the situation has markedly improved in recent years, and the desired level of uncertainty, $\approx$10\%, may be in sight. In this review the current understanding of this critical reaction is summarized. The emphasis is placed primarily on the experimental work and interpretation of the reaction data, but discussions of the theory and astrophysics are also pursued. The main goal is to summarize and clarify the current understanding of the reaction and then point the way forward to an improved determination of the reaction rate.
\end{abstract}

\pacs{Valid PACS appear here}
\maketitle
\tableofcontents

\vspace{5mm}
\noindent {\it ``There are three kinds of lies: lies, damned lies, and statistics."} \\ \\
-Mark Twain
\\ \\
{\it ``Consider the subtleness of the sea; how its most dreaded creatures glide under water, unapparent for the most part, and treacherously hidden beneath the loveliest tints of azure."} \\ \\
-Herman Melville, {\it Moby Dick}
\\ \\
\section{\label{sec:intro} Introduction}

The baryonic matter that is a product of the Big Bang takes the form of hydrogen, helium, and very small amounts of lithium. This is the seed material that has fueled the chemical evolution of our Universe. Through the many generations of stars their life cycles have been governed by myriads of microscopic interactions driven by the short range strong and weak forces and the long range electromagnetic force. Chemical reactions define the molecular configurations of elements in our environment, while nuclear processes are responsible for the formation of the chemical elements themselves. The history and environments where the formation processes occur dictate the elemental and isotopic abundance distributions that we observe today.

The nuclear reactions necessary for the formation of the elements can only take place at conditions of high density and temperature. These conditions occur only in special settings in the universe, such as the center of stars and during stellar explosions. Tens of thousands of nuclear reactions can participate in a specific nucleosynthesis scenario, depending on the various environmental conditions. However, only a small fraction of these reactions have a strong impact on the overall chemical evolution of the elements. These few reactions have far reaching consequences for the chemistry and the subsequent molecular evolution of baryonic matter. There is one reaction of particular relevance, $^{12}$C($\alpha$,$\gamma$)$^{16}$O, that influences the $^{12}$C/$^{16}$O ratio in our universe. This reaction, together with the 3$\alpha$-process, the fusion of three $^4$He nuclei into one $^{12}$C nucleus, defines the carbon and oxygen abundance that is the fundamental basis for all organic chemistry and for the evolution of biological life in our universe. As Willy Fowler wrote in his 1983 Nobel Prize lecture \cite{Fowler23111984}, ``The human body is 65\% oxygen by mass and 18\% carbon with the remainder mostly hydrogen. Oxygen (0.85\%) and carbon (0.39\%) are the most abundant elements heavier than helium in the sun and similar Main Sequence stars. It is little wonder that the determination of the ratio $^{12}$C/$^{16}$O, produced during helium burning, is a problem of paramount importance in nuclear astrophysics." As a consequence, the reaction has been dubbed ``the holy grail of nuclear astrophysics".

The significance of the $^{12}$C$(\alpha,\gamma)^{16}$O reaction for energy production and nucleosynthesis in stars is closely tied to that of the 3$\alpha$ process. The simultaneous fusion of three $\alpha$ particles was discussed by \textcite{PhysRev.55.434}, but it was not until preliminary measurements of the long life-time of the $^8$Be were made that it was realized by \textcite{1952ApJ...115..326S} that a much more efficient two-step reaction was possible. Finally it was \textcite{1954ApJS....1..121H} who deduced that there must be an actual resonance in the $^8$Be$(\alpha,\gamma)^{12}$C reaction, the famous Hoyle state in $^{12}$C, that enhances the cross section even further \cite{Salpeter2002}. The experimental work of \textcite{PhysRev.107.508} rather quickly established the rate of the 3$\alpha$ process since it depends mainly on the strength of the Hoyle state (see \textcite{Freer20141} for a recent review). Current estimates of the uncertainty in the 3$\alpha$ rate are at about the 10\% level over the regions of typical astrophysical interest. However, at lower temperatures ($<$~0.1~GK), the uncertainty is likely much larger, because other reaction mechanisms become significant (see, e.g. \cite{PhysRevC.94.054607} and references therein).   

Nature is not so kind to us with the $^{12}$C$(\alpha,\gamma)^{16}$O reaction. Here the cross section enhancement is not the result of a single narrow resonance or even several such resonances, but stems from the very delicate (and seemingly devious!) interferences between broad overlapping resonances and nonresonant reaction components, properties which are much more difficult to determine accurately. Originally only the $E1$ contribution to the cross section was thought to dominate and the reaction rate and estimates were based purely upon the properties of the 1$^-$ subthreshold state in $^{16}$O. At that time, since no direct measurements had been made, only very rough predictions of the cross section, based on theory and indirect measurements, were possible. For example, in the seminal work of \textcite{RevModPhys.29.547} (B$^2$FH), the ground state $\gamma$ width of the 1$^-$ subthreshold level at $E_x$~=~7.12 MeV had been measured by \textcite{SwM56} as $\Gamma_{\gamma_0}$~=~130$^{+90}_{-80}$ meV but no experimental information was available for the reduced $\alpha$ width, which had to be calculated based on rudimentary nuclear theory. The significance of the comparable $E2$ contribution was not realized for another 30 years \cite{Redder1987385}.

This review will provide an overview of the astrophysical significance of the $^{12}$C$(\alpha,\gamma)^{16}$O reaction, the particular role it plays in nuclear physics and a review of the interpretation and analysis of the experimental nuclear physics data that provide the basis for the presently used nuclear reaction rate in astrophysical simulations. In this work we seek to employ as comprehensive a study of the $^{12}$C($\alpha,\gamma$)$^{16}$O reaction as possible by including other measurements that provide important information on the $^{16}$O  compound nucleus for our interpretation of the reaction mechanism. This is implemented using a state of the art $R$-matrix analysis, whose theoretical basis and implementation is explored in detail. Based on this complementary information a reaction rate analysis is performed that includes all available reaction and decay data associated with the $^{16}$O compound nucleus. The goal is to investigate the uncertainties associated with the low energy extrapolation of the existing laboratory data into the stellar energy range. The uncertainties in the reaction rate are determined by Monte Carlo simulation techniques and a detailed investigation of the systematic uncertainties in both data and model. Finally, the impact of these uncertainties will be investigated in the framework of stellar model simulations.

\section{\label{sec:astro} Helium Burning and its Astrophysical Significance}

The $^{12}$C($\alpha$,$\gamma$)$^{16}$O reaction plays a major role in
key nuclear burning phases driving the evolution and the associated
nucleosynthesis in low mass and massive stars. This includes
main-sequence hydrogen burning, where $^{12}$C and $^{16}$O formed by the
$^{12}$C($\alpha$,$\gamma$)$^{16}$O reaction in previous generations of stars
can play a critical role.  On the main-sequence, hydrogen burning
fuses four hydrogen nuclei into helium releasing about 25~MeV of
energy. This energy release generates the internal pressure conditions
for maintaining the stability of the stellar core against
gravitational contraction. For low mass stars with initial masses M~$\lesssim$~1.5 M$_\odot$ 
the fusion process is facilitated by the $pp$-chains, a
sequence of light ion capture reactions building upon the fusion of
two protons into a deuteron by the weak interaction. In more massive
stars M~$\gtrsim$~1.5~M$_\odot$, the importance of the $pp$-chains is
diminished and the fusion process is dominated by a catalytic reaction
sequence, the CNO cycles that are characterized by four proton capture reactions and two $\beta^+$ decays on carbon and oxygen forming a cycle by emitting an $\alpha$ particle. The result of CNO nucleosynthesis is the
conversion of hydrogen to $^4$He and enrichment of $^{14}$N based on
the depletion of the initial $^{12}$C and $^{16}$O nuclei.

With the depletion of hydrogen in the stellar core, hydrogen
burning continues only in a shell surrounding the inert core.  The hydrogen depleted
core contracts gravitationally, increasing the density
and temperature of the core matter. This contraction is halted with
the ignition of helium burning as a new energy source. Helium burning
is triggered by the 3$\alpha$-process releasing 7.5~MeV in fusion energy
and producing  $^{12}$C.
This is a rather unique process, setting stringent conditions
for the ignition of helium burning in stars.
The 3$\alpha$-process is
followed by the subsequent $\alpha$ capture reaction
$^{12}$C$(\alpha,\gamma)^{16}$O, converting the $^{12}$C into
$^{16}$O. These two isotopes are the principal products of helium
burning. The ratio of these products affects not only their own
nucleosynthesis but the future evolution of the star in its subsequent
burning phases. The ratio of $^{12}$C/$^{16}$O is determined by the competition between the
3$\alpha$ and $^{12}$C($\alpha$,$\gamma$)$^{16}$O reaction rates at a given temperature. The
time evolution of the molar abundances Y($^{12}$C) and Y($^{16}$O) can be
calculated as a function of the helium seed abundance Y($^4$He), the
reaction rates $\lambda_{\rm{reaction}}$ of the helium burning processes,
and the density $\rho$ using the following equations:
\begin{equation}\label{eq:12Cnetwork}
\begin{split}
\frac{dY(^{12}{\rm C})}{dt} = & \frac{1}{3!} \ Y^3(^4{\rm He})\cdot\rho^2\cdot\lambda_{(3\alpha)} \\
                            & - Y(^4{\rm He})\cdot Y(^{12}{\rm C})\cdot \rho\cdot\lambda_{^{12}{\rm C}(\alpha,\gamma)^{16}{\rm O}} \\
\frac{dY(^{16}{\rm O})}{dt} = & Y(^4{\rm He})\cdot Y(^{12}{\rm C})\cdot \rho\cdot\lambda_{^{12}{\rm C}(\alpha,\gamma)^{16}{\rm O}} \\
                              & -Y(^4{\rm He})\cdot Y(^{16}{\rm O})\cdot \rho\cdot\lambda_{^{16}{\rm O}(\alpha,\gamma)^{20}{\rm Ne}},
\end{split}
\end{equation}
where the stoichiometric factor of 1/3! accounts for indistinguishable $\alpha$-particles.

\begin{figure}
\includegraphics[width=1.0\columnwidth]{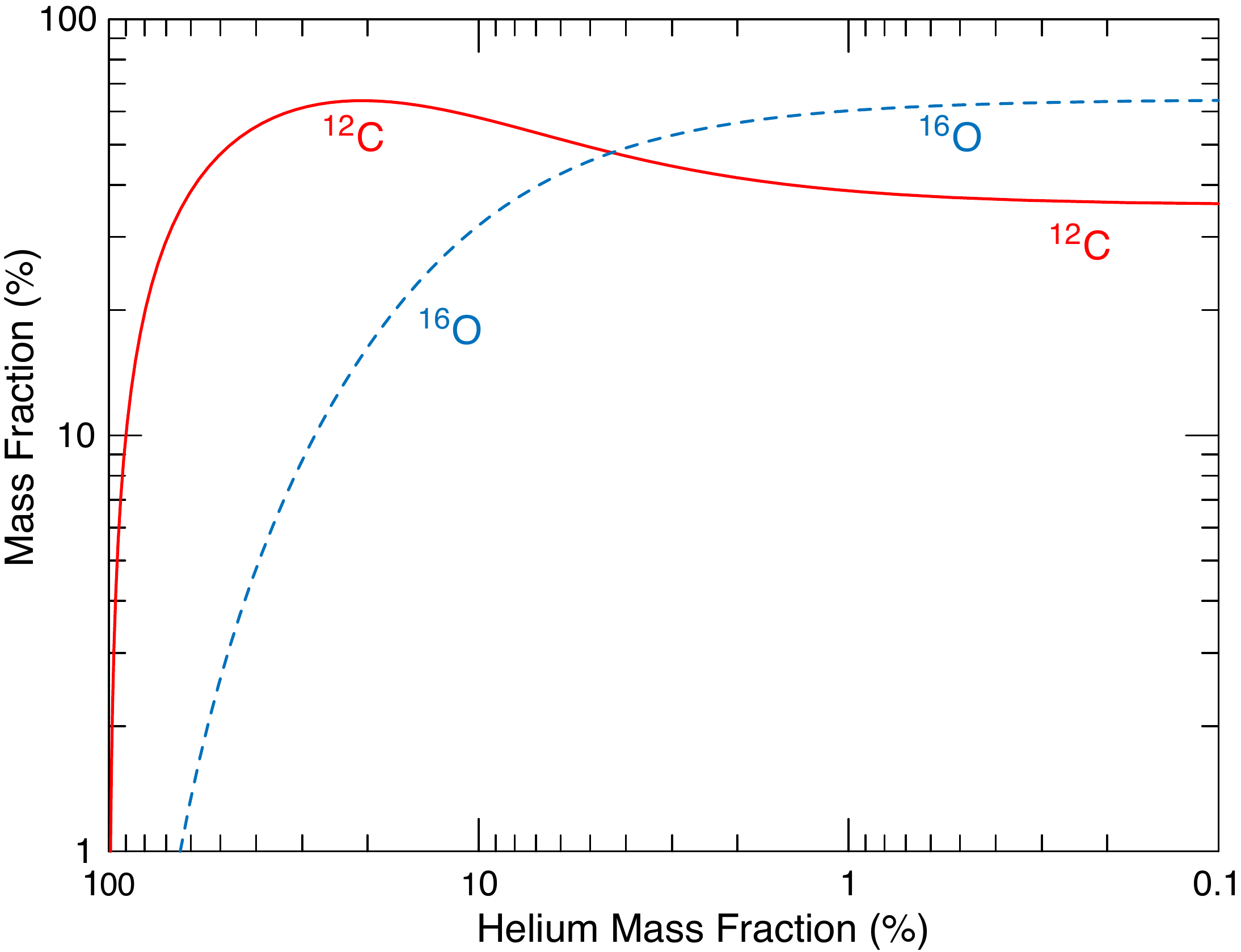}
\caption{(Color online) Typical evolution of $^{12}$C and $^{16}$O mass fractions as a function
of the $^{4}$He mass fraction at constant temperature and density. A mass fraction X$_i$ of isotope $i$
is related to the molar abundance Y$_i$ of Eq.~\ref{eq:12Cnetwork} by X$_i$=W$_i$$\cdot$Y$_i$,
where  W$_i$ the atomic weight. The oxygen mass fraction rises above the carbon mass fraction
only when the helium abundance is relatively small.
}
\label{fig:heburn}
\end{figure}

Fig.~\ref{fig:heburn} illustrates a typical evolution of
Eq.~(\ref{eq:12Cnetwork}) at constant temperature and density. Putting helium on
the x-axis instead of time makes the evolution largely independent of the exact
thermodynamic conditions. The feeding of $^{12}$C, driven by the
3$\alpha$-process, occurs early in the evolution when the abundance of carbon is
low and helium is high. Oxygen production occurs later by $\alpha$
capture on the freshly produced $^{12}$C. This shows the sensitivity of the
$^{12}$C/$^{16}$O ratio to the strengths of the
$^{12}$C($\alpha$,$\gamma$)$^{16}$O and $^{16}$O($\alpha,\gamma$)$^{20}$Ne
rates in addition to the 3$\alpha$-process that facilitates the feeding of
these isotopes. Both reactions are therefore critical for our understanding of
the emergence of $^{12}$C and the evolution of its abundance.

In typical helium burning environments the reaction rate
$\lambda_{^{16}{\rm O}(\alpha,\gamma)^{20}{\rm Ne}}$ is considerably
smaller than $\lambda_{^{12}{\rm C}(\alpha,\gamma)^{16}{\rm O}}$. Both
the 3$\alpha$-process and the $^{12}$C($\alpha,\gamma$)$^{16}$O
reaction burn with high efficiency through pronounced resonance
mechanisms. In contrast, the $^{16}$O($\alpha,\gamma$)$^{20}$Ne reaction lacks any such resonance enhancement in the stellar energy range making its rate comparatively much lower \cite{PhysRevC.82.035802}. This essentially
prohibits further helium burning beyond $^{16}$O and maintains the
$^{12}$C/$^{16}$O balance as we observe it today. This effect of a
sensitive balance between these three reactions is frequently
discussed as an example for the anthropic principle, a prerequisite
for the evolution of biological life as we know it in our universe
\cite{Carr_Nature1979}. These deliberations dominated the discussion
of the importance of the interplay between these three reactions in
the early second half of the 20$^{\text{th}}$ century
\cite{Kragh_DOI:_10.1007/s00407-010-0068-8}.

With the emergence of more sophisticated stellar modeling and
nucleosynthesis simulation techniques, a number of more intricate
questions emerged that underlined the importance of the
$^{12}$C$(\alpha,\gamma)^{16}$O reaction. It plays a crucial role for
stellar evolution and the associated nucleosynthesis during later
stages. The aspects and consequences of the $^{12}$C($\alpha,\gamma$)$^{16}$O
rate have been investigated in detail by performing extensive modeling
of the evolution and nucleosynthesis patterns in stars over a wide
range of stellar masses. These simulations have been performed using
tabulated reaction rates \cite{Caughlan1988283}, \cite{nacre}, or
later \cite{Buchmann2006254} as reference, with variations based on
the predicted uncertainty ranges. There are pronounced differences
with respect to the role of the reaction rate for nucleosynthesis
in low- and intermediate-mass stars, M~$\le$~8~M$_{\odot}$, that
develop into Asymptotic Giant Branch (AGB) stars with subsequent mass
loss, ending as white dwarfs and massive stars, M $\ge$ 8 M$_{\odot}$,
that develop towards their final fate as core-collapse supernova. The
outcome of these studies is discussed in the following sections.

\subsection{\label{sec:LMS}Helium Burning in Low- and Intermediate Mass Stars}

When a single star on the main sequence exhausts the supply of
hydrogen in its core, the core contracts and its temperature
increases, while the outer layers of the star expand and cool. The
star becomes a red giant
\citep[e.g.,][]{iben_1991_aa,stancliffe_2009_aa,PAS:9305903}.
The subsequent onset of helium burning in the core,
for stars with initial masses M $\gtrsim$ 0.5~$M_{\odot}$,
causes the star to
populate the horizontal branch in the Hertzsprung-Russell diagram
for more metal-poor stars or the red clump for more metal-rich stars
\citep{cannon_1970_aa,faulkner_1973_aa,seidel_1987_aa,castellani_1992_aa,girardi_1999_aa}.
After the star depletes the supply of helium in its core, the
carbon-oxygen (CO) core contracts while the envelope once again expands and
cools along a path that is aligned with its previous red-giant track. The
star becomes an asymptotic giant branch (AGB) star
\citep[e.g.,][]{hansen_2004_aa,herwig_2005_aa,kippenhahn_2012_aa,salaris_2014_aa,fishlock_2014_aa}.

A variation of the $^{12}$C($\alpha,\gamma$)$^{16}$O rate affects the core helium burning
lifetime which in turn impacts the mass of the resulting He-exhausted core at
the onset of the AGB phase. This mass is an important quantity that affects
many of the star's properties during the AGB phase. Low-mass and
intermediate-mass stars enter the AGB phase with hydrogen and helium
fusion continuing in shells around a hot core composed primarily of
carbon and oxygen and a trace amount of the neutron-rich isotope
$^{22}$Ne.  The precise influence of the
$^{12}$C($\alpha,\gamma$)$^{16}$O rate on the nucleosynthesis in AGB
stars is challenging to evaluate accurately in the framework of
present models due to multiple uncertainties (e.g., mixing
processes). Yet, the reaction does play a key role for the nucleosynthesis
during the AGB phase affecting the ejected abundances as well as the
$^{12}$C/$^{16}$O ratio in the white dwarf remnant.

During the AGB phase helium ignites in the He-shell under electron
degenerate conditions, whose energy release triggers a sequence of
convective thermal pulses, often called He-shell flashes.
Depending on mass and composition, there
may be a few to several hundred thermal pulses. During a He-shell
flash the 3$\alpha$-process is the dominant source of energy and a
producer of $^{12}$C.  The $^{12}$C($\alpha,\gamma$)$^{16}$O reaction
creates $^{16}$O, whose mass fraction increases with depth. However,
the duration of a He-shell flash is relatively short. Simulations
suggest the $^{16}$O mass fraction in the intershell rises to
$\simeq$0.2, the $^{12}$C mass fraction to 0.2$-$0.45, and the
remaining material is mainly $^4$He
\cite{10.1086/500443,0004-637X-827-1-30}.

The energy release from the thermal pulses also temporarily reduces or
extinguishes H-burning in the layers beneath the stellar envelope, and
causes convection to pull material from the central regions of the
star towards its surface \citep[e.g.,][]{herwig:05,
  straniero:06, PAS:9305903}. This dredged up material is enriched in
carbon, oxygen and $s$-process elements from the helium intershell,
modifying the surface composition \citep[e.g.,][]{gallino:98}.  This
phenomenon is confirmed by spectroscopic analysis of AGB stars
\citep[e.g.,][]{abia:01,zamora:09}, post-AGB stars
\citep[e.g.,][]{Delgado-Inglada11052015}, measurements of presolar
grains \citep[e.g.,][]{lugaro:03}, and planetary nebula as the
dredged-up material is blown into the interstellar medium by stellar
winds \citep[e.g.,][]{vanwinckel:00}.

For stars with an initial mass less than $\simeq$6~M$_{\odot}$ the
temperature in the stellar core is too low to ignite $^{12}$C fusion,
and the post-AGB evolution leads to a white dwarf.  The
$^{12}$C($\alpha,\gamma$)$^{16}$O reaction rate has a large influence
on the mass fraction profiles of $^{12}$C and $^{16}$O in the white
dwarf remnant \citep[e.g.,][]{PhysRevC.73.025802,fields_2016_aa}.
For example, properties of white dwarf models derived from Monte Carlo
stellar evolution surveys suggest variation of
$^{12}$C($\alpha,\gamma$)$^{16}$O within the experimental
uncertainties causes a $\simeq$50\% spread in the central carbon and
oxygen mass fractions.  Surveys of planetary nebulae find cases of
oxygen enrichment by nearly a factor of two relative to carbon
\cite{Delgado-Inglada11052015}, which may be due to an enhanced
$^{12}$C($\alpha,\gamma$)$^{16}$O rate or mixing processes
\cite{0067-0049-225-2-24}. Asteroseismology studies of white dwarfs
seek to determine the $^{12}$C/$^{16}$O profile and infer the reaction
rate from the transition of the white dwarfs inner oxygen-rich region to
the carbon-rich region in the outer layers \cite[e.g.,][]{1538-4357-587-1-L43}.

For stars with an initial mass above $\simeq$7~$M_{\odot}$ the
temperatures in the stellar core are high enough to ignite carbon,
and the minimum mass for neon ignition is $\simeq$10~$M_{\odot}$.
Stars with initial masses between $\simeq$7~$M_{\odot}$ and $\simeq$10~$M_{\odot}$
are designated as super-AGB stars.
Depending primarily on the initial mass, $^{12}$C/$^{16}$O profile,
and composition mixing model, the ignition of carbon may not occur
at all (for stars $\lesssim$ 7~$M_{\odot}$ ), occur at the center of the star (for
stars $\gtrsim$ 10~$M_{\odot}$), or occur somewhere off-center
\citep{siess:07,poelarends_2008_aa,siess_2009_aa,0004-637X-772-2-150, doherty:15, farmer_2015_aa}
In the off-center case, ignition is followed by the inward propagation of a
subsonic burning front \citep{nomoto_1985_aa,timmes_1994_aa,garcia-berro_1997_aa,lecoanet_2016_aa}.
The ignition conditions depend on the $^{12}$C/$^{16}$O ratio determined by the
$^{12}$C($\alpha,\gamma$)$^{16}$O rate, and may vary by a factor of
$\simeq$13 at ignition \textcite{0004-637X-583-2-878}.

The $^{12}$C/$^{16}$O profile in the remnant white dwarf impacts
the ignition of Type Ia supernovae, one of the premier probes for
measuring the cosmological properties of the Universe
\citep[e.g.,][]{riess_1998_aa,perlmutter_1999_aa}.  The carbon mass
fraction impacts the overall energy release, expansion velocity,
silicon-group and iron-group ejecta profiles
\cite{0004-637X-495-2-617, Ropke2006, calder_2007_aa, raskin_2012_aa, seitenzahl_2016_aa, miles_2016_aa} in progenitor systems involving
either one white dwarf (single degenerate channel) or two white dwarfs
(double degenerate channel).

\subsection{\label{sec:MS}Helium Burning in Massive Stars}

Most of a main sequence star's initial metallicity comes from the CNO
and $^{56}$Fe nuclei inherited from its ambient interstellar medium.
The slowest step in the hydrogen-burning CNO cycle is proton
capture onto $^{14}$N. This results in all the CNO catalysts piling up
into $^{14}$N when hydrogen burning is completed.  During the early
onset of core helium burning, the reaction sequence
$^{14}$N($\alpha$,$\gamma$)$^{18}$F($\beta^{+}$$\nu_e$)$^{18}$O($\alpha$,$\gamma$)$^{22}$Ne
converts all of the $^{14}$N into $^{22}$Ne. For the first time, the
stellar core has a net neutron excess. As detailed below, this neutronization is
important for the the slow neutron capture (\emph{s}-)process in massive stars.

Helium burning in massive stars with initial masses M~$\gtrsim8$~M$_{\odot}$
has a lifetime of $\simeq$10$^6$ years. Typical core temperatures and densities
in solar metallicity stellar models are $\simeq$2$\times10^8$~K and
$\simeq$1$\times10^3$~g~cm$^{-3}$, respectively
\citep{iben_1966_aa,RevModPhys.74.1015,limongi_2003_aa,nomoto_2013_aa}. As the
convective helium core evolves the temperature and density rise significantly,
and thus so does the energy generation due to the 3$\alpha$-process and
$^{12}$C($\alpha,\gamma$)$^{16}$O reactions.  Carbon production is favored by
larger densities and smaller $\lambda_{^{12}{\rm C}(\alpha,\gamma)^{16}{\rm                                                                            
O}}$, while oxygen production is favored for smaller densities (see
Eq.~(\ref{eq:12Cnetwork})) and larger $\lambda_{^{12}{\rm C}(\alpha,\gamma)^{16}{\rm O}}$. Fig.~\ref{fig:heburn} shows that the last remnants of helium fuel are the most important in setting the final
$^{12}$C/$^{16}$O ratio.

In addition, during the final stages of helium burning, when the
temperature and density are larger, the three reactions
$^{12}$C($\alpha,\gamma$)$^{16}$O, $^{22}$Ne($\alpha$,n)$^{25}$Mg,
and $^{16}$O($\alpha,\gamma$)$^{20}$Ne compete for those
last few $\alpha$-particles.  The
$^{22}$Ne($\alpha$,n)$^{25}$Mg reaction is the dominant neutron source
the $s$-process in massive stars
\citep[e.g.,][]{raiteri:91,the:07,pignatari:10,tur_2009_aa, kaeppeler:94}, producing about 8 neutrons per iron seed nuclei. Therefore, the
$^{12}$C($\alpha,\gamma$)$^{16}$O reaction has an impact on
$s$-process nucleosynthesis, where a larger rate translates
into a smaller production of neutrons.
For temperatures larger than $\simeq$4$\times$10$^{8}$~K, the
$^{16}$O($\alpha,\gamma$)$^{20}$Ne rate becomes larger
than the $^{12}$C($\alpha,\gamma$)$^{16}$O rate, converting some of
the abundant $^{16}$O into $^{20}$Ne.

The neutronization, entropy profile, and $^{12}$C/$^{16}$O ratio that
emerges from core helium-burning influences the subsequent evolution
of a massive star. The uncertainty of the
$^{12}$C($\alpha,\gamma$)$^{16}$O rate propagates to the subsequent
carbon, neon, oxygen, and silicon burning stages
\citep[e.g.,][]{imbriani:01,eleid:04}.  For example, upon helium
depletion the core again contracts and heats to conditions conducive
to carbon burning by the $^{12}$C + $^{12}$C reaction. Ignition of carbon
depends on the fusion cross section of this heavy-ion reaction and on the
square of the carbon fuel abundance that is chiefly set by the
$^{12}$C($\alpha,\gamma$)$^{16}$O reaction. 
\citet{tur_2007_aa,tur_2010_aa} considered the influence of uncertainties in the
3$\alpha$ and $^{12}$C$(\alpha,\gamma)^{16}$O reactions on the
evolution and nucleosynthesis of massive stars.  Using a reference
$^{12}$C$(\alpha,\gamma)^{16}$O rate of 1.2 times that of
\citep{PhysRevC.54.393}, they concluded that variations of this rate
induced variations in the final abundances ejected by the
supernova explosion including $^{12}$C, the key radionuclides $^{26}$Al, $^{44}$Ti,
and $^{60}$Fe, and the final mass of the remnant.

These later evolutionary phases are a rich site of fascinating
challenges that include the interplay between
nuclear burning \citep{couch_2015_aa,mueller_2016_aa,jones_2017_aa,farmer_2016_aa},
convection \citep{meakin_2007_ab,viallet_2013_aa},
rotation \citep{heger_2000_aa,rogers_2015_aa,chatzopoulos_2016_aa},
radiation transport \citep{jiang_2015_aa,jiang_2016_aa},
instabilities \citep{garaud_2015_aa,wheeler_2015_aa},
mixing \citep{maeder_2012_aa,martins_2016_aa},
waves \citep{rogers_2013_aa,fuller_2015_aa,aerts_2015_ab},
eruptions \citep{humphreys_1994_aa,kashi_2016_aa,quataert_2016_aa},
and binary partners \citep{justham_2014_aa, marchant_2016_aa,pavlovskii_2017_aa}.
This bonanza of physical puzzles is closely linked with
compact object formation by core-collapse supernovae \citep[e.g.,][]{timmes_1996_ac,eldridge_2004_aa,ozel_2010_aa,suwa_2015_aa,
perego_2015_aa,sukhbold_2016_aa}
and the diversity of observed massive star transients \citep[e.g.,][]{van-dyk_2000_aa,ofek_2014_aa,smith_2016_aa}.
Recent observational clues that challenge conventional wisdom
\citep{zavagno_2010_aa,vreeswijk_2014_aa,boggs_2015_aa,jerkstrand_2015_aa,strotjohann_2015_aa},
coupled with the expectation of large quantities of data
from upcoming surveys \citep[e.g.,][]{creevey_2015_aa,papadopoulos_2015_aa,sacco_2015_aa,yuan_2015_aa},
new measurements of key nuclear reaction rates and techniques for assessing
reaction rate uncertainties \citep{wiescher_2012_aa,sallaska_2013_aa,iliadis_2016_aa},
and advances in 3D pre-SN modeling \citep{couch_2015_aa,mueller_2016_aa,jones_2017_aa},
offer significant improvements in our quantitative understanding of the end states of massive stars.

\subsection{\label{sec:FS}Helium Burning in First Stars}

After the photons of the cosmic microwave background were released,
the Universe exhibited a uniform structure with no point sources of
light. As gravitational perturbations grew, dark matter and gas
aggregated.  No metals existed to facilitate the cooling and further
condensation of gas into stars, as in later generations.
Primordial star formation was instead driven by cooling through
molecular hydrogen line emission \citep[e.g.,][]{palla_1983_aa}.  The
first stars -- referred to as H-He, pregalactic, population III, or
zero metallicity stars -- are thought to have initially formed at
redshifts $z \simeq$ 20 in small dark matter haloes of mass
$\simeq$10$^{6}$~M$_{\odot}$
\citep[e.g.,][]{abel_2002_aa,turk_2009_aa,0004-637X-681-2-771}.
Simulations suggest that fragmentation of the central gas
configuration allow for a range of stellar masses, 1~M$_{\odot}$
$\lesssim$~M~$\lesssim$~1000~M$_{\odot}$, depending on the
dimensionality, spatial resolution, and local physics used in the
simulations \citep[e.g.,][]{truran_1971_aa,stacy_2016_aa, hosokawa_2016_aa, 1986ApJ...307..675F}.

The $^{12}$C($\alpha$,$\gamma$)$^{16}$O reaction impacts the
early nucleosynthesis steps of the first generation of stars.
In sufficiently massive first generation stars, M $\gtrsim$
10~M$_{\odot}$, the $pp$-chains have too weak a temperature dependence
to provide enough energy generation to halt gravitational
contraction. Such stars continue to get denser and hotter until the
central temperature reaches $\simeq$10$^{8}$~K, where the 3$\alpha$
reaction synthesizes $^{12}$C \citep{ezer_1971_aa}.  This
self-production of carbon is followed by the
$^{12}$C($\alpha$,$\gamma$)$^{16}$O reaction to produce oxygen.  The
zero metallicity star thus makes enough of its own CNO elements to
power the catalytic, hydrogen burning CNO cycles, halt
gravitational contraction, and proceed onto the main sequence. The evolution
of the first stars from the main sequence to their final fate
continues to be investigated across the entire initial mass spectrum
\citep[e.g.,][]{dantona_1982_aa,guenther_1983_aa,el-eid_1983_aa,bond_1984_aa,
weiss_2000_aa,umeda_2000_aa,marigo_2001_ab,heger_2002_aa,heger_2010_aa,ritter_2016_aa}.
In stars sufficiently massive to burn helium, the $^{12}$C($\alpha$,$\gamma$)$^{16}$O reaction
establishes the $^{12}$C/$^{16}$O profile which impacts the subsequent evolution.

Indeed, the most metal poor stars that we observe today carry signatures
of the first core-collapse supernovae, characterized by enhancements of
carbon and oxygen relative to iron, [C/Fe] $\sim$ [O/Fe] $\sim$ + 3.0
\citep[e.g.,][]{bond_1981_aa,bessell_1984_aa,beers_1985_aa,beers_1992_aa,
christlieb_2003_aa,frebel_2005_aa,keller_2014_aa,Frebel_AJL2015,yoon_2016_aa,hansen_2016_aa}.
A large fraction of these stars show [C/Fe] and [O/Fe] ratios larger than
those in the Sun \citep[e.g.,][]{hansen:15,bonifacio:15}.
The full potential of stellar archaeology can likely be reached in
ultrafaint dwarf galaxies, where the simple formation history may
allow for straightforward identification of second-generation stars.
\citep[e.g.,][]{Alexander_MNRAS2015}. These observations confirm the important role
of the $^{12}$C($\alpha$,$\gamma$)$^{16}$O reaction for interpreting the
onset of nucleosynthesis in the first stars.

\subsection{\label{sec:UC} Uncertainty considerations}

The reliability of nucleosynthesis predictions depends on the quality of the stellar models and the nuclear reaction input parameters. The interplay between these two components defines the overall uncertainty in the model predictions. The quality of stellar model simulations has seen a rapid improvement over the last two decades due to the enormous increase in computational power. This has effectively reduced the traditional uncertainties associated with the model parameters, putting larger demands on the uncertainties associated with the reaction cross section. 

An unprecedented effort has also been invested into improved experimental data and extrapolation techniques. While there have been significant advances, cross section measurements towards lower energies represent a staggering experimental challenge. The exponential decline of the cross section can translate a few 10's of keV step towards lower energies into an order of magnitude reduction in reaction yield. This needs to be compensated by either a significant increase in beam current, a significant increase in detection efficiency and/or a significant decrease in the experimental background. Past experiments have pushed the measurements to the limit of the practical amount of time and effort that is achievable with current resources. However, advances in detector technology and high current, low background accelerator facilities offer renewed chances to move forward. 

Improvements in the extrapolation technique are also possible. One major step forward can be made be making a more complete and consistent treatment of all reaction parameters and data using $R$-matrix theory. The overall uncertainty in the cross section evaluations, however, is difficult to assess. In addition, the propagation of the rate through the stellar models is also open to interpretation. An attempt based on statistical means has been suggested by \citet{0954-3899-42-3-034007} (and references therein). This has been adopted by \citet{0004-637X-823-1-46} to provide uncertainty ranges for different quantities predicted by a stellar model (e.g. density or mass fraction of $^{12}$C) as they are sensitive to the uncertainty in the rate of the $^{12}$C($\alpha,\gamma$)$^{16}$O reaction given by \citep{0004-637X-567-1-643}.

The long-standing large uncertainties associated with the $^{12}$C($\alpha,\gamma$)$^{16}$O reaction rate and the difficulties in providing a reliable extrapolation of laboratory data to the stellar energy range triggered initiatives to deduce the reaction rate from nucleosynthesis simulations for massive stars and the comparison with observational abundance distributions. The first attempt utilized a set of massive star models ranging from 12 to 40~$M_{\odot}$, following nuclear burning through all phases of stellar evolution up to the
point of iron core collapse \cite{Weaver199365}. Within the uncertainties of the model simulations, the results indicated a reaction rate that is in good agreement with that suggested later by \textcite{Buchmann2006254} on the basis of an $R$-matrix analysis. A similar approach was taken by \textcite{Garnett199727}, who used the C/O abundance ratio in the ionized interstellar gas of galaxies, with very low heavy element abundances, to constrain the $^{12}$C($\alpha,\gamma$)$^{16}$O rate. This study confirmed the results of the former analysis by \textcite{Weaver199365}.

A similar analysis on the impact of the $^{12}$C($\alpha,\gamma$)$^{16}$O reaction rate on the nucleosynthesis of heavier element yields during pre-supernova evolution and supernova explosions was performed by \textcite{0004-637X-671-1-821}. They considered the nucleosynthesis in stars with initial masses ranging from 13 to 27~$M_{\odot}$ calculated from the implicit, one-dimensional, hydrodynamical stellar evolution code \texttt{KEPLER} \citep[e.g.,][]{RevModPhys.74.1015}. They varied the $^{12}$C($\alpha,\gamma$)$^{16}$O reaction rate by scaling the $S$-factor of $S$(300 keV)~=~146 keV~barn as suggested by \citep{Buchmann2006254} by a factor of 0.6 to 1.9, probing the impact on the production factors of light elements and in particular the carbon/oxygen mass fraction at carbon ignition at the center of these massive stars. The results again suggested good agreement with the prediction by \textcite{Buchmann2006254}, confirming the earlier results of \textcite{Weaver199365, Garnett199727}. The study concluded that for a reliable nucleosynthesis simulation for massive stars, the $^{12}$C($\alpha,\gamma$)$^{16}$O reaction rate needs to be known to an uncertainty of 10\%. 

This work was followed more recently by a more expanded study \cite{0004-637X-769-1-2} where the sensitivity of presupernova evolution and supernova nucleosynthesis yields of massive stars was considered in dependence of variations in the $3\alpha$ and the $^{12}$C($\alpha,\gamma$)$^{16}$O rates. These variations were kept within an uncertainty range of $\pm$2$\sigma$. A grid of twelve initial stellar masses between 12 and 30 $M_{\odot}$, using 176 models per stellar mass, were computed to explore the effects of the two independently varying rates on the production of intermediate mass elements $A$~=~16-40 and the $s$-only isotopes produced efficiently by the weak $s$-process ($^{70}$Ge, $^{76}$Se, $^{80}$Kr, $^{82}$Kr, $^{86}$Sr, and $^{87}$Sr) in comparison to the solar abundance distribution. The study found a close correlation between the two rates for an optimal fit of the abundances, as to be expected by Eq.~(\ref{eq:12Cnetwork}), indicating that an increase of the $^{12}$C($\alpha,\gamma$)$^{16}$O rate requires an increase in the $3\alpha$ rate.

\subsection{\label{sec:RR_intro} The Nuclear Reaction Rate}

While these model based studies are certainly of great interest, they depend on the reliability of the stellar models, the model parameters, and the numerical treatment of the hydrodynamic aspects of stellar evolution. Depending on the internal burning conditions in the specific environments the stellar reaction rate needs to be well known over a wide energy range. 

The nuclear reaction rate can be calculated from the total reaction cross section $\sigma(E)$ by integration over the Maxwell-Boltzmann distributions of the interacting particles in a stellar environment of temperature $T$. The reaction rate per particle pair is given by
\begin{equation} \label{eq:rr}
N_A\langle\sigma v\rangle =
\left(\frac{8}{\pi\mu}\right)^{1/2}\frac{N_A}{(k_B T)^{3/2}}\int^\infty_0\sigma(E)Ee^{-E/k_B T}dE,
\end{equation}
where $\mu$ is the reduced mass, $E=\mu v^2/2$ is the center-of-mass energy, $N_A$ is Avogadro's number, and $k_B$ is Boltzmann's constant. The energy-dependent cross section is the key input for determining the reaction rate. This is determined by various reaction contributions and mechanisms. 

Traditionally, the charged-particle reaction cross section is expressed in terms of the astrophysical $S$~factor
\begin{equation} \label{eq:sfactor}
S(E)=\sigma(E) \, E \, e^{2\pi\eta}.
\end{equation}
The exponential term $e^{2\pi\eta}$ approximately accounts for the influence of the Coulomb barrier on the cross section, where $\eta$ is the Sommerfeld parameter ($\sqrt{\frac{\mu}{2E}}Z_1Z_2\frac{e^2}{\hbar^2}$). Therefore $S(E)$ essentially describes the nuclear and centrifugal barrier components of the reaction mechanism and is also more convenient for plotting and extrapolation. The reaction rate scales with the $S$~factor at the stellar energy range, and the literature therefore often quotes the $S$~factor at a typical stellar energy. For the $^{12}$C($\alpha,\gamma$)$^{16}$O reaction this is at $E_\text{c.m.}$~=~300~keV. Thus the value of $S$(300~keV) is often given for ease of comparison of the impact of the nuclear reaction data on the extrapolation.

Equation~(\ref{eq:rr}) can be approximated when either of two extreme cases dominate the $S$~factor. First, some $S$~factors are dominated by non-resonant processes (e.g., direct capture) and are often characterized by a nearly energy-independent $S$~factor. In this case, the energy range of interest for a specific burning temperature $T$ is traditionally defined in terms of the Gamow window, which is defined by the integrand in Eq.~(\ref{eq:rr}). For a constant $S$-factor, the integrand can be approximated by a Gaussian distribution around the mean center-of-mass energy $E_0$ (in units of MeV) of
\begin{equation}\label{eq:Gamow}
E_0 = 0.122\cdot(Z_1^2Z_2^2 \hat{\mu} T_9^2)^{1/3},
\end{equation}
with $Z_1$ and $Z_2$ being the charges of the interacting particles, $\hat{\mu}$ the reduced mass in atomic mass units, and $T_9$ the temperature in GK. Using the same notation, the width $\Delta E$ of the Gamow range is given by
\begin{equation}\label{eq:Gamow_delta}
\Delta E = 0.236\cdot(Z_1^2Z_2^2 \hat{\mu} T_9^5)^{1/6}.
\end{equation}
The reader is cautioned that this Gaussian distribution concept may break down in the case of resonances or as one moves above the Coulomb barrier (i.e., at higher temperatures); see Fig.~\ref{fig:RR_limit} below.

This simple formalism facilitates the quick identification of the energy range over which the reaction cross section needs to be determined to provide a reaction rate for stellar burning simulations. Table~\ref{tab:astro_sites_and_temps} summarizes the energy ranges corresponding to the characteristic temperatures of the various stellar environments discussed above in Sec.~\ref{sec:astro}. A purely experimentally determined reaction rate would require experimental cross section data covering the full range of these energies (0.15~$<E_\text{c.m.}<$~3.4~MeV). Since this has not been achieved for the lower energies, the reaction rate for $^{12}$C($\alpha,\gamma$)$^{16}$O has to rely on the extrapolation of experimental data obtained at higher energies.

\begin{table*}
\caption{Astrophysical environments and burning stages where the $^{12}$C$(\alpha,\gamma)^{16}$O reaction plays an important role. The temperatures of these environments dictate the energy ranges where the $^{12}$C$(\alpha,\gamma)^{16}$O cross section must well known for an accurate calculation of the reaction rate. \label{tab:astro_sites_and_temps}}
\begin{ruledtabular}
\begin{tabular}{c c c c}
Burning Stages & Astro. Sites & Temp. Range (GK) & Gamow Energy Range (MeV) \\
\hline
Core Helium Burning & AGB stars and Massive Stars & 0.1-0.4 & 0.15-0.65\\
Core Carbon and Oxygen Burning & Massive Stars & 0.6-2.7 & 0.44-2.5 \\
Core Silicon Burning & Massive Stars & 2.8-4.1 & 1.1-3.4 \\
Explosive Helium Burning & Supernovae and X-Ray Bursts & $\approx$1 & 0.6-1.25 \\
Explosive Oxygen and Silicon Burning & Supernovae & $>$5 & $>$1.45 \\
\end{tabular}
\end{ruledtabular}
\end{table*} 

The second case is when the $S$-factor is dominated by narrow isolated resonances (i.e., such that the resonance width is small compared to the resonance energy and interference effects can be neglected). Ignoring all energy dependences except the Lorentzian approximation of the Breit-Wigner cross section, Eq.~(\ref{eq:rr}) can then be integrated analytically. This yields an expression for the reaction rate in terms of resonance strengths $\omega\gamma_i$
\begin{multline} \label{eq:rres}
N_A\langle\sigma v\rangle = 1.5394\times 10^{11}(\hat{\mu} T_9)^{-3/2}\\ 
\times\sum\limits_i \omega\gamma_i \,\exp\left(\frac{-11.605\cdot E_{R_i}}{T_9}\right) \left[   \frac{\text{cm}^3} {\text{sec} \cdot \text{mol}} \right] ,
\end{multline} 
where $\omega\gamma_i$ and $E_{R_i}$ are the resonance strength and resonance energy of the $i^{th}$ resonance in MeV, respectively.

The resonance strengths are proportional to the production and decay widths, $\Gamma_\text{in}$ and $\Gamma_\text{out}$:
\begin{equation}
\omega\gamma = \frac{(2J+1)}{(2J_1+1)(2J_2+1)} \frac{\Gamma_\text{in}\Gamma_\text{out}}{\Gamma},
\end{equation}
where $J$ is the spin of the resonance, $J_1$ and $J_2$ are the spins of the nuclei in the entrance channel, and $\Gamma$ is the total width of the resonance. In case of radiative $\alpha$ capture reactions, $\Gamma_\text{in}$ and $\Gamma_\text{out}$ correspond to the $\alpha$ and $\gamma$ partial widths $\Gamma_{\alpha}$ and $\Gamma_{\gamma}$, respectively. 

Complicating matters, the energy dependence of the $^{12}$C($\alpha,\gamma$)$^{16}$O reaction does not  fall into either of these two specialized categories. Instead, the $S$-factor is dominated by broad resonances which interfere with one another, a regime in between the two extreme cases discussed above. Therefore the reaction rate must be determined through numerical integration of Eq.~(\ref{eq:rr}). However, in addition to these broad resonances there are also a few narrow resonances that are superimposed upon them. Because of practical experimental considerations (i.e. target thickness and accelerator energy resolution), the strengths of these narrow resonances are much easier quantities to measure accurately than the individual widths or actual cross sections over them. Therefore, in practice, numerical integration of Eq.~(\ref{eq:rr}) is used in conjunction with the narrow resonance specific form of Eq.~(\ref{eq:rres}) to calculate the total rate of the $^{12}$C($\alpha,\gamma$)$^{16}$O reaction. This process is described in more detail in Sec.~\ref{sec:RR}.

The following section is dedicated to outlining our present knowledge of the reaction mechanisms and the underlying nuclear structure and reaction phenomena that are needed for an accurate calculation of the reaction rate. For an informed extrapolation it is important to treat and determine the $^{12}$C($\alpha,\gamma$)$^{16}$O cross section as a nuclear physics problem that can only be solved by understanding the complex quantum mechanics of the reaction mechanism. Further, nuclear theory, as discussed in Secs.~\ref{sec:nuc_phys} and \ref{sec:r_matrix}, calculates the different multipolarities of the reaction independently. Thus for the extrapolation to be made, it is necessary to not only measure the total cross section as a function of energy $\sigma(E)$, but also to understand its composition in terms of photon multipolarities and  ${}^{16}{\rm O}$ final states.


\section{\label{sec:nuc_phys} Nuclear Physics Aspects}

The reaction mechanism of $^{12}$C($\alpha,\gamma$)$^{16}$O, and therefore its cross section or $S$-factor, is characterized by strong resonant and non-resonant contributions and the interference effects between these components. The strength of these components is directly associated with the nuclear structure of the $^{16}$O nucleus. Being doubly magic it has been the subject of numerous studies and its unique level structure has provided a long standing challenge for theoretical descriptions.

\begin{figure*}
\includegraphics[width=2\columnwidth]{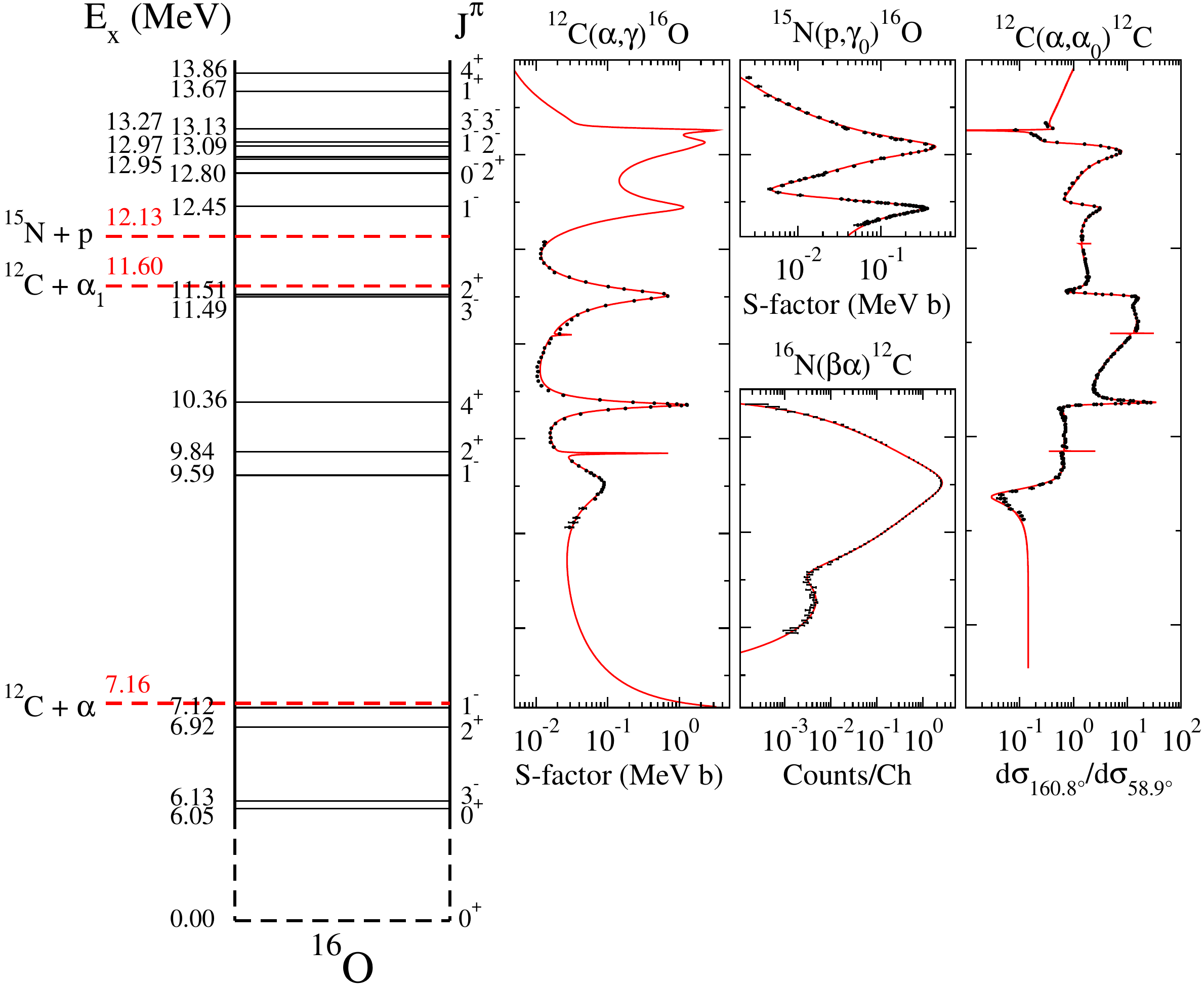}
\caption{(Color online) Level diagram of the $^{16}$O compound nucleus. Levels that are irrelevant to the analysis are omitted. For example, unnatural parity states below the proton separation energy. Several reactions that populate the CN are shown where their energy axis has been converted to excitation energy. At low energy only $^{12}$C+$\alpha_0$, $^{16}$O+$\gamma$ and $^{16}$N($\beta\alpha$) partitions are considered. At higher energies, the $^{15}$N+$p$ and $^{12}$C+$\alpha_1$ partitions are also included. Representative experimental cross section measurements that populate the CN are shown on the right. The total cross section data for the $^{12}$C$(\alpha,\gamma)^{16}$O reaction are those of \textcite{schurmann2005}, the $^{15}$N$(p,\gamma_0)^{16}$O those of \textcite{PhysRevC.82.055804}, the $^{16}$N$(\beta\alpha)^{12}$C spectrum is from \textcite{PhysRevLett.70.726}, and the $^{12}$C$(\alpha,\alpha_0)^{12}$C data are from \textcite{PhysRevLett.88.072501}. The solid red curve represents the phenomenological $R$-matrix fit described in this work. \label{fig:level_diagram}}
\end{figure*}

The $^{16}$O compound nucleus is represented schematically in Fig.~\ref{fig:level_diagram}. It has four particle bound excited states at excitation energies: $E_x$~=~6.05, 6.13, 6.92, and 7.12~MeV. As an even-even nucleus, the spin of the ground state is $J^\pi$ = 0$^+$ and the four excited states are 0$^+$, 3$^-$, 2$^+$, and 1$^-$ respectively. The two odd parity states are considered to be single particle configurations that can be described well in the framework of the shell model, while the two of even parity have been characterized as cluster configurations that require a microscopic potential or cluster model approach \cite{Langanke_Friedrich_1986}. From the following cluster model discussions in Sec.~\ref{sec:cluster_models}, one might expect that the separation energy $S_{\alpha_0}$ of the $^{16}$O CN into an $\alpha$ particle and the ground state configuration of $^{12}$C is at $E_x$ = 7.16~MeV. It will become of utmost relevance for the reaction rate that $S_{\alpha_0}$ is only a few hundred keV above the 2$^+$ and 1$^-$ bound states. It is useful to note that all of the excited bound states that $\gamma$ decay, do so to the ground state with nearly 100\% probability. Angular momentum and spin selection rules dictate that if the $^{16}$O compound nucleus is formed by a $^{12}$C+$\alpha_0$ reaction (intrinsic spins both equal to 0), then only states with $J$ = $l$ and $\pi$ = (-1)$^l$ (natural parity states), where $J$ is the total spin and $l$ is the relative orbital angular momentum, can be populated. With the limitation to only natural parity states, the $\gamma$-ray decay selection rules give that only electric transitions to the 0$^+$ ground state can occur. Further, $\gamma$-ray decays from 0$^+$ to 0$^+$ states are strictly forbidden.

The $^{12}$C($\alpha,\gamma$)$^{16}$O cross section is greatly influenced by the isospin of the states in $^{16}$O. The two 1$^-$ levels that most influence the low energy cross section, those at $E_x$~=~7.12~MeV (bound) and 9.59~MeV (unbound) are $T=0$, for which $E1$ $\gamma$-ray decays would be strictly forbidden to the ground state if the states were isospin pure. However, the Coulomb interaction breaks isospin symmetry, causing the states to become isospin mixed, allowing for such transitions to take place, albeit at a reduced strength. This is the primary reason that the $E1$ and $E2$ multipolarity components of the $^{12}$C($\alpha,\gamma_0$)$^{16}$O cross section are of nearly equal strength. In fact, the earliest studies of the $^{12}$C($\alpha,\gamma$)$^{16}$O reaction were made primarily to study the effects of isospin symmetry breaking. At higher energies, the next $T$~=~0 state is at $E_x$~=~12.45~MeV and the first $T$~=~1 state is that at $E_x$~=~13.09~MeV. Reproducing the properties of these states, especially the $\gamma$-ray decay widths, has proven very challenging for nuclear models as will be discussed further in Sec.~\ref{sec:cluster_models}.   

The level structure of $^{16}$O results in very different reaction mechanisms favored by the $\gamma$-ray de-excitations to the ground state versus those to higher-lying excited states. As will be discussed in more detail in Sec.~\ref{sec:radiative_capture}, $E1$ direct capture to the ground state is greatly suppressed. On the other hand, the large $Q$-value for the ground state transition favors resonance decays. This results in resonances, including those of the subthreshold states, dominating over the direct capture. It should also be noted that while the $E1$ direct capture strength is negligible, there is the possibility that $E2$ direct capture could be a significant component to the cross section in off-resonance regions. In contrast, resonant de-excitations to the high-lying excited states in $^{16}$O are suppressed because of their smaller $Q$-values. This then puts the strength of resonance decays on par with that of the direct capture, with direct capture even dominating in several cases (see Sec.~\ref{sec:cascades_12Cag}). Therefore, the cascade transition cross sections are expected to be small compared with that of the ground state. This then makes the determination of asymptotic normalization coefficients for the corresponding final states critical for modeling the external capture component of the capture cross sections for these transitions (see Secs.~\ref{sec:radiative_capture} and \ref{sec:cascades_12Cag}).

The essential goal is to obtain an accurate value of the cross section over the energy range that contributes significantly to the reaction rate calculation (see Sec.~\ref{sec:RR}). Ideally the cross section could simply be measured directly in the laboratory, but this not a viable option since the Coulomb repulsion between the two charged particles makes the cross sections over the Gamow energy region extremely small. For the $^{12}$C$(\alpha,\gamma)^{16}$O reaction, the cross section at $E_{c.m.}$~=~300~keV is estimated to be about 2$\times$10$^{-17}$~barns. This is still about 5 orders of magnitude below the sensitivity, a few picobarns, achieved by the most state of the art measurements. For this reason, nuclear reaction theories must be used to aid in the {\it extrapolation} of the cross section to the astrophysically relevant region. This is the crux of the problem.


Several different approaches have been investigated. The cluster model approach provides guidance for interpreting the level structure of the $^{16}$O compound nucleus, but so far they are not a sufficiently reliable method for predicting reaction cross sections or for extrapolating existing experimental data from laboratory studies into the stellar energy range. Phenomenological models, fit to experimental data and extrapolated to low energy, are more accurate and have been the mainstay for many years. The remainder of this section is devoted to an introduction to the experimental data and the status of the cluster and phenomenological models used to interpret it.

\subsection{The Experimental Situation}

A host of experimental measurements have been made to study the $^{12}$C$(\alpha,\gamma)^{16}$O reaction over the years. Because of its incredible importance to the field of nuclear astrophysics and the extreme challenge of its measurement, nearly every kind of technique in the experimental nuclear physicist's tool box has been brought to bear. Experiments have ranged from the most sophisticated, brute force, high beam current, direct measurements, to techniques as indirect as the study of the $\beta$ delayed $\alpha$ emission of $^{16}$N and Coulomb excitation. These extensive and diverse efforts have aimed either at the direct study of the low energy cross section or at the study of the nuclear properties of the levels near the $\alpha$ threshold in the $^{16}$O compound nucleus.

Direct methods have evolved from measuring reaction yields in close geometry where only faint signals were observed, and are difficult to interpret, to measuring high statistics detailed angular distributions in far geometry that approach direct observation of the differential cross sections. Most experiments have focused on a limited low energy region from 1~MeV~$< E_\text{c.m.} <$~3~MeV for two primary reasons. First, a broad 1$^-$ resonance (corresponding to the level at $E_x$~=~9.59~MeV in Fig.~\ref{fig:level_diagram}) enhances the cross section in this region making measurements more viable. This state then serves as a touch stone for measurements toward lower energies. Second, measurements are greatly hindered by the increasing, and very high cross section, $^{13}$C$(\alpha,n)^{16}$O background reaction. The large amount of neutrons it creates causes several serious experimental difficulties. The neutrons themselves cause damage to the delicate lattice structure of solid state detectors. Further, secondary $\gamma$-rays are created through inelastic neutron scattering and neutron capture on both surrounding material and the detectors themselves. They create $\gamma$-rays over a wide energy range that hinder the measurements of all the transitions of the $^{12}$C$(\alpha,\gamma)^{16}$O reaction (see, e.g., \textcite{Makii2005411}).

In recent years, the vast improvements in experimental techniques have expanded the accessible energy range, both to lower and higher energies. This effort, coupled with substantial improvements in the phenomenological description of the reaction contributions and the overall reaction mechanism through $R$-matrix theory, lead to a substantially improved confidence in the low energy extrapolation of the cross section. Before entering into a detailed discussion of the experimental data and its phenomenological interpretation in Secs.~\ref{sec:data} and \ref{sec:R-matrix_analysis}, the following sections provide a summary of past utilization of both cluster models and $R$-matrix theory for interpreting the $^{12}$C($\alpha,\gamma$)$^{16}$O cross sections.  


\subsection{Cluster Models}\label{sec:cluster_models}

Since the pioneering work of \textcite{PhysRev.52.1083}, it has been argued that individual nucleons should often be found in tightly bound $\alpha$ particle cluster configurations (see, e.g., \cite{Beck_ClustersVol1, Beck_ClustersVol2, Beck_ClustersVol3} for a series of recent reviews). It follows then that for nuclei that are integer multiples of the $\alpha$ particle, many of the nuclear excitations in the compound nucleus can be interpreted as the molecular configurations of $\alpha$ particle clusters. This then strongly influences the strength of the associated resonance states and the strength of direct capture transitions, which usually dominate the reaction mechanism. This idea is particularly interesting when played out using the theoretical framework of the Ikeda model \cite{Ikeda01071968}, which predicts pronounced $\alpha$ clustering configurations near the $\alpha$ threshold. Thus the effect of $\alpha$ clustering is particularly critical for reactions in stellar helium burning, where the 3$\alpha$-process and the subsequent $^{12}$C$(\alpha,\gamma)^{16}$O reaction dominate \cite{0034-4885-70-12-R03}.

For modeling the radiative capture cross section of the $^{12}$C$(\alpha,\gamma)^{16}$O reaction, a number of cluster models of varying levels of sophistication have been applied to investigate the impact of these effects on the various transition components. This is, in particular, important for the $E2$ ground state transition due to the fact that the subthreshold 2$^+$ state at $E_x$~=~6.92~MeV represents a high degree of $\alpha$ clustering \cite{Brown1966401}. In the $E1$ ground state component, the cluster contribution of the $E_x$~=~7.12 MeV subthreshold state and its interference with the resonance corresponding to the broad $E_x$~=~9.59 MeV cluster state, add significantly to the reaction strength.

Most of the cluster models applied to the study of this reaction have their roots in the Resonating Group Method (RGM) originally proposed by \textcite{PhysRev.52.1107}. Microscopic potential models have been presented by \textcite{Langanke1983334,Langanke1985384}, and \textcite{FUNCK198511} and are summarized by \textcite{Langanke_Friedrich_1986}. Single-channel and multi-channel Generator Coordinate Method (which is equivalent to the RGM) calculations, have also been developed \cite{Descouvemont1984426, PhysRevC.36.1249, PhysRevC.47.210, PhysRevC.78.015808}. Many of these works have also included calculations of the strengths of the $^{12}$C$(\alpha,\gamma)^{16}$O cascade transitions \cite{Descouvemont1984426, Langanke1985384, PhysRevC.36.1249, PhysRevC.78.015808}. Since none of these models allow for all of the degrees of freedom associated with 16 nucleons, the use of effective interactions is required along with some phenomenological adjustment of parameters to agree with experimental inputs such as separation thresholds and resonance energies.

Despite the longstanding theoretical development, precision $\alpha$ cluster modeling is still very challenging. As a recent example, the $\alpha$ cluster configuration of the $^{16}$O nucleus was studied using a modified shell model approach built on a cluster-nucleon configuration interaction model with advanced realistic shell-model Hamiltonians. The model was constructed in order to investigate the strength of clustering phenomena in the harmonic oscillator basis \cite{PhysRevC.91.044319}. This study provides a comprehensive description of the $\alpha$ cluster structure of the $^{16}$O nucleus up to a very high excitation energy range based on the $^{12}$C ground state configuration. The study demonstrates the possible existence of pronounced $\alpha$ cluster configurations. In particular for the energy range near the $\alpha$ threshold where large $\alpha$ spectroscopic factors are predicted for natural parity resonance and subthreshold levels between 6.0~MeV~$< E_x< $~8.5~MeV. These theoretical results are in general agreement with the tabulated values obtained by $\alpha$ transfer and capture reactions \cite{Tilley19931}.

In principle these models are highly constrained and consequently have great predictive power. However, the cluster models are challenged to describe all of the available experimental data with the precision required for nuclear astrophysics applications. This is largely because they only take certain cluster components in the reaction mechanism into account. For example, there are difficulties in correctly describing the width of the narrow $2^+$ state at $E_x$~=~9.84 MeV with this approach, as it is predominantly not an $\alpha$-cluster state~\cite{PhysRevC.78.015808}. The new approach of coupling the cluster model with modern shell model techniques offers new opportunities for a more comprehensive theoretical description of the level structure of the $^{16}$O compound nucleus that is necessary for a reliable theoretical prediction of the $^{12}$C($\alpha,\gamma$)$^{16}$O reaction cross section \cite{PhysRevC.91.044319}. Another state-of-the-art example is the calculation of \textcite{PhysRevLett.112.102501}, where binding energies, charge radii, quadrupole moment (for the $2^+$ state), and electromagnetic transition strengths are provided for the even-parity bound states of $^{16}$O. As theoretical approaches become truly {\em ab initio}, they will significantly further our understanding of the $^{12}$C$(\alpha,\gamma)^{16}$O reaction. Already cluster models can be very useful for making theoretical calculation of Asymptotic Normalization Coefficients (ANCs) for the bound states of $^{16}$O and this should see further attention. However, because of the limitations described, phenomenological approaches must still be employed. 

\subsection{Phenomenological Models} \label{sec:phenom_models}

A strong point of phenomenological reaction models is that, while remaining ignorant of more fundamental nuclear physics (i.e. internal nuclear wave functions), well-established quantum-mechanical symmetries and conservation laws, such as angular momentum conservation and unitarity, can still be enforced. This allows the model to remain flexible, while still providing many stringent constraints. On the other hand, phenomenological models have little predictive power without data and the quality of their constraints or predictions are very much dependent upon the quality of the supporting experimental data.

The most long standing and pervasive phenomenological model for the resolved resonances region is the $R$-matrix theory of \textcite{PhysRev.72.29}, which was further developed by \textcite{RevModPhys.30.257} and \textcite{Bloch1957503}. It has been used by many in the field for the analysis of the $^{12}$C$(\alpha,\gamma)^{16}$O reaction as well as many other reactions. The $R$-matrix model, as described in more detail in Sec.~\ref{sec:r_matrix}, offers the best approach to phenomenological analysis of the $^{12}$C$(\alpha,\gamma)^{16}$O reaction at this time. Some of the other alternatives, and justification for this assertion, are discussed below.

One extension of the $R$-matrix method that has been applied to the
$^{12}$C$(\alpha,\gamma)^{16}$O reaction is the ``hybrid'' $R$-matrix-optical
model~\cite{PhysRevC.7.561,Koonin1974221}. In this approach, the
broad $1^-$ state at $E_x$~=~9.85~MeV and higher energy background levels
are modeled using an optical potential, with the subthreshold $1^-$
state being introduced as a separate $R$-matrix level.
Refinements have been provided in subsequent publications
\cite{Langanke1983334,Langanke1985384}. This method has the attractive
feature that the optical model should reproduce many of the higher lying broad resonances. If this can replace the need for additional background pole terms, the number of free parameters could be greatly reduced and an improved constraint on the extrapolated $S$~factor could be achieved. However, these models have difficulties describing, over a broad energy range, the high-precision $^{12}$C$(\alpha,\alpha)^{12}$C elastic scattering data that are now available \cite{Plaga1987291, PhysRevC.79.055803}. These difficulties are not present in the standard $R$-matrix approach due to the greater flexibility. Nevertheless, the hybrid model may still be interesting for investigating the effects of the nuclear interaction beyond the channel radius.

An alternative to the $R$-matrix is the $K$-matrix, as developed by
Jean Humblet \cite{Humblet1976210,PhysRevC.42.1582}. $K$-matrix theory
is based upon a pole expansion (Mittag-Leffler series) of a meromorphic
function, rather than the properties of eigenfunctions satisfying boundary
conditions as in $R$-matrix theory. Some advantages of the $K$-matrix approach are that there is no channel radius and the computation of Coulomb wavefunctions is not needed. The fitting of experimental data is in general quite similar to the
$R$-matrix approach, with equal numbers of free parameters leading
to similar quality fits, $S$-factor extrapolations, and uncertainties.
A detailed comparison of the $K$- and $R$-matrix approaches for
$^{12}$C$(\alpha,\gamma)^{16}$O has been given by \textcite{PhysRevC.50.1194}.
A disadvantage of the $K$-matrix method is that the background remaining in
addition to the explicitly-included levels has a complicated and
generally uncertain functional form, including the possibility of sub-threshold
poles (echo poles) and complex-energy poles: see \textcite{Barker1994361, Brune1996122, Humblet1998714}. The situation is much clearer in the $R$-matrix approach, where the remaining background can only consist of real pole terms at higher energies.

Recently, several researchers have investigated new approaches for obtaining bound-state ANCs from elastic scattering data \cite{4252797920090601, PhysRevC.81.011601, 0954-3899-40-4-045106}, but it is not clear if these methods offer any advantages over the phenomenological $R$-matrix approach.

With the above considerations, a phenomenological $R$-matrix approach will be used to interpret the experimental data and perform the interpolation and extrapolation of the $^{12}$C$(\alpha,\gamma)^{16}$O cross section over the entire range of astrophysical interest. The following section details the critical aspects of this theoretical framework.

\section{\texorpdfstring{$R$}{R}-matrix Theory} \label{sec:r_matrix}

Since the first analysis of the $^{12}$C$(\alpha,\gamma)^{16}$O reaction by \textcite{Barker1971}, $R$-matrix theory has been used to model the experimental data. Over the intervening years many different approaches have been used (see Table~\ref{tab:S-extrap_history} below), but for the reasons discussed above, $R$-matrix has been the most common method and is the choice adopted for the present analysis. In many previous works, the $R$-matrix formalism has been specialized to the $^{12}$C$(\alpha,\gamma)^{16}$O case in order to simplify the formulas. In this more global analysis, that considers several other reaction partitions in addition to $^{12}$C+$\alpha$ and $^{16}$O+$\gamma$, the complete formalism is required. Subsections \ref{subsec:rm_general}$-$\ref{subsec:rmatrix_beta_channels} below cover $R$-matrix theory in a general manner, keeping in mind that the $R$-matrix approach is a useful tool for many applications besides the phenomenological analysis of nuclear reactions. Subsection \ref{subsec_rm_phenom} discusses considerations that are specific to the phenomenological analysis of nuclear reaction data and finally Subsec.~\ref{theory_Rmatrix_12Cag} covers our specific application to $^{12}$C$(\alpha,\gamma)^{16}$O.

\subsection{General \texorpdfstring{$R$}{R}-matrix Theory \label{subsection:rmatrixtheory}}
\label{subsec:rm_general}

$R$-matrix theory has been explained in detail in previous reviews~
\cite{RevModPhys.30.257,Hal80,0034-4885-73-3-036301,azure}, and only certain
details will be described here. For the most part, we utilize
the approach and notation of \cite{RevModPhys.30.257} (LT); an important
alternative is the Bloch operator formalism
\cite{Bloch1957503,PhysRev.151.774}.

The most basic premise of $R$-matrix theory divides the nuclear
configuration space into two distinct regions:
the nuclear \textit{interior} where the many-body nuclear
interactions are complicated, and the \textit{exterior} where it is
assumed there are two clusters in each channel that can be treated as separate and non-interacting (except for the Coulomb potential).
Channels are assumed to be orthogonal and are labeled by the indices $\alpha sl \equiv c$, where $\alpha$ defines a particular pair of nuclei, $s$ is the coupled spin (channel spin) of the pair, and $l$ is the relative orbital angular momentum.
We will work in the nuclear center-of-mass system;
the quantities $\mu_\alpha$, $k_\alpha$, $v_\alpha$, and
$r_\alpha$ are the reduced mass, wavenumber, relative velocity,
and radial separation for pair $\alpha$.
Channel spin wave functions are defined by
\begin{equation}
|\psi\rangle_{\alpha s\nu}=[|\psi\rangle_{\alpha 1} \otimes |
  \psi\rangle_{\alpha 2}]_{s\nu}
\end{equation}
where $|\psi\rangle_{\alpha 1}$ and $|\psi\rangle_{\alpha 2}$ are the internal
wavefunctions of the nuclei $1$ and $2$ making up pair $\alpha$,
and $\nu$ is the channel spin projection.

For each channel, the the dividing surface between the regions is taken
to be a sphere of radius $r_\alpha=a_c$.
This radius is known as the {\em channel radius} and may be different
for different channels.
It is convenient to utilize the ``channel surface functions'' introduced
by \textcite{Hal80} which have total angular momentum $J$ with component $M$:
\begin{eqnarray}
|{\mathcal S}cJM\rangle &=& \left(\frac{\hbar^2}{2\mu_\alpha a_c}
\right)^{1/2} \frac{\delta(r_\alpha-a_c)}{a_c} \\
\nonumber && \times\left[ |\psi\rangle_{\alpha s\nu} \otimes
i^l Y_{lm}(\hat{r}_\alpha)\right]_{JM},
\end{eqnarray}
where $Y_{lm}$ are the spherical harmonics.
These functions can be used to project a total wavefunction into a
well-defined angular momentum state on a particular channel surface.

In the internal region, we take the basis vectors $|\lambda JM\rangle$ to be the solutions to the nuclear Hamiltonian with
energy eigenvalues $E_\lambda$ that satisfy boundary conditions (LT, Eq.~V.2.1)
\begin{equation}
\langle {\mathcal S}cJM|\frac{\partial}{\partial r_\alpha}
r_\alpha|\lambda JM\rangle = B_c \langle {\mathcal S}cJM|\lambda JM\rangle
\end{equation}
where $B_c$ are real energy-independent boundary condition constants.
They orthogonal and normalized over the internal region such that
\begin{equation}
\langle\lambda'J'M'|\lambda JM\rangle=\delta_{\lambda'\lambda},
\delta_{J'J}\delta_{M'M} .
\label{eq:norm_interior}
\end{equation}
They are also understood to be complete, provided that all basis vectors satisfying the boundary condition (an infinite number) are included. Under time reversal these basis vectors transform according to (LT, Eq.~III.3.4)
\begin{equation} \label {eq:eigen_t_rev}
K|\lambda JM\rangle = (-1)^{J-M}|\lambda J\,{-}M\rangle,
\end{equation}
where $K$ is the time-reversal operator.
The real $M$-independent reduced-width amplitudes are given by
(LT, Eq.~III.4.8a)
\begin{equation}
\gamma_{\lambda cJ} = \langle {\mathcal S}cJM|\lambda JM\rangle,
\end{equation}
i.e., by the amplitude of the eigenfunction at the nuclear surface.

In the external region wavefunctions can be expressed in terms of
(LT, Eq.~III.2.19)
\begin{subequations}
\begin{eqnarray}
|{\mathcal I}cJM\rangle&=& \frac{I_{\alpha l}(r_\alpha)}{v_\alpha^{1/2} r_\alpha}
\left[|\psi\rangle_{\alpha s\nu}\otimes i^l
Y_{lm}(\hat{r}_\alpha)\right]_{JM} \\
|{\mathcal O}cJM\rangle&=& \frac{O_{\alpha l}(r_\alpha)}{v_\alpha^{1/2} r_\alpha}
\left[|\psi\rangle_{\alpha s\nu}\otimes i^l
Y_{lm}(\hat{r}_\alpha)\right]_{JM}
\end{eqnarray} \label{eq:coul_in_out}
\end{subequations}
where the incoming and outgoing Coulomb functions $I_{\alpha l}$ and $O_{\alpha l}$
are defined by (LT, Eq.~II.2.13). For closed channels the outgoing
solution $O_{\alpha l}$ is taken to be the exponentially-decaying
Whittaker function (LT, Eq.~II.2.17) and $v_\alpha$ for negative
energies is a positive real quantity as defined in (LT, Sec.~III.1).
In addition one defines
\begin{equation}
L_c=\left(\frac{r_\alpha}{O_{\alpha l}}\frac{\partial O_{\alpha l}}
{\partial r_\alpha} \right)_{a_c} =S_c+iP_c
\end{equation}
where the shift factor $S_c$ and and penetration factor $P_c$
are real quantities. Other Coulomb surface quantities are given by
$O_c=O_{\alpha l}(a_c)$, $I_c=O_{\alpha l}(a_c)$, and
$\Omega_c=(I_c/O_c)^{1/2}$. The relative Coulomb phase shift
$\omega_{\alpha l}$ is defined by (LT, Eq. III.2.13c).
From this point forward, we will suppress the angular momentum
labels $J$ and $M$ where it introduces no ambiguity and denote
$|{\mathcal S}cJM\rangle$ by $|c\rangle$ in order to simplify the
presentation.

By expanding an arbitrary wavefunction $|\Psi\rangle$ in the internal
region in terms of the basis $|\lambda\rangle$ and applying Green's theorem,
it can be shown (LT, Eq.~V.2.7) that
\begin{equation}
\langle c|\Psi\rangle_{\rm int} = \sum_{c'} R_{cc'}
  \langle c'|\frac{\partial}{\partial r_{\alpha'}}r_{\alpha'}-B_{c'}
  |\Psi\rangle_{\rm int},
\label{eq:psi_int}
\end{equation}
where $R_{cc'}$ are elements of {\em the} ${\bm R}$~matrix.
It is a function of the energy $E$
and can be expressed in terms of the reduced-width amplitudes and
energy eigenvalues as
\begin{equation} \label{eq:rmatrix}
R_{c'c}=\sum_\lambda \frac{\gamma_{\lambda c'}\gamma_{\lambda c}}
  {E_\lambda -E}.
\end{equation}
Essentially, ${\bm R}$ defines the logarithmic derivative of the
radial wavefunction at the channel surface(s) as a function of energy.

The general wavefunction $|\Psi\rangle$ may be expanded in the external
region (i.e., outside the channel radii) via
\begin{equation} \label{eq:psi_ext}
|\Psi\rangle_{\rm ext} = \sum_{cJM} z_{cJM}\left[ |{\mathcal I}cJM\rangle-
  \sum_{c'} U^J_{c'c}|{\mathcal O}c'JM\rangle\right],
\end{equation}
where the expansion coefficients $z_{cJM}$ specify the incoming flux which can
only be non-zero for open channels and $U_{cc'}$ are elements of
the scattering matrix ${\bm U}$ (also called the collision matrix).
By evaluating $\langle c|\Psi\rangle_{\rm ext}$ and
$\langle c| \frac{\partial}{\partial r_{\alpha}}r_{\alpha}|\Psi\rangle_{\rm ext}$
and comparing the results to Eq.~(\ref{eq:psi_int}) considering the
continuity of the logarithmic radial derivative at the channel radius,
${\bm U}$ can be related to ${\bm R}$ and the external Coulomb
functions evaluated at the channel radii. The result is (LT, Eq.~VII.1)
\begin{equation} \label{eq:scat_mat_channel}
{\bm U} = {\bm\Omega}\left[ {\bm 1} + 2i{\bm P}^{1/2}
  [{\bm 1}-{\bm R}({\bm L}-{\bm B})]^{-1}{\bm R}{\bm P}^{1/2} \right]{\bm\Omega},
\end{equation}
where ${\bm\Omega}$, ${\bm P}$, ${\bm L}$, and ${\bm B}$ are are purely
diagonal with elements $\Omega_c$, $P_c$, $L_c$, and $B_c$, respectively;
${\bm 1}$ is the unit matrix.
Alternatively, the elements of the scattering matrix can be expressed as
\begin{equation} \label{eq:scat_mat_level}
U_{c'c}=\Omega_{c'}\Omega_c \left[ \delta_{c'c}+2i(P_{c'} P_c)^{1/2}
\sum_{\lambda\mu}A_{\lambda\mu}\gamma_{\lambda c'}\gamma_{\mu c} \right]
\end{equation}
where $A_{\lambda\mu}$ are elements of the \textit{level matrix} ${\bm A}$ that is defined in level space by its inverse:
\begin{equation} \label{eq:ainv}
[{\bm A}^{-1}]_{\lambda\mu}=(E_\lambda-E) \delta_{\lambda\mu}-\sum_c
\gamma_{\lambda c}\gamma_{\mu c}(L_c-B_c).
\end{equation}
We also define a matrix ${\bm M}$ that is closely related to the scattering
matrix ${\bm U}$:
\begin{eqnarray} \label{eq:m_matrix}
{\bm U} &=& {\bm\Omega}\left[ {\bm 1} + 2i{\bm P}^{1/2}{\bm M}
  {\bm P}^{1/2} \right]{\bm\Omega} \label{eq:U_M}\\
M_{c'c} &=& \left\{ [{\bm 1}-{\bm R}({\bm L}-{\bm B})]^{-1}{\bm R}
  \right \}_{c'c} \\
&=& \sum_{\lambda\mu}A_{\lambda\mu}\gamma_{\lambda c'}\gamma_{\mu c}.
\end{eqnarray}
The matrix ${\bm M}$ may be interpreted physically as the projection of the
outgoing-wave Green's function onto the the channel surfaces
\cite[Eq.~(65)]{PhysRev.151.774}.

If the scattering matrix is diagonal, the nuclear phase shifts $\delta_c$
may be defined via
\begin{equation} \label{eq:phase_shift}
U_{cc} = e^{2i\delta_c}.
\end{equation}
The $\Omega_{c'}\Omega_c\delta_{c'c}$ term provides so-called hard-sphere
contribution to $U_{cc'}$; it is only present for elastic scattering.
While the phase shift corresponding to this term is mathematically
identical to that resulting from an infinite repulsive core at the
channel radius, one should avoid placing too much physical significance on
it since the total phase shift has contributions from both this term and
from the $R$-matrix. Note also that the hard-sphere term is present even
if the nuclear interactions vanish.

The solution corresponding to Eq.~(\ref{eq:psi_ext}) in the internal region
can be found using (LT, Eq.~IX.1.31):
\begin{equation} 
|\Psi\rangle_{\mathrm int}=-i\sum_{cJM}
\Omega_c(2\hbar P_c)^{1/2} z_{cJM}
\sum_{\lambda\mu}A_{\lambda\mu} \gamma_{\mu c}|\lambda JM\rangle .
\label{eq:psi_int_in}
\end{equation}
With the particular choice
\begin{equation}
z_{\xi JM}=
i \left[\frac{\pi\hbar(2l+1)}{\mu_\alpha k_\alpha}\right]^{1/2} (sl\nu 0|JM)
\end{equation}
where $\xi\equiv\alpha sl$ and $z_{c JM}=0$ for $c\neq\xi$,
$|\Psi\rangle_{\mathrm ext}$
is asymptotically equal to
$\exp[i(k_\alpha z_\alpha +\eta_\alpha\log k(r_\alpha-z_\alpha)]|\psi\rangle_{\alpha s\nu}$
plus \textit{outgoing} waves.
In this case, noting that $M=\nu$ only, $|\Psi\rangle$ becomes
\begin{subequations} \label{eq:outwave}
\begin{eqnarray} \label{eq:out_ext}
\lefteqn{ |\alpha s\nu \rangle_{\rm ext} =
\sum_{lJ} i \left[\frac{\pi\hbar(2l+1)}{\mu_\alpha k_\alpha}\right]^{1/2}
  } \hspace{0.15in} \\
\nonumber && \times(sl\nu 0|J\nu) \left[|{\mathcal I}\xi J\nu\rangle-\sum_c
U^J_{c\xi}|{\mathcal O}cJ\nu\rangle\right] \\ \label{eq:out_int}
\lefteqn{ |\alpha s\nu \rangle_{\rm int} =
\sum_{lJ} \hbar\Omega_\xi
\left[\frac{2\pi P_\xi(2l+1)}{\mu_\alpha k_\alpha}\right]^{1/2}}\hspace{0.15in} \\
\nonumber && \times(sl\nu 0|J\nu)
\sum_{\lambda\mu}A_{\lambda\mu}\gamma_{\mu \xi}|\lambda J\nu\rangle .
\end{eqnarray}
\end{subequations}
The wavefunction $|\alpha s\nu \rangle$ corresponds to an incident plane wave in partition $\alpha$ with channel spin $s$ and projection $\nu$;
the asymptotic form of Eq.~(\ref{eq:out_ext}) may be used to define scattering amplitudes. These equations are also useful for calculating radiative capture in perturbation theory, as described below. In addition, the plane
wave states can be expressed in terms of partial waves:
\begin{equation} \label{eq:plane_partial}
|\alpha s \nu\rangle = \frac{i\pi^{1/2}}{k_\alpha}\sum_{lJ} (2l+1)^{1/2}
  (sl\nu0|J\nu) |\alpha slJ\nu\rangle,
\end{equation}
where the internal and external representations of $|\alpha slJ\nu\rangle$
can be read off by inspection of Eq.~(\ref{eq:outwave}).

For the calculation of observables, formulas from general reaction
theory may be utilized. Defining the transition matrix to be
\begin{equation} \label{eq:t-matrix}
T_{cc'}=e^{2 i \omega_c}\delta_{cc'} - U_{cc'},
\end{equation}
the angle-integrated cross section can then be computed via (LT, Eq.~VIII.3.2b)
\begin{equation}\label{equation:crosssection}
\sigma_{\alpha \alpha'} = \frac\pi{k_\alpha^2} \sum_{J l l' s s'} g_J \left| T_{cc'} \right|^2,
\end{equation}
where the case of elastic scattering of charged particles is excluded.
The statistical factor is given by
\begin{equation}
g_J = \frac{2J+1}{(2J_{\alpha 1}+1)(2J_{\alpha 2}+1)},
\end{equation}
where $J_{\alpha 1}$ and $J_{\alpha 2}$ are the individual particle spins for the pair $\alpha$.

While the angle integrated cross section is related in a rather simple way to the transition matrix via Eq.~(\ref{eq:t-matrix}), the unpolarized differential cross section takes on a more complicated form because different partial waves
may interfere:
\begin{equation} 
\begin{split}
\frac{d\sigma_{\alpha,\alpha'}}{d\Omega_{\alpha'}} = &
  \frac{1}{(2J_{\alpha 1}+1)(2J_{\alpha 2}+1)} \\
& \times \sum_{ss'}(2s+1)\frac{d\sigma_{\alpha s, \alpha's'}}{d\Omega_{\alpha'}}
\end{split}
\end{equation}
where
\begin{equation} \label{eq:diff_XS_per_s}
\begin{split}
&(2s+1) \frac{k^2_\alpha}{\pi}\frac{d\sigma_{\alpha s, \alpha's'}}{d\Omega_{\alpha'}}
  =  (2s+1)|C_{\alpha'}(\theta_{\alpha'})|^2\delta_{\alpha s,\alpha's'} \\
& +\frac{1}{\pi}\sum_LB_L(\alpha s,\alpha' s') P_L(\cos\theta_{\alpha'})
  +\delta_{\alpha s,\alpha's'}(4\pi)^{-1/2} \\
& \times \sum_{Jl}(2J+1) 2\text{Re}\left[i(T^J_{c'c})^*C_{\alpha'}(\theta')
   P_l(\cos\theta_{\alpha'})\right],
\end{split}
\end{equation}
with
\begin{equation}
\begin{split}
B_L&(\alpha s, \alpha' s') =
\frac{(-1)^{s-s'}}{4}\sum_{J_1J_2l_1l_2l_1'l_2'} \text{\={Z}}(l_1J_1l_2J_2,sL) \\
&\times  \text{\={Z}}(l_1'J_1l_2'J_2,s'L)(T^{J_1}_{\alpha's'l_1',\alpha sl_1})(T^{J_2}_{\alpha's'l_2',\alpha s l_2})^*
\end{split}
\end{equation}
and
\begin{equation}
\begin{split}
\text{\={Z}}&(l_1J_1l_2J_2,sL) = \\
& ((2l_1+1)(2l_2+1)(2J_1+1)(2J_2+1))^{1/2} \\
& \times (l_1l_200|L0)W(l_1J_1l_2J_2;sL).
\end{split}
\end{equation}
Here the $P_L(\cos\theta_{\alpha})$ are the Legendre Polynomials and $W$ is the Racah coefficient. The $C_{\alpha}(\theta_{\alpha})$ are the Coulomb amplitudes that are only present for charged-particle elastic scattering. They are given by
\begin{equation}
\begin{split}
C_{\alpha}(\theta_{\alpha})&=(4\pi)^{-1/2}\eta_{\alpha}\csc^2\left(\frac{\theta_{\alpha}}{2}\right) \\&\times\exp\left\{-2i\eta_{\alpha}\ln\left[\sin\left(\frac{\theta_{\alpha}}{2}\right)\right]\right\},
\end{split}
\end{equation}
where $\eta_{\alpha}$ is the Sommerfeld parameter for the partition $\alpha$. The differential cross section for polarized particles, which is not utilized in this analysis, can be found in \textcite{nuc_pol}, for example.

\subsection{Physical Interpretation of the \texorpdfstring{$R$}{R}-Matrix Parameters}

Following \citet{PhysRev.81.148}, it is instructive to make a one level approximation to the level matrix ${\bm A}$. This leads (ignoring the hard-sphere contribution if $c=c'$) to
\begin{equation}
|T_{cc'}|^2 =  \frac{\Gamma_{\lambda c}\Gamma_{\lambda c'}}{(E_\lambda-
E+\Delta_{\lambda})^2+\frac14\left(\sum_{c''}\Gamma_{\lambda c''}\right)^2},
\label{eq:one_level_formal}
\end{equation}
where $\Gamma_{\lambda c} = 2P_c\gamma_{\lambda c}^2$ is the formal partial width for channel $c$ and $\Delta_\lambda$ is the energy-dependent level shift:
\begin{equation}
\Delta_\lambda = -\sum_c \gamma_{\lambda c}^2[S_c(E)-B_c].
\end{equation}
This form is functionally quite similar to the expression of \citet{PhysRev.49.519} for a single resonance level, with the exception of the level shift.
Considering that the boundary condition constants $B_c$ are arbitrary
real parameters, the correspondence to the Breit-Wigner formula may be made closer by choosing
\begin{equation}
B_c=S_c(E_\lambda),
\label{eq:bc_equal_sc}
\end{equation}
i.e., such that the level shift vanishes at $E_\lambda$. In this situation, we may associate $E_\lambda$ with the resonance energy. When the boundary conditions satisfy Eq.~(\ref{eq:bc_equal_sc}), we will denote the corresponding $R$-matrix parameters with tildes, i.e., as $\tilde{E}_\lambda$ and $\tilde{\gamma}_{\lambda c}$.

Further following \citet{PhysRev.81.148}, one may make a linear approximation to the level shift
\begin{equation} \label{eq:narrow_res_approx}
\Delta_\lambda \approx  (\tilde{E}_\lambda-E) \sum_c \tilde{\gamma}_{\lambda c}^2 \frac{dS_c}{dE}(\tilde{E}_\lambda)
\end{equation}
and Eq.~(\ref{eq:one_level_formal}) becomes (LT, Eq.~VII.3.2)
\begin{equation}
|T_{cc'}|^2 =  \frac{\tilde{\Gamma}_{\lambda c} \tilde{\Gamma}_{\lambda c'}}{(\tilde{E}_\lambda-
E)^2+\frac14\left(\sum_{c''}\tilde{\Gamma}_{\lambda c''}\right)^2},
\label{eq:one_level_observed}
\end{equation}
where (LT, Eqs.~XII.3.5 and~XII.3.6)
\begin{equation}\label{eq:width_convert}
\tilde{\Gamma}_{\lambda c} = \frac{2P_c\tilde{\gamma}_c^2}{1+\sum_{c'}
\tilde{\gamma}_{\lambda c'}^2 \frac{dS_{c'}}{dE}(\tilde{E}_\lambda)}.
\end{equation}
With this definition, Eq.~(\ref{eq:one_level_observed}) is now formally identical to the Breit-Wigner expression. One may expect the Breit-Wigner formula to be a particularly good approximation to $R$-matrix theory in the case of an isolated narrow resonance, such that the importance of other resonances and any non-linear energy dependence of $S_c(E)$ is minimal.

The partial width for decay into channels which are closed (or bound) is zero. In this  case, one defines instead the ANC, which  are real quantities that can be related to the reduced width via (LT, Eqs.~IV.7.1-IV.7.4), \cite[Eqs.~(8) and~(16)]{Barker1995693}, \cite[Eqs.~(60) and~(63)]{PhysRevC.59.3418}:
\begin{equation} \label{eq:ANC_to_rw}
\begin{split}
C_{\lambda c} = & \frac{(2\mu_\alpha a_c)^{1/2}}{\hbar W_c(a_c)} \\
  & \times\frac{\tilde{\gamma}_{\lambda c}}{\left[1+\sum_{c'}
  \tilde{\gamma}_{\lambda c'}^2 \frac{dS_{c'}}{dE}(\tilde{E}_\lambda) \right]^{1/2} },
\end{split}
\end{equation}
where $ W_c(a_c)$ is the exponentially-decaying Whittaker function
evaluated at the channel radius.
Note that the square of this equation, $C_{\lambda c}^2$,
is very similar in structure to Eq.~(\ref{eq:width_convert}) that
describes the partial widths in unbound channels.
If the level in question is bound in all channels, it can be shown (LT, Eq.~A.29) that the factor of $[1+\sum_{c'}
  \tilde{\gamma}_{\lambda c'}^2 \frac{dS_{c'}}{dE}(\tilde{E}_\lambda)]^{1/2}$ in the denominator of Eq.~(\ref{eq:ANC_to_rw}) is exactly what is required to change the normalization volume of the eigenfunction from the interior region (see Eq.~(\ref{eq:norm_interior})) to all space.

Based on the the correspondence of Eq.~(\ref{eq:one_level_observed}) to the Breit-Wigner formula, the quantities $\tilde{E}_\lambda$ and $\tilde{\Gamma}_{\lambda c}$ defined above are often called the observed resonance energy and partial widths corresponding to an $R$-matrix level; in addition, \cite{azure} used the terminology ``physical $R$-matrix parameters'' for $\tilde{E}_\lambda$ and $\tilde{\gamma}_{\lambda c}$. The reader is cautioned, however, that many conventions have been used in the past to define ``resonance energies'' and ``partial widths''. Some workers also go on to define ``observed'' reduced widths, which provides an additional opportunity for confusion (such parameters are not used in this work). In addition,  $\tilde{E}_\lambda$ and $\tilde{\Gamma}_{\lambda c}$ are somewhat dependent upon the channel radius that is used.
A more fundamental and unambiguous definition of resonance energies and partial widths is provided by the poles and residues of the scattering matrix which may be extracted from an $R$-matrix parametrization \cite{PhysRevLett.59.763}. This approach brings with it complications, including the fact that these poles and residues are generally complex quantities. However, for the case of a bound state, the relation is simple: $\tilde{E}_\lambda$ is a pole of the scattering matrix and its residues are proportional to $C_{\lambda c} C_{\lambda c'}$.

In order to give a more intuitive measure of the strength of the reduced width, it is often divided by the Wigner limit \cite{PhysRev.72.29} to give the dimensionless reduced width
\begin{equation}\label{eq:drw}
\theta^2_{\lambda c} = \frac{\tilde{\gamma}_{\lambda c}^2}{\gamma_{\text{W}}^2}, \quad \text{where}
\end{equation}
\begin{equation}
\gamma_{\text{W}}^2 = \frac{3}{2}\frac{\hbar^2}{\mu_\alpha a_c^2} \text{ or } \frac{\hbar^2}{\mu_\alpha a_c^2}
\end{equation}
is the Wigner limit. Unfortunately different conventions have been used for the Wigner limit that has led to some confusion in the literature.
The quantity $\gamma_{\text{W}}^2$ may be thought of as a crude estimate for the reduced width corresponding to a ``single-particle'' assumption for channel $c$. Consequently, $\theta^2_{\lambda c}$ is similar to the spectroscopic factor and can be interpreted physically as a dimensionless measure of the strength of a level relative to the single-particle case. We generally avoid the use of $\theta^2_{\lambda c}$ in this work because of its ambiguous definition and dependence on channel radius. We have, however, included some discussion in order to allow comparison with previous work.

\subsection{Parameter Transformations}
\label{sec:param_trans}

The tilde notation implies that the parameters are relative to the choice of boundary condition given by Eq.~(\ref{eq:bc_equal_sc}), i.e. that the level shift vanishes at the resonance energy. It is only for this choice of $B_c$ that the $R$-matrix parameters have a simple physical interpretation. Since the $B_c$ are energy independent, this implies that only one level of a given spin and parity can satisfy Eq.~(\ref{eq:bc_equal_sc}) and thus the parameters corresponding to other levels will not have a simple physical interpretation. \citet{Bar1972} has shown that the scattering matrix is invariant with respect to changes in the $B_c$, provided the $R$-matrix parameters are adjusted using a transformation that was also given. That this result holds even when the number of levels is finite is rather remarkable and unexpected, and is the likely reason why it took nearly 30 years after the formulation of $R$-matrix theory for this to be noticed. It is possible via iterative searching to find the transformations which yield Eq.~(\ref{eq:bc_equal_sc}) for all channels simultaneously and thus deduce a physical interpretation for all of the levels.

\citet{PhysRevC.66.044611} showed that the number of independent transformations yielding Eq.~(\ref{eq:bc_equal_sc}) in all channels is equal to the original number of $R$-matrix levels, provided that $dS_c/dE>0$ (which appears to be true in practice, although it remains in general unproven). This work further showed that the $R$-matrix formalism could be cast in a form such that the scattering matrix is given directly in terms of the $\tilde{E}_\lambda$ and $\tilde{\gamma}_{\lambda c}$ for all of the levels. In this approach, the $B_c$ do not appear and all of the parameters have a simple physical interpretation. We call $\tilde{E}_\lambda$ and $\tilde{\gamma}_{\lambda c}$ the alternative $R$-matrix parameters. The alternative level matrix may be defined via
\begin{equation} \label{eq:alta_inv}
\begin{split}
(&\tilde{\bm A}^{-1})_{\lambda\mu} = (\tilde{E}_\lambda-E)\delta_{\lambda\mu}-\sum_c
\tilde{\gamma}_{\lambda c} \tilde{\gamma}_{\mu c} (S_c+{\rm i}P_c) \\
& + \sum_c \left\{ \begin{array}{ll}
\tilde{\gamma}_{\lambda c}^2 S_{\lambda c} & \lambda=\mu \\
\tilde{\gamma}_{\lambda c} \tilde{\gamma}_{\mu c}
\frac{S_{\lambda c}(E-\tilde{E}_\mu) - S_{\mu c}(E-\tilde{E}_\lambda)}
{\tilde{E}_\lambda-\tilde{E}_\mu} & \lambda\neq\mu \end{array} \right. ,
\end{split}
\end{equation}
where $S_{\lambda c}\equiv S_c(\tilde{E}_\lambda)$.
The ${\bm M}$ matrix, which via Eq.~(\ref{eq:U_M}) determines the scattering matrix and thus the observables, is then given by
\begin{equation} \label{eq:M_alt_A}
M_{c'c} = \sum_{\lambda\mu}\tilde{A}_{\lambda\mu}\tilde{\gamma}_{\lambda c'}
  \tilde{\gamma}_{\mu c}.
\end{equation}
It is important to note that this formalism is mathematically equivalent to the
original $R$-matrix theory. This approach is used exclusively in the present analysis.

\subsection{Radiative Capture} \label{sec:radiative_capture}

$R$-matrix theory as described above does not include reactions involving photons and the channel label $c$ used in the previous equations do not include such channels. As is the case for most theoretical treatments, we include photon channels in $R$-matrix calculations via perturbation theory, where the transition matrix is given by the matrix element of the electromagnetic interaction Hamiltonian evaluated between initial and final nuclear states. The interaction Hamiltonian is decomposed into a sum of transition operators corresponding to particular multipolarities which are classified as electric ($EL$) or magnetic ($ML$). The $R$-matrix formalism is then used to define the nuclear states.  We assume here that the final state is bound in all nuclear decay channels, although an extension to unbound final states is possible. The matrix elements can be evaluated in coordinate space by considering separately the contributions from inside and outside the channel radii of the initial scattering state. In the internal region, the key quantities are the matrix elements of the transition operators between the $R$-matrix basis states $|\lambda\rangle$ and the final state, which are defined to be the (internal) reduced widths for photons. In the external region, the Coulomb functions can be used for both the initial and final state.  Importantly, a bound final state may be parametrized completely in the external region by its ANCs. In the external region, we only consider electric transitions and utilize the simple Siegert form of the transition operators in the long-wavelength approximation. This contribution to the transition matrix has been traditionally referred to as {\it external capture}, and depends only on the $R$-matrix parameters for nuclear channels and the final-state ANCs.

There are different nomenclatures that have been used in the literature to describe the direct (one-step) process that can occur for a capture reaction. Since this has led to considerable confusion, a moment is taken here to better define the terminology surrounding the direct capture process. In this work ``direct capture'' always refers to transitions from an initial state to a final state in a non-resonant manner, where no obvious intermediate compound nucleus resonance is populated. It does \textit{not} refer to any particular model, including the direct capture \textit{model} developed by \textcite{Rolfs197329}. In fact, it should be noted that in \textcite{Rolfs197329} careful distinction is made between the direct capture model that was being used and the more general concept of a direct capture process.  From a quantum-mechanical point of view there is no completely unambiguous way to separate resonant and non-resonant processes, which implies that the concept of ``direct capture'' is likewise somewhat ambiguous. In the $R$-matrix approach, the physical process of direct capture is described by a combination of external capture \textit{and} background poles. From this point of view, the ambiguity is related to the somewhat arbitrary choice of channel radius, which affects the strength of background poles and the division between internal and external capture. Another point which has caused confusion is that ``direct capture'' is not synonymous with {\em external capture} in the $R$-matrix approach where resonances also contribute to external capture. However, the two concepts do have considerable overlap and much of the important early work in this area used ``direct capture''
models that only included {\em external capture} (or extranuclear capture) \cite{CHRISTY196189,PhysRev.131.2582}.

The importance of external radiative capture in the $R$-matrix approach was first considered by \citet{PhysRev.88.1109}. The general formalism has been presented by \citet{azure}, which is based upon the work of \cite{RevModPhys.30.257,LANE1960563,Lyn68,PhysRevC.18.1962,BarkerKajino1991,Angulo2001755}. We we largely follow \citet{azure} which employs the general notation of \citet{BarkerKajino1991} but utilizes ANCs to parametrize the strength of the final states.

We will utilize the label $p\equiv\epsilon L\lambda_f$ for photon channels,
where $\epsilon$ indicates the transition type
($\epsilon=0$ for magnetic, $\epsilon=1$ for electric),
$L$ is the multipolarity, and $\lambda_f$ characterizes the
final nuclear state by its total angular momentum $J_f$, parity, energy,
and possibly its ANCs. Note that $\lambda_f$ is analogous to the label $\alpha$
used for nuclear partitions.

The differential radiative capture cross section may be calculated in first-order perturbation theory via \cite{PhysRevC.59.2152}
\begin{equation} \label{eq:goldenrule}
\begin{split}
\frac{d\sigma_{\alpha\rightarrow\lambda_f}}{d\Omega_\gamma} & = \frac{k_\gamma}{2\pi\hbar v_\alpha}
  \frac{1}{(2J_{\alpha 1}+1)(2J_{\alpha 2}+1)} \\
  & \sum_{s\nu q M_f} \left| \langle \lambda_f M_f
  |H_e(\vec{k}_\gamma,q)|\alpha s \nu\rangle \right|^2,
\end{split}
\end{equation}
where $|\lambda_f M_f\rangle$ is the final-state wavefunction with total
angular momentum projection $M_f$;
$\vec{k}_\gamma$ is the photon wave vector with magnitude $k_\gamma=(E-E_f)/\hbar c$ with $E_f$ being the final state energy;
$H_e(\vec{k}_\gamma,q)$ is the photon emission Hamiltonian,
with $q$ the photon circular polarization;
and $|\alpha s \nu\rangle$ are plane wave states with outgoing boundary conditions, normalized to unit magnitude, and are described within the $R$-matrix approach by Eq.~(\ref{eq:outwave}).
The final state wavefunction is normalized over all space such that
$\langle\lambda_f M_f|\lambda_f M_f'\rangle=\delta_{M_f M_f'}$
and it behaves under time reversal according to
\begin{equation} \label{eq:final_t_rev}
K|\lambda_f M_f\rangle = (-1)^{J_f-M_f}|\lambda\,{-}M_f\rangle.
\end{equation}
The photon emission Hamiltonian is given by $H_e(\vec{k}_\gamma,q) = [H_a(\vec{k}_\gamma,q)]^\dagger$, where $H_a(\vec{k}_\gamma,q)$ is the photon absorption Hamiltonian \cite{RevModPhys.39.306,PhysRevC.59.2152,PhysRevC.88.024602}
\begin{equation}
H_a(\vec{k}_\gamma,q) = -\sum_{\epsilon L\mu} q^{1-\epsilon}
  \alpha_{\epsilon L} \mathcal{M}^{\epsilon L}_{\mu}
  D^{L}_{\mu q}(-\phi_\gamma,-\theta_\gamma,0),
\end{equation}
with
\begin{equation}
\alpha_{\epsilon L} = -\left[\frac{2\pi(L+1)(2L+1)}{L}\right]^{1/2}
  \frac{i^{L+1-\epsilon} k_\gamma^L}{(2L+1)!!}.
\end{equation}
Here, $D^L_{\mu q}$ is the Wigner rotation matrix,
$\theta_\gamma$ and $\phi_\gamma$ describe the photon emission angles,
$\mathcal{M}^{\epsilon L}_{\mu}$ are the multipole operators,
and $\mu$ is the projection of $L$.

The transition matrix connecting nuclear and photon channels may
be defined as
\begin{equation} \label{eq:photon_tmatrix}
\begin{split}
T^J_{c\rightarrow p} = & \left[\frac{8\pi(L+1)}{\hbar v_\alpha L}\right]^{1/2}
  \frac{k_\gamma^{L+1/2}}{(2L+1)!!} \times \\
  & \langle\alpha slJ||i^{L+1-\epsilon}\mathcal{M}^{\epsilon L}||\lambda_f \rangle^*,
\end{split}
\end{equation}
where the definition of the reduced matrix element is
\begin{equation} \label{eq:reduced_M}
\begin{split}
\langle\alpha slJM|& i^{L+1-\epsilon} \mathcal{M}^{\epsilon L}_{\mu} |
  \lambda_f M_f\rangle \equiv \\
  & (LJ_f \mu M_f|JM) \langle\alpha slJ|| i^{L+1-\epsilon}\mathcal{M}^{\epsilon L}||
  \lambda_f \rangle,
\end{split}
\end{equation}
and
$|\alpha slJ\nu\rangle$ are the partial-wave components of
$|\alpha s\nu\rangle$ defined by Eq.~(\ref{eq:plane_partial}).
With this definition of the transition matrix,
the angle-integrated radiative capture cross section corresponding to
Eq.~(\ref{eq:goldenrule})
can be written as
\begin{equation} \label{eq:sigma_photon_total}
\sigma_{\alpha\rightarrow\lambda_f} = \frac\pi{k_\alpha^2}
  \sum_{J l s L \epsilon} g_J \left| T^J_{c\rightarrow p} \right|^2,
\end{equation}
which is analogous to Eq.~(\ref{equation:crosssection}) for the
cross section connecting nuclear partitions.
The expression for the differential cross section in terms of
$T^J_{c\rightarrow p}$ is given by Eq.~(36) of \citet{azure}.

We now specify our approach to $R$-matrix theory, where the internal
and external contributions to the matrix element
$\langle \alpha slJ || \mathcal{M}^{\epsilon L} || \lambda_f \rangle$
are considered separately and we can
define the total transition matrix to be the sum of internal and
external contributions:
\begin{equation} \label{eq:t-matrix_gammas}
T^J_{c\rightarrow p} =  T^J_{c\rightarrow p}({\rm int})+T^J_{c\rightarrow p}({\rm ext}).
\end{equation}
Using Eqs.~(\ref{eq:out_int}) and~(\ref{eq:plane_partial}) the internal
contribution to the matrix element is given by
\begin{equation}
\begin{split}
\langle\lambda_f & || i^{L+1-\epsilon}\mathcal{M}^{\epsilon L} ||
  \alpha s lJ\rangle_{\rm int}^* =
  -i\Omega_c(2\hbar v_\alpha P_c)^{1/2}  \\
  & \times \sum_{\lambda\mu} A_{\lambda\mu}\gamma_{\mu c}
  \langle\lambda || i^{L+1-\epsilon} \mathcal{M}^{\epsilon L} ||
  \lambda_f \rangle_{\rm int}^*,
\end{split}
\end{equation}
where the $J$ index is suppressed in the r.h.s.\@ and the reduced matrix
element is defined as in Eq.~(\ref{eq:reduced_M}).
We thus have
\begin{equation} \label{eq:photon_T_internal}
T^J_{c\rightarrow p}({\rm int)} = -2i\Omega_c (P_c k_\gamma^{2L+1})^{1/2}
  \sum_{\lambda\mu} A_{\lambda\mu} \gamma_{\mu c}\gamma_{\lambda p},
\end{equation}
where the photon reduced-width amplitude is given by
\begin{equation} \label{eq:photon_red_width}
\gamma_{\lambda p} = \left[\frac{4\pi(L+1)}{L}\right]^{1/2} \\
  \frac{\langle\lambda ||i^{L+1-\epsilon} \mathcal{M}^{\epsilon L}||
  \lambda_f \rangle_{\rm int}}{(2L+1)!!}.
\end{equation}
Due to the time-reversal properties of $|\lambda\rangle$ and $|\lambda_f\rangle$
given by Eqs.~(\ref{eq:eigen_t_rev}) and~(\ref{eq:final_t_rev}), as well as of the multipole operators, these reduced-width amplitudes are
real quantities~\cite{RevModPhys.30.257,PhysRevC.18.1962,PhysRevC.59.2152} and hence the complex conjugation symbol on the reduced matrix element has been dropped. We assume there is no residual photon-energy dependence of
the multipole operators, as is the case
for the long-wavelength approximation, such that the $\gamma_{\lambda p}$ are constants. Note also that the form of $T^J_{c\rightarrow p}({\rm int})$ given by Eq.~(\ref{eq:photon_T_internal})
has the same structure as the transition matrix connecting
nuclear channels, with the exception that photon channels
{\em do not} contribute to $A_{\lambda\mu}$ (or equivalently, to the
resonance denominators).

In the external region, explicit forms for the wavefunctions
and multipole operators will be utilized.
Here, the final-state wavefunction is assumed to consist of two clusters in each channel and may be written as
\begin{equation} \label{eq:final_external}
\begin{split}
|\lambda_f M_f\rangle_{\rm ext} &= \sum_{c}
  C_{c}\frac{W_{c}(r_{\alpha})}{r_{\alpha}} \\ & \times
  \left[ |\psi\rangle_{\alpha s \nu} \otimes i^l
  Y_{lm}(\hat{r}_{\alpha})\right]_{J_f M_f},
\end{split}
\end{equation}
where the sum is over final-state channels $c$,
$W_{c}(r_{\alpha})$ is the Whittaker function
which describes the radial dependence of the final channel $c$,
and $C_{c}$ is the ANC describing the channel's asymptotic strength.

Only the electric multipole operators will be considered in the external
region. Assuming that each cluster is represented by a point charge, using the Siegert form of the operators, and making the long-wavelength approximation, the electric multipole operators in partition $\alpha$ become
\begin{equation} \label{eq:multipole_external}
\mathcal{M}^{1L}_\mu = \bar{e}^L_\alpha r_\alpha^L Y_{L\mu}(\hat{r}_\alpha),
\end{equation}
where the effective charge is
\begin{equation} \label{eq:eff_charge}
\bar{e}^L_\alpha = e\left[Z_{\alpha 1}\left(\frac{M_{\alpha 2}}{M_\alpha}\right)^L
  + Z_{\alpha 2}\left(-\frac{M_{\alpha 1}}{M_\alpha}\right)^L \right]
\end{equation}
with $eZ_{\alpha i}$ and $M_{\alpha i}$ the charges and masses of
partition $\alpha$ and $M_\alpha=M_{\alpha 1}+M_{\alpha 2}$. The effective charge factor plays a critical role in ${}^{12}{\rm C}(\alpha,\gamma){}^{16}{\rm O}$ $E1$ capture. Due to the nearly equal charge-to-mass ratios of ${}^4{\rm He}$ and ${}^{12}{\rm C}$ nuclei, $\bar{e}^L_\alpha$ nearly vanishes for this case and external capture is strongly surpressed.

The external contribution to the transition matrix can now be calculated using
Eqs.~(\ref{eq:out_ext}) and~(\ref{eq:plane_partial}) for the initial state,
Eq.~(\ref{eq:final_external}) for the final state,
Eqs.~(\ref{eq:scat_mat_channel}) and~(\ref{eq:m_matrix})
for the nuclear scattering matrix,
Eq.~(\ref{eq:multipole_external}) for the multipole operators,
and Eq.~(\ref{eq:photon_tmatrix}) for the transition matrix.
Note that matrix elements of these simple electric multipole operators
vanish unless $\alpha=\alpha_f$ and $s=s_f$. One thus obtains
\begin{equation} \label{eq:photon_T_external}
\begin{split}
&T^J_{c\rightarrow p}({\rm ext}) = -2i\Omega_c
  (P_c k_\gamma^{2L+1})^{1/2} \\
& \times \sum_{c'l_f}  \frac{\bar{e}^L_{\alpha'}}{\hbar}
  (\mu_{\alpha'} a_{c'})^{1/2}
   V(Ll_fJs';l'J_f) C_{\alpha' s' l_f} \\
& \times \left[ \delta_{cc'} P_c^{-1} J_{cl_f L}'
  + M_{cc'}(J_{c'l_f L}'' + iJ_{c'l_f L}') \right],
\end{split}
\end{equation}
where the real function $V$ contains angular momentum factors:
\begin{equation}
\begin{split}
V(Ll_fJs; & lJ_f) = \frac{1}{(2L+1)!!}
  \left[\frac{2(L+1)(2L+1)}{L}\right]^{1/2} \\
& \times i^{l+L-l_f} (lL00|l_f0) (2l+1)^{1/2} \\
& \times (2J_f+1)^{1/2} W(Ll_fJs;lJ_f).
\end{split}
\end{equation}
When $c$ is an open channel, the radial integrals $J_{cl_f L}'$ and $J_{cl_f L}''$
are given by
\begin{equation}
\begin{split}
J_{cl_f L}' = \int_{a_c}^\infty & \frac{ G_c(a_c) F_c(r_\alpha) -
  F_c(a_c)G_c(r_\alpha) }{ F_c^2(a_c)+G_c^2(a_c)} \\
& \times W_{c_f}(r_\alpha) \, r_\alpha^L \, dr_\alpha
\end{split}
\end{equation}
and
\begin{equation}
\begin{split}
J_{cl_f L}'' = \int_{a_c}^\infty & \frac{ F_c(a_c) F_c(r_\alpha) +
  G_c(a_c)G_c(r_\alpha) }{ F_c^2(a_c)+G_c^2(a_c)} \\
& \times W_{c_f}(r_\alpha) \, r_\alpha^L \, dr_\alpha ,
\end{split}
\end{equation}
where $c_f\equiv\alpha s l_f$ and $F_c(r_\alpha)$ and $G_c(r_\alpha)$ are the
regular and irregular Coulomb wavefunctions, respectively.
If channel $c$ is closed, we take $J_{cc_f L}'=0$ and
\begin{equation}
J_{cl_f L}'' = \int_{a_c}^\infty
\frac{W_{c}(r_\alpha)}{W_c(a_c)}W_{c_f}(r_\alpha) \, r_\alpha^L \, dr_\alpha .
\end{equation}
Our radial integrals are very similar to those introduced by
\citet{BarkerKajino1991}, but use a different normalization and a
different convention for closed channels.

As expressed by Eq.~(\ref{eq:photon_T_external}), the external photon
transition matrix consists of two types of terms. The first is
proportional to $\delta_{cc'} J_{cc_f L}'$ and thus only receives
contributions from the entrance channel. It is also non-resonant and
independent of the $R$-matrix parameters of the nuclear channels, except
for the channel radii. In addition, the radial scattering wavefunction in the
$J_{cc_f L}'$ integral corresponds to elastic scattering by a hard sphere.
For these reasons, this contribution has been called hard-sphere capture.
The second term is proportional to the ${\bm M}$ matrix and
consequently does depend on the nuclear $R$-matrix parameters
and exhibits resonances along with the nuclear channels.
This contribution has been called channel capture in the literature.
Note that the division of capture strength between the internal contribution,
hard-sphere capture, and channel capture is dependent upon the choice
of channel radii.

It is possible to write the internal and external contributions to
$T^J_{c\rightarrow p}$ in a different form that emphasizes another aspect of
the underlying physics that they share.
Considering radiative captures to a particular photon channel $p$ from
nuclear channels of total spin $J$, we can write
Eq.~(\ref{eq:photon_T_internal}) as
\begin{equation}
{\bm T}^J_p({\rm int}) = -2i{\bm \Omega}{\bm P}^{1/2} k_\gamma^{L+1/2}
  {\bm \gamma}^T{\bm A}{\bm \gamma}_p,
\end{equation}
where ${\bm T}^J_p({\rm int})$ is a column vector in channel space
with elements $T^J_{c\rightarrow p}$,
${\bm \gamma}$ is a matrix (in general, rectangular) with elements
$\gamma_{\lambda c}$, ${\bm A}$ is the level matrix defined by
Eq.~(\ref{eq:ainv}), and ${\bm \gamma}_p$ is a column
vector in level space with elements $\gamma_{\lambda p}$.
Using the method described in the appendix of \citet{PhysRevC.66.044611},
this equation can also be written as
\begin{equation}
{\bm T}^J_p({\rm int}) = -2i{\bm \Omega}{\bm P}^{1/2} k_\gamma^{L+1/2}
[{\bm 1}-{\bm R}({\bm L}-{\bm B})]^{-1}{\bm R}_p,
\end{equation}
where ${\bm R}_p$ is a column vector in channel space with components
\begin{equation}
[{\bm R}_p]_c=\sum_\lambda \frac{\gamma_{\lambda c}\gamma_{\lambda p}}{E_\lambda-E}.
\end{equation}

Considering now the external region, one can define column vectors ${\bm x}$
and ${\bm y}$ in channel space with components
\begin{eqnarray}
x_c &=& \sum_{l_f}  \frac{\bar{e}^L_{\alpha}}{\hbar}
  (\mu_{\alpha} a_{c})^{1/2} V(Ll_fJs;lJ_f) C_{\alpha s l_f} J_{cl_f L}' \\
y_c &=& \sum_{l_f}  \frac{\bar{e}^L_{\alpha}}{\hbar}
  (\mu_{\alpha} a_{c})^{1/2} V(Ll_fJs;lJ_f) C_{\alpha s l_f} J_{cl_f L}'' .
\end{eqnarray}
Equation~(\ref{eq:photon_T_external}) can then be written as
\begin{eqnarray} \label{eq:T_ext_first}
{\bm T}^J_p && ({\rm ext}) = -2i{\bm \Omega}{\bm P}^{1/2} k_\gamma^{L+1/2}
  \left\{ {\bm P}^{-1}{\bm x} \right. \nonumber \\
&& \quad \left. +[{\bm 1}-{\bm R}({\bm L}-{\bm B})]^{-1}
  {\bm R}({\bm y}+i{\bm x}) \right\} \nonumber \\
&& = -2i{\bm \Omega}{\bm P}^{1/2} k_\gamma^{L+1/2}
  [{\bm 1}-{\bm R}({\bm L}-{\bm B})]^{-1} \\
&& \quad \times \left\{  [{\bm 1}-{\bm R}({\bm L}-{\bm B})]{\bm P}^{-1}{\bm x}
  +{\bm R}({\bm y}+i{\bm x}) \right\}.  \nonumber
\end{eqnarray}
We define $P_c^{-1}x_c\equiv 0$ for closed channels, since from
Eq.~(\ref{eq:photon_T_external}) it is clear that these values only
affect $T^J_{c\rightarrow p}$ when $c$ is closed, which are channels we
are not interested in.
Defining ${\bm S}$ to be a diagonal matrix in channel space
with elements consisting of the shift function $S_c$, the quantity in braces
in Eq.~(\ref{eq:T_ext_first}) simplifies to
\begin{equation}
{\bm R}_p({\rm ext}) \equiv
[{\bm 1}-{\bm R}({\bm S}-{\bm B})]{\bm P}^{-1}{\bm x}+{\bm R}{\bm y} ,
\end{equation}
which is a real quantity as the complex pieces have canceled.

The total transition matrix can now be written as
\begin{equation} \label{eq:T_photon_watson}
\begin{split}
{\bm T}^J_p =& -2i{\bm \Omega}{\bm P}^{1/2} k_\gamma^{L+1/2}
[{\bm 1}-{\bm R}({\bm L}-{\bm B})]^{-1} \\
& \times \left[ {\bm R}_p + {\bm R}_p({\rm ext})\right].
\end{split}
\end{equation}
The important result is that both ${\bm R}_p$ and
${\bm R}_p({\rm ext})$ are real quantities,
which implies the complex phases of the
$T^J_{c\rightarrow p}$ are determined entirely by Coulomb interactions
and the $R$-matrix parameters for nuclear channels -- and
thus not by the photon emission Hamiltonian.
\citet{PhysRevC.59.2152} has pointed out that results such as this
are a manifestation of Watson's Theorem \cite{PhysRev.95.228},
which is more general than $R$-matrix theory.
The primary assumptions required for Watson's Theorem
are first-order perturbation theory for photon emission and
time-reversal invariance.
A derivation of the analogous result for the single-channel $R$-matrix case
has been given by \citet{BarkerKajino1991}, Eqs.~(25-27), where the
complex phase of the capture matrix element is found to be simply given by the sum
of the Coulomb and nuclear elastic-scattering phase shifts.
As shown by Eq.~(\ref{eq:sigma_photon_total}),
complex phases do not affect the total cross section.
They do, however, significantly impact angular distributions,
which has important implications for the
${}^{12}{\rm C}(\alpha,\gamma){}^{16}{\rm O}$ reaction, as $\gamma$-ray
angular distributions are used to separate the $E1$ and $E2$ multipole
transitions to the ${}^{16}{\rm O}$ ground state.
It has been found that elastic scattering
data, which can precisely fix the nuclear $R$-matrix parameters,
are very helpful for improving the accuracy of the extracted
radiative capture multipoles. See \citet{PhysRevC.64.055803} and
\citet{PhysRevC.88.062801} for further discussion.

In the case of a narrow resonance or bound state, a single-level
approximation again provides for a physical interpretation.
We will assume the level shift vanishes for level $\lambda$ and
consequently describe it with $\tilde{E}_\lambda$, $\tilde{\gamma}_{\lambda c}$,
and $\tilde{\gamma}_{\lambda p}$.
We make a single-level approximation to the matrix ${\bm M}$ appearing
in Eq.~(\ref{eq:photon_T_external}) and ignore the hard-sphere capture term.
The cross section can then be put into the Breit-Wigner form
as shown before for nuclear channels.
It is also useful to define a channel contribution to the $\gamma$-ray
reduced width amplitude:
\begin{equation}
\begin{split}
\tilde{\gamma}_{\lambda p}({\rm ch}) = & \sum_{cl_f}
  \frac{\bar{e}^L_{\alpha}}{\hbar} (\mu_{\alpha} a_{c})^{1/2} V(Ll_fJs;lJ_f) \\
& \times  \gamma_{\lambda c} C_{\alpha s l_f} (J_{cl_f L}'' +iJ_{cl_f L}').
\end{split}
\end{equation}
The resulting partial $\gamma$-ray width can then be written as
\begin{equation} \label{eq:G_to_g_gammas}
\tilde{\Gamma}_{\lambda p} = \frac{2k_\gamma^{2L+1}\left|\tilde{\gamma}_{\lambda p} +\tilde{\gamma}_{\lambda p}({\rm ch})\right|^2}{1+\sum_{c}
\tilde{\gamma}_{\lambda c}^2 \frac{dS_{c}}{dE}(\tilde{E}_\lambda)}.
\end{equation}
Note that $\tilde{\gamma}_{\lambda p}$ and $\tilde{\gamma}_{\lambda p}({\rm ch})$,
which are the internal and external contributions,
are combined coherently and that $\tilde{\gamma}_{\lambda p}({\rm ch})$ is
in general a complex quantity. We should also emphasize here that
photon channels do not contribute to the total width in the denominator
of the Breit-Wigner formula in this approach.

The radiative capture formalism presented above is easily adapted to the
alternative $R$-matrix parametrization. The alternative photon reduced widths
$\tilde{\gamma}_{\lambda p}$ are defined by the replacements
$\gamma_{\lambda p}\rightarrow\tilde{\gamma}_{\lambda p}$ and
$\langle\lambda|\rightarrow\langle\tilde{\lambda}|$ in
Eq.~(\ref{eq:photon_red_width}).
The internal contribution to the transition matrix is calculated
using the replacement
\begin{equation}
\sum_{\lambda\mu}A_{\lambda\mu}\gamma_{\mu c}\gamma_{\lambda p}\rightarrow
\sum_{\lambda\mu}\tilde{A}_{\lambda\mu}\tilde{\gamma}_{\mu c}\tilde{\gamma}_{\lambda p}
\end{equation}
in Eq.~(\ref{eq:photon_T_internal}) \cite{PhysRevC.66.044611}.
For the external contribution, one only needs
the ${\bm M}$ matrix, which is already defined in terms of the alternative
parameters by Eq.~(\ref{eq:M_alt_A}).

\subsection{\texorpdfstring{$\beta$}{beta}-Delayed Particle Emission} \label{subsec:rmatrix_beta_channels}

As in the case of radiative capture, $\beta$-delayed particle spectra can be modeled in the $R$-matrix approach using first-order perturbation theory. General formulas have been given by
\textcite{Barker1988269} and formulas specific to the $\beta$-delayed $\alpha$-particle spectrum
from ${}^{16}{\rm N}$ are given, for example, by \textcite{PhysRevC.50.1194}. It should be noted that only allowed transitions are considered, which is roughly analogous
to only considering $E1$ transitions for radiative capture. In addition, we do not consider any external contribution to the transition matrix element. Such contributions are not thought to be significant for the ${}^{16}{\rm N}$ $\beta$ decays, although they have been found to be important for understanding the $\beta$-delayed deuteron spectrum from ${}^6{\rm He}$ decay~\cite{BARKER199417}. The reader should be aware that $\beta$~decays into unbound states has received considerably less theoretical attention, in the $R$-matrix context or otherwise, than radiative capture.

The rate of $\beta$-delayed particle decay may be written as \cite{Barker1988269}
\begin{equation} \label{eq:beta_decay_life_time}
\frac{\ln 2}{t_{1/2}} = \int \sum_{Jc} w^J_c(E)dE
\end{equation}
where differential decay rate $w^J_c$ is summed over the final-state angular momenta $J$ and channels $c\equiv\alpha sl$ and $t_{1/2}$ is the partial half life for $\beta$-delayed particle emission. For example, the total half-life of $^{16}$N is 7.13(2) s \cite{Tilley19931}, but the branching ratio for $\beta$-delayed $\alpha$ emission is only 1.2$\times$10$^{-5}$. Therefore the $\beta$-delayed $\alpha$ emission half-life is 5.9$\times$10$^5$~s.


The $w^J_c(E)$ describe the $\beta$-delayed particle energy spectrum (or possibly spectra) components that do not interfere. These quantities may be written in the $R$-matrix formalism as \cite{Barker1967,Barker1969b,Barker1988269}
\begin{equation} \label{eq:w_j_c}
w^J_c(E) = C^2f_\beta P_c\sum_x\left| \sum_{\lambda\mu} g_{\lambda x}
  \gamma_{\mu c}A_{\lambda\mu} \right|^2,
\end{equation}
where $C^2$ is a constant factor, $f_\beta$ is the integrated Fermi function (the $\beta$-decay phase space factor), and the $g_{\lambda x}$ are the $\beta$-decay feeding factors. Here, $x$ is used to indicate either Fermi or Gamow-Teller transitions. The $J$-dependence of the $R$-matrix quantities $g_{\lambda x}$, $P_c$, $\gamma_{\mu c}$, and $A_{\lambda\mu}$ in the r.h.s.\@ of Eq.~(\ref{eq:w_j_c}) has been suppressed. It should also be noted that the $\beta$-decay feeding factors are assumed here to take only real values.

In practice it is often convenient to rewrite Eq.~(\ref{eq:w_j_c}) as
\begin{equation} \label{eq:n_j_c}
n^J_c(E) = f_\beta P_c\sum_x\left| \sum_{\lambda\mu} B_{\lambda x}
  \gamma_{\mu c}A_{\lambda\mu} \right|^2,
\end{equation}
where the feeding factor is now defined to be
\begin{equation}
B_{\lambda x} = C(Nt_{1/2}/\ln 2)^{1/2}g_{\lambda x},
\end{equation}
$n^J_c(E)$ is the number of counts per unity energy, and $N$ is the total number of counts in the spectrum. Further variations of this formula exist in the literature, including dividing Eq.~(\ref{eq:n_j_c}) by $N$ and redefining $B_{\lambda x}$ so that it is independent of $N$. In addition, some workers, such as  \textcite{PhysRevC.50.1194}, absorb the reduced width $\gamma_{\mu c}$ into the definition of the feeding factor.

In the case of $\beta$~decay to a narrow unbound level, the single-level approximation may be used to relate the measured transition strength to the feeding factor for that level. This approximation results in a Breit-Wigner energy spectrum for the particle decay. We will assume the level shift vanishes for the level $\lambda$ and that its parameters are $\tilde{E}_\lambda$, $\tilde{\gamma}_{\lambda c}$, and $\tilde{B}_{\lambda x}$. If one then ignores the energy dependences of $f_\beta$ and $P_c$ and assumes $S_c$ is linear in energy, the integral over the resulting Lorentzian energy distribution can be performed analytically to obtain \cite{Barker1988269}
\begin{equation} \label{eq:logft}
(ft_{1/2})_\lambda = \frac{N_\lambda t_{1/2}[1+\sum_{c} \tilde{\gamma}_{\lambda c}^2 \frac{dS_{c}}{dE}(\tilde{E}_\lambda)]}{\pi\sum_x\left|\tilde{B}_{\lambda x}\right|^2},
\end{equation}
where $N_\lambda$ is the total number of counts observed for the transition. This formula may also be used to define $\log (ft_{1/2})_\lambda$ values. Also note that this equation becomes exact (in the sense of the $R$-matrix approach) if the final state is bound. See also appendix A of \textcite{Riisager2014112} for further discussions of this topic. We utilize such calculations in this work to define the feeding factors for bound states and to compare our results with previous studies in the literature. It should also be noted that  $\beta$-delayed particle emission is easily implemented using the alternate $R$-matrix parameterization -- all that is required is the replacement
\begin{equation}
\sum_{\lambda\mu} B_{\lambda x}\gamma_{\mu c}A_{\lambda\mu} \rightarrow
\sum_{\lambda\mu} \tilde{B}_{\lambda x}\tilde{\gamma}_{\mu c}\tilde{A}_{\lambda\mu},
\end{equation}
where the relation of the alternative feeding factors $\tilde{B}_{\lambda x}$ to the $B_{\lambda x}$ is given by \citet{PhysRevC.66.044611}. Finally, since only Gamow-Teller transitions are allowed for $^{16}$N$(\beta\alpha)^{12}$C, we drop the $x$ index from the labeling of the $\beta$-decay feeding parameters for this case.

\subsection{\texorpdfstring{$R$}{R}-matrix Phenomenology}
\label{subsec_rm_phenom}

$R$-matrix theory can be used for the phenomenological analysis of nuclear
reaction data by adjusting the parameters to optimize the agreement with
experimental data. More specifically, this means adopting channel radii and
adjusting the parameters $E_\lambda$ and $\gamma_{\lambda c}$ that determine
the scattering matrix in nuclear channels.
These parameters can also define the energies and ANCs of final
states in radiative capture.
If radiative capture data and/or $\beta$-delayed particle data are included
in the analysis, then the photon reduced widths $\gamma_{\lambda p}$
and/or the feeding factors $B_{\lambda x}$ would also be adjusted.

One significant approximation in phenomenological $R$-matrix approach is
that the sums over levels must be truncated. Typically, the known levels up to
a certain excitation energy are included and the remainder of the spectrum
is modeled with one or more ``background'' pole terms for each spin and parity.
This has been the standard technique for some time
(e.g., \citet{PhysRev.58.1068}), but the exact implementation varies.
This approach is utilized in the present work
and its effect on the fit is discussed in Sec.~\ref{sec:systematic_more}.
In addition, it is also generally necessary to truncate the sums over
channels in a phenomenological $R$-matrix analysis.  Channels that are
strongly closed energetically are typically neglected, as they
are expected to have very little influence (LT, subsection X.2).
Channels with large orbital angular momenta are likewise typically excluded,
as their influence is suppressed by the angular momentum barrier.

The choice of channel radius warrants some discussion. According to formal
$R$-matrix theory, the channel radius should be large enough so that at and
beyond the channel radius, nuclear forces are negligible and
Coulomb wavefunctions are a good approximation.
However, increasing the channel radius increases the density of
background poles, as can be seen from (LT, IV.3.3b) for the case of
zero nuclear potential. In a phenomenological analysis, choosing too
large of a radius leads to problems with the background poles becoming
overly complicated. For example, multiple background poles might be required
to cancel most of the large hard-sphere elastic scattering phase shift (which increases along with with the channel radius).

In practice, phenomenological $R$-matrix fits must use channel radii which
enclose most but not all of the nuclear interactions.
As an example, the {\it ab initio} calculation of
\textcite{PhysRevLett.99.022502} using realistic nuclear forces
found that a radius of 9~fm was required
for nuclear interactions in the neutron+$\alpha$ system to become
negligible. However, a radius this large would be impractical for
phenomenological fitting; see, for example, \citet{PhysRevLett.59.763}
where 3~fm was used for this radius in a phenomenological description.
As a consequence of some nuclear interactions beyond the channel radius,
the phenomenological reduced width amplitudes must be considered to be
in some sense to be re-normalized quantities.
$R$-matrix fits should, however, be fairly insensitive to the specific
value of the channel radius for a reasonable range of values.
As the radius increases, the penetrability factor becomes larger and
the reduced width amplitudes decrease to preserve the physical width.
It is thus good practice to explore the sensitivity of the
phenomenological fit to the channel radius (or radii).
If a strong variation in the fit quality exists, this can often
indicate that background poles have not been sufficiently considered.

The phenomenological $R$-matrix approach derives much of its power from the
fact that it automatically produces a scattering matrix that is unitary and
symmetric, even with the truncations mentioned above. Unitarity is
a particularly powerful constraint when data are available from multiple
reaction channels. A related statement is that
a single set of $R$-matrix parameters should be able to simultaneously
describe essentially all low-energy nuclear reaction and nuclear structure data relating
to a given compound nucleus.
Our implementation of a phenomenological $R$-matrix analysis of
data relevant to the ${}^{12}{\rm C}(\alpha,\gamma){}^{16}{\rm O}$ reaction
is discussed below.

\subsection{\texorpdfstring{$R$}{R}-matrix Strategy} \label{theory_Rmatrix_12Cag}

As described in Sec.~\ref{sec:astro}, the energy range of the $^{12}$C$(\alpha,\gamma)^{16}$O cross section that is needed to calculate the reaction rate for astrophysical environments is very low ($E_\text{c.m.}$~=~300~keV), well below the limits of current experimental sensitivity ($\sigma\approx$~2$\times$10$^{-17}$~barns). Therefore, while there is a significant amount of data at higher energies, an extrapolation to low energy must be made. This is the primary reason that the phenomenological model must be employed. Further, from a more fundamental theory stand point, the different contributions to the cross section are calculated independently (see Secs.~\ref{sec:cluster_models} and \ref{subsection:rmatrixtheory}). Therefore, it provides further constraint to the phenomenological model if each of the individual contributions, as well as their sum, are measured independently as well. Further, as will be emphasized in Sec.~\ref{sec:R-matrix_analysis}, measurements over a wide energy range, up to several MeV above the $\alpha$ threshold, are also very useful since they help constrain both the interference patterns and the background contributions of the different components, both of which can result in a large source of uncertainty in the cross section extrapolation.

Because low energy measurements of the $^{12}$C$(\alpha,\gamma)^{16}$O reaction are greatly hindered by the Coulomb barrier, indirect techniques are extremely valuable. In particular these techniques can be used to deduce the level parameters (i.e. energies, ANCs, lifetimes) that can then be used in $R$-matrix or other reaction models. These types of measurements have proven the most useful in constraining the contributions to the cross section from the subthreshold states. In particular, it is the 1$^-$ level at $E_x$ = 7.12~MeV ($E_\text{c.m.}$ = -45~keV) and the $E_x$ = 2$^+$ level at 6.92~MeV ($E_\text{c.m.}$ = -245~keV) that have the greatest contribution to the total cross section at $E_\text{c.m.}$~=~300~keV. The energies and lifetimes (or $\gamma$-widths) are well known for these states, but the ANCs (or reduced widths) have proven difficult to determine accurately until recently (see Sec.~\ref{sec:subthreshold}).

So far, the most successful indirect methods include measurements of the $\alpha$-spectrum from $^{16}$N$(\beta\alpha)^{12}$C decay, the differential cross section of $^{12}$C$(\alpha,\alpha_0)^{12}$C elastic scattering, and $\alpha$-transfer reactions. All can be used to determine or constrain one or both of the ANCs of the 1$^-$ and 2$^+$ subthreshold states. One limitation of elastic scattering is that, as shown by Eq.~(\ref{eq:diff_XS_per_s}), the Coulomb amplitude dominates elastic scattering at low energies. Thus for energies below $E_{\rm c.m.}\approx 2.0$~MeV, the elastic scattering cross section is essentially indistinguishable from Rutherford scattering. In the case of $^{16}$N$(\beta\alpha)^{12}$C, the spectrum is surpessed at low $\alpha+{}^{12}{\rm C}$ relative energies by the Coulomb barrier but the Coulomb amplitude is not present. This spectrum has been measured to below $E_{\rm c.m.}=1$~MeV, i.e. closer to the subthreshold states.

As compound nucleus reactions, the data from the $^{16}$N$(\beta\alpha)^{12}$C decay and $^{12}$C$(\alpha,\alpha_0)^{12}$C reaction can be fit directly in the $R$-matrix analysis. These data have the added benefit that they give constraints on other important level parameters as well. On the other hand, as a direct reaction, the $\alpha$-transfer data is analyzed using a distorted wave Born approximation analysis. The ANCs are deduced from a DWBA analysis, then the values and associated uncertainties can be used in the $R$-matrix model (see, e.g., \cite{PhysRevC.59.3418, PhysRevC.63.024612}). 

However, greatly complicating the issue, the subthreshold resonances interfere with other higher lying broad resonances. These interferences are implemented in the $R$-matrix by the relative signs of the reduced width amplitudes in Eqs.~(\ref{eq:rmatrix}) and~(\ref{eq:photon_T_internal}); note the relation between the reduced width and the ANC is given in Eq.~(\ref{eq:ANC_to_rw}). The relative signs determine if the amplitudes of the cross section from the different resonances will add or subtract, which can give drastically different values for the cross section in off-resonance regions. This is because when two components of the cross section ($\sigma_1$ and $\sigma_2$) interfere with one another the magnitude goes as
\begin{equation}
\sigma_\text{interference} \propto 2\sqrt{\sigma_1\sigma_2}.
\end{equation}
Therefore even if one of the cross section components is small, the interference term can still be significant compared to the total. It is into just such an off-resonance region where the extrapolation must be made to reach the stellar energy range. Therefore a reliable and precise extrapolation hinges on the determination of both the magnitude of the level parameters and their relative signs (see Sec.~\ref{sec:interferences}). This means that detailed measurements of the $^{12}$C$(\alpha,\gamma)^{16}$O cross section over, experimentally accessible, off-resonance regions at higher energies are vary valuable in constraining the extrapolation to low energy. 

It should also be emphasized that there are two different general types of interference effects seen in nuclear reactions. One type is when levels of the same $J^\pi$ combine to produce energy-dependent interference effects. Another type is when processes with different $J^\pi$ values combine to produce angle-dependent effects (see Eq.~(\ref{eq:diff_XS_per_s})). Both types of interference are important for understanding the $^{12}$C$(\alpha,\gamma)^{16}$O reaction. The former being particular critical for the low-energy extrapolation of the cross section as just discussed. The latter are critical in disentangling the $E1$ and $E2$ contributions to the cross section (see Sec.~\ref{sec:gs_12Cag}). Practical experimental considerations may allow for only an angle integrated and differential cross section measurement in a single setup. This reemphasizes the need to combine many different kinds of experimental results since different types of data are critical to the $R$-matrix analysis and have different types of uncertainties associated with them.

The general $R$-matrix strategy is then to utilize as much experimental data as possible in order to provide as much physical constraint as possible to the phonological model. While low energy measurements of the $^{12}$C$(\alpha,\gamma)^{16}$O cross section are critical, so too are indirect measurements and those at higher energies. It is only by combining this wide array of experimental data that the phenomenological model can be constrained to the point that it can yield an extrapolated cross section approaching the desired accuracy of nuclear astrophysics applications. With this clearly in mind, a summary of these many and diverse experimental endeavors is in order.   

\section{\label{sec:data} Experimental Measurements}

The study of the $^{12}$C$(\alpha,\gamma)^{16}$O reaction can be naturally divided into three eras: first measurements, the push to low energies, and a return to indirect methods. The division of these eras is marked by some drastic improvement or new discovery in the experimental measurements.

Early measurements sought to investigate the low energy cross section, not for nuclear astrophysics motivations, but to study the effects of isospin breaking of $T$ = 0 transitions (as discussed in Sec.~\ref{sec:intro}). In these investigations many different experimental techniques, nearly all of the indirect methods used today, were developed to study the properties of the compound nucleus. Many of these experiments simply suffered from the immaturity of the field, both in experimental techniques and theoretical interpretation. The capstones for this first period were the unprecedented measurement of the low energy capture cross section around the 1$^-$ resonance at $E_\text{c.m.}$ = 2.68 MeV by \textcite{Dyer1974495} and the multilevel-multichannel $R$-matrix analysis of \textcite{Barker1971} that utilized capture, scattering, and $\beta$ delayed $\alpha$ emission data. 

Once the capture cross section was actually measured, a race began to push the measurements to lower energies, closer to the range of astrophysical interest ($E_\text{c.m.} \approx$~300~keV). A host of experimental improvements and new techniques were developed, including highly $^{13}$C depleted and stable targets, high purity target chambers, recoil separators, inverse kinematic measurements with pure helium gas targets, and high energy resolution detectors. Despite the extraordinary efforts, the rapid drop in the low energy cross section made lower energy measurements hard won. The major discovery of this period was that not only $E1$, but also $E2$ multipolarity, perhaps even in almost equal amplitudes, make up the dominating ground state transition cross section at stellar energies. Further theoretical methods to interpret the higher precision data were also more thoroughly explored. As it became more apparent that direct techniques would be extremely difficult to improve upon, there was a return to indirect methods. While the transition to the next period is not so clear cut, the works of \textcite{PhysRevLett.70.726, PhysRevLett.70.2066} serve as a reasonable division point, as they mark one of the early re-measurements of the $\beta$ delayed $\alpha$ emission spectrum of $^{16}$N and \textcite{PhysRevC.50.1194} made one of the most detailed global analyses of the time. These measurements would dramatically decrease the uncertainty in the $E1$ cross section.

While measurements of the low energy capture cross section continued, attempting improved measurements as new detectors or techniques were developed, many efforts have been made to revive the original indirect methods of transfer, scattering, and $\beta$ delayed $\alpha$ emission of $^{16}$N. Transfer reaction studies have probably benefited the most from theoretical and experimental developments, allowing new measurements to achieve an unprecedented level of consistency. New measurements of the $\beta$ delayed $\alpha$ emission spectra also continued in an effort to reach greater sensitivity and achieve improved accuracy. New scattering and recoil separator measurements were made that covered a wide energy range with high precision, providing a strong underpinning for the $R$-matrix analyses. Increases in computational power also brought about improvements in the sophistication of analysis methods, allowing large amounts of data to be utilized simultaneously to better constrain phenomenological fits and making Monte Carlo uncertainty methods viable. Because of its complexity, additional efforts are always underway to tackle this difficult problem. New indirect methods such as photo-disintegration and Coulomb excitation are underway. New theoretical models are under development, with {\it ab initio} calculations on the horizon (e.g. \citet{Elhatisari2015}).

\begin{figure*}
\includegraphics[width=2.0\columnwidth]{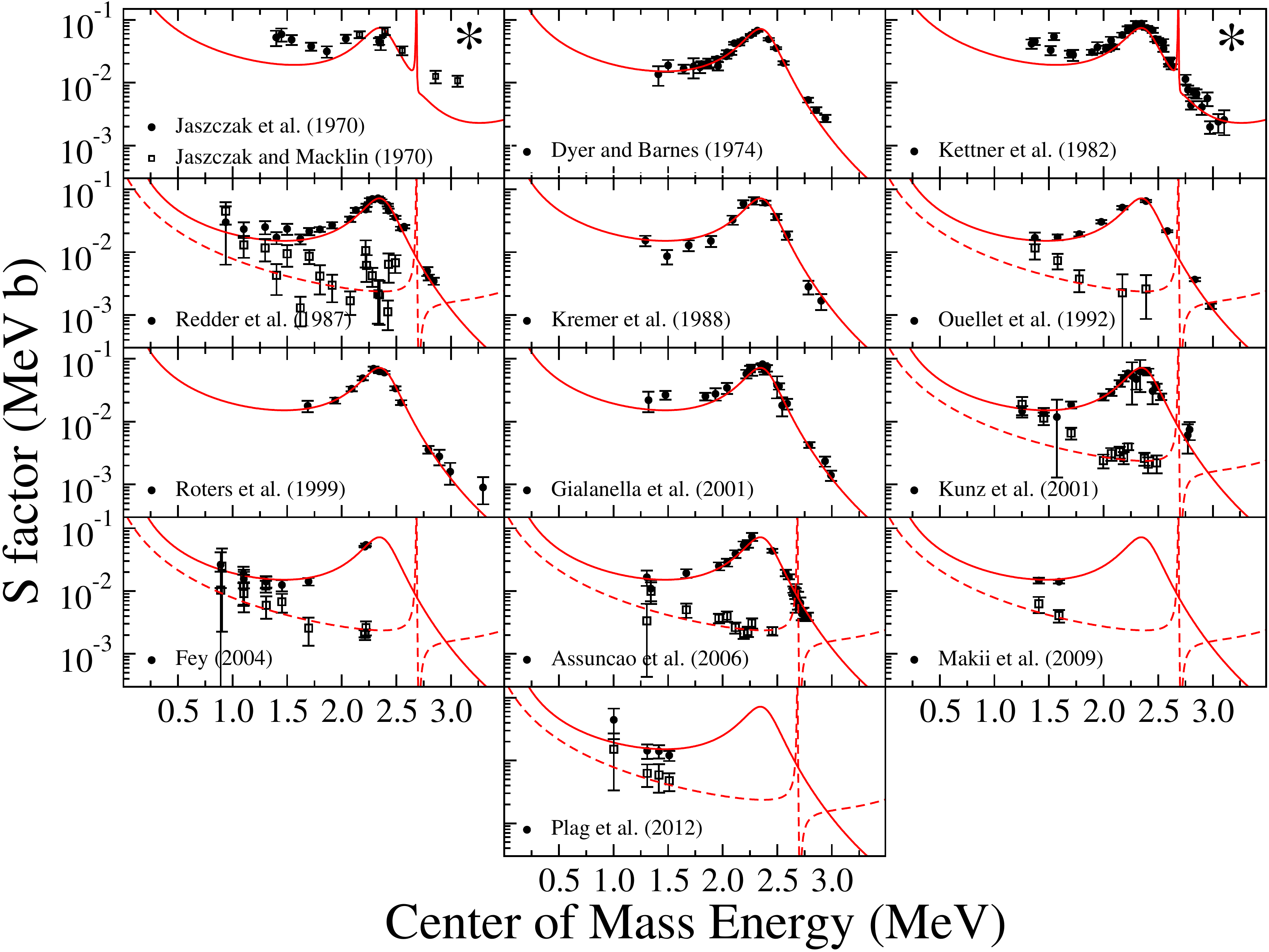}
\caption{(Color online) Comparison of all $E1$ and $E2$ cross sections measured to date. The early works of \textcite{PhysRevC.2.63, PhysRevC.2.2452, springerlink:10.1007/BF01415851} give only total cross sections. These are demarcated by an (*). As a standard for comparison, the $R$-matrix fit described later in this work is also shown. The solid red line shows the $E1$ contribution (except where only the total is given) while the red dashed line gives the $E2$ contribution. No normalization factors have been applied to the data. It should be noted that the region of astrophysical interest is at roughly $E_\text{c.m.} \approx$~300~keV, far below the lowest energy measurements at $E_\text{c.m.} \approx$~1~MeV.  \label{fig:E1_and_E2_comparisons}}
\end{figure*}

To aid the following discussions, Fig.~\ref{fig:E1_and_E2_comparisons} compares all the $E1$ and $E2$ $^{12}$C$(\alpha,\gamma)^{16}$O ground state cross section data reported over the low energy range. While the $R$-matrix fit, described later in this work, represents one of the more detailed phenomenological analyses to date, its use in this section is to simply provide a standard for comparison of the different data sets. This is most helpful when the data are difficult to compare on a one-to-one basis. For example, when experimental effects are significant or the cross section data are presented using different representations. Since this figure is intended to illustrate an unbiased comparison between the different data sets, no scaling factors have been applied to the data.

Experimental techniques have improved significantly over the years and several different techniques have been explored. One of the most significant improvements has been target quality and stability or the use of a helium gas target for inverse kinematics. As a summery for the reader, Table~\ref{tab:targets} collects this information for the capture measurements. 

\begin{table*}
\caption{Summary of target details for different $^{12}$C$(\alpha,\gamma)^{16}$O experiments. \label{tab:targets}}
\begin{ruledtabular}
\begin{tabular}{ c c c c c } 
Ref. & Target & Backing & Thickness & $^{13}$C depletion or gas purity \\
\hline
\textcite{Larson1964497} & cracking acetylene & Ta (0.025 cm) & 96 $\mu$g/cm$^2$ and thinner & factor of 10 $^{13}$C depletion \\
\textcite{PhysRevC.2.63} & cracking of acetylene & Ta (0.025 cm) & 98-178 $\mu$g/cm$^2$ & 99.94\% $^{12}$C \\
\textcite{Dyer1974495} & cracking of methyl alcohol & Ta (0.008 cm) & 150-200 $\mu$g/cm$^2$ & 99.945\% $^{12}$C \\
\textcite{springerlink:10.1007/BF01415851} & He gas target &  & 10 Torr & $<$1 ppm \\ 
\textcite{Redder1987385} & ion implantation & Au & 80~keV at 2.68~MeV & $^{13}$C/$^{12}$C$\approx$10$^{-4}$ \\ 
\textcite{PhysRevLett.60.1475} & He gas target &  & 3.6(2) $\mu$g/cm$^2$  & recoil separator \\ 
\textcite{PhysRevLett.69.1896, PhysRevC.54.1982} & ion implantation & Au & 3-5$\times$10$^{18}$ atoms/cm$^2$ & factor of 10$^3$ $^{13}$C depletion  \\ 
\textcite{Roters1999} & He gas target &  & 9.1 Torr & 0.0001\% \\
\textcite{Gialanella2001} & He gas target & & 20 Torr & 0.0001\% \\
\textcite{PhysRevLett.86.3244} & ion implantation & Au & 2-3$\times$10$^{18}$ atoms/cm$^2$ & factor of 10$^3$ $^{13}$C depletion \\
\textcite{Fey2004} & ion deposition & Au & $\approx$2$\times$10$^{18}$ atoms/cm$^2$ & \\
\textcite{schurmann2005} & He gas target &  & 4.21(14)$\times$10$^{17}$ atoms/cm$^2$ & recoil separator \\
\textcite{PhysRevC.73.055801} & ion implantation & Au & 0.5-11$\times$10$^{18}$ atoms/cm$^2$ & factor of 10$^3$ $^{13}$C depletion \\
\textcite{PhysRevLett.97.242503} & He gas target &  & 4-8 Torr & recoil separator \\ 
\textcite{PhysRevC.80.065802} & cracking of methane gas & Au & 250-400 $\mu$g/cm$^2$ & 99.95\% $^{12}$C  \\
\textcite{Schurmann2011557} & He gas target &  & 4$\times$10$^{17}$ atoms/cm$^2$ & recoil separator \\
\textcite{PhysRevC.86.015805} & ion deposition & Au & 30-120 $\mu$g/cm$^2$ & $^{13}$C/$^{12}$C$<$10$^{-4}$ \\
\end{tabular}
\end{ruledtabular}
\end{table*} 

\subsection{\label{sec:classical_era} First Measurements ('55-'74)} 

\begin{figure}
\includegraphics[width=1.0\columnwidth]{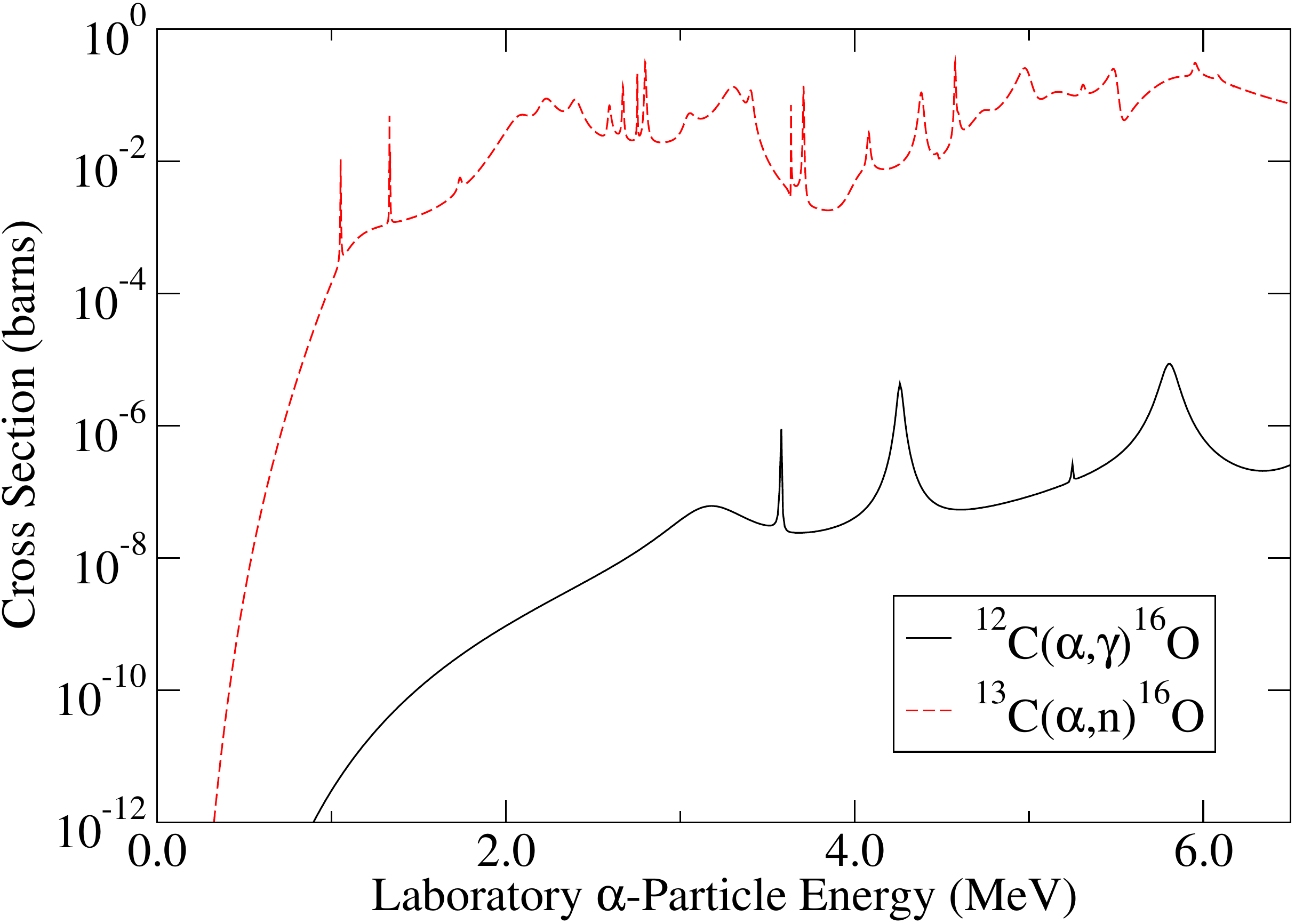}
\caption{(Color online) Comparison of the cross section of the $^{12}$C$(\alpha,\gamma)^{16}$O reaction (this work) to that of the $^{13}$C$(\alpha,n)^{16}$O reaction. Because of the large difference in cross sections, even trace amounts of $^{13}$C in target materials and beam line elements can create large backgrounds in the $\gamma$-ray spectra of $^{12}$C$(\alpha,\gamma)^{16}$O measurements. These backgrounds are chiefly the result of $(n,n'\gamma)$ and $(n,\gamma)$ reactions on the detector materials themselves and nearby beam line components. Level parameters used to calculate the $^{13}$C$(\alpha,n)^{16}$O cross section have been taken from \textcite{doi:10.1080/00223131.2002.10875047} and \textcite{refId0}. \label{fig:12Cag_vs_13C_an}}
\end{figure}

The first published attempt at a direct measurement of the low energy $^{12}$C$(\alpha,\gamma)^{16}$O cross section was made by \textcite{0370-1298-68-6-410} at Imperial College in London with the sole goal of simply detecting a signal from the capture reaction. The experiment was performed with an $\alpha$ beam of 1.6 MeV, a thick target (of unspecified thickness) made of natural carbon (98.9\% $^{12}$C, 1.1\% $^{13}$C by mole fraction), and a NaI detector. Like all subsequent experiments using forward kinematics, it was greatly hindered by background produced from the high cross section $^{13}$C$(\alpha,n)^{16}$O reaction. Indeed several studies were made simply to characterize this reaction (see e.g. \textcite{0370-1298-66-12-415}), which is a background for all $\alpha$ induced reaction studies. A comparison of the cross sections is shown in Fig.~\ref{fig:12Cag_vs_13C_an} where it can be seen that that of the $^{12}$C$(\alpha,\gamma)^{16}$O reaction, on top of the lowest energy 1$^-$ resonance, is more than six orders of magnitude smaller than that of the $^{13}$C$(\alpha,n)^{16}$O reaction. Only upper limits were determined by \textcite{0370-1298-68-6-410}, not surprising in hindsight, as it is now known that the capture cross section at $E_\alpha$ = 1.6 MeV is about 0.2 nbarns!

An estimate for the astrophysical $^{12}$C$(\alpha,\gamma)^{16}$O cross section was given soon after by \textcite{RevModPhys.29.547}, using the most basic kind of resonance theory: a single-level Breit-Wigner \cite{PhysRev.49.519}. The analysis was limited to only the contribution from the $1^-$ subthreshold state at $E_x$ = 7.12~MeV with the $\gamma$-ray width measured by \textcite{SwM56} and assuming $\theta^2_\alpha(7.12)$ = 0.1. This result was later updated by \textcite{fcz1967} using an improved resonance energy, $\gamma$-ray width \cite{PhysRev.108.982}, and a theoretical calculation of $\theta^2_\alpha(7.12)$ \cite{Stephenson1966} that was in reasonable agreement with the result of the first $\alpha$-transfer reaction experiment~\cite{Loebenstein1967481}.

\textcite{doi:10.1080/14786435708231722} (at Brookhaven National Laboratory) were the first to resolve a signal from the $^{12}$C$(\alpha,\gamma)^{16}$O reaction. This was done by subtracting out the large background produced by the $^{13}$C$(\alpha,n)^{16}$O reaction. A thick target (450~$\mu$g/cm$^2$) technique was used and measurements were made over an $\alpha$ energy range from 3.00 to 3.45 MeV. These measurements were associated with decays of $\gamma$ rays from the $E_x$ = 9.59 MeV state, whose $\gamma$ decay width was of great interest at the time to test theoretical predictions of isospin mixing of $T$ = 1 contributions into the predominately $T$ = 0 state. The main result was the measurement of the ground state $\gamma$ width of the state as $\Gamma_{\gamma_{0}} \approx$ 6 meV, about a factor of 2-3 smaller than the accepted value today. 

Measurements were then extended to higher energies by \textcite{0370-1328-75-2-312} (at the Atomic Energy Research Establishment in Harwell, UK) who were the first to study the ground state $\gamma$ decay widths of the 2$^+$ levels at $E_x$ = 9.85 and 11.50 MeV, comparing their Weisskopf widths to that of the $E_x$ = 6.92 MeV state measured by \textcite{PhysRev.108.982}. The experiment was also the first to use targets depleted in $^{13}$C in order to suppress the neutron induced background. Great effort was also made to limit additional carbon build up on the target, resulting from contamination in the beam line, by both target heating and the use of a cold trap. Angular distribution measurements were made for the first time to verify the multipolarities of the transitions.

A very ambitious measurement campaign was then carried out by \textcite{Larson1964497} at the California Institute of Technology (Cal Tech) who measured the $^{12}$C$(\alpha,\gamma)^{16}$O excitation function over an unprecedented energy range from $E_\alpha$ = 2.8 to 8.3 MeV. While this experiment was motivated by further structure studies, for the first time it also sought to investigate the cross section for nuclear astrophysics purposes. Building on the experience of the previous studies, the experiment utilized a depleted $^{13}$C target, a cold trap, and oil free pumps to limit background. While very successful at higher energies, even measuring $\gamma$ ray angular distributions, yields at low energies were still insufficient to map the resonance corresponding to the 1$^-$ state at $E_x$~=~9.59 MeV. The state's properties were still investigated but a thicker target (96~$\mu$g/cm$^2$) was necessary. A new larger value for the $\gamma$ width of $\Gamma_\gamma$ = 22(5) meV was found, in good agreement with current measurements! A more detailed account of the experiment can be found in James Larson's thesis \cite{Larson_thesis}. It is interesting to note the acknowledgment section, which reads like a who's who of nuclear astrophysics at the time. The thesis project was suggested by Willie Fowler, with advising from Charles Lauritsen, Ward Whaling, Charlie Barnes, and Ralph Kavanagh, and with additional discussions with Tom Tombrello and Fred Barker. 

A rather unique measurement testing time reversal invariance was made by \textcite{VonWimmersperg1970291} using measurements of the $^{12}$C$(\alpha,\gamma_0)^{16}$O reaction and its inverse $^{16}$O$(\gamma,\alpha_0)^{12}$C over a $E1/E2$ mixed region at $E_x \approx$~13.1~MeV. Detailed balance was used to test the consistency of the forward/backward asymmetry of the angular distribution at this energy. No significant deviation was observed. 

The first excitation curve measurement of the lowest energy 1$^-$ resonance in the $^{12}$C$(\alpha,\gamma)^{16}$O reaction was reported by \textcite{PhysRevC.2.63} at Oak Ridge National Laboratory ORNL (although preliminary measurements had also been reported at Cal Tech \cite{Adams1968}). Highly $^{13}$C depleted targets, for the time, were utilized (99.94\% $^{12}$C) with thicknesses ranging from 98 to 178 $\mu$g/cm$^2$. The now typical precautions were taken to avoid carbon build up on the target and limit background. The experiment was further notable in that it was one of the first to use a bunched helium beam for time-of-flight (\textcite{Adams1968} had also used this technique) to separate the $\gamma$ rays from the neutron background signals. The measurements were limited to the low energy side of the resonance ranging from $E_\alpha$ = 1.86 to 3.20 MeV but were extended up to $E_\alpha$ = 4.2 MeV in \textcite{PhysRevC.2.2452}. No attempt was made to extrapolate the cross section to stellar energies.

With the extreme difficulty of measuring the capture cross section directly, indirect studies pursued the determination of the reduced $\alpha$ widths. \textcite{Loebenstein1967481}, also at Cal Tech, made $\alpha$ transfer measurements covering the ground state and first five excited states of $^{16}$O using the $^6$Li$(^{12}$C,$d)^{16}$O reaction. While the experiments were performed at relatively low energies of $E_\text{c.m.}$ = 7~MeV, they were still not low enough to avoid the effects of compound nucleus contributions to the cross section. Because the cross sections were known to be a mixture of direct and compound nucleus formation processes, the data were difficult to interpret with theory, and the uncertainties of the extracted $\theta^2_\alpha$ values were difficult to quantify. A range of values were given for many of the low lying states in $^{16}$O including the $E_x$~=~6.05~MeV 0$^+$ (0.14 $< \theta^2_\alpha <$ 0.30), the $E_x$~=~6.92 MeV 2$^+$ (0.15 $< \theta^2_\alpha <$ 0.27) and the $E_x$~=~7.12 MeV 1$^-$ states (0.06 $< \theta^2_\alpha <$ 0.14), but this did not include any contributions to the uncertainties from theory. \textcite{Puhlhofer1970258} made similar measurements except using the $^{12}$C$(^7$Li$,t)^{16}$O reaction, but encountered similar complications in the interpretation of the data. 

Just as in the capture data, the $E_x$ = 7.12 MeV state appears as a subthreshold state in $\alpha$ scattering on $^{12}$C. While the Rutherford cross section masks the low energy subthreshold compound nucleus contributions, it was realized that there should be a measurable effect even at higher energies if $\theta^2_\alpha$ was large enough. A detailed scattering measurement was performed at Australia National University by \textcite{Clark1968481} and the effect of the subthrehold state was subsequently analyzed by \textcite{Clark1969} using a multilevel $R$-matrix analysis. While, the effect of the subthreshold state was shown to contribute significantly to the scattering cross section, the uncertainty in the extracted phase shifts, and the need for a large background pole in the $R$-matrix analysis, resulted in a large uncertainty in $\theta^2_\alpha$(7.12 MeV) of 0.71$^{+0.37}_{-0.18}$. Further, the results differed greatly from those of the transfer measurements.

Another compound nucleus reaction that can populate the $E_x$~=~7.12 and $E_x$~=~9.59 MeV states is $\beta$ delayed $\alpha$ emission from $^{16}$N. This decay almost exclusively populates the $E_x$~=~9.59 MeV state through an allowed Gamow-Teller transition, but should also weakly populate the high energy tail of the $E_x$~=~7.12 MeV subthreshold state just as in the $E1$ component of the capture reaction. This reaction had already been investigated carefully in order to observe the weak parity forbidden decay to the 2$^-$ state in $^{16}$O at $E_x$~=~8.87~MeV by \textcite{PhysRevLett.25.941, Hattig1969144} at the Max-Plank-Institute in Mainz. \textcite{PhysRevC.4.1591} subsequently analyzed the spectrum using an $R$-matrix fit and showed that the data were sensitive to contributions from the subthreshold state constraining $\theta^2_\alpha$(7.12 MeV). In a similar manner as the scattering data, it was found that unconstrained contributions from background states resulted in a large uncertainty. However, the range of 0.013~$< \theta^2_\alpha <$~0.105 was found to be in good agreement with values determined from the transfer reaction data, but in disagreement with those of the scattering. The data sets resulting from these measurements were never published and only a subset of the data have survived, they are commonly referred to in the literature as the ``W\"{a}ffler data''.

As a culmination of these early measurements, \textcite{Barker1971} performed the first comprehensive $R$-matrix analysis by fitting iteratively the scattering phase shifts of \textcite{Clark1968481}, \textcite{MillerJones19621} and \textcite{Morris196897}, the $\beta$ delayed $\alpha$ data of \textcite{PhysRevLett.25.941}, and the capture cross section data of \textcite{PhysRevC.2.2452}. The main goal was to re-analyze all of the data within a self consistent analysis in an effort to resolve the inconsistent determinations of $\theta^2_\alpha$(7.12 MeV). \textcite{Barker1971} found that the large value deduced by \textcite{Clark1969} was in error because of the invalid approximation of using a single level $R$-matrix and an improper treatment of the boundary conditions. The analysis found that in fact a general consistency could be obtained for the value of $\theta^2_\alpha$(7.12 MeV) (see Table~\ref{tab:drwa}). The uncertainty estimate resulted in a range of the extrapolated capture $S$-factor at $E_{c.m.}$~=~300~keV of 50~keV~b~$<$~$S$(300~keV)~$<$~330~keV~b, with a best fit value of $S$(300~keV)~=~150~keV~b. It should be noted that this uncertainty band includes both interference solutions for the low energy $E1$ capture cross section, which the data could not differentiate between. A similar analysis was soon performed by \textcite{Weisser1974460}, which included the much more accurate capture data of \textcite{Dyer1974495}. A similar best fit value of $S$(300~keV)~=~170~keV~b was obtained. Although model uncertainties were investigated throughly, no over all uncertainty range was given.

\begin{table*}
\caption{Summary of subthreshold state reduced $\alpha$ widths (prior to the convention of using asymptotic normalization coefficients). Here $E1$ and $E2$ refer to the ground state transition. Note that the reduced widths are radius dependent, which has caused some confusion in the past. Values of $r_\alpha\approx$~5.5~fm are typical. \label{tab:drwa}}
\begin{ruledtabular}
\begin{tabular}{ c c c c c } 
Ref. & $\theta^2_{\alpha, 6.92}$ & $\theta^2_{\alpha,7.12}$ & $\theta^2_{\alpha,7.12}$/$\theta^2_{\alpha,9.59}$ & data considered \\
\hline
\textcite{Loebenstein1967481} & 0.15 - 0.27 & 0.06 - 0.14 & 0.07 - 0.16\footnotemark[2] & $^6$Li$(^{12}$C$,d)^{16}$O \\ 
\textcite{Clark1969} & & 0.71\footnotemark[1] & & $^{12}$C$(\alpha,\alpha)^{12}$C \\
\textcite{Puhlhofer1970258} & 0.18 & 0.025 & & $^7$Li$(^{12}$C$,t)^{16}$O \\
\textcite{PhysRevC.4.1591} & & 0.013 - 0.105 & & $^{16}$N$(\beta\alpha)^{12}$C \\
\textcite{Barker1971} & & 0.047 - 0.176 & & $^{12}$C$(\alpha,\gamma)^{16}$O ($E1$), $^{12}$C$(\alpha,\alpha)^{12}$C, $^{16}$N$(\beta\alpha)^{12}$C \\
\textcite{Weisser1974460} & & 0.11 & & $^{12}$C$(\alpha,\gamma)^{16}$O ($E1$), $^{12}$C$(\alpha,\alpha)^{12}$C \\
\textcite{Koonin1974221} & & 0.18$^{+0.14}_{-0.10}$ & 0.19$^{+0.16}_{-0.11}$ & $^{12}$C$(\alpha,\gamma)^{16}$O ($E1$), $^{12}$C$(\alpha,\alpha)^{12}$C \\
\textcite{PhysRevC.14.491} & & & 0.1-0.2\footnotemark[2] & $^{12}$C$(^7$Li$,t)^{16}$O \\
\textcite{Becchetti1978293} & & & 0.35(13) & $^{12}$C$(^7$Li$,t)^{16}$O \\
\textcite{Becchetti1978313} & & & $\sim$0.4\footnotemark[2] & $^{12}$C$(^6$Li$,d)^{16}$O \\
\textcite{Becchetti1980336} & & & 0.3-0.6 & $^{12}$C$(^6$Li$,d)^{16}$O \\
\textcite{springerlink:10.1007/BF01415851} & 1.0$^{+0.4}_{-0.3}$ & 0.19$^{+0.14}_{-0.08}$ & & $^{12}$C$(\alpha,\gamma)^{16}$O ($E1$\&$E2$ \& 6.92), $^{12}$C$(\alpha,\alpha)^{12}$C \\
\textcite{Descouvemont1984426} & 0.10(2) & 0.09(2) & & $^{12}$C$(\alpha,\gamma)^{16}$O ($E1$, $E2$ \& 6.92) \\
\textcite{Langanke1985384} & $\approx$0.17 & & & $^{12}$C$(\alpha,\gamma)^{16}$O ($E1$, Total, $\sigma_{E2}/\sigma_{E1}$, 6.92) \\
\multirow{ 2}{*}{\textcite{BarkerKajino1991}} &  \multirow{ 2}{*}{0.730} & \multirow{ 2}{*}{0.114} & \multirow{ 2}{*}{0.14} & $^{12}$C$(\alpha,\gamma)^{16}$O ($E1$, $E2$, 6.92, 7.12) \\
& & & & $^{12}$C$(\alpha,\alpha)^{12}$C, $^{16}$N$(\beta\alpha)^{12}$C \\

\end{tabular}
\end{ruledtabular}
\footnotetext[1]{Corrected by \textcite{Barker1971} to 0.11.}
\footnotetext[2]{Recalculated by \textcite{Essays_in_Nuc_Astro}.}
\end{table*} 

\begin{figure}
\includegraphics[width=1.0\columnwidth]{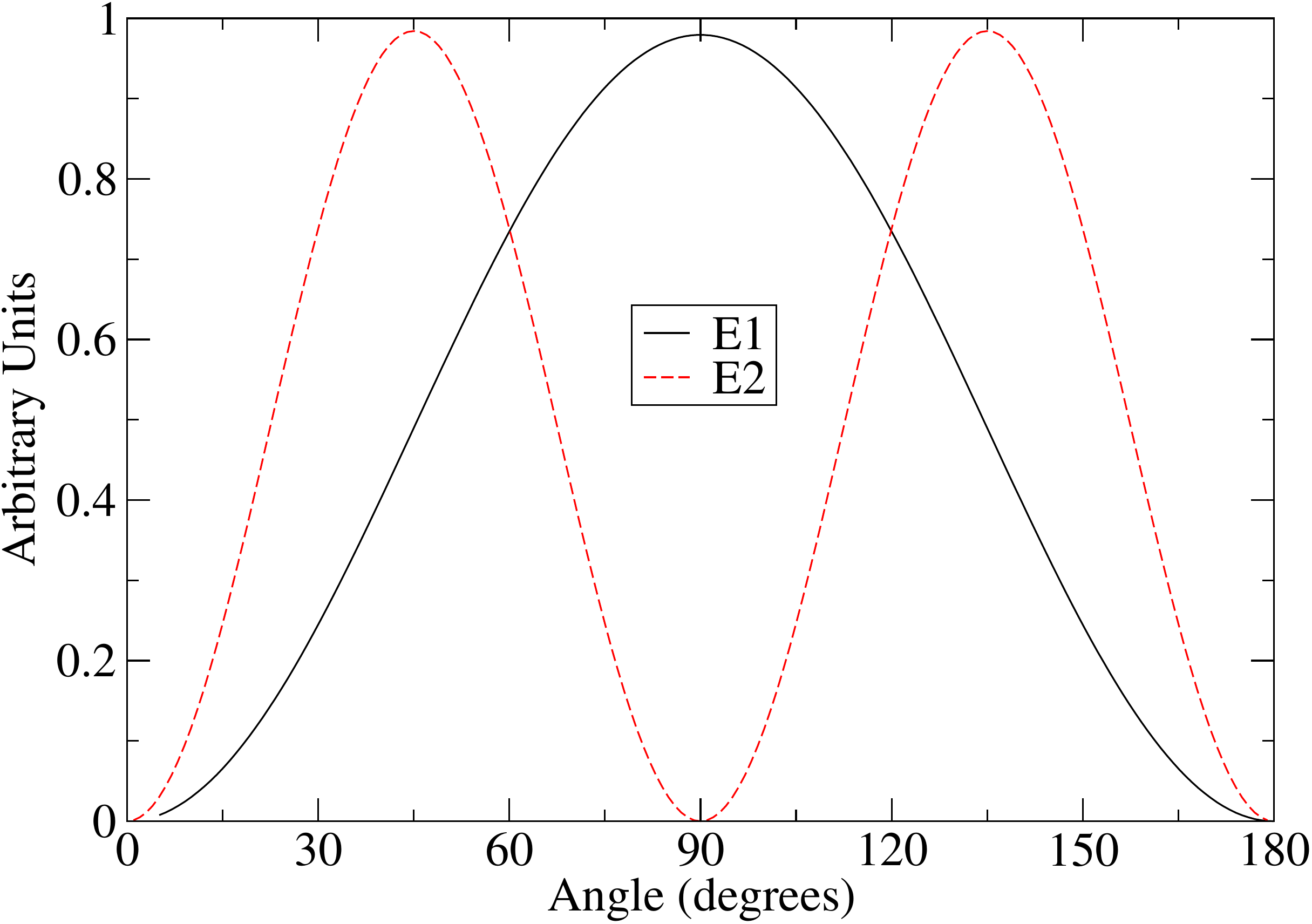}
\caption{(Color online) Theoretical calculation comparing the $E1$ and $E2$ angular distributions for the ground state transition of the $^{12}$C$(\alpha,\gamma_0)^{16}$O reaction. \label{fig:E1_E2_theory}}
\end{figure}

At long last, the first accurate low energies measurement of the $^{12}$C$(\alpha,\gamma)^{16}$O ground state cross section was made by \textcite{Dyer1974495} at Cal Tech. The experiment was also notable because it was the first observation of interference between a subthreshold and unbound 1$^-$ state, a phenomenon predicted several years before by \textcite{1957ApJ...125..221M}. The experiment utilized a target very similar to that of \textcite{PhysRevC.2.63}, and a clean target chamber setup. Only $E1$ and $E2$ multipolarities are allowed from the decays of the 1$^-$ and 2$^+$ excited states to the 0$^+$ ground state of $^{16}$O, therefore these multipolarities are expected to dominate the cross section. To simplify the interpretation of the data, measurements were made primarily at 90$^\circ$ to the beam axis because the $E1$ and $E2$ angular distributions are such that the $E1$ cross section is both maximum and the $E2$ cross section is zero at this angle as shown by Fig.~\ref{fig:E1_E2_theory}. The $E1$ cross section can be written in the simple form as \cite{Dyer1974495} 
\begin{equation} \label{eq:E1_simple}
\sigma_{E1} = 4\pi\left(\frac{2}{3}\right)\left(\frac{d\sigma}{d\Omega}\right)_{90^\circ}.
\end{equation}
Angular distributions were also measured for the first time over this low energy region. The angular distribution data are critical in extracting the $E2$ cross section (as described in Sec.~\ref{sec:gs_12Cag}) since, as shown in Fig.~\ref{fig:E1_E2_theory}, there is no angle where the $E2$ cross section can be isolated. A detailed discussion of the different contributions to the $E1$ cross section is given, noting in particular the large uncertainty that is found from the interferences of higher energy states, modeled using a single background pole, with two explicitly defined $E_x$~=~7.12~MeV subthreshold state and the $E_x$~=~9.59~MeV unbound state.

It was found that $\theta^2_\alpha$(7.12 MeV) was not well constrained by the capture data, strongly reinforcing the motivation for indirect studies. The technique of a three level $R$-matrix calculation to model the $E1$ capture used by \textcite{Dyer1974495} would become the standard for many years to follow. The direct capture contribution (dominated by $E2$ multipolarity) was modeled using the formalism found in e.g. \textcite{PhysRev.122.224} and \textcite{PhysRev.131.2582}, which would later be expanded by \textcite{BarkerKajino1991} into the external capture model. In addition, a hybrid $R$-matrix model was also investigated following the work of \textcite{Koonin1974221}. In this case, the potential model gives the contribution from single particle states, including those at higher energies, which could drastically decrease the uncertainty in the extrapolation of the cross section, but this depends again on the reliability of the potential model and its likewise phenomenologically determined parameters. With limited angular distribution data, it was incorrectly assumed that the small $E2$ component observed in the angular distributions came from direct capture. No contribution from the 2$^+$ subthreshold state at $E_x$ = 6.92 MeV was considered.

The significant progress in studying the level structure and compound nucleus cross sections of $^{16}$O was not limited to these very low energies. Continuing the work of \textcite{Larson1964497}, the $^{12}$C$(\alpha,\gamma)^{16}$O cross section was investigated above the proton separation energy $S_p$~=~12.13 MeV by \textcite{Mitchell1964529}, \textcite{Kernel1971352}, and \textcite{Brochard1973}. These studies were complemented by other $^{12}$C+$\alpha$ reaction measurements by \textcite{Mitchell1965553} and \textcite{Morris196897}. The motivation for most of these studies was to understand the increasingly complicated level structure of $^{16}$O at these higher energies. However, it was also realized that these measurements provided an indirect way of studying the $^{15}$N$(p,\gamma)^{16}$O and $^{15}$N$(p,\alpha)^{12}$C reactions, which \cite{PhysRev.55.434} had pointed out as being of great interest for nucleosynthesis as they form the branch point of the CNO cycle. A study of the $^{15}$N$(p,\gamma)^{16}$O reaction had been made by \textcite{Hebbard1960289} and several measurements of the $^{15}$N$(p,\alpha_0)^{12}$C and $^{15}$N$(p,\alpha_1)^{12}$C reactions by \textcite{doi:10.1139/p52-068}, \textcite{PhysRev.86.527}, \textcite{doi:10.1080/14786441108520389}, \textcite{Neilson1953}, \textcite{PhysRev.108.1015} and \textcite{PhysRev.114.1543}. \textcite{PhysRev.108.735} and \textcite{PhysRev.114.1543} had also studied the proton scattering cross section. These measurements confirmed that the $^{12}$C$(\alpha,\gamma)^{16}$O cross section over the energy range just above $S_p$ is dominated by two broad 1$^-$ resonances ($E_x$~=~12.45 and 13.09~MeV) and angular distributions hinted at a weak contribution from a broad 2$^+$ state ($E_x$~=~12.95 MeV). While these resonances are a few MeV above the astrophysical energy range of interest for the $^{12}$C$(\alpha,\gamma)^{16}$O reaction, their large widths can produce significant interference effects that impact the cross section even at low energies.

\begin{table*}
\caption{Extrapolations of the $^{12}$C$(\alpha,\gamma)^{16}$O $S$-factor to $E_\text{c.m.}$ = 300 keV categorized by either cluster model calculations are phenomenological fits. The abbreviations used below are for the generalized coordinate method (GCM) and potential model (PM) for the theoretical works and Breit-Wigner (BW), $R$-matrix ($R$), and $K$-matrix ($K$) for the phenomenological calculations. Hybrid $R$-matrix (H$R$) models have also been used in an effort to connect the phenomenological calculations more closely to more fundamental theory. \label{tab:S-extrap_history}}
\begin{ruledtabular}
\begin{tabular}{ c c c c c c } 
& \multicolumn{5}{c}{$S$(300 keV) keV~b} \\ \cline{2-6}
Ref. & $E1$ & $E2$ & Cascades & Total & Model\\
\hline
\textbf{Cluster Models} & & & & & \\
\textcite{Descouvemont1984426} & 300 & 90 & & & GCM\\
\textcite{Langanke1985384} & 160-280 & 70 & $<$10\footnotemark[3] & 230-350 & H$R$\&PM \\
\textcite{FUNCK198511} & & 100 & & & PM \\
\textcite{Redder1987385} & 140$^{+120}_{-80}$ & 80$\pm$25 & 7$\pm$3\footnotemark[3] 1.3$^{+0.5}_{-1.0}$\footnotemark[4] & & $R$\&PM \\
\textcite{PhysRevC.36.1249} & 160 & 70 & & & GCM\\ 
\textcite{PhysRevLett.69.1896} & 1$^{+6}_{-1}$ & 40$\pm$7 & & & $R$\&PM \\
\textcite{PhysRevC.47.210} & & 90 & & & GCM \\
\textcite{PhysRevC.54.1982} & 79$\pm$16 & 36$\pm$6 & & 120$\pm$40 & $R$,$K$,PM \\
\textcite{PhysRevC.78.015808} & & 42$\pm$2 & & & GCM \\
\textcite{0004-637X-745-2-192} & $\approx3$ & 150$^{+41}_{-17}$ & 18.0$\pm$4.5\footnotemark[5] & 171$^{+46}_{-22}$ & PM \\
\textcite{Xu201361} (NACRE2) & 80$\pm$18 & 61$\pm$19 & 6.5$^{+4.7}_{-2.2}$\footnotemark[5] & 148$\pm$27 & PM \\
\hline
\textbf{Phenomenological Fits} & & & & & \\
\textcite{RevModPhys.29.547} & 340 & & & 340 & BW \\
\textcite{Barker1971} & 50-330 & & & 50-330 & $R$ \\
\textcite{Koonin1974221} & 80$^{+50}_{-40}$ & & & 80$^{+50}_{-40}$ & H$R$ \\
\textcite{Dyer1974495} & 140$^{+140}_{-40}$ & & & 140$^{+140}_{-40}$ & $R$\&H$R$ \\
\textcite{Weisser1974460} & 170 & & & 170 & $R$ \\
\textcite{Humblet1976210} & 80$^{+140}_{-70}$ & & & 80$^{+140}_{-70}$ & $K$ \\
\textcite{springerlink:10.1007/BF01415851} & 250 & 180 & 12(2)\footnotemark[3]$^,$\footnotemark[4] & 420$^{+160}_{-120}$ & BW \\
\textcite{Langanke1983334} & 150 or 340 & $<$4\% of $E1$ & & 150 or 340 & H$R$ \\
\textcite{Barker1987} & 150$^{+140}_{-60}$ & 30$^{+50}_{-30}$ & & & $R$ \\
\textcite{PhysRevLett.60.1475} & 0-140 &  & & & $R$\&H$R$ \\
\textcite{PhysRevC.40.515} & 0-170 & 5-28 & & 0-170 & $K$ \\
\textcite{BarkerKajino1991} & 150$^{+170}_{-70}$ or 260$^{+140}_{-160}$ & 120$^{+60}_{-70}$ & 10\footnotemark[3] 1-2\footnotemark[4] & 280$^{+230}_{-140}$ or 390$^{+200}_{-230}$ & $R$ \\
\textcite{PhysRevC.44.2530} & 43$^{+20}_{-16}$ & 7$^{+24}_{-5}$ & & 50$^{+30}_{-20}$ & $K$ \\
\textcite{PhysRevC.48.2114} & 45$^{+5}_{-6}$ & & & & $K$ \\
\textcite{PhysRevC.50.1194} & 79$\pm$21 or 82$\pm$26 & & & & $R$\&$K$ \\
\textcite{PhysRevC.54.393} & 79$\pm$21 & 70$\pm$70 & 16$\pm$16\footnotemark[1]$^,$\footnotemark[3]$^,$\footnotemark[4]&  165$\pm$75 & $R$\&$K$ \\
\textcite{Hale1997177} & 20 & & & & $R$ \\
\textcite{Trautvetter1997161} & 79 & 14.5 & & & BW \\
\textcite{PhysRevLett.83.4025} & 101$\pm$17 & 42$^{+16}_{-23}$ & & & $R$ \\
\textcite{Roters1999} & 79$\pm$21 & & & & $R$ \\
\textcite{PhysRevC.61.064611} & & 190-220 & & & $R$ \\
\textcite{Gialanella2001} & 82$\pm$16 or 2.4$\pm$1.0 & & & & $R$ \\
\textcite{PhysRevLett.86.3244} & 76$\pm$20 & 85$\pm$30 & 4$\pm$4\footnotemark[5] & 165$\pm$50 & $R$ \\
\textcite{PhysRevLett.88.072501} & & 53$^{+13}_{-18}$ & & & $R$ \\
\textcite{Hammer2005514} & 77$\pm$17 & 81$\pm$22 & & 162$\pm$39 & $R$ \\
\textcite{Buchmann2006254} & & & 5$^{+7}_{-4.5}$\footnotemark[1] 7$^{+13}_{-4}$\footnotemark[3] & & $R$ \\
\textcite{PhysRevLett.97.242503} & & & 25$^{+16}_{-15}$\footnotemark[1] & & $R$ \\
\textcite{PhysRevC.78.065801} & & & 7.1$\pm$1.6\footnotemark[3] & & $R$ \\
\textcite{PhysRevC.81.045809} & 86$\pm$22 & & & & $R$ \\
\textcite{Schurmann2011557} & & & $<$1\footnotemark[1] & & $R$ \\
\textcite{Schurmann201235} & 83.4 & 73.4 & 4.4\footnotemark[5] & 161$\pm$19$_\text{(stat)}$$^{+8}_{-2}$$_\text{(syst)}$ & $R$ \\
\textcite{PhysRevC.85.035804} & 100$\pm$28 & 50$\pm$19 & & 175$^{+63}_{-62}$ & $R$ \\
\textcite{PhysRevLett.109.142501} & & 62$^{+9}_{-6}$ & & & $R$ \\
\multirow{ 2}{*}{\textcite{PhysRevLett.114.071101}} & & & 1.96$\pm$0.30 or 4.36$\pm$0.45\footnotemark[1]  & & \multirow{ 2}{*}{$R$} \\
 & & & 0.12$\pm$0.04 or 1.44$\pm$0.12\footnotemark[2] & & \\
\textcite{PhysRevC.92.045802} & 98.0$\pm$7.0 & 56$\pm$4.1 & 8.7$\pm$1.8\footnotemark[5] & 162.7$\pm$7.3 & $R$ \\
this work & 86.3 & 45.3 & 7\footnotemark[5] & 140$\pm$21$_\text{(MC)}$$^{+18}_{-11 \text{(model)}}$ & $R$ \\
\end{tabular}
\end{ruledtabular}
\footnotetext[1]{6.05 MeV transition}
\footnotetext[2]{6.13 MeV transition}
\footnotetext[3]{6.92 MeV transition}
\footnotetext[4]{7.12 MeV transition}
\footnotetext[5]{sum of all cascade transitions}
\end{table*} 

\subsection{\label{sec:low_energy_race} The Push to Lower Energies ('74-'93)}

Despite the very successful work by \textcite{Dyer1974495}, their hard won success marks the beginning of a gap in capture cross section measurements of nearly a decade. When they were finally picked up again in the early 1980's, a more focused set of experiments would emerge, with the primary goal of pushing the cross section data to ever lower energies. 

In an attempt to avoid the large background problems from the $^{13}$C$(\alpha,n)^{16}$O reaction that plagued earlier measurements, the experiment of \textcite{springerlink:10.1007/BF01415851} (Claus Rolfs' M\"{u}nster group, experiment performed at Bochum) was performed for the first time in inverse kinematics using a high intensity $^{12}$C beam (50 $\mu$A) on a windowless extended helium gas target. The gas target allowed for the use of higher beam intensities, since the destruction of the target was no longer an issue, and avoided the issue of carbon build up on the target. However, a new set of complications presented themselves that were largely related to interpretation of the yield from the extended geometry gas target. The end result was a measurement of the low energy total ground state capture cross section down to $E_{c.m.}$ = 1.34 MeV, about 0.1 MeV below the measurements of \textcite{Dyer1974495} but not as low as \textcite{PhysRevC.2.63}. It was found that the cross section at low energies was in better agreement with that of \textcite{PhysRevC.2.63} than that of \textcite{Dyer1974495}, but the comparison was not straightforward since the data of \textcite{Dyer1974495} only represents the $E1$ component while that of \textcite{springerlink:10.1007/BF01415851} represented an angle integrated cross section. The most notable result of the experiment was the realization that the ground state $E2$ cross section could make a sizable contribution to the low energy cross section through the high energy tale of the 2$^+$ subthreshold state at $E_x$~=~6.92~MeV. Perhaps even with an amplitude equal to that of the $E1$ cross section at $E_\text{c.m.}$~=~300~keV. 

Because of the successful suppression of the background from the $^{13}$C$(\alpha,n)^{16}$O reaction, \textcite{springerlink:10.1007/BF01415851} achieved another first, measurements of the cascade transitions at low energies. An excitation curve of the $E_x$~=~6.92~MeV transition is given, where it is assumed that the $E_x$~=~6.92~MeV transition dominates over the $E_x$~=~7.12 MeV transition (the two individual $\gamma$ lines could not be resolved in the NaI detectors used). It was found later that this was not a good assumption as both transitions have comparable cross sections over the reported energy range \textcite{Redder1987385}. This highlights a continued issue with the observation of the individual cascade transitions. The $E_x$~=~6.05 and 6.13~MeV first and second excited states in $^{16}$O are only 80 keV apart and the $E_x$~=~6.92 and 7.12~MeV third and fourth excited states are only separated by 100 keV. Thus either a detector with high $\gamma$-ray energy resolution or a detailed unfolding simulation is required in order to extract their individual contributions.

Measurements were then made by \textcite{Redder1987385} (again Rolfs' M\"{u}nster group, this time at Suttgart), but now switching back to forward kinematics. Major experimental improvements included implanted targets and the first use of high energy resolution germanium detectors Ge(Li). The targets were made by implanting $^{12}$C ions into a gold backing, drastically decreasing the amount of $^{13}$C contamination. The gold backing was further soldered onto a copper back plate that allowed for better cooling, through its enhanced thermal conductivity. This was combined with flowed water cooling. These targets were estimated to now be depleted in $^{13}$C by two orders of magnitude (earlier experiments were about one order of magnitude). Up until this point all previous experiments had been performed with NaI detectors, but the improved low level of neutrons allowed for the use of Ge(Li) detectors. The improved energy resolution over NaI detectors allowed for better separation of background peaks and for the $\gamma$ ray secondary peaks of the $E_x$~=~6.92 and~7.12 MeV cascade transitions to be clearly distinguished. The measurements were made over the course of three different experimental campaigns using different accelerators and different arrangements of NaI and Ge(Li) detectors. Angular distributions were measured at an unprecedented eight angles from $E_{c.m.}$~=~1.7 to 2.84~MeV. The lowest energy measurements were extended to $E_{c.m.}$ = 0.94 MeV, now the record for the lowest energy, with the cross section at a minuscule 48 picobarns. Even at the lowest energies, angular distributions were measured, but with only three Ge(Li) detectors. Most significantly, the angular distribution measurements showed a substantial $E2$ component to the cross section, confirming that this multipolarity is quite significant at stellar energies. Again, the energy dependence of the cross section at low energies was found to be higher than that of \textcite{Dyer1974495}.

The improved angular distribution measurements by \textcite{Redder1987385} provided more sensitivity to the $E2$ ground state cross section triggered several theoretical calculations to model this previously neglected component of the cross section. For the first time \textcite{Descouvemont1984426} made use  of a microscopic model using the generator coordinate method. \textcite{Langanke1983334} updated their hybrid $R$-matrix calculations taking into account the new capture data and then refining the calculations again in \textcite{Langanke1985384}, correcting some previous errors. Now including the $E2$ cross section, \textcite{Barker1987} updated his calculations as well, using purely $R$-matrix calculations for both the $E1$ and $E2$ cross sections. Further, several calculations were made for the $E_x$~=~6.92 MeV cascade contribution. All agreed that its contribution to the total capture $S$-factor should be small ($<$15 keV b) at stellar energies. The general result was that the $^{12}$C$(\alpha,\gamma)^{16}$O cross section at stellar energies should be significantly larger, 2-5 times of the value estimated by \textcite{Dyer1974495}, but the recommended values varied widely as summarized in Table~\ref{tab:S-extrap_history}. History gives us a valuable lesson here. While several experiments were in apparent contradiction to \textcite{Dyer1974495}, later measurements would find that these measurements were in fact erroneously large, perhaps the result of insufficient background subtraction.

The Cal Tech \cite{PhysRevLett.60.1475} group now looked to re-investigate the $^{12}$C$(\alpha,\gamma)^{16}$O reaction but this time using another novel technique for the first time: a recoil separator. By detecting the $\gamma$ rays in coincidence with the recoiling $^{16}$O, nearly background free spectra could be obtained. Because of the acceptance of the CTAG separator, the 90$^\circ$ placement of the NaI detectors, and the much different angular distributions of $E1$ and $E2$ radiation (see Fig.~\ref{fig:E1_E2_theory}), the efficiency for detecting the $E2$ component was only 50\%-65\% of that for the $E1$ component. Theoretical values of the $E1$/$E2$ cross section ratio from \textcite{Langanke1983334} and \textcite{Langanke1985384} were necessary to extract the $E1$ cross section over the range from $E_{c.m.}$ = 1.29 to 3.00 MeV so the results were somewhat theory dependent. The $E1$ cross section was found to be in good agreement with that of \textcite{Dyer1974495}, reinforcing the tension between the different measurements.

To try to resolve these differences, a new measurement was then performed by the Queen's University group. The data was first reported in \textcite{PhysRevLett.69.1896} but results were subsequently revised in \textcite{PhysRevC.54.1982}. The experiment was performed in forward kinematics using water cooled implanted targets very similar to those of \textcite{Redder1987385}.  The beam was also wobbled over the target surface to insure even beam coverage lessening the sensitivity of the experiment to any target inhomogeneity. With the low neutron background, six germanium detectors were used to measure angular distributions over an energy range from $E_\text{c.m.}$~=~1.37 to 2.98~MeV. The data turned out to split the difference between those of \textcite{Redder1987385} and \textcite{PhysRevLett.60.1475}, providing no solution to the issue. In the analysis of \textcite{PhysRevC.54.1982}, the $E1$ data, along with that of \textcite{Dyer1974495}, \textcite{Redder1987385}, and \textcite{PhysRevLett.60.1475} were fit simultaneously using a three level $R$-matrix fit along with the newly measured $^{16}$N$(\beta\alpha)^{12}$C data of \textcite{PhysRevC.50.1194} and the elastic scattering phase shift data of \textcite{Plaga1987291}. The $E2$ data was fit separately using a cluster model method. It should be noted that in \textcite{PhysRevLett.69.1896}, the destructive solution between the 1$^-$ subthreshold state ($E_x$~=~7.12 MeV) and the broad resonance corresponding to the 1$^-$ level at $E_x$ = 9.59 MeV was reported to produce the best $\chi^2$ fit. However, in the revised analysis of \textcite{PhysRevC.54.1982}, the authors concluded that, considering all the capture data, the constructive solution was in fact favored and the destructive one was statistically ruled out.

While it has been neglected in all analyses, a high energy measurement of the ground state transition cross section, from 12~$<E_x<$~28~MeV, was made at this time by \textcite{PhysRevLett.32.1061}. The cross section data were obtained in order to measure the $E2$ strength of the giant dipole resonance and hence were decomposed into $E1$ and $E2$ components. While the measurement was made for purely structure motivations, this data could provide valuable upper limits on the high energy background contributions of phenomenological $R$-matrix fits at lower energies for the dominant ground state transition. This topic will be revised in Sec.~\ref{sec:systematic_more}.

Several measurements continued to study the properties of the $^{16}$O compound nucleus in the energy region just above $S_p$. Several measurements were made to continue in the investigation of the astrophysically important CNO branch point reactions $^{15}$N$(p,\alpha)^{12}$C \cite{Bray1977334, Zyskind1979404, springerlink:10.1007/BF01419081, Pepper1976163} and $^{15}$N$(p,\gamma)^{16}$O \cite{Rolfs1974450}. Improved measurements of $^{12}$C+$\alpha$ scattering were also made by \textcite{dagostino}.   

\subsection{\label{sec:modern_era} Return to Indirect Techniques ('93-present)}

The last 20 years has witnessed a continued, and even increasingly, intense effort to study the $^{12}$C$(\alpha,\gamma)^{16}$O reaction. With the seeming impasse of a capture experiment reaching the stellar energy range, there has been a renewed interest in indirect techniques. While many direct measurements continue to be made, the development of improved theoretical and experimental methods for interpreting transfer reaction data, continued development of more accurate $^{16}$N$(\beta\alpha)^{12}$C measurements, improved recoil separators, and more sophisticated analyses have arguably produced the greatest impact.

\textcite{Plaga1987291} (Rolfs' group) measured the scattering cross section at 35 angles covering a wide angular range from $\theta_{\text{lab}}$ = 22$^\circ$ to 163$^\circ$ at 51 energies between $E_{c.m}$ = 0.75 and 5.0 MeV. Phase shifts were extracted for angular momentum $l$ = 0 to 6 using a multilevel $R$-matrix fit. One of the main findings was that the reduced $\alpha$ widths were highly correlated with the background pole parameters, a problem observed before, resulting in large model uncertainties. This issue seems to be a limiting factor in the determination of subthreshold reduced widths from scattering data in general.

A major step forward was the measurements of the $\alpha$-particle energy spectrum from $^{16}$N$(\beta\alpha)^{12}$C in the 1990s. The first results were reported from an experiment performed at  TRIUMF by \textcite{PhysRevLett.70.726}, with a more complete description given in \textcite{PhysRevC.50.1194}. The $^{16}$N$(\beta\alpha)^{12}$C spectrum was concurrently measured by the Yale group, \textcite{PhysRevLett.70.2066}. These measurements were subsequently extended by \textcite{FranceIII1997165}. Another $^{16}$N$(\beta\alpha)^{12}$C measurement was performed by the Seattle group shortly after but the results were not published. The spectrum can be found in the later work of \textcite{PhysRevC.75.065802}.

These measurements were highly motivated by the theoretical calculations of \textcite{Baye1988445}, \textcite{PhysRevC.41.1736} and \textcite{PhysRevC.44.2530} that predicted a characteristic interference pattern at low energies. This is the result of the interference between the 1$^-$ levels at $E_x$~=~7.12 and $E_x$~=~9.59 MeV. It is very sensitive to the relative values of the reduced $\alpha$ widths and the $\beta$ decay branching ratios of the two states. However, the interpretation of the spectrum is complicated by the presence of an $l$~=~3 component coming from the 3$^-$ subthreshold state at $E_x$~=~6.13~MeV. Never the less, the interference pattern was in fact observed and would mark a drastic improvement in the constraint of the $E1$ ground state capture cross section.

The $\beta$-delayed $\alpha$-particle spectrum provided for a high level of constraint on the reduced $\alpha$ width of the 1$^-$ subthreshold state. A detailed global analysis was presented by \textcite{PhysRevC.50.1194} where the TRIUMF $^{16}$N$(\beta\alpha)^{12}$C data were fit simultaneously with the scattering phase shifts of \textcite{Plaga1987291} and the $E1$ capture data of \textcite{Dyer1974495}, \textcite{Redder1987385}, \textcite{PhysRevLett.60.1475}, and \textcite{PhysRevLett.69.1896}. It was found that the $^{16}$N$(\beta\alpha)^{12}$C data significantly improved the constraint on the $E1$ cross section, by way of the 1$^-$ subthreshold state's ANC, over the capture data. It was noted however that the general shape of the capture data were still critical because only they can determine the interference pattern between the two 1$^-$ resonances, which greatly influences the low energy cross section.  


A global analysis of the capture, scattering, and $^{16}$N$(\beta\alpha)^{12}$C data was performed by \textcite{PhysRevC.54.393} and several important conclusions were made. One focus of the analysis was to look at biases that had developed because of the historical convention of dividing the ground state cross section into $E1$ and $E2$ cross sections. Two general techniques have been used. The first is to measure the angular distributions and perform a fit to a theoretically motivated angular distribution function. While this technique has been widely used, it also has its pitfalls. One issue is that it can be difficult to measure differential cross sections at several angles given the very low yields. These low yields are often influenced by systematic uncertainties that can be difficult to quantify and can be easily overlooked. Further, the fitting also requires has a phase that can either be left free in the fitting or can be constrained by scattering data. The second method uses a large diameter detector also centered at 90$^\circ$ but placed in very close geometry to the target to measure the angle integrated cross section over approximately 2$\pi$, effectively measuring $\sigma_\text{total}$/2. Then the $E2$ cross section can be deduced as $\sigma_\text{total} - \sigma_{E1} = \sigma_{E2}$. Both of these techniques require assumptions and simulations, allowing more opportunities for errors to be made. For these reasons, \textcite{PhysRevC.54.393} advocated that global analysis should instead rely on ``primary" data, meaning either the actual differential cross sections that were measured or the total cross section for a close geometry setup. Unfortunately, many of the early measurements only reported the deduced $E1$ and $E2$ cross sections, not the differential data. Another important conclusion was that while the $E1$ cross section seemed to be fairly consistent over different measurements, the $E2$ cross section showed large fluctuations. This seems to be because analyses are attempting to extract a small $E2$ contribution over most of the experimentally accessible energy region, i.e., over the broad low energy 1$^-$ resonance. The uncertainties of this process seem to have often been underestimated. 

\textcite{PhysRevC.54.393} also predicted through simulation that high precision scattering measurements could be used to improve the constraint of the fits. In direct response, \textcite{PhysRevLett.88.072501} performed a very detailed measurement of the scattering cross section. Measurements were made from $E_\text{c.m.}$~=~2.0 to 6.1~MeV in energy steps of approximately 10~keV and at 32 angles ranging from $\theta_\text{lab}$ = 24$^\circ$ go 166$^\circ$. The measurement sought to place stronger constraints on both the $\alpha$ widths of the unbound states and the reduced widths of the subthreshold states. In particular the goal was an improved constraint on the reduced width of the 2$^+$ subthreshold state, which is not constrained by the $^{16}$N$(\beta\alpha)^{12}$C reaction. However, because of the issues of background pole contributions in the $R$-matrix analysis, the constraint was not as great as expected. Additionally, because of issues with the target thickness varying due to carbon build up on the target, the data were analyzed as ratios of the yields instead of as absolute cross sections. It was however demonstrated that this still provides significant constraint on the $R$-matrix fit while greatly reducing systematic uncertainties that are difficult to quantify. A more complete description of the experiment and analysis, together with an extraction of the phase shifts, was later given in \textcite{PhysRevC.79.055803}. An $R$-matrix fit including data above $S_p$, which included $^{12}$C$(\alpha,\alpha_1)^{12}$C and $^{12}$C$(\alpha,p)^{15}$N data, was subsequently given in \textcite{deboer2012}.

As noted already, transfer reactions can in provide information about reduced widths. In general, the interpretation of these experiments is subject to uncertainties in the optical potentials and the reaction mechanism (direct transfer versus multi-step processes and/or compound-nuclear fusion). Sub-Coulomb measurements, where the energies in the entrance \textit{and} exit channels are below the Coulomb barrier, provide a powerful way to minimize these uncertainties. For sub-Coulomb kinematics, other reaction mechanisms are supressed relative to direct transfer and and the Coulomb potentials dominate, leading to little dependence on the nuclear parts of the optical potentials. Due to the poximity of the 6.92-MeV $2^+$ state and 7.12-MeV $1^-$ state to the $\alpha$ threshold, these states are ideal for application of the sub-Coulomb $\alpha$ transfer technique. While not quite as ideal, the 6.05-MeV $0^+$ state and 6.13-MeV $3^-$ state are still well-suited for it. For the ${}^{16}{\rm O}$ ground state, it is unfortunately impossible to realize the kinematics required for sub-Coulomb transfer to be applicable, due to the large positive $Q$~value for $\alpha$-transfer reactions to this state. 

\textcite{PhysRevLett.83.4025} performed the first sub-Coulomb $^{12}$C$(^6$Li$,d)^{16}$O and $^{12}$C$(^7$Li$,t)^{16}$O experiments. Further, it was realized that analyzing the transfer cross section to determine the model-independent ANC, rather than the spectroscopic factor, removed un-necessary model dependence from the results. The ANC can be related to the reduced width for a particular channel radius by Eq.~(\ref{eq:ANC_to_rw}). The experiment determined the ANCs of the 1$^-$ and 2$^+$ subtheshold states with greatly reduced uncertainties and an $R$-matrix fit was used to deduce the impact on the capture extrapolation. The result was a greatly reduced uncertainty on the $E2$ cross section and a value for the $E1$ cross section that was roughly consistent with that deduced from the high-precision $^{16}$N$(\beta\alpha)^{12}$C spectrum. 

Another low energy cross section measurement was made by Rolfs' group at Bochum \cite{Roters1999}. The experiment was performed again in inverse kinematics on a helium gas target similar to that used in \textcite{springerlink:10.1007/BF01415851} but also used bismuth germanate detectors for the first time. The BGO detectors were three times as efficient compared to NaI detectors of equal size, allowing a farther geometry measurement with the same statistics for a given beam time, reducing angular resolution effects. The setup was used to measured the $E1$ cross section in far geometry at $\theta_\text{lab}$~=~90$^\circ$ and the angle integrated cross section by placing a larger BGO in close geometry. An $R$-matrix fit was performed and the extrapolation predicted about an equal contribution from $E1$ and $E2$ multipolarities at $E_\text{c.m.}$~=~300~keV.

\textcite{Gialanella2001} performed a measurement similar to that of \cite{Roters1999}. The main result of these work was that a detailed Monte Carlo uncertainty analysis was preformed for the $R$-matrix fit for the first time. The analysis highlighted the systematic differences between the different $E1$ data sets. The main result was, that depending on which low energy data were included in the fit, the destructive $E1$ solution could not be statistically ruled out.

The first in a series of detailed angular distribution studies at Stuttgart was performed by \textcite{PhysRevLett.86.3244}. The measurements covered an energy range from $E_\text{c.m.}$~=~0.95 to 2.8~MeV at 20 energies and was notable because measurements were made at up to nine different angles and used high purity germanium detectors HPGe for the first time. The experiment benefited greatly from high background suppression provided by a BGO array allowing for reasonable statics with less beam time than similar previous setups. Subsequent experiments were performed using the EUROGRAM and GANDI arrays. The EUROGRAM measurements covered an energy range from $E_\text{c.m.}$~=~1.3 to 2.78~MeV and are published in \textcite{PhysRevC.73.055801}. The ``turntable experiment" data is available in full only in the PhD thesis of Michael Fey \cite{Fey2004}, but some details and data are given in \textcite{Hammer2005363, Hammer2005514}. These measurements represent the largest set of angular distribution data currently available, but their limited peer-review publication and apparently underestimated systematic errors \cite{Brune201349} has brought their validity into question.

\textcite{Fleurot2005167} are developing a new indirect approach using Coulomb dissociation of the reaction $^{208}$Pb$(^{16}$O,$^{16}$O*$)^{208}$Pb for the first time. A preliminary experiment was performed at Kernfysisch Versneller Instituut KVI using the big bite spectrometer. The method should be more sensitive to the $E2$ cross section, offering a complementary indirect approach to the $\beta$ delayed $\alpha$ emission measurements. Reminiscent of high energy transfer reaction studies, the reaction mechanism is quite complicated, requiring models for both the nuclear and Coulomb amplitudes of the cross section. The 2$^+$ states at $E_x$~=~9.84 and 11.52~MeV were populated and angular distributions were extracted. However, some of the angular distributions showed large systematic deviations from their expected values at certain angles. The results are encouraging but significant development in the technique and theory is likely required before reliable data can be obtained.


A detailed recoil separator measurement was made at DRAGON \cite{Hutcheon2003190} at the TRIUMF-ISAC facility by \textcite{PhysRevC.78.065801}. The experiment covered a wide energy range from $E_\text{c.m.}$~=~2.22 to 5.42~MeV. This measurement focused on the $E_x$~=~6.05~MeV cascade transition and an $R$-matrix analysis of the data reported that this contribution was much larger (25$^{+15}_{-16}$ keV b) at $E_\text{c.m.}$~=~300~keV than previously estimated. However, the interpretation of the data was later found to be in error and the later measurements of \textcite{Schurmann2011557} and \textcite{PhysRevLett.114.071101} have confirmed a smaller value ($\sim$ 2-5 keV b). In addition to the $E_x$~=~6.05~MeV transition, the total cross section was evaluated but is only available in the thesis of \textcite{Matei_thesis}. Other cascade transition data were observed in the spectra but remain unanalyzed. 

A re-measurement of the $^{16}$N$(\beta\alpha)^{12}$C spectrum was made by \textcite{PhysRevC.75.065802} (Yale group) in an attempt to clarify the inconsistency issues in the different data sets. By convoluting the $R$-matrix fits of the previous data with the experimental resolution functions, it was reasserted that the TRIUMF data \cite{PhysRevC.50.1194} were inconsistent with both their measurement and that of the Seattle measurement. In addition, the data from the previously unpublished experiments at Mainz \cite{Hattig1969144, PhysRevLett.25.941, PhysRevC.10.320} and Seattle were made available, a very valuable service to the community. 

Another measurement of the $^{16}$N$(\beta\alpha)^{12}$C spectrum was made soon after at Argonne National Laboratory by \textcite{PhysRevC.81.045809}. This experiment attempted to lessen the effects of $\beta$ background and contaminant reactions by using the in-flight technique \cite{Harss2000} to create the $^{16}$N beam. To minimize the energy convolution of the spectrum by the catcher, thin carbon foils were used with thicknesses of only 17(2) $\mu$g/cm$^2$. The resulting spectrum is similar to that of \textcite{PhysRevC.50.1194}, but there are some very significant differences as discussed further in Sec.~\ref{sec:beta_delayed_alpha}.   

A low energy measurement of the $^{12}$C$(\alpha,\gamma)^{16}$O reaction was made at the Research Laboratory for Nuclear Reactors at the Tokyo Institute of Technology by \textcite{PhysRevC.80.065802}. The experiment concentrated on a very low energy range, measuring at just two energies of $E_\alpha$~=~2.000 and 2.270~MeV. However, the goal of the experiment was a high accuracy measurement of the $E1$ and $E2$ cross sections at these energies where past experiments had showed considerable disagreement, especially in the $E2$ cross section. This was accomplished by measuring at three critical angles, $\theta_\text{lab}$~=~40, 90, and 130$^\circ$, and in far geometry with small solid angles (as reflected by the $Q$ coefficients) using time-of-flight. Compton suppressed NaI detectors were utilized, targets were obtained by cracking $^{13}$C depleted methane gas. Additionally, the time-of-flight capability facilitated a very detailed study of the different sources of background. These were primarily found to be secondary $(n,\gamma)$ and $(n,n'\gamma)$ reactions induced by neutrons from the $^{13}$C$(\alpha,n)^{16}$O reaction (see also \cite{Makii2005411}). The deduced $\sigma_{E1}$ and $\sigma_{E2}$ cross sections were found with the smallest uncertainties to date in this region. They also showed significantly less scatter than many previous measurements, and in general are somewhat lower in overall cross section. Their energy dependence, albeit with only two data points, is in excellent agreement with previous $R$-matrix fits.    

\textcite{PhysRevLett.109.142501} used a novel method \cite{PhysRevC.64.055803} of determining the $E2$ interferences by measuring the energy integrated differential yield over the narrow low energy 2$^+$ resonance, corresponding to the state at $E_x$~=~9.85~MeV, in the $^{12}$C$(\alpha,\gamma)^{16}$O reaction. The result was that the number of possible interference solutions with this resonance from the 2$^+$ subthreshold and the next higher energy state at $E_x$~=~11.51~MeV could be reduced to two. These interference solutions will be discussed further in Sec.~\ref{sec:interferences}.  

The measurement of \textcite{PhysRevC.86.015805} investigated the low energy cross section of $^{12}$C$(\alpha,\gamma)^{16}$O using a standard forward kinematics setup but using a nearly 4$\pi$ BaF$_2$ detector array for the first time. The BaF$_2$ detectors have the advantage that they are less sensitive to neutrons than HPGe detectors and are more efficient. Their disadvantage is a decreased energy resolution compared to HPGe's. The array is segmented in such a way that angular distributions at twelve angles can be extracted. The angular information was used to separate the $E1$ and $E2$ components using the traditional procedure of fitting to Legendre polynomials. In addition, the measurement also reported the sum of the cascade transition and therefore could give the total capture cross section. 

A dedicated experimental campaign is ongoing at the Kyushu University Tandem Laboratory (KUTL) \cite{Ikeda2003558} to measure the $^{12}$C$(\alpha,\gamma)^{16}$O reaction. The experiment aims at a direct measurement of the capture cross section down to $E_\text{c.m.}$~=~0.7~MeV using an inverse kinematics setup, time-of-flight, and a recoil separator. A windowless helium gas target is used with a pressure of $\approx$25 Torr and beam intensities in excess of 10~p$\mu$A. The experimental development has steadily progressed with an ever improving setup. Total cross section measurements have been made at $E_\text{c.m.}$~=~2.4 and 1.5~MeV, and measurements are ongoing for $E_\text{c.m.}$~=~1.2~MeV \cite{:/content/aip/proceeding/aipcp/10.1063/1.4874074}.

Significantly improved measurements have not been limited to the $^{12}$C$(\alpha,\gamma)^{16}$O reaction. The $^{15}$N$(p,\gamma)^{16}$O reaction has been the subject of several recent measurements at the LUNA facility and the University of Notre Dame's nuclear science laboratory \cite{0954-3899-36-4-045202, PhysRevC.82.055804, Caciolli2011, imbriani2012}. This was highly motivated by new measurements of the bound state proton ANCs in $^{16}$O \cite{PhysRevC.78.015804}, which gave strong evidence that the measurement of \textcite{Rolfs1974450} over estimated the low energy cross section, a common theme. These complementary measurements resulted in a significant improvement in the uncertainty of this reaction at stellar energies (now at the $\approx$5\% level). The $^{15}$N$(p,\alpha)^{12}$C reaction has also been re-investigated using the Trojan Horse method \cite{PhysRevC.76.065804}. Additional proton scattering data have been measured by \textcite{PhysRevC.85.038801}. The work of \textcite{PhysRevC.87.015802} combined the vast majority of the data above $S_p$ and obtained a combined fit for all open reaction channels up to $E_x\approx$14~MeV. Preliminary fits were also made to a very limited set of $^{12}$C$(\alpha,\gamma)^{16}$O data.

The sub-Coulomb transfer reaction experiment of \textcite{PhysRevLett.114.071101} has reconfirmed the earlier measurements of the $\alpha$ ANCs for the levels at $E_x$~=~6.92 (2$^+$) and 7.12 (1$^-$)~MeV and has additionally measured those of the $E_x$~=~6.05 (0$^+$) and 6.13 (3$^-$)~MeV states for the first time. These measurements reconfirmed the assertion of \textcite{Schurmann2011557} that the large value for the low energy $S$-factor of the $E_x$~=~6.05 MeV transition given in \textcite{PhysRevLett.97.242503} was incorrect. However, the value found in \textcite{PhysRevLett.114.071101} is also in disagreement with that of \textcite{Schurmann2011557} since their assumed ANC was significantly smaller than that measured by \textcite{PhysRevLett.114.071101}. These issues are discussed in detail in Sec.~\ref{sec:cascades_12Cag}.


Another study has been recently performed at KVI where the goal was to determine the total $\beta\alpha$ branching ratio \cite{Refsgaard2016296} for $^{16}$N$(\beta\alpha)^{12}$C decay. A value of (1.49~$\pm$~0.05(stat)$^{+0.0}_{-0.10}$(sys))$\times$10$^{-5}$ was obtained, a 24\% increase over the literature value of 1.20(5)$\times$10$^{-5}$. If correct, this could have an effect on the analysis of the $^{16}$N$(\beta\alpha)^{12}$C spectrum. The implications have not yet been fully explored.


Four recent comprehensive analysis of the $^{12}$C$(\alpha,\gamma)^{16}$O reaction are conspicuously absent from this section, those of \textcite{Schurmann201235}, \textcite{PhysRevC.85.035804}, \textcite{Xu201361} (NACRE2), and \textcite{PhysRevC.92.045802}. A review of each of these analyses has been reserved for later in this work, Sec.~\ref{sec:discussion}, so that more detailed comparisons can be made with the present global analysis presented in Sec.~\ref{sec:R-matrix_analysis}.

\subsection{\label{sec:future_exp} Up Coming Experiments}

An experiment long under development is the measurement of the inverse photo-disintegration reaction $^{16}$O$(\gamma_0,\alpha)^{12}$C. While this method has the limitation of only being sensitive to the ground state transition, this is the most important transition for the $^{12}$C$(\alpha,\gamma)^{16}$O reaction at stellar energies. Two independent groups have attempted to tackle the measurement using quite different measurement apparatuses but so far at the same beam facility, the High-Intensity $\gamma$-ray Source (HI$\gamma$S) \cite{Weller2009257}. One setup uses a bubble chamber to detect the recoiling $\alpha$ particles \cite{DiGiovine201596}. The super-heated liquid used in the chamber acts as both target and detection medium. A successful proof of principle experiment has been performed using a C$_4$F$_{10}$ liquid by \cite{Ugalde201374} for the $^{19}$F$(\gamma_0,\alpha)^{15}$N reaction. Continued measurements are now taking place at Jefferson laboratory. 

Another experiment has been proposed using an optical time projection chamber as described by \textcite{1742-6596-337-1-012054}. A successful experiment has been performed to extract cross section data for the $^{12}$C$(\gamma_0,\alpha)^8$Be reaction as reported in \textcite{PhysRevLett.110.152502}. One great advantage of this setup is that it allows for the extraction of very detailed angular distributions. Both types of experiments are limited in their energy resolution by available $\gamma$ ray beams (resolution at HI$\gamma$S $\sim$200 keV at these energies for example). If the experimental techniques can be further developed, inverse measurements may be the best way to probe the low energy cross section since the photo-disintegration cross section is about 50 times larger than the capture cross section. Plans are also underway to perform these kinds of experiments at the upcoming ELI-NP facility \cite{1742-6596-590-1-012005} where a significantly higher $\gamma$ flux will be available.

Several recoil separator measurements are also planned for the $^{12}$C$(\alpha,\gamma)^{16}$O reaction. Further measurements have already begun at TRIUMF's DRAGON facility \cite{Hutcheon2003190}. The European recoil separator for nuclear astrophysics ERNA recoil separator has also been recommissioned at the center for isotopic research on the cultural and environmental heritage CIRCE laboratory in Caserta, Italy. A reinvestigation of the $^{12}$C$(\alpha,\gamma)^{16}$O reaction is planned in order to improve upon the previously successful experimental campaigns \cite{schurmann2005, Schurmann2011557} in Bochum, Germany. In addition, the St. George recoil separator \cite{Couder200835} is currently in the commissioning phase and $^{12}$C$(\alpha,\gamma)^{16}$O is a reaction of primary interest. 

Underground experimental facilities have also yet to weigh in. At LUNA \cite{0034-4885-72-8-086301}, $\alpha$ particle beams have been prohibited in the past due to the risk of creating background signals in other nearby experiments. This ban has been lifted however and measurements of the $^{12}$C$(\alpha,\gamma)^{16}$O reaction are in the planning phase. However, the current LUNA facility can only create helium beams of up to $E_\alpha$~=~400~keV, a very difficult point at which to start the measurements. Therefore the first measurements may have to wait until the instillation of the new higher-energy LUNA MV facility is completed (scheduled for operation in 2019). 

The compact accelerator system for performing astrophysical research CASPAR \cite{Robertson2016} is also nearing completion. This new underground facility located at the Sanford underground research facility in Lead, South Dakota will be the first underground nuclear astrophysics facility in the United States. At present a 1~MV KN accelerator has been installed and commissioning is underway. While the neutron producing reactions $^{13}$C$(\alpha,n)^{16}$O and $^{22}$Ne$(\alpha,n)^{25}$Mg are the planned flagship experiments, the $^{12}$C$(\alpha,\gamma)^{16}$O reaction will certainly be investigated in the future.

The Jinping underground laboratory is also currently under construction in Sichuan, China. The facility will have a high current (pmA) 400~kV accelerator with an ECR source able to accelerate $^4$He$^{++}$ beams up to $E_\alpha$~=~800~keV. One of the flagship experiments for the facility is to measure the $^{12}$C$(\alpha,\gamma)^{16}$O reaction at this energy ($E_\text{c.m.}$~=~600~keV). The current goal is to perform this measurement by the end of 2019 \cite{doi:10.7566/JPSCP.14.011101}.

As a final point, as experimental data are obtained with greater precision and at higher energies, experiments may soon become sensitive to less probable second order reaction channels. Some candidate decay channels are forbidden $\beta$, ($e^-e^+$) $\pi$, internal conversion, and simultaneous multiple $\gamma$ emission. While it is likely that most of these processes remain below current experimental detection thresholds, care should be taken to not forget they are possible. Some of these reactions could be used to further indirectly constrain the capture cross section. For example, detection of the $\pi$ decay of the $E_x$~=~6.05~MeV transition is already being planned at the CIRCE laboratory \cite{Guerro2014, Tabassam2015}.

\subsection{\label{sec:world_data} World Data Set} 

The previous sections described the many experimental endeavors that have provided a wealth of data for the understanding of the $^{12}$C$(\alpha,\gamma)^{16}$O reaction. Fig.~\ref{fig:E1_and_E2_comparisons} shows how the capture data have evolved over time. Because the $E1$ cross section dominates over much of the experimentally accessible region, these measurements are in reasonable agreement (barring some of the earliest measurements). While the agreement between the different $E2$ data sets have certainty improved, there are still large disagreements both between different data sets and with theory. There are two major trends, a large scatter in the data in regions where the cross section should be smooth, and an increase in the cross section at low energy. Both of these phenomena are very common to measurements that push the limits of the experimental techniques employed. What is encouraging to see is that both these problems have lessened with more recent measurements. The trend of the data shows that attention should be made in determining accurate measurements of the $E2$ cross section, since the $E1$ is fairly well established. If the $E2$ data can be improved to a level of consistency similar to the $E1$, this could lead to a significant reduction in the overall uncertainty in the extrapolation to low energy.

On reflecting back over the experimental data, two issues stand out clearly. The first is that of the overall normalization of the data. Future measurements will likely attempt to push to lower energies, yet it is important to remember that new measurements should not be limited to the lowest energy ranges. In particular, it is always useful to have at least one measurement near the maximum of the broad resonance at $E_\text{c.m.}$~=~2.2~MeV. In this way, so long as all the data can be considered to share the same overall systematic uncertainties, the normalization and shape of the data provide significantly better constraint on the extrapolation. The second issue is the splitting of the $E1$ and $E2$ data into separate cross sections. While this makes sense from a theory point of view, it leads to further assumptions in the analysis of the experimental data. This procedure may be responsible for the large scatter of the $E2$ data. It also emphasizes the point that the differential cross section measurements (and even spectra) should be retained, since if this data where available, it would likely provide a better understanding of what caused these problems.

For this review, a global $R$-matrix analysis has been performed in order to facilitate a better comparison of the different capture data sets and combine several recent results that have not yet been considered relative to other data. The method provides a standard framework to interpret the impact and gauge the level of agreement between all the types of measurements, both direct and indirect, on the extrapolation of the capture cross section. An attempt has been made to consider as much of the relevant experimental data as possible. However there are a few cases in which the experimental data are clearly in very poor agreement, and are therefore excluded from the analysis. 

Most of the past data sets are found to be in reasonable agreement. The few that are excluded are the ground state capture data of \textcite{PhysRevC.2.63, PhysRevC.2.2452}, both the ground state and cascade capture data of \textcite{springerlink:10.1007/BF01415851}, the $E2$ data of \textcite{Redder1987385}, and the $E_x$ = 6.05 MeV transition data of \textcite{PhysRevLett.97.242503}. Further, the $\beta$ delayed $\alpha$ data sets of \textcite{PhysRevC.10.320, PhysRevC.75.065802}, that is the Yale, Mainz, and Seattle data, are also excluded because not enough information regarding the target effect corrections have been given to perform a reanalysis of these data sets. This is discussed in more detail in Sec.~\ref{sec:beta_delayed_alpha}. There have been several $\alpha$ scattering experiments that have all been largely consistent only with improved uncertainties. For this reason only the most comprehensive data set of \textcite{PhysRevC.79.055803} is considered here. Further, the data reported in \textcite{PhysRevC.73.055801} are replaced by the corrected data presented in \textcite{Brune201349}.

Table~\ref{tab:data_sources} summarizes the data considered in the energy region below $S_p$ in $^{16}$O, but the higher energy data given in Table~I of \textcite{PhysRevC.87.015802} are also included. The table also summarizes where the actual numerical values of the data for each of the measurements were obtained. It is fortuitous that most of the $E1$ and $E2$ data, below $S_p$, have been made available in tabular form and therefore few had to be digitized from figures. It should be noted that this was not true for the data from \textcite{PhysRevC.2.63, PhysRevC.2.2452}, but this data have been excluded from the analysis as mentioned above. Unfortunately much of the angular distribution data did have to be digitized. The figures were these data were obtained are listed in Table~\ref{tab:data_sources}. For ease of reference, Table~\ref{tab:targets} also summarizes the different kinds of targets that have been used in the various capture measurements. 

\begin{table*}
\caption{Summary of experimental data considered in the present analysis. The data included are in addition to that already presented in Table~I of \textcite{PhysRevC.87.015802}. The reaction, the multipolarity of the $\gamma$ rays measured, the source of the data, and the systematic uncertainty for each data set is noted. For several of the data sets, the absolute cross section was found by normalizing to previous data, so these entries are left blank. \label{tab:data_sources}}
\begin{ruledtabular}
\begin{tabular}{ c c c c c } 
Ref. & Reaction(s) & Source & $\sigma_\text{syst}$ (\%) \\
\hline
\textcite{Dyer1974495} & $^{12}$C$(\alpha,\gamma_0)^{16}$O, $E1$ & Table~2, Fig.~6 & 10 \\
\textcite{Redder1987385} & $^{12}$C$(\alpha,\gamma_{0,6.92,7.12})^{16}$O, $E1$ and $E2$ & Tables~1,2,3, Fig.~5 & 6 \\ 
\textcite{PhysRevLett.60.1475} & $^{12}$C$(\alpha,\gamma_0)^{16}$O, $E1$ & Private Communication\footnotemark[1] & 15 \\ 
\textcite{PhysRevLett.70.726} & \multirow{ 2}{*}{$^{16}$N$(\beta\alpha)^{12}$C} & \multirow{ 2}{*}{Table~1} & \multirow{ 2}{*}{6} \\
\& \textcite{PhysRevC.50.1194} & & & \\
\textcite{PhysRevLett.69.1896, PhysRevC.54.1982} & $^{12}$C$(\alpha,\gamma_0)^{16}$O, $E1$ and $E2$ & Table~3,4  & $-$ \\ 
\textcite{Roters1999} & $^{12}$C$(\alpha,\gamma_0)^{16}$O, $E1$ and $E2$ & Table~1 & $-$ \\
\textcite{Gialanella2001} & $^{12}$C$(\alpha,\gamma_0)^{16}$O, $E1$ & Table~2 & 9 \\
\textcite{PhysRevLett.86.3244} & $^{12}$C$(\alpha,\gamma_0)^{16}$O, $E1$ and $E2$ & Table~1, Fig.~3 & $-$ \\
\textcite{Fey2004} & $^{12}$C$(\alpha,\gamma_0)^{16}$O, $E1$ and $E2$ & Tables~E1,E2, Figs.~D1 to D13 & $-$ \\
\textcite{schurmann2005} & $^{12}$C$(\alpha,\gamma)^{16}$O & Table~1 & 6.5 \\
\textcite{PhysRevC.73.055801}, & \multirow{ 2}{*}{$^{12}$C$(\alpha,\gamma_0)^{16}$O,$E1$ and $E2$} & \multirow{ 2}{*}{Table~1 (3 params.),Fig.~10} & \multirow{ 2}{*}{$-$} \\
\& \textcite{Brune201349} & & & \\
\textcite{PhysRevC.80.065802} & $^{12}$C$(\alpha,\gamma_0)^{16}$O, $E1$ and $E2$ & Table~7 & $-$\footnotemark[2] \\
\textcite{PhysRevC.79.055803} & $^{12}$C$(\alpha,\alpha)^{12}$C & EXFOR\footnotemark[3] & $-$\\ 
\textcite{PhysRevC.81.045809} & $^{16}$N$(\beta\alpha)^{12}$C & Table~1 & 2 \\
\textcite{Schurmann2011557} & $^{12}$C$(\alpha,\gamma_{all})^{16}$O & Table~1 & 6.5 \\
\textcite{PhysRevC.86.015805} & $^{12}$C$(\alpha,\gamma_{all})^{16}$O & Table~III &  $<$10 \\
\end{tabular}
\end{ruledtabular}
\footnotetext[1]{Data available in Supplemental Material.}
\footnotetext[2]{Statistical and systematic uncertainties combined.}
\footnotetext[3]{Data available as both yields (\href{http://www.nndc.bnl.gov/exfor/servlet/X4sGetSubent?reqx=34022&subID=121461002}{C1461002}) and cross sections (\href{http://www.nndc.bnl.gov/exfor/servlet/X4sGetSubent?reqx=34022&subID=121461014}{C1461014}).}
\end{table*} 

\section{\texorpdfstring{$R$-matrix analysis of $^{12}$C$(\alpha,\gamma)^{16}$O}{R-matrix analysis of 12C(a,g)16O}} \label{sec:R-matrix_analysis}

As described in Sec.~\ref{sec:r_matrix}, the phenomenological $R$-matrix method is currently the preferred method for the analysis of the $^{12}$C$(\alpha,\gamma)^{16}$O reaction. This method has been used to simultaneously fit the experimental measurements that populate the $^{16}$O compound nucleus at energies below $E_x \approx$~14~MeV (see Fig.~\ref{fig:level_diagram}). A modified version of the $R$-matrix code \texttt{AZURE2} \cite{azure, azure2} has been used. The code implements the generalized mathematical formalism that has been described in Sec.~\ref{subsection:rmatrixtheory} including the alternate $R$-matrix formalism of \textcite{PhysRevC.66.044611} in order to more conveniently utilize level parameters from the literature. Another convince of this alternate parameterization is that boundary conditions are eliminated. However, the channel radii still need to be specified. In principle a different radius can be chosen for each channel, but it is common practice to only choose different radii for different partitions. For the best fit, values of $a_{\alpha_0}$~=~$a_{\alpha_1}$~=~5.43~fm and $a_{p_0}$~=~5.03~fm were found. Discussions of the sensitivity of the fit to the choice of channel radii is given in Sec.~\ref{sec:systematic_more}.

Several physical quantities have uncertainties much smaller than those from other sources and are treated as constants in the analysis. These are summarized in Table~\ref{tab:r_const}. Entrance channel angular momenta were considered up to $l$~=~8. Unless specifically labeled otherwise, all quantities are given in the center-of-mass reference frame.  

\begin{table}
\caption{Masses and particle separation energies used in the $R$-matrix calculation. The quantities $S_\alpha$, $S_p$, and $S_{\alpha_{1}}$ represent the separation energies of an $\alpha$ particle, a proton, and an $\alpha$ particle with $^{12}$C in its first excited state respectively. Masses are in atomic mass units. All values are taken from \textcite{Audi2003337}. \label{tab:r_const}}
\begin{ruledtabular}
\begin{tabular}{ c c }
Parameter & Value \\
\hline
$S_\alpha$ & 7.16192(1)~MeV \\
$S_p$ & 12.12741(1)~MeV \\
$S_{\alpha_{1}}$ & 11.60083(31)~MeV \\ 
$m_p$ & 1.00782503207(10)  \\
$m_\alpha$ & 4.00260325415(6)  \\
$m(^{12}\text{C})$ & 12  \\ 
$m(^{15}\text{N})$ & 15.00010889823(15)  \\
$m(^{16}\text{N})$ & 16.006101658(2815)  \\
$m(^{16}\text{O})$ & 15.99491461956(16)  \\ 
\end{tabular}
\end{ruledtabular}
\end{table}  

The present work is an extension of the $R$-matrix analysis given in \textcite{PhysRevC.87.015802}. In that work, a global $R$-matrix fit was achieved for data belonging to all the open channels above the first excited state $\alpha$ particle ($S_{\alpha_1}$~=~11.60~MeV) and proton ($S_p$~=~12.13~MeV) separation energies and below $E_x \approx$~14.0 MeV in $^{16}$O (see Fig.~\ref{fig:level_diagram}). In the present analysis, all the previous channels and data are again considered, but, in addition, the lower energy data for the $\alpha$ capture reaction and the $^{16}$N$(\beta\alpha)^{12}$C spectra are included. This global, simultaneous, analysis considers over 15,000 data points, the majority of the data available in the literature. In addition to the primary aim of facilitating a comparison between the different data sets, this global analysis has the potential to place more stringent constraints on the extrapolation of the $^{12}$C$(\alpha,\gamma)^{16}$O reaction to stellar energies. This is mainly the result of the inclusion of the higher energy data and the extension of the phenominological model to those energies. While the resonances at higher energy do not have a strong impact on the low energy cross section, an explicit fitting to these higher energies places much more stringent limits on possible low energy tail contributions of even higher lying resonances. As will be described, this has been, and remains, one of the largest uncertainties in the extrapolation of the cross section to stellar energies. 

\subsection{``Best Fit" Procedure} \label{sec:fitting}

One of the main reasons that the $^{12}$C$(\alpha,\gamma)^{16}$O cross section has such large uncertainties in the extrapolation is that there are different possible fit solutions, corresponding to the different relative interference patterns of the different resonances, depending on different assumptions and interpretations of the data. However, as more and more measurements have been undertaken, these ambiguities have steadily decreased. Here the assumptions of this analysis are described and a large amount of the remainder of this work details what happens when these assumptions are bent or broken in order to more fully gauge the uncertainties. 

As with most analyses of this kind, the path to the final solution was not a straight forward procedure but was an iterative process. In this section the details of the ``best fit" are given. This was not the fit with the lowest over all $\chi^2$, nor was it the fit that allowed all possible parameters to vary freely, but it is believed to be the most physically reasonable one. In principle a $\chi^2$ minimization should lead to the best solution, but this assumes that all of the uncertainties have been correctly quantified in the data, and this is certainly not the case. While the $\chi^2$ minimization also includes a term for a constant systematic uncertainty, and often this is a dominant contribution, energy and angular dependent systematic uncertainties are also present, which may or may not have been quantified. These usually have a smaller effect, but in some cases, especially when the statistical uncertainties become very small, it is very likely that these are the cause of poorer quality fits. 

The assumptions that brought about the ``best fit'' were 
\begin{itemize} 
\item The ANCs, as determined from modern transfer reactions, are reliable and their values are fixed in the fit
\item $\gamma$ ray widths of the subthreshold states are reliable and are fixed in the fit
\item Fits are unacceptable if the normalization factors of all data sets in a given channel systematically deviate (the exception to this rule is the $E2$ ground state transition data)
\end{itemize}
and important conclusions brought about by a thorough review of the data and the $R$-matrix fits were
\begin{itemize}
\item The $^{12}$C$(\alpha,\gamma)^{16}$O $E2$ ground state transition cross section data show large deviations between one another 
\item Background pole contributions are negligible for the $^{12}$C$(\alpha,\gamma)^{16}$O capture data for all transitions
\item For the ground state $^{12}$C$(\alpha,\gamma)^{16}$O data, the low energy measurements yielding larger cross sections are more likely to be affected by un-reported systematic uncertainties.
\end{itemize}
These assumptions and conclusions were in most cases not assumed a priori, but came about through the iteration of many test calculations. Of course these assumptions, although logically motivated, are still somewhat subjective and may have (in fact have greatly) varied from one evaluator to the next over the years. Therefore it is of the utmost importance that these assumptions are tested rigorously and the sensitivity of the fit must be gauged for each one. This is what much of Sec.~\ref{sec:unc_analysis} is devoted to, a quantification of these assumptions into uncertainties on the extrapolation of the cross section to stellar energies. Throughout this work there are discussions of these different assumptions but a brief discussion of all of them is first given here. 

While ANCs can be determined, in theory, through compound nucleus reactions like scattering, capture, and $\beta$ delayed particle emission, there are nearly always complications. These analyses are usually performed using a phenomenological $R$-matrix analysis but are often complicated by the presence of broad resonances contributions and the need for background contributions. It is the experience of the authors that unless a capture cross section can be described well using only the external capture model (i.e. no broad resonances are present) it can be difficult to extract a reliable value for ANCs from this type of data. For scattering it is nearly always the case that many background poles are needed to compensate for the potential phase shift that is only approximately reproduced by the hard sphere phase shifts. The situation is similar for $^{16}$N$(\beta\alpha)^{12}$C data were again background contributions are often required. Further, even small errors in the corrections of the data for experimental effects can effect the ANC determinations.  

Transfer reactions have their own issues, namely that their are theoretical uncertainties that can be difficult to quantify, but recent Sub-Coulomb measurements seem to have succeed in limiting these effects so that ANCs can be extracted reliably to about the 10\% uncertainty level. This has been confirmed by measurements by different groups using different experimental setups that yield consistent results. Certainly more experiments need to be performed to better verify this claim, but at this time they seem to be the most reliable and accurate method.

The only $\gamma$ ray widths of subthreshold states that have a large impact on the cross section determination are the ground state widths of the $E_x$~=~6.92 and 7.12~MeV states which make strong subthreshold contributions to the ground state $E2$ and $E1$ cross sections respectively. These widths are known to better than 5\% total uncertainty and have been verified by several different measurements, although they have not been studied recently.

Unless there is some reason to suspect that all data for a specific reaction suffer from a shared systematic uncertainty, it seems reasonable to assume that the weighted average of different measurements should be very close to one. This has been found to be true for all of the reaction channels studied here except for the $^{12}$C$(\alpha,\gamma)^{16}$O $E2$ ground state data. This data shows both large scatter, that is not reflected by the experimental error bars, and has been found to by systematically too large in value. These issues have lessened with more recent measurements. The issue here has always been the attempt to determine a small $E2$ component from a cross section that is dominated by the broad $E1$ resonance at $E_x$~=~9.59 MeV.

With the addition of the higher energy states in this analysis, a reasonable fit can be achieved for both the $E1$ and $E2$ ground state capture transition data without any background contributions. This does not mean that the present $R$-matrix analysis does not have any background poles, several are still required to fit the $^{12}$C$(\alpha,\alpha_0)^{12}$, $^{15}$N$(p,p_0)^{15}$N, and $^{16}$N$(\beta\alpha)^{12}$C reaction data. The physical justification for this is clearly seen in the higher energy $^{12}$C$(\alpha,\gamma)^{16}$O data of \textcite{PhysRevLett.32.1061} where it has been shown that the cross section, for both $E1$ and $E2$ multipolarities, becomes weak at high energies and does not show any resonances that compare in strength to those that correspond to the 1$^-$ levels $E_x$~=~12.45 and 13.09~MeV. Since these two strong higher energy states are now included explicitly, they should account for the majority of the higher energy background contributions.

For the $E2$ ground state transition data, more recent measurements have achieved significantly more consistent measurements (e.g. \cite{PhysRevC.80.065802, PhysRevC.86.015805}). This is also true for the $E1$ ground state transition data, but to a lesser extreme. This can be seen clearly in Fig.~\ref{fig:E1_and_E2_comparisons}. From \textcite{PhysRevC.2.63} to \textcite{PhysRevC.86.015805}, there has been a general decrease in the values given for the cross sections on the low energy side of the 1$^-$ resonance. In general, background contaminations are often underestimated for the yield extraction in low statistics measurements and this trend is quite prevalent in the literature.

\subsubsection{Systematic Uncertainty $\chi^2$ Term}

Every data set has some systematic uncertainty, which represents the experimentalist's best estimate of contributions to the uncertainty that come from sources that are not statistical. In many cases these uncertainties are approximated by a constant factor (e.g. target thickness, beam intensity, efficiency, etc.) that affects the entire data set. However, as measurements become more precise, the quantification of the systematic uncertainties often become increasingly complicated. 

As a first step, it is critical to separate the uncertainties into contributions from point-to-point and over all systematic uncertainties in order to perform the $\chi^2$ minimization accurately. For analyses that consider multiple data sets, each with independent systematic uncertainties, this becomes even more crucial. This has been shown some time ago, e.g. by \citet{PhysRevC.15.518}, yet has been neglected even in some relatively recent and comprehensive analyses (e.g. \citet{Hammer2005514}). Further, it has been shown that the method of introducing the systematic uncertainty term into the $\chi^2$ fitting solves Peelle's pertinent puzzle, or at least makes the effect negligible \cite{Carlson20093215, Hale_2004}.

For the approximation that the systematic uncertainty of an individual data set can be treated as constant, it is included in the $\chi^2$ fit using the method described in, e.g., \textcite{DAgostini1994306, PhysRevC.15.518, Schurmann201235} and is given by
\begin{equation} \label{eq:standard_chi_squared}
\chi^2 = \sum_i\left(\sum_j R_{ij}^2 + \frac{(n_i-1)^2}{\sigma^2_{\text{syst},i}}\right)
\end{equation}
\begin{equation}
R_{ij} = \frac{f(x_{i,j}) - n_iy_{i,j}}{n_i\sigma_{i,j}}
\end{equation}
\noindent where $n_i$ is the normalization factor of an individual data set, $f(x_{i,j})$ is the value of the cross section from the $R$-matrix fit, $y_{i,j}$ is the experimental cross section of a given data point, $\sigma_{i,j}$ is the (hopefully mostly) statistical uncertainty of the data point, and $\sigma^2_{\text{syst},i}$ is the over all fractional systematic uncertainty of the data set. A summary of the systematic uncertainties of the different experiments is listed in Table~\ref{tab:data_sources}. Since several of the capture measurements were normalized to earlier measurements, several lack independent normalizations (indicated by the $-$ in Table~\ref{tab:data_sources}). In these cases, the normalizations of the data are allowed to vary freely in the fitting. Table~\ref{tab:normalizations} lists the normalization factors for the excitation curve data resulting from the $R$-matrix fit and compares them to the experimental systematic uncertainties.    

\begin{table*}
\caption{Normalization factors ($n_i$), $\chi^2$, and number of data points (N) for excitation curve data resulting from the $R$-matrix fit. Note the reasonable scattering in the normalizations of the $E1$ data compared to the much larger scatter present in the $E2$ data. Additionally, it should be noted that the reduced $\chi^2$ values for the $^{16}$N$(\beta\alpha)^{12}$C and $^{12}$C$(\alpha,\alpha_0)^{12}$C data sets are significantly greater than one. The effect of this on the uncertainty estimate of the extrapolated cross section will be discussed in Sec.~\ref{sec:total_unc}. \label{tab:normalizations}}
\begin{ruledtabular}
\begin{tabular}{ c c c c c c c } 
Reaction & Ref. & $\sigma_{\text{syst},i}$ \% & Norm. ($n_i$) & $\chi^2$ & $L^{-1}$ & N \\
\hline
$^{12}$C$(\alpha,\gamma_0)^{16}$O & \textcite{Brune201349} & 9 & 1.12 & 20.3 & 14.0 & 16 \\ 
\hline
$^{12}$C$(\alpha,\gamma_0)^{16}$O (28$^\circ$) & \cite{PhysRevC.54.1982} & - & 1.200 & 18.4 & 14.9 & 15 \\
(60$^\circ$) & \textcite{PhysRevC.54.1982} & - & 0.970 & 91.7 & 20.7 & 16 \\
(90$^\circ$) & \textcite{PhysRevC.54.1982} & - & 1.046 & 78.6 & 18.0 & 16 \\
(90$^\circ$) & \textcite{PhysRevC.54.1982} & - & 0.957 & 86.7 & 18.7 & 16 \\
(120$^\circ$) &\textcite{PhysRevC.54.1982}  & - & 1.068 & 45.3 & 14.5 & 15 \\
(143$^\circ$) & \textcite{PhysRevC.54.1982} & - & 1.117 & 28.7 & 15.6 & 15 \\
(40$^\circ$) & \textcite{PhysRevC.80.065802} & - & 1.085 & 1.55 & 2.23 & 2 \\
(90$^\circ$) & \textcite{PhysRevC.80.065802} & - & 0.926 & 0.527 & 1.81 & 2 \\
(130$^\circ$) & \textcite{PhysRevC.80.065802} & - & 0.842 & 1.982 & 2.50 & 2 \\ 
\hline
$^{12}$C$(\alpha,\gamma_0)^{16}$O ($E1$) & \textcite{Dyer1974495} & 10 & 1.031 & 55.9 & 25.2 & 24 \\
 & \textcite{Redder1987385} & 6 & 1.006 & 72.8 & 29.3 & 26 \\
 & \textcite{PhysRevLett.60.1475} & 15 & 1.110 & 18.4 & 13.1 & 12 \\ 
 & \textcite{PhysRevC.54.1982} & - & 0.957 & 44.8 & 11.6 & 9 \\ 
 & \textcite{Roters1999} & - & 1.092 & 15.4 & 12.0 & 13 \\
 & \textcite{Gialanella2001} & 9 & 0.913 & 27.3 & 18.5 & 20 \\
 & \textcite{PhysRevLett.86.3244} & - & 1.011 & 12.4 & 15.6 & 19 \\ 
 & \textcite{Fey2004} & - & 1.000 & 16.9 & 8.3 & 11 \\
 & \textcite{PhysRevC.80.065802} & - & 0.959 & 0.43 & 1.47 & 2 \\
 & \textcite{Schurmann2011557} & 6.5 & 0.994 & 0.605 & 0.95 & 1 \\
 & \textcite{PhysRevC.86.015805} & 12-21 & 1.017 & 2.01 & 3.25 & 4 \\
\hline
$^{12}$C$(\alpha,\gamma_0)^{16}$O ($E2$) & \textcite{PhysRevC.54.1982} & - & 0.883 & 3.42 & 5.70 & 5 \\
& \textcite{Roters1999} & - & 1.698 & 0.246 & 1.54 & 2 \\
& \textcite{PhysRevLett.86.3244} & - & 1.065 & 23.2 & 13.6 & 11 \\
& \textcite{Fey2004} & - & 1.364 & 15.1 & 12.8 & 12 \\
& \textcite{PhysRevC.80.065802} & - & 1.095 & 0.46 & 1.63 & 2 \\ 
& \textcite{Schurmann2011557} & 6.5 & 0.958 & 29.4 & 8.0 & 7 \\
& \textcite{PhysRevC.86.015805} & 30-61 & 1.016 & 0.342 & 2.93 & 5 \\
\hline
$^{12}$C$(\alpha,\gamma_{6.05})^{16}$O ($E1)$ & \textcite{Schurmann2011557} & 6.5 & 1.00 & 1.71 & 1.08 & 1 \\
 ($E2$) & \textcite{Schurmann2011557} & 6.5 & 1.19 & 16.5 & 9.4 & 6 \\
\hline
$^{12}$C$(\alpha,\gamma_{6.13})^{16}$O & \textcite{Schurmann2011557} & 6.5 & 1.03 & 8.9 & 5.2 & 7 \\
\hline
$^{12}$C$(\alpha,\gamma_{6.92})^{16}$O & \textcite{Redder1987385} & - & 0.261 & 26.4 & 42.8 & 25 \\
 & \textcite{Kunz_thesis} & - & 0.644 & 9.0 & 21.0 & 12 \\
 & \textcite{Schurmann2011557} & 6.5 & 0.993 & 18.1 & 12.7 & 7 \\
\hline
$^{12}$C$(\alpha,\gamma_{7.12})^{16}$O  & \textcite{Redder1987385} & - & 0.265 & 52.4 & 29.5 & 24 \\
 & \textcite{Kunz_thesis} & - & 0.469 & 8.3 & 16.7 & 12 \\
 & \textcite{Schurmann2011557} & 6.5 & 1.00 & 3.64 & 6.6 & 7 \\
\hline
$^{12}$C$(\alpha,\gamma_{\text{total}})^{16}$O & \textcite{schurmann2005} & 6.5 & 0.926 & 301 & 136 & 76 \\
 & \textcite{PhysRevC.86.015805} & 8-21 & 1.08 & 4.90 & 4.18 & 4 \\
 & \textcite{:/content/aip/proceeding/aipcp/10.1063/1.4874074} & - & 0.972 & 0.982 & 6.2 & 3 \\
\hline
$^{16}$N$(\beta\alpha)^{12}$C & \textcite{PhysRevC.50.1194} & 5 & 0.91 & 496 & 122 & 87 \\
 & \textcite{PhysRevC.81.045809} & 2 & 1.13 & 545 & 135 & 88 \\
\hline
$^{12}$C$(\alpha,\alpha)^{12}$C & \textcite{PhysRevC.79.055803} & - & - & 56021 & 11775 & 9728 \\
\end{tabular}
\end{ruledtabular}
\end{table*}

Ideally all parameters of the $R$-matrix fit could be varied simultaneously to achieve the best fit, but this situation could not be realized. The current analysis has 64 fit parameters not including the normalization parameters. These parameters correspond to the partial widths and energies of the 12 particle unbound states in $^{16}$O that were used to describe the broad energy structure of the reaction cross sections. An additional 7 background poles were necessary, primarily to reproduce the scattering reaction data. However, the best fit contains no background poles for the $^{12}$C$(\alpha,\gamma)^{16}$O reaction, for any of the transitions. Five subthreshold states and five levels corresponding to narrow resonances were also included, with their ANCs or partial particle widths and $\gamma$ decay widths fixed to values from the literature. Tables~\ref{app:fit_parameters} and \ref{tab:fit_params_gamma} give the $R$-matrix parameters necessary to reproduce the best fit of this analysis. The parameters in bold indicate those actually used for the fitting. Other parameters were fixed at values taken from the literature. Additionally, Table~\ref{tab:fit_params_grwa} gives the reduced width amplitudes associated with the $\gamma$-ray widths, subdivided into their internal end channel contributions. Note that these are not additional fitting parameters.

It should be highlighted here that many different fitting combinations where investigated and are discussed, but for clarity and practicality only the details of the best fit are given in Tables~\ref{app:fit_parameters} and \ref{tab:fit_params_gamma}. For example, in the subsequent sections that investigate the uncertainties in the fitting, background poles for the $^{12}$C$(\alpha,\gamma)^{16}$O reaction are introduced, but these are absent from the parameter tables since they are not included in the best fit.

With such a complicated fitting, it is reasonable to question whether the fit is unique. That is, can a similar quality fit be obtained but with a vary different parameter set? For levels in the $R$-matrix that correspond to physical levels, it is believed that these values are in fact unique and well defined, at least to within their uncertainties. This can be said with some confidence because of the unitarity condition of the $R$-matrix theory, but this does only apply for the particle channels. Further, it is often the case that only the partial width that corresponds to the lowest orbital angular momentum channel is used in the fitting. Higher orbital angular momentum channels have been investigated but had values consistent with zero for the present experimental data. When more precise data are obtained, this assumption should be re-investigated.    

For the $\gamma$-ray channels, unitarity is not enforced, as this reaction mechanism is introduced into the theory as a perturbation. This is well satisfied in the current case because the capture cross sections are always much smaller than the nuclear ones. However, because the particle width(s) of a given level are always much larger than the $\gamma$-ray widths, they determine the resonance's total width while the $\gamma$-ray widths effectively only determine the height of the resonance. In this way the capture data uniquely constrain the total $\gamma$-ray widths. However, it is sometimes possible that more than one $\gamma$-ray multipolarity can contribute and may be of similar intensity. This is especially true for the well known case when $E2$ and $M1$ decays are both possible and it is often true that the data may not uniquely determine the multipolarity. Therefore the value of the total $\gamma$-ray widths are likely unique, but their multipolarities should be viewed as tentative assignments. Additionally, like the particle channels, higher order multipolarities may also be possible. The fit was tested for sensitivities to these higher order terms but again they were found to not be significant considering the uncertainties of the current data.

What are more ambiguous are the interference combinations that are possible for the capture data. The ground state inferences patterns appear to now be well known as discussed in Sec.~\ref{sec:interferences}, but some of these assignments are based on only a single measurement or only a few data points. The ambiguity is much greater for the cascade transitions. Very few data exist for these transitions and often the interference signs have been deduced based on a few number of data points and the constraint imposed by the total capture cross section measurements. The situation is even worse above $S_p$, here the interference signs are only constrained by their effects on the lower energy data. However, since the cascade transitions are dominated by external capture at low energies, these different solutions have a negligible effect on the low energy extrapolation.

Finally, the values of the background poles are certainly not unique. This is perfectly acceptable since their energies are rather arbitrary. The only condition is that these energies should be significantly larger than the highest energy data points so that they can provide an approximately energy-independent underlying background that represents the sum of the low energy tails of all higher lying resonances. What is unique is the magnitude of the underlying background provided by the background poles, and this contribution can be produced many different equivalent ways. For example, the background poles are often placed at $E_x$~=~20~MeV for the best fit, but an investigation of the extension of the capture cross section to higher energies necessitates moving the background poles up to $E_x$~=~40~MeV. While the values of the partial widths for these poles are of course much different than those at lower energy, the cross section that they produce at low energy is relatively unchanged.

\subsubsection{Corrections for Experimental Effects} \label{sec:target_effects}

It is a simple fact that measurements made in the laboratory are never actually the true cross sections, statistical variations aside. Even for arguably the simplest of experimental data, for example $^{12}$C$(\alpha,\alpha_0)^{12}$C, the reported quantities are often expressed as yields instead of actual cross sections. Even quantities labeled as cross sections in the literature are often only normalized yields, which may or may not have been subjected to any number of different deconvolution techniques and other corrections for experimental effects. A general formula relating the experimental yield $Y(E_b)$ at mean beam-particle energy $E_b$ to true cross section $\sigma(E)$ is
\begin{eqnarray} \label{eqn:target_effs}
Y(E_b)=\int_{E_b-\Delta}^{E_b}\int^{\infty}_{0} \frac{\sigma(E)}{\epsilon(E_b')} g(E-E_b') dEdE_b'. 
\end{eqnarray}
\noindent Here, the function $g(E-E_b)$ describes the spreading of the beam particle energy around the mean energy, $\epsilon(E_b)$ is the stopping power that describes the energy loss of the beam particles as they passes through the target material, and $\Delta$ is the total energy loss in the target. Other effects, such as energy straggling, may also be important depending on the experimental conditions, but Eq.~(\ref{eqn:target_effs}) serves as a general enough example. As is commonly implemented in the case of charged particle beams, the spreading function is approximated by a Gaussian function
\begin{eqnarray} \label{eqn:conv}
g(E-E_b) = \frac{1}{\sqrt{2\pi\sigma_b^2}}\exp\left[-\frac{(E-E_b)^2}{2\sigma_b^2}\right] ,
\end{eqnarray}
\noindent where $\sigma_b$ defines the energy width of the beam. 


In this analysis, Eq.~(\ref{eqn:target_effs}) is used to approximately correct for the resolution of the experimental measurements, as most of the data under analysis assumes this sort of convolution function. It should be kept in mind that this is an approximate method and that for data with very small uncertainties this simple method may not prove accurate enough. In this analysis a good example are the $\beta$ delayed $\alpha$ emission data. Possible failings of the deconvolution method have been discussed recently by \textcite{PhysRevC.80.045803} and are described further in Sec.~\ref{sec:beta_delayed_alpha}.
 
\subsection{\label{sec:gs_12Cag} Ground State Transition}

\begin{figure}
\includegraphics[width=1.0\columnwidth]{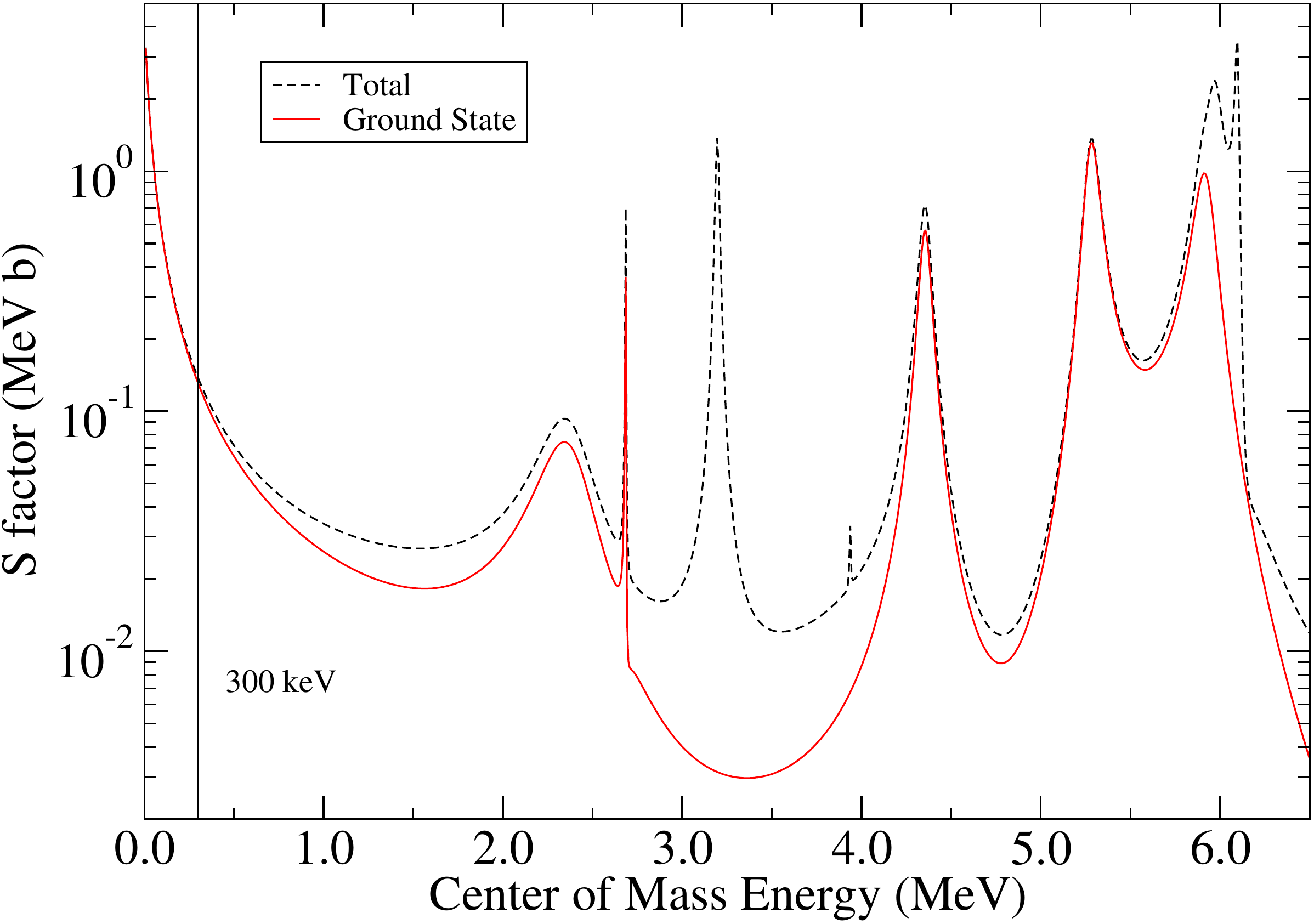}
\caption{(Color online) Illustration of the total $^{12}$C$(\alpha,\gamma)^{16}$O $S$-factor (black dashed line) compared to the ground state transition (red line) based on the $R$-matrix analysis of this work. The ground state transition dominates at stellar energies. $E_\text{c.m.}$ = 300 keV is indicated by the vertical black line. Experimental measurements of the ground state transition have reached as low as $E_\text{c.m.} \approx$ 1 MeV. \label{fig:gs_compare}}
\end{figure}

The largest contribution to the $^{12}$C$(\alpha,\gamma)^{16}$O cross section at low energy ($E_\text{c.m.} \approx$~300 keV) is the ground state transition. This is illustrated in Fig.~\ref{fig:gs_compare}\footnote{While the $E1$ constructive solution is shown here, this statement is true even for a destructive $E1$ solution, since the $E2$ cross section still dominates over the cascade transition contributions. There is no $E2$ interference pattern that has been considered viable that makes its contribution of similar magnitude or smaller than the cascade contributions.}. The $E1$ and $E2$ multipolarites dominate the low energy cross section in nearly equal amplitudes as discussed in Sec.~\ref{sec:nuc_phys}. At higher energies, high order multipolarities could become significant, although this has not yet been observed. A prime candidate is the ground state $E3$ decay of the broad 3$^-$ level at $E_x$~=~11.49~MeV.   

\begin{figure}
\includegraphics[width=1.0\columnwidth]{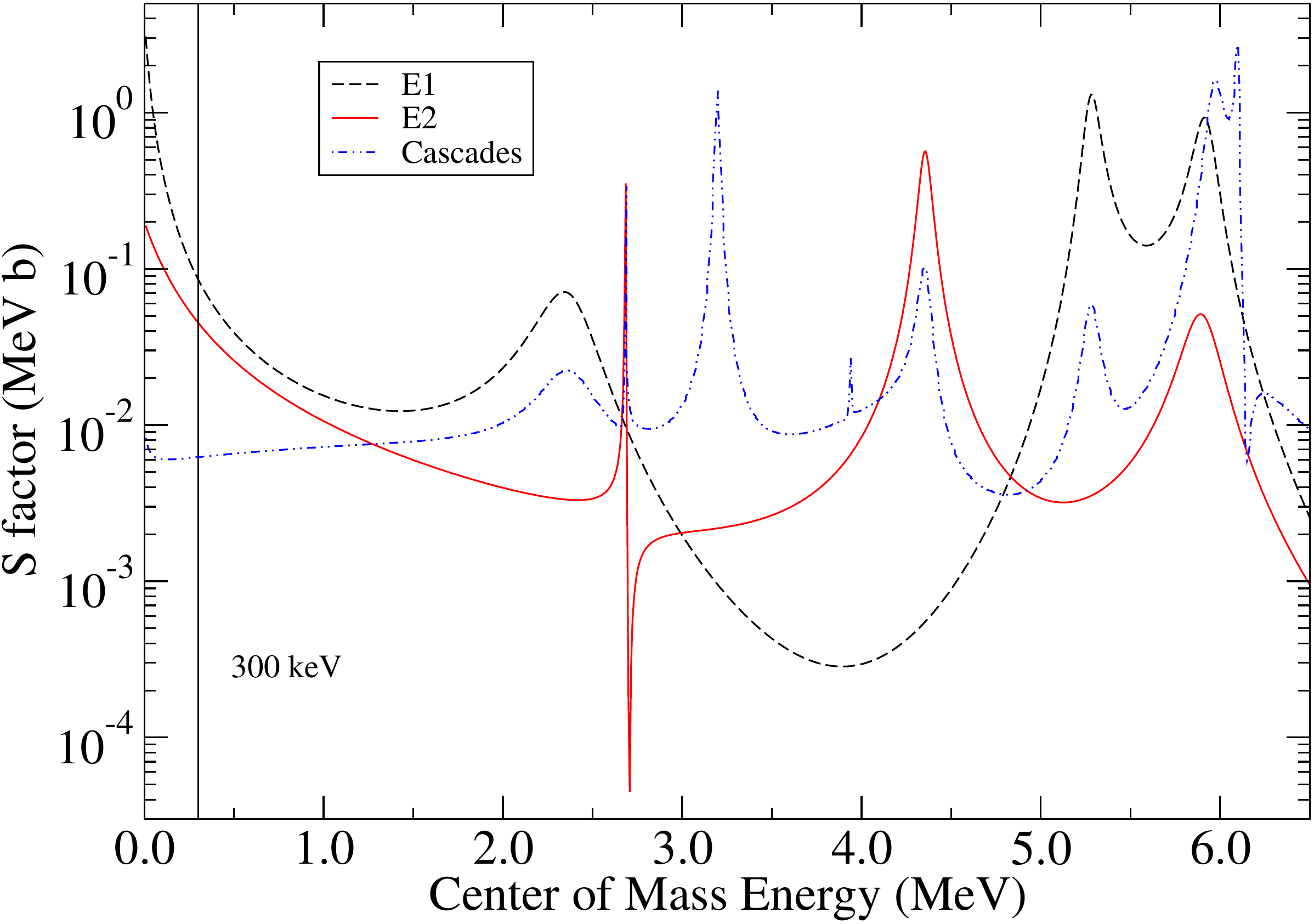}
\caption{(Color online) Comparison of predicted $E1$ (dashed black line), $E2$ (solid red line), and the sum of the cascade transitions (dot-dot-dashed blue line) cross sections. Over much of the low energy range covered by most measurements, the broad resonance corresponding to the $E1$ decay of the 1$^-$ level at $E_x$ = 9.59 MeV dominates over the $E2$ contribution. This has made the experimental determination of the $E2$ cross section extremely challenging. \label{fig:E1vsE2}}
\end{figure}

The separation of the ground state capture cross section into $E1$ and $E2$ multipolarities ($\sigma_{E1}$ and $\sigma_{E2}$) dates back to \textcite{Dyer1974495}. As discussed in Sec.~\ref{sec:data}, at that time $\sigma_{E1}$ was thought to dominate the low energy cross section, which was determined by decay through the 1$^-$ subthreshold state and its interference with the unbound level at $E_x$~=~9.59~MeV. The $E1$ cross section was also easily isolated experimentally by measuring at 90$^\circ$ where $\sigma_{E2}$ and the interference terms are zero (see Sec.~\ref{sec:classical_era}). This then also greatly simplifies the mathematics of the analysis, which at the time was usually a three level $R$-matrix fit. Complications arose when it was found that $\sigma_{E2}$ was also significant (see Sec.~\ref{sec:low_energy_race}). From an experimental standpoint, the immediate difficulty was that there is no angle where $\sigma_{E1}$ is zero and $\sigma_{E2}$ is not, therefore $\sigma_{E2}$ must be deduced indirectly. The traditional technique is to measure the differential cross section at several angles, spanning a wide angular range, and then perform a fit to a theory motivated function representing the angular distribution. If only $E1$ and $E2$ multipolarities contribute to the cross section, the differential cross section can be written as \cite{Dyer1974495}
\begin{equation} \label{eq:E1_E2_diff_XS}
\begin{split}
&4\pi\left(\frac{d\sigma}{d\Omega}\right)(E,\theta_\gamma) = \\ & 
\begin{split}& \hspace{3mm}\sigma_{E1}(E)[1-Q_2P_2(\cos\theta_\gamma)] \\ &
\hspace{3mm}+\sigma_{E2}(E)\left[1+\frac{5}{7}Q_2P_2(\cos\theta_\gamma)-\frac{12}{7}Q_4P_4(\cos\theta_\gamma)\right] \\ &
\hspace{3mm}+ 6\cos\phi(E)\sqrt{\frac{\sigma_{E1}(E)\sigma_{E2}(E)}{5}}[Q_1P_1(\cos\theta_\gamma) \\ & 
\hspace{52mm}-Q_3P_3(\cos\theta_\gamma)]\end{split}
\end{split}
\end{equation}
where $P_n\cos(\theta_\gamma)$ are the Legendre polynomials, $Q_n$ are the geometric correction factors \cite{PhysRev.91.610}, and $\phi$ is the difference in phase between the $E1$ and $E2$ transition matrix elements. The phase difference can be written as 
\begin{equation}
\cos\phi = \cos[\delta_{\alpha 1}-\delta_{\alpha 2}-\tan^{-1}(\eta/2)]
\label{eq:e1e2phase}
\end{equation}
where $\delta_{\alpha 1,2}$ are the angular momentum $l$~=~1 and 2 $\alpha$ scattering phase shifts (see Eq.~(\ref{eq:phase_shift})) and $\eta$ is the Sommerfeld parameter. As discussed above in Sec.~\ref{sec:radiative_capture}, Eq.~(\ref{eq:e1e2phase}) is very general and is a consequence of Watson's theorem. It is also fully consistent with the $R$-matrix formalism used here. This simply illustrates the connection between the scattering cross section, from which the phase shifts can be extracted, and the capture cross section. Since the scattering cross section is large, the phase shifts can be easily and accurately measured and used to constrain $\cos\phi$ up to an overall sign~\cite{PhysRevC.64.055803}. If $\sigma_{E1}$ is then determined from measurements at 90$^\circ$, then $\sigma_{E2}$ is essentially the only undetermined quantity. In principle this provides a straightforward way of obtaining $\sigma_{E2}$ but there are complications. The main issue is that $\sigma_{E1}$ is much larger than $\sigma_{E2}$ over much of the experimentally accessed low energy range because the cross section is dominated by the broad resonance corresponding to the 1$^-$ level at $E_x$~=~9.59~MeV as illustrated in Fig.~\ref{fig:E1vsE2}. The fact that the interference term is proportional to $\sqrt{\sigma_{E1}\sigma_{E2}}$ increases the sensitivity to the small $E2$ component, but in practice this approach has yielded a large scatter in the $\sigma_{E2}$ data as shown in Figs.~\ref{fig:E1_and_E2_comparisons} and \ref{fig:angle_int_12Cag}. The scatter is far outside the acceptable statistical range and suggests that systematic errors in the radiative capture measurements have been underestimated.

One way to understand the sign ambiguity in $\cos\phi$ is that the nuclear phase shifts and arctangent function in Eq.~(\ref{eq:e1e2phase}) are ambiguous by multiples of $\pi$. However, in our $R$-matrix formalism, the sign of $\cos\phi$ is determined by the phases of the transition matrix elements which are ultimately determined by the signs of the reduced width amplitudes (phase shifts are inherently ambiguous and are not used in the calculations). Note also that a generalization of Watson's theorem still applies when inelastic scattering and reaction channels are open, such as is the case shown below in Fig.~\ref{fig:E1_E2_int}. In this situation, the phases of the radiative capture transition matrix elements are determined by the $R$-matrix parameters for the nuclear channels, as shown by Eq.~(\ref{eq:T_photon_watson}), which can be constrained if sufficient data in the nuclear channels are available.

\begin{figure}
\includegraphics[width=1.0\columnwidth]{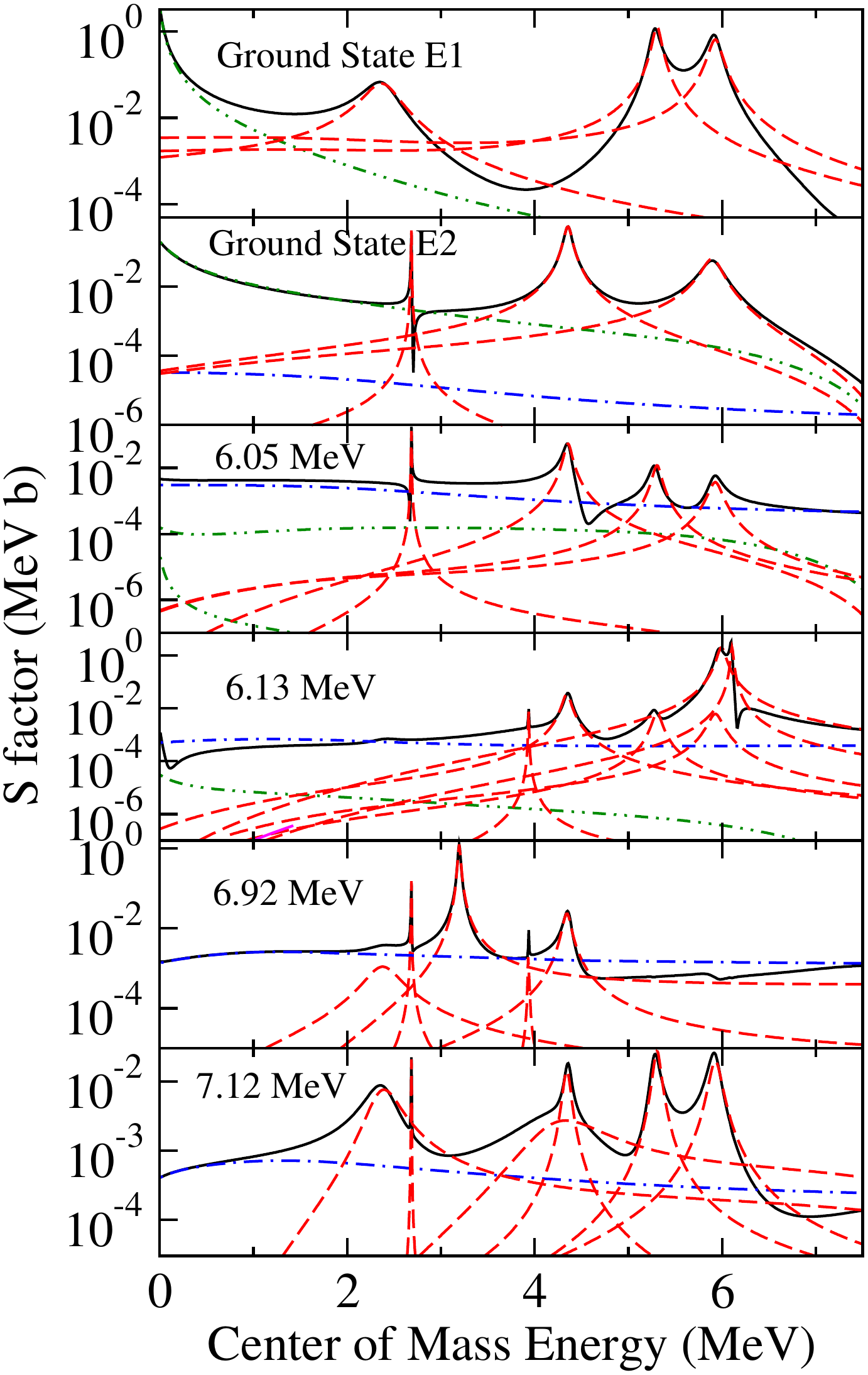}
\caption{(Color online) $S$-factors of the different transitions that make up the total $^{12}$C$(\alpha,\gamma)^{16}$O cross section. The ground state is further divided into $E1$ and $E2$ multipolarities. Red dashed lines indicate individual resonance contributions (single level calculation), blue dashed-dot lines are the hard-sphere contributions to the external capture, green dashed-dot-dot lines are subthreshold contributions, and the solid black lines represent the total with interferences included (i.e. individual contributions summed coherently). \label{fig:components}}
\end{figure}
\begin{figure}
\includegraphics[width=1.0\columnwidth]{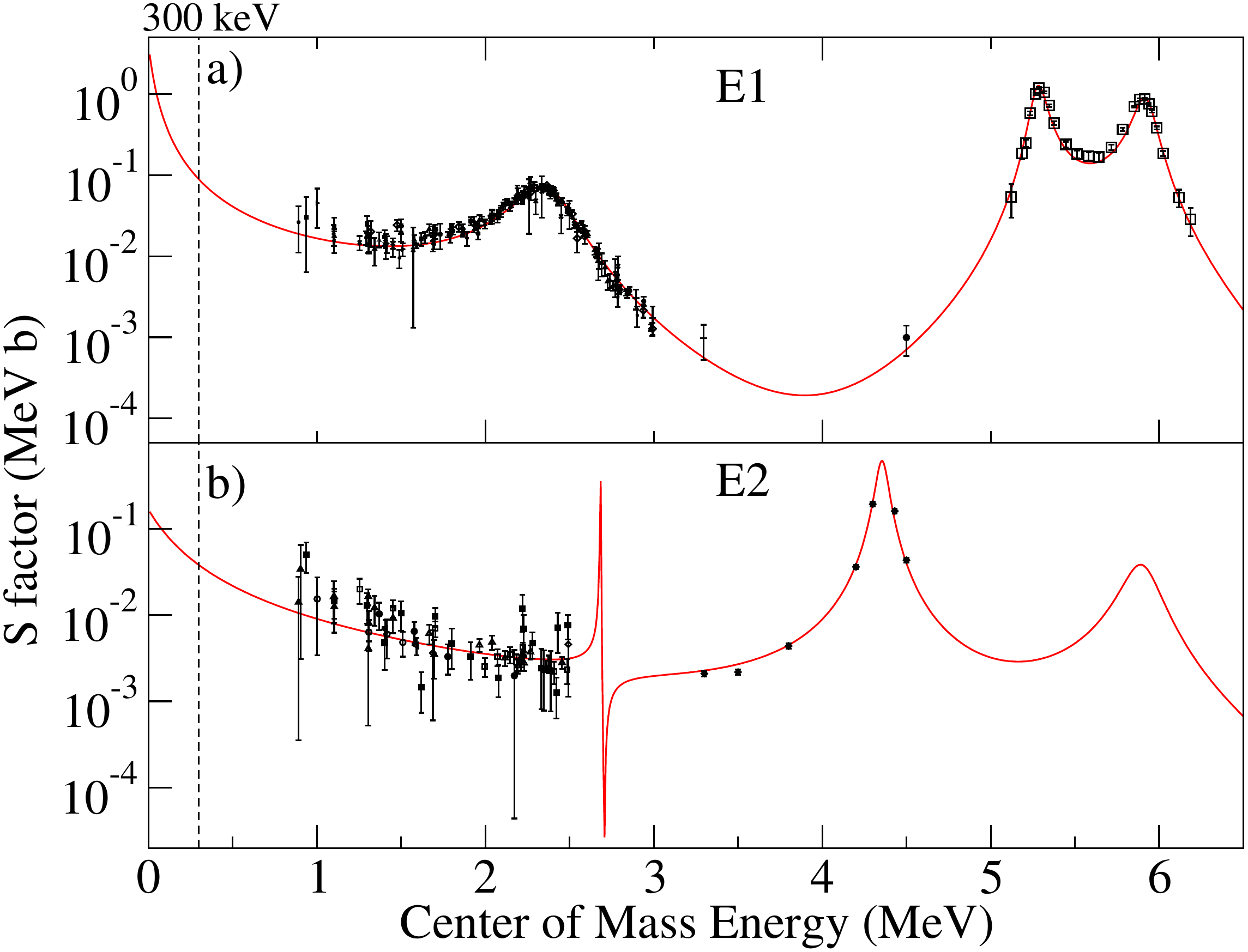}
\caption{(Color online) Fit to the $^{12}$C$(\alpha,\gamma_0)^{16}$O cross section. The $E1$ contribution from \textcite{Dyer1974495, Redder1987385, Gialanella2001, Roters1999, PhysRevLett.86.3244, PhysRevC.73.055801, PhysRevLett.60.1475, PhysRevC.54.1982, Fey2004, PhysRevC.80.065802, Schurmann2011557, PhysRevC.86.015805} is shown in Fig.\ref{fig:angle_int_12Cag} a), and the $E2$ contribution from \textcite{PhysRevC.73.055801, PhysRevC.54.1982, PhysRevLett.86.3244, Redder1987385, Roters1999, Schurmann2011557, Fey2004, PhysRevC.80.065802, PhysRevC.86.015805} is shown in Fig.\ref{fig:angle_int_12Cag} b). In Fig.\ref{fig:angle_int_12Cag} a), the angle integrated cross section data of \textcite{Brochard1973} are also shown at high energy for comparison as they are dominated by $E1$ capture. Note that the data have been subjected to overall normalizations as determined by the fitting procedure.  \label{fig:angle_int_12Cag}}
\end{figure}

For the $E1$ data, only decays from $1^-$ levels can contribute because the spins of the entrance channel particles and the final state are all zero. The different levels that are considered are the subthreshold state at $E_x$~=~7.12~MeV and the unbound states at $E_x$~=~9.59, 12.45, and 13.10~MeV. While the two higher lying states are 3 to 4~MeV above the lowest energy resonance, their large total widths and ground state $\gamma$-ray decay widths make their contributions significant even at low energies. The different contributions used to reproduce the cross section data are shown in Fig.~\ref{fig:components}. This $E1$ ground state cross section is unique in that it has a very weak external capture contribution (see Sec.~\ref{sec:radiative_capture}, Eq.~\ref{eq:eff_charge}). It does however have a strong subthreshold contribution from the $E_x$~=~7.12~MeV state. The ANC$_{\alpha}$ together with the energy and the $\Gamma_{\gamma_0}$ of this state characterize the strength of the subthreshold contribution. A discussion of the subthreshold state parameters is deferred to Sec.~\ref{sec:subthreshold}. It is also of note that the magnitude of the contributions from the subthreshold state and all three unbound states are similar at $E_\text{c.m.} \approx$ 1.2~MeV. The cross section that is shown in the figures of this work is the result of the choice of the constructive $E1$ interference solution. A discussion of why this particular solution has been chosen is given in Sec.~\ref{sec:interferences}. 

The situation is similar for the $E2$ data where only $2^+$ states can contribute. The states that are considered explicitly are the subthreshold at $E_x$~=~6.92~MeV and two unbound ones at $E_x$~=~11.51 and 12.96~MeV. The narrow state at $E_x$~=~9.84~MeV is also included but its energy and partial widths are all fixed to the values in the literature \cite{Tilley19931}. Further, as suggested in \textcite{PhysRevLett.109.142501}, experimental data in the vicinity of this state (2~MeV~$<E_\text{c.m.}<$~3~MeV) have been excluded because of the deconvolution of this data from yield to cross sections is unreliable (see Figs.~\ref{fig:E1_and_E2_comparisons} and \ref{fig:angle_int_12Cag}). The interference sign is also fixed to the one determined in \textcite{PhysRevLett.109.142501}. The different reaction components used for the fit are shown in Fig.~\ref{fig:components}. The subthrehold state completely dominates at low energy and is a slowly varying function of energy. Note that it is dominant all the way up to $E_\text{c.m.} \approx$~3.5~MeV. Only one experimental measurement has been made at these higher energies, that of \textcite{Schurmann2011557}. There is a small $E2$ external capture contribution that is present but because of interference terms it can have a significant effect on the $E2$ cross section in certain isolated regions. In fact it is just in the region near $E_\text{c.m.} \approx$~3.5~MeV that the effect is maximal since the other resonances make their smallest relative contributions here (see Fig.~\ref{fig:components}). Therefore the lowest energy $E2$ ground state transition data of \textcite{Schurmann2011557} have the somewhat unexpected ability to constrain the ground state external capture contribution. However, it is only the two lowest energy data points that have any significant sensitivity. Further discussions of the subthreshold state parameters are given in Sec.~\ref{sec:subthreshold}.

Ideally, the $R$-matrix fit would be made directly to primary data\footnote{That is, the data that are most closely related to the yields that are measured.}. This is done when the data are available, but in several instances only the derived $E1$ and $E2$ cross sections are given, and the original angular distributions are not reported. Differential cross section measurements below $S_p$ are available only for the ground state transition and only in the limited energy range around the broad lowest energy 1$^-$ resonance that corresponds to the level at $E_x$~=~9.59~MeV \cite{Dyer1974495, Redder1987385, PhysRevC.73.055801, Fey2004, PhysRevC.54.1982, PhysRevC.80.065802}. These data are used to determine the relative interference between the $E1$ and $E2$ components of the cross section, but it is possible that measurements over other regions, where the two components are closer in magnitude, would provide better constraint. Above $S_p$, measurements are available in \textcite{Larson1964497, Kernel1971352} over the broad states at $E_x$~=~12.45~(1$^-$), 12.96~(2$^+$), and 13.10~(1$^-$)~MeV. The $Q$ coefficients \cite{PhysRev.91.610, Longland2006452} used to correct for the extended geometry of the $\gamma$ ray detectors are listed in Table~\ref{tab:Q_coeffs}.

\begin{table*}
\caption{Summary of $Q$ coefficients for extended detector geometry corrections. In cases where the coefficients were not reported they have been approximated using a \texttt{GEANT4} simulation and the details of the geometry presented in the reference; the source for these cases is indicated as ``this work.''. \label{tab:Q_coeffs}}
\begin{ruledtabular}
\begin{tabular}{ c c c c c c c } 
Ref. & Det. & $Q_1$ & $Q_2$ & $Q_3$ & $Q_4$ & Source \\
\hline
\textcite{Larson1964497} & & 0.897 & 0.719 & 0.509 & 0.311 & this work \\
\hline
\textcite{Kernel1971352} & & 0.989 & 0.968 & 0.937 & 0.896 & this work \\
\hline
\textcite{Dyer1974495} & & 0.955 & 0.869 & 0.750 & 0.610 & Table~5.5 of \textcite{sayre_diss} \\ \hline
\textcite{Ophel1976397} & & 0.990 & 0.969 & 0.948 & 0.900 & this work \\ 
\hline
\textcite{PhysRevC.54.1982} & 28$^\circ$ & 0.9719 & 0.9173 & 0.8395 & 0.7431 & Table~1 \\
 & 60$^\circ$ & 0.9675 & 0.9047 & 0.8162 & 0.7061 & \\
 & 90$^\circ$ & 0.9541 & 0.8670 & 0.7474 & 0.6068 & \\
 & 90$^\circ$ & 0.9543 & 0.8675 & 0.7486 & 0.6091 & \\
 & 120$^\circ$ & 0.9762 & 0.9296 & 0.8672 & 0.7787 & \\
 & 143$^\circ$ & 0.9831 & 0.9500 & 0.9017 & 0.8400 & \\ \hline
\textcite{Redder1987385} & & & 0.92 & & 0.75 & in text \\ \hline
\textcite{PhysRevC.73.055801} &  & 0.989(2) & 0.968(4) & 0.936(8) & 0.895(14) & in text \\  
\hline
\textcite{PhysRevC.80.065802} & 40$^\circ$ & 0.980 & 0.947 & 0.898 & 0.837 & Table~VI \\
 & 90$^\circ$ & 0.980 & 0.946 & 0.897 & 0.835 & \\
 & 130$^\circ$ & 0.980 & 0.948 & 0.901 & 0.841 & \\
\hline
\textcite{PhysRevC.86.015805} & & 0.948 & 0.927 & 0.862 & 0.775 & Eq.~2 \\
 
\end{tabular}
\end{ruledtabular}
\end{table*}

\begin{figure*}
\includegraphics[width=1.9\columnwidth]{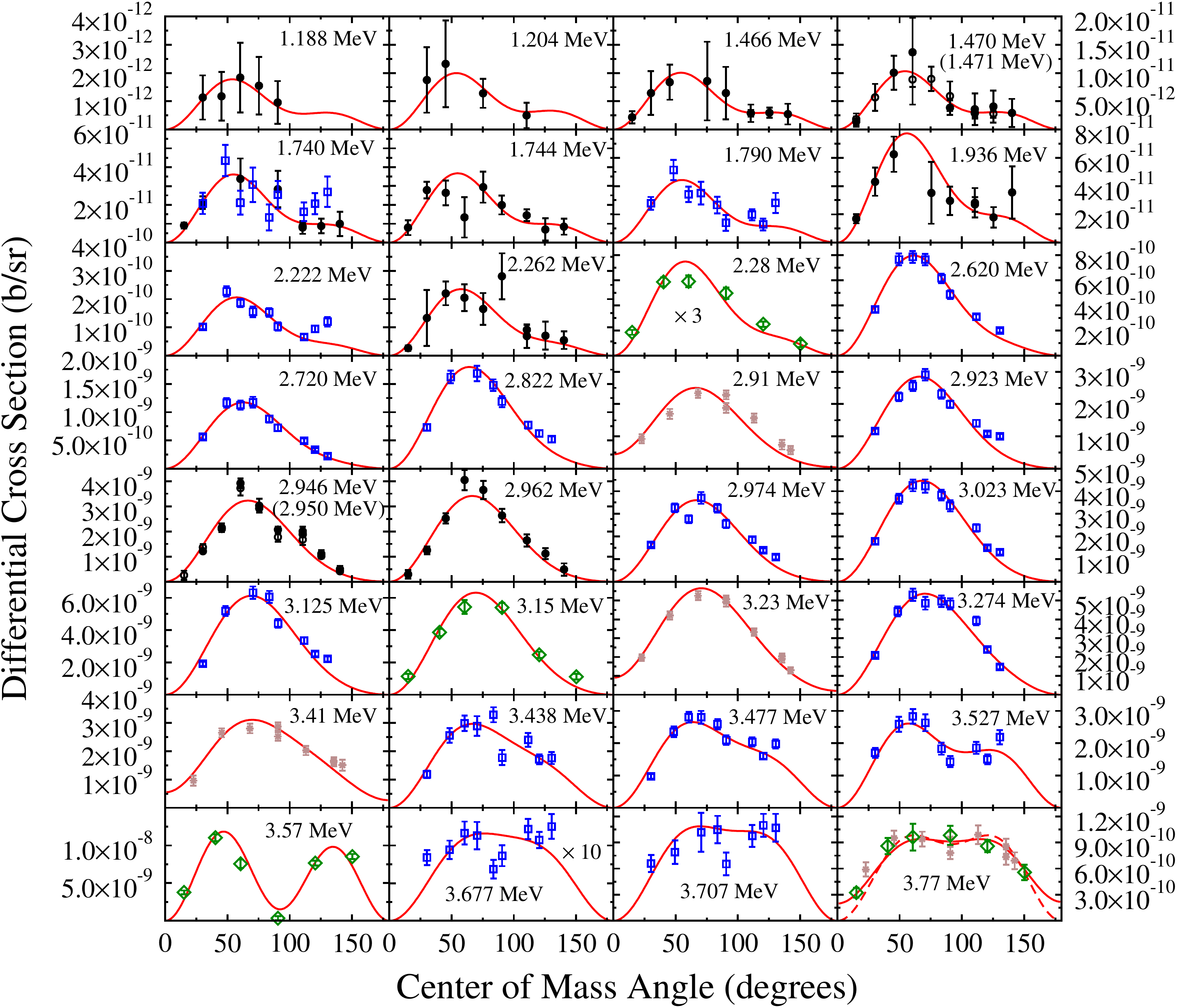}
\caption{(Color online) Comparison of the $R$-matrix fit to the angular distribution data of \textcite{Dyer1974495} (green diamonds), \cite{Redder1987385} (brown stars), \cite{PhysRevC.73.055801} (black circles), and \cite{Fey2004} (blue squares). Note that the data at $E_\alpha$~=~2.28 and 3.677~MeV have been scaled for plotting convenience. The data from \textcite{Fey2004} show a systematic deviation from the $R$-matrix fit and other data sets at backward angles. This is most clearly visible at $E_\alpha$~=~1.740~MeV where the data from \textcite{PhysRevC.73.055801, Fey2004} were measured at the same energy.  \label{fig:ang_dist_big_plot}}
\end{figure*}

\begin{figure}
\includegraphics[width=0.7\columnwidth]{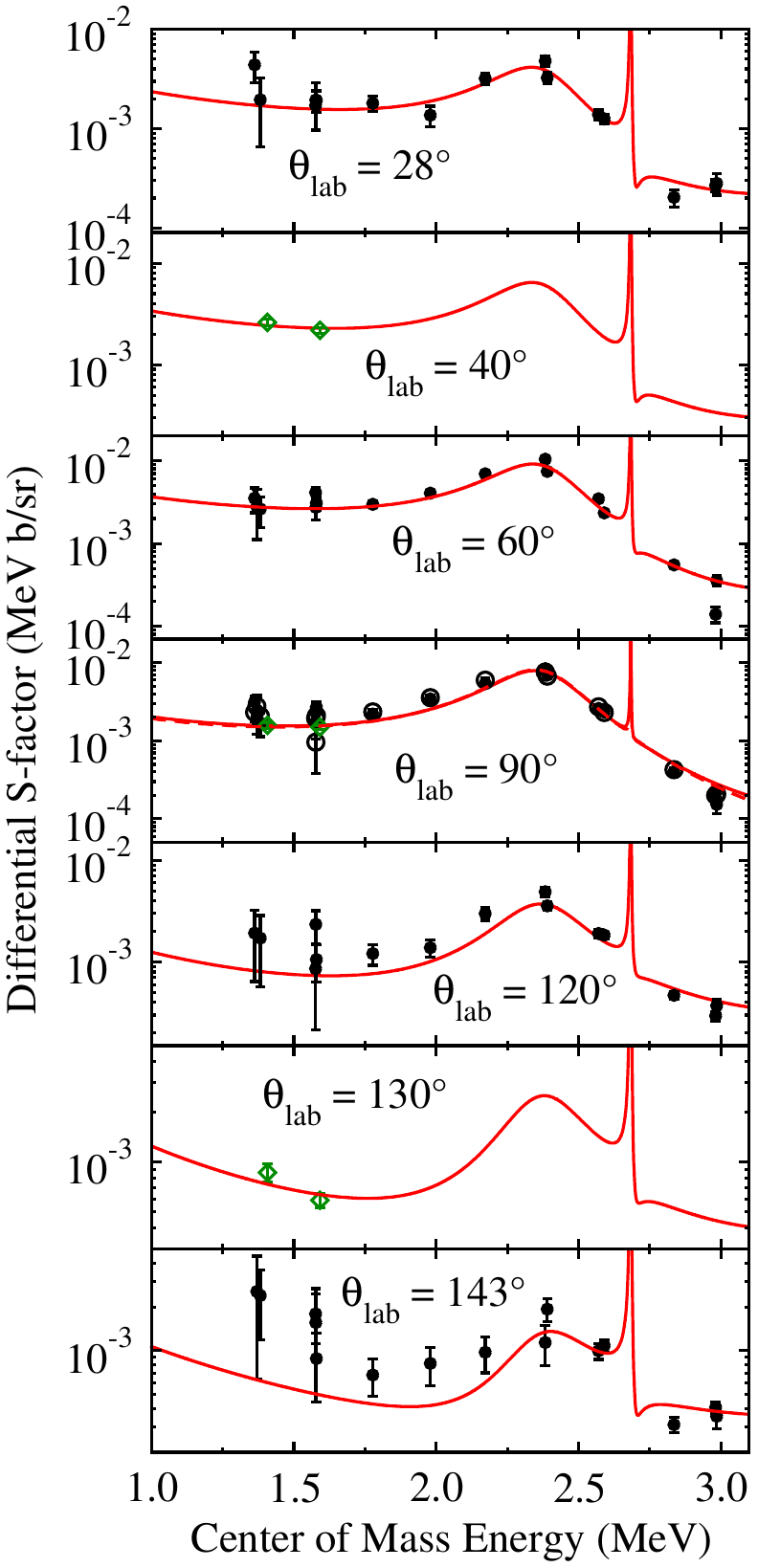}
\caption{(Color online) Fits to the differential excitation curve data of \textcite{PhysRevC.54.1982} (solid black) and \cite{PhysRevC.80.065802} (green diamonds). At $\theta_{\text{lab}}$ = 90$^\circ$ two measurements were made by \textcite{PhysRevC.54.1982} but with different detectors on opposite sides of the target. This second data set is represented by open black points. This is also the only angle where both experiments measured the cross section at the same angle. \label{fig:differential_excitation_functions}}
\end{figure}

The best fit to the $^{12}$C$(\alpha,\gamma_0)^{16}$O angle integrated data of \textcite{Dyer1974495, Redder1987385, Gialanella2001, Roters1999, PhysRevLett.86.3244, PhysRevC.54.1982, Schurmann2011557, PhysRevC.73.055801, Fey2004, PhysRevC.80.065802, springerlink:10.1007/BF01415851, Brochard1973, PhysRevLett.60.1475, PhysRevC.86.015805} is shown in Fig.~\ref{fig:angle_int_12Cag}. The simultaneous fit to the ground state angular distribution differential cross section data \cite{Dyer1974495, Redder1987385, PhysRevC.73.055801, Fey2004} is shown in Fig.~\ref{fig:ang_dist_big_plot} and the differential excitation curves of \textcite{PhysRevC.54.1982, PhysRevC.80.065802} are shown in Fig.~\ref{fig:differential_excitation_functions}. 

\subsection{\label{sec:cascades_12Cag} Cascade Transitions}

While the cascade cross sections make a small contribution to the total low energy cross section ($\approx$5\% at $E_\text{c.m.}$~=~300~keV), at higher energies they can dominate as shown in Figs.~\ref{fig:gs_compare} and \ref{fig:E1vsE2}. However, another compelling reason for their accurate measurement would be to constrain the ANCs of the subtheshold states, in particular those of the $E_x$~=~6.92 and 7.12~MeV states, through their external capture contributions. The $E_x$~=~6.13 MeV transition capture cross section, which is external capture dominated, is also connected to the $\beta$ delayed $\alpha$ emission spectrum through its ANC as discussed further in Sec.~\ref{sec:beta_delayed_alpha}.

\begin{figure}
\includegraphics[width=0.9\columnwidth]{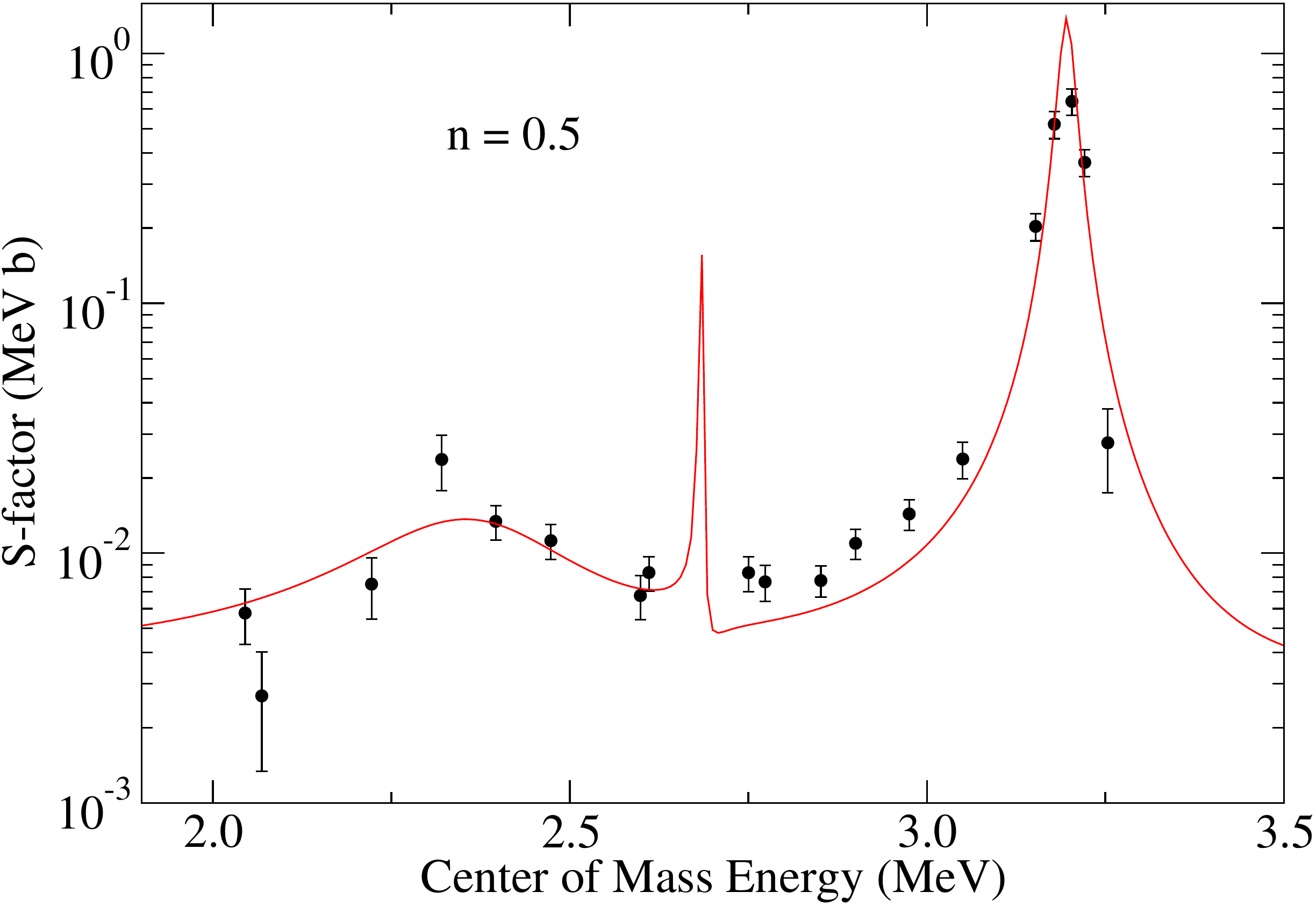}
\caption{(Color online) Data for the sum of the $E_x$ = 6.92 and 7.12 MeV transitions for the $^{12}$C$(\alpha,\gamma)^{16}$O data of \textcite{springerlink:10.1007/BF01415851}. The data have been normalized by a factor of 0.5 to match the $R$-matrix fit. The agreement is reasonably good at low energies but above the narrow resonance corresponding to the 2$^+$ level at $E_x$ = 9.845 MeV the agreement worsens. These data were not included in the global fit. \label{fig:sum_cascades}}
\end{figure}

Cascade transition excitation curves for the $^{12}$C$(\alpha,\gamma)^{16}$O reaction have been measured by \textcite{springerlink:10.1007/BF01415851, Redder1987385, PhysRevLett.97.242503, Schurmann2011557}. The measurement of these transitions are complicated experimentally by the close energy spacing of the bound states at $E_x$~=~6.05~(0$^+$) and 6.13~(3$^-$)~MeV and those at $E_x$~=~6.92~(2$^+$) and 7.12~(1$^-$)~MeV. This can make separating the individual contributions difficult when using Sodium Iodide, Barium Fluoride, or Lanthanum Bromide detectors, which have poorer energy resolution than a Germanium detector. For example, \textcite{springerlink:10.1007/BF01415851} first reported measurements of the transition to the $E_x$~=~6.92~MeV state but in \textcite{Redder1987385}, where Ge(Li) detectors were used, it was clarified that the measurement was actually the sum of the de-excitations through the $E_x$~=~6.92 and 7.12~MeV states. The combined cross section, $E_x$~=~6.92~+~7.12~MeV transition, of \textcite{springerlink:10.1007/BF01415851} is shown in Fig.~\ref{fig:sum_cascades} compared to the sum of the cross sections for each transition from the $R$-matrix fit. While the data is in rough agreement, the shape is distorted in the vicinity of the 4$^+$ level at $E_x$~=~10.36~MeV. Further, while the $E_x$~=~6.92 and 7.12~MeV data of \textcite{Redder1987385} are in agreement as far as their energy dependence, their absolute scale is about a factor of two larger than recent measurements by \textcite{Kunz_thesis} (see Table~\ref{tab:normalizations} and \textcite{Schurmann201235}). Note that the data of \textcite{springerlink:10.1007/BF01415851} are excluded from the fitting. Further measurements of these cascade transitions at low energy are needed in order to understand these discrepancies.

There has been a great deal of interest and contradictory results regarding the cascade cross section measurements of the $E_x$~=~6.05 MeV transition. The measurement of \textcite{PhysRevLett.97.242503} reports an excitation curve for this transition over a wide energy range. Further, in that work an $R$-matrix fit was made to this data that gave a much larger low energy contribution than previously estimated. A subsequent measurement by \textcite{Schurmann2011557} over a more limited higher energy range claimed that their data was inconsistent with that of \textcite{PhysRevLett.97.242503} and their $R$-matrix fit predicted the opposite extreme, that the transition's contribution was lower than previously estimated. A reanalysis of the data of \textcite{Schurmann2011557} by \textcite{PhysRevC.87.015802} found a value in between the two, which was closer to previous estimates. Finally, \textcite{PhysRevLett.114.071101} measured the ANCs of the $E_x$~=~6.05 and 6.13~MeV transitions and arrived at values very close to those deduced by \textcite{PhysRevC.87.015802}. Several details of the analysis of this transition have not been explained fully in the literature. For this reason they are addressed here in some detail. 

The first issue is that the $R$-matrix fit presented by \textcite{PhysRevLett.97.242503} is erroneous. An error was made in the code used for the fit that allowed for a larger $E1$ contribution. Proper calculations show that the large $E1$ component is impossible to reproduce.

\begin{figure}
\includegraphics[width=1.0\columnwidth]{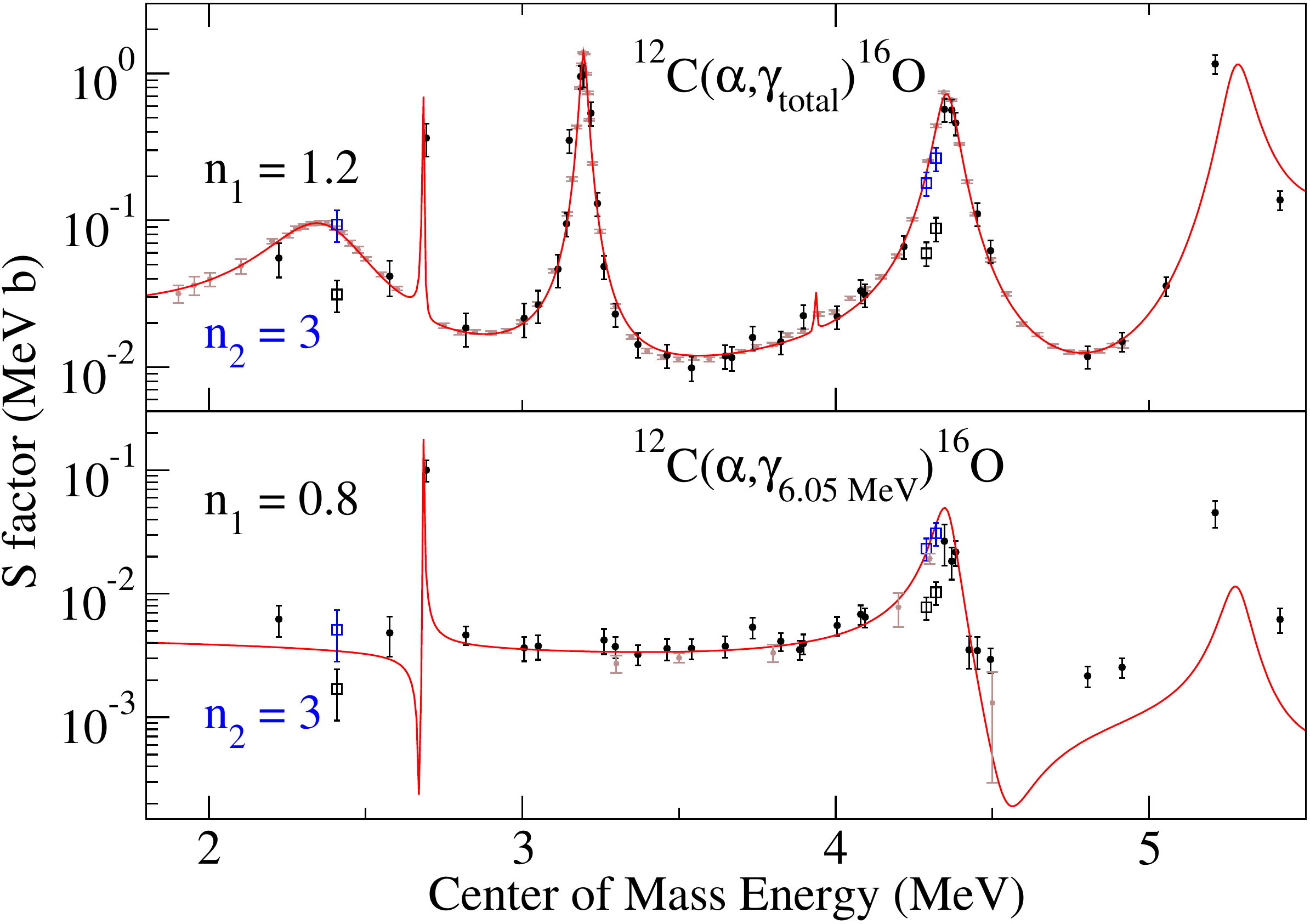}
\caption{(Color online) Comparison of the $^{12}$C$(\alpha,\gamma_{\text{total}})^{16}$O and the $^{12}$C$(\alpha,\gamma_{\text{6.05 MeV}})^{16}$O data of \textcite{PhysRevLett.97.242503, Matei_thesis} (solid black points) and \textcite{schurmann2005, Schurmann2011557} (brown stars) with the global $R$-matrix fit from this work (solid red line). The data have been renormalized by the factors n$_1$ to match the fit. The blue square points represent the further renormalization of data points that are all systematically low by the same amount as discussed in the text. \label{fig:TRIUMF_data}}
\end{figure}

Second, in \textcite{Schurmann2011557} it was shown that their $E_x$~=~6.05~MeV cascade data is inconsistent with that of \textcite{PhysRevLett.97.242503}. This is true if the data are taken at face value. However, the experiment of \textcite{PhysRevLett.97.242503} measured not only the cascade transition to the $E_x$~=~6.05 MeV state, but also the total cross section. Unfortunately this data have never been published and are available only in the thesis of \textcite{Matei_thesis}. A comparison of the fit from this work with both the total and the $E_x$~=~6.05~MeV transition cross section is shown in Fig.~\ref{fig:TRIUMF_data}. The total cross section data from \textcite{Matei_thesis} is not in immediately good agreement with the fit, but becomes so if multiplied by a factor of $\approx$1.2 (n$_1$ in Fig.~\ref{fig:TRIUMF_data}). This normalization factor is well within the quoted experimental systematic uncertainty of +31/-25\%. The data are in excellent agreement with the fit over the entire energy range except for three points at $E_\text{c.m.}$~=~2.41, 4.29, and 4.32~MeV. However, if the cross sections of these three points are all multiplied by a factor of 3 (n$_2$ in Fig.~\ref{fig:TRIUMF_data}), they are in excellent agreement with the total cross section data of \textcite{Schurmann2011557}. 

\begin{figure}
\includegraphics[width=1.0\columnwidth]{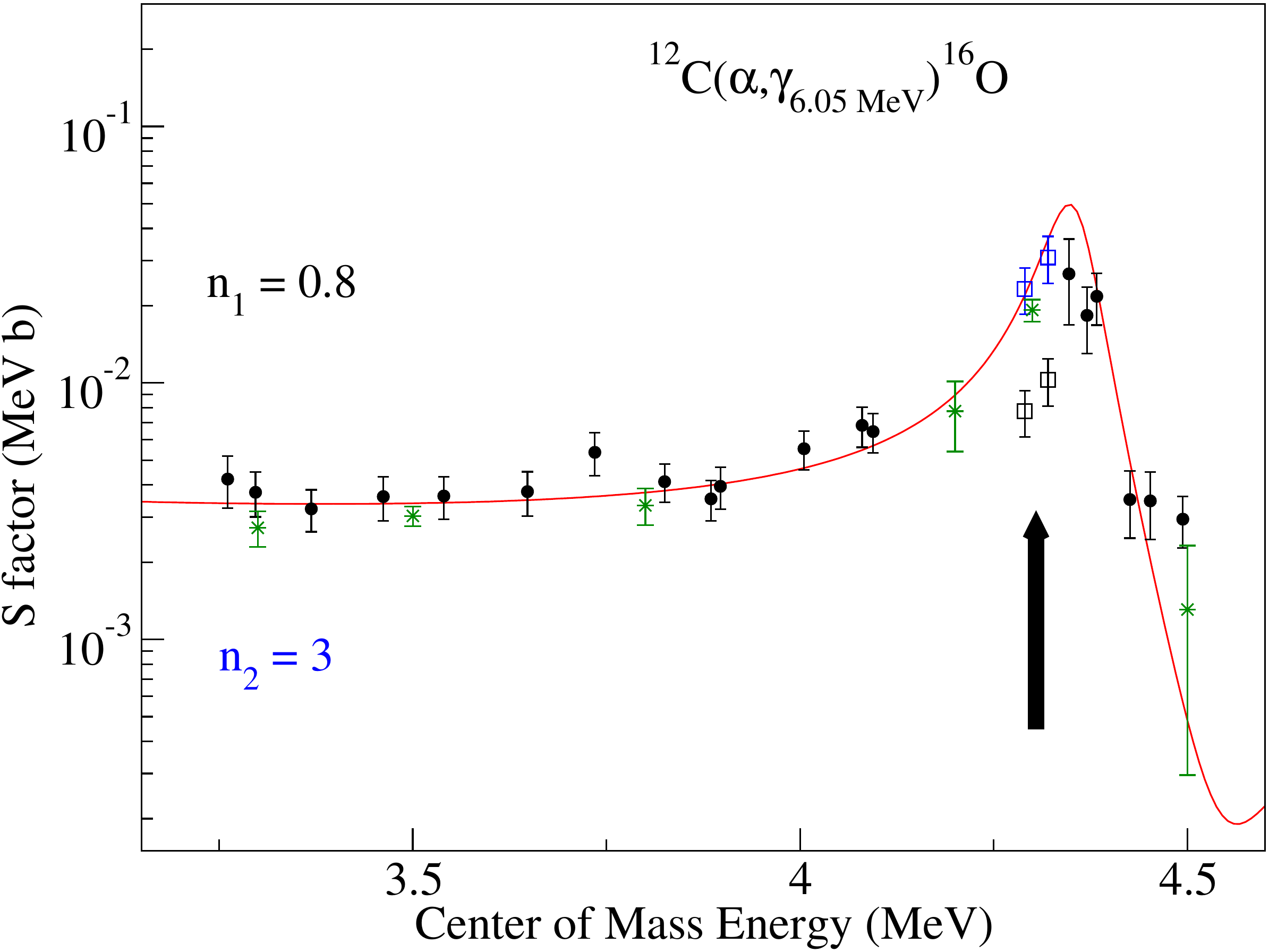}
\caption{(Color online) Comparison between the $E_x$ = 6.05 MeV transition data of \textcite{PhysRevLett.97.242503} and \textcite{Schurmann2011557} over the smaller energy range of the \textcite{Schurmann2011557} data. The data of \textcite{PhysRevLett.97.242503} have been normalized to those of \textcite{Schurmann2011557} with a factor of n$_1$ = 0.8. The cross section of the two data points at $E_\text{c.m.}$ = 4.29 and 4.32 MeV have been scaled up by the same factor of 3 used to normalized the corresponding points in the total cross section as shown in Fig.~\ref{fig:TRIUMF_data}. The original values are indicated by the open black squares, the scaled values by the open blue squares. \label{fig:TRIUMF_zoom-in}}
\end{figure}

\begin{figure}
\includegraphics[width=1.0\columnwidth]{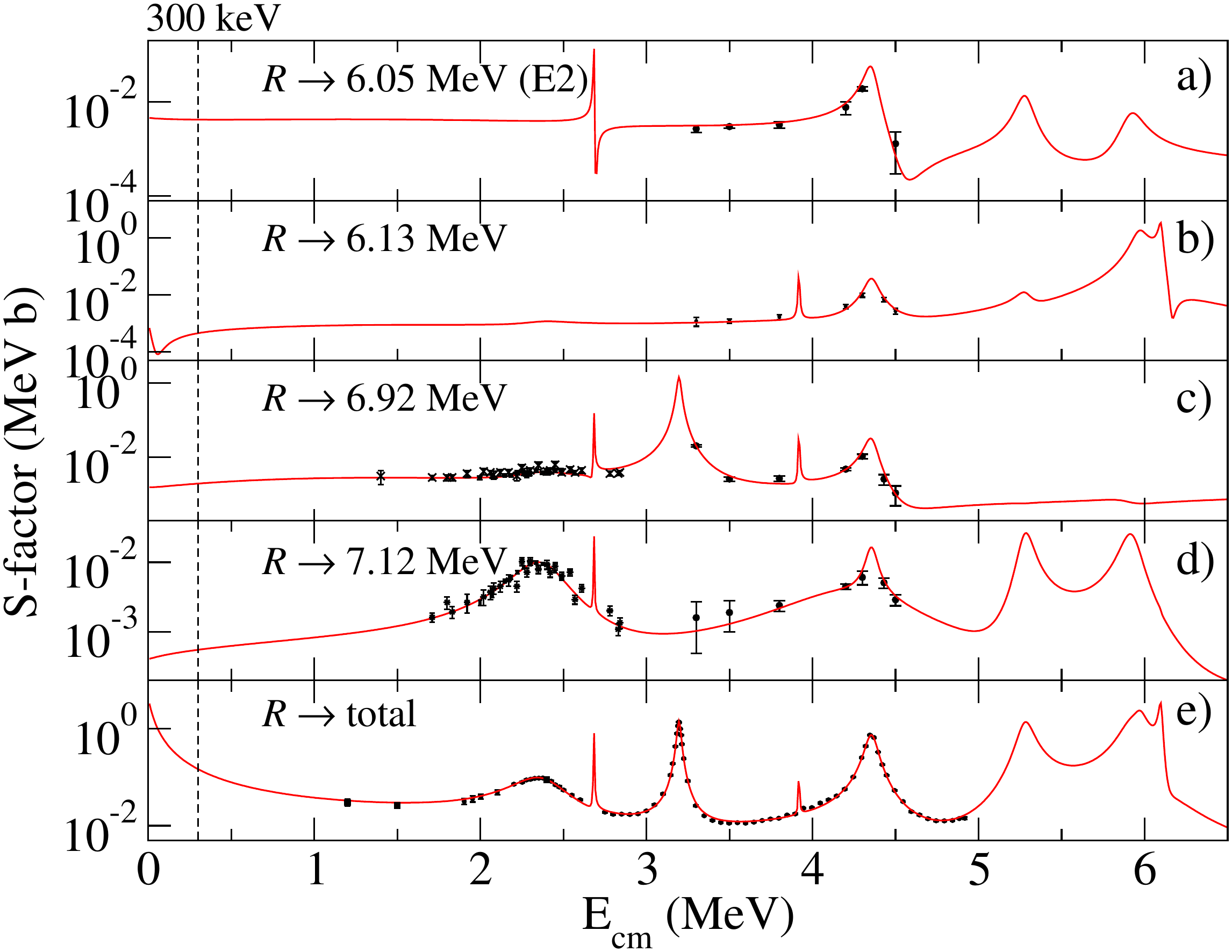}
\caption{(Color online) Transitions for primary $\gamma$ rays to excited states in $^{16}$O for the $^{12}$C$(\alpha,\gamma)^{16}$O reaction are given in Figs.~\ref{fig:cascades} a) through d). The data include those from \textcite{schurmann2005, Schurmann2011557, Schurmann201235, Redder1987385, PhysRevC.86.015805, :/content/aip/proceeding/aipcp/10.1063/1.4874074}. The total $S$ factor is shown in Fig.~\ref{fig:cascades} e). \label{fig:cascades}}
\end{figure}

Turning back to the $E_x$~=~6.05 MeV transition, it was found that the data of \textcite{PhysRevLett.97.242503} deviated from those of \textcite{Schurmann2011557} not only in normalization but also in energy dependence. Albeit the comparison could only be made over the limited energy range of the data of \textcite{Schurmann2011557} (3.5~$<E_\text{c.m.}<$~4.5~MeV, see Fig.~\ref{fig:cascades}). Further, the $R$-matrix analysis of \textcite{Schurmann2011557} indicated a difference in energy dependence over an even wider energy range. Both of these issues can now be addressed. As for the $R$-matrix fit, the energy dependence does not match over the broad energy range because a value of $\gamma_{6.05}$~=~0.1$^{+0.05}_{-0.01}$ (corresponding to an ANC$_{\alpha,6.05\text{ MeV}}$~=~44$^{+270}_{-44}$~fm$^{-1/2}$ ) was chosen, which has subsequently been shown to be much too small by \cite{PhysRevLett.114.071101} (ANC$_{\alpha,6.05\text{ MeV}}$~=~1560$\pm$100~fm$^{-1/2}$). Since external capture dominates the $E_x$~=~6.05 MeV transition cross section over much of the energy region (see Fig.~\ref{fig:components}), this caused a significant difference in the cross section. In \textcite{Schurmann2011557}, the large systematic uncertainty of \textcite{PhysRevLett.97.242503} was not taken into account and instead the difference in the two data sets was attributed to the data of \textcite{PhysRevLett.97.242503} being the sum of the $E_x$~=~6.05 and 6.13~MeV transitions. However, the experimental technique used in \textcite{PhysRevLett.97.242503} uses a coincidence of $\gamma$ rays with recoils. Since the $\gamma$ ray decay from the $E_x$~=~6.05~MeV state to the ground state (0$^+ \rightarrow$ 0$^+$) is strictly forbidden, this explanation seems impossible. If instead the data of \textcite{PhysRevLett.97.242503} are normalized to the data of \textcite{Schurmann2011557} in the off-resonance region over the range from 3.25 $<E_\text{c.m.}<$ 4.10 MeV (normalization factor of 0.8), the agreement is much improved. The remaining deviation occurs for just two points, which happen to fall at $E_\text{c.m.}$~=~4.29 and 4.32~MeV, the vary same points that were found to be too low in the total cross section data. If these points are subjected to the same normalization factor as those in the total cross section, the data are in much better agreement as shown in Fig.~\ref{fig:TRIUMF_zoom-in}. Regarding the low energy point at $E_\text{c.m.}$~=~2.41~MeV, its agreement is also much improved with the energy dependence of the $R$-matrix fit if subjected to this same normalization factor.

Summarizing the situation for the $E_x$~=~6.05 MeV transition, the data of \textcite{PhysRevLett.97.242503} and \textcite{Schurmann2011557} are in generally good agreement if the systematic uncertainties are considered. There are a few points in the data of \textcite{PhysRevLett.97.242503} that appear to have some unconsidered systematic shift in their absolute scale compared to the rest of the data set. Attempts were made to re-examine the log books of the experiment but no correlation between these data points could be established. It was found that no significant $E1$ contribution was necessary to fit the data and that {\bf the low energy cross section is dominated by $E2$ external capture} (see Fig.~\ref{fig:components}) in contradiction to the recent reassertion in \textcite{PhysRevC.92.045802} that this cross section is $E1$ dominated. Finally, the ANCs measured in \textcite{PhysRevLett.114.071101} are found to be in good agreement with the capture data. 


The $R$-matrix fit and the cascade data included in the global fit are shown in Fig.~\ref{fig:cascades}. It was found that the cascade data of \textcite{Kunz_thesis} ($E_x$~=~6.92 and 7.12~MeV transitions) require normalization factors of $\approx$0.5 while those of \textcite{Redder1987385} require values of $\approx$0.25 (see Table~\ref{tab:normalizations}). The normalization is somewhat unexpected since this was not required in the fit of \textcite{Schurmann201235}. This may be the result of the different ANCs used in this analysis. In addition, the value of the absolute normalization for the cascade data is highly sensitive to the normalization factor of the total cross section data. If the normalization of the total cross section data are increased by a few percent, well with in their systematic uncertainty of 6.5\%, more consistent normalizations can be obtained for the $E_x$~=~6.92 and 7.12~MeV transition data of \textcite{Kunz_thesis}. However, at low energies the external capture dominates the cross section for both of these transitions and the data are only over the resonance region corresponding to the $E_x$~=~9.59 MeV state. Hence the fit to the data over this energy region have little effect on the extrapolation to stellar energies.  

A detail that is only briefly mentioned in \textcite{Schurmann201235} is that their data gives the first evidence for $\gamma$ ray decays from the 3$^-$ state at $E_x$~=~11.49 MeV. In this analysis evidence is found for decays to the $E_x$~=~6.13~($M1$/$E2$), 6.92~($E1$), and 7.12~($E2$) MeV final states. The lowest order multipolarity has been assumed except for the $E_x$~=~6.13~MeV transition. No evidence for the $E3$ decay through the $E_x$~=~6.05 MeV transition is observed in the data. The widths of these states can be found in Table~\ref{tab:fit_params_gamma}.

\subsection{\label{sec:beta_delayed_alpha} \texorpdfstring{$\beta$ Delayed $\alpha$ Emission}{beta Delayed alpha Emission}}

There have been several measurements of the $\beta$ delayed $\alpha$ emission spectrum of $^{16}$N as discussed in Sec.~\ref{sec:data}. However, the different spectra are can not be compared directly since all of them suffer from experimental resolution effects to different degrees. This point was the subject of an analysis by \textcite{0954-3899-24-8-042} where it was determined that the spectra of the Yale \cite{FranceIII1997165}, Seattle (unpublished), and Mainz \cite{PhysRevLett.25.941} groups were inconsistent with that of TRIUMF \cite{PhysRevC.50.1194}. This began a long debate regarding the consistency between the data sets that still remains unresolved. 

More recently, \textcite{PhysRevC.80.045803} performed another comparison between the different data sets where detailed \texttt{GEANT} simulations were created for several of the past setups. The Monte Carlo simulations were used to simulate the observed $\alpha$ spectra. Many different deconvolution effects and background sources were investigated for the Mainz, Seattle, Yale, Argonne and TRIUMF data sets. In the end, the main conclusion was that the different spectra were very sensitive to the deconvolution technique that was used. In particular, energy loss effects through the catcher foils were found to be a main factor. The claim was also made that the deconvolution technique used by \textcite{PhysRevC.75.065802} was incorrect and that this data set may in fact be in agreement with those of \textcite{PhysRevC.50.1194, PhysRevLett.25.941}. Unfortunately another conclusion of the work was that not enough experimental information is available to properly analyze much of the data with the degree of confidence that is desired. It remains a challenge for future measurements to produce consistent results for the measurement of the $^{16}$N$(\beta,\alpha)^{12}$C spectrum.

For the current analysis, because of the above issues, the data are limited to that of \textcite{PhysRevC.50.1194, PhysRevC.81.045809}. These data sets have the most detailed documentation of how to simulate their remaining resolution effects and claim that they can be accurately done so using the simple method given by Eq.~(\ref{eqn:conv}). In addition to the usual convolution term, it has also been recommended that each spectrum be shifted in energy in order to agree with the more easily determined energy calibration of the scattering data. These correction factors are summarized in Table~\ref{tab:beta_conv}.

\begin{table}
\caption{\label{tab:beta_conv} Summary of convolution and energy shift corrections for $^{16}$N$(\beta\alpha)^{12}$C spectra for \textcite{PhysRevC.50.1194, PhysRevC.81.045809}. For the other measurements, the convolution parameters are estimated from the best fits to the data of \textcite{PhysRevC.50.1194, PhysRevC.81.045809}.}
\begin{ruledtabular}
\begin{tabular}{ c c c }
Ref. & $\sigma$ (keV) & $\Delta E$ (keV) \\ \hline 
\textcite{PhysRevC.50.1194} (TRIUMF) & 30 & -5 \\
\textcite{PhysRevC.81.045809} (Argonne) & 40 & -3.75 \\ \hline
\textcite{PhysRevC.75.065802} (Yale) & 100 & 0 \\
\textcite{PhysRevLett.25.941} (Mainz) & 40 & -10 \\
unpublished (Seattle) & 40 & -5 \\
\end{tabular}
\end{ruledtabular}
\end{table} 

\begin{figure}
\includegraphics[width=1.0\columnwidth]{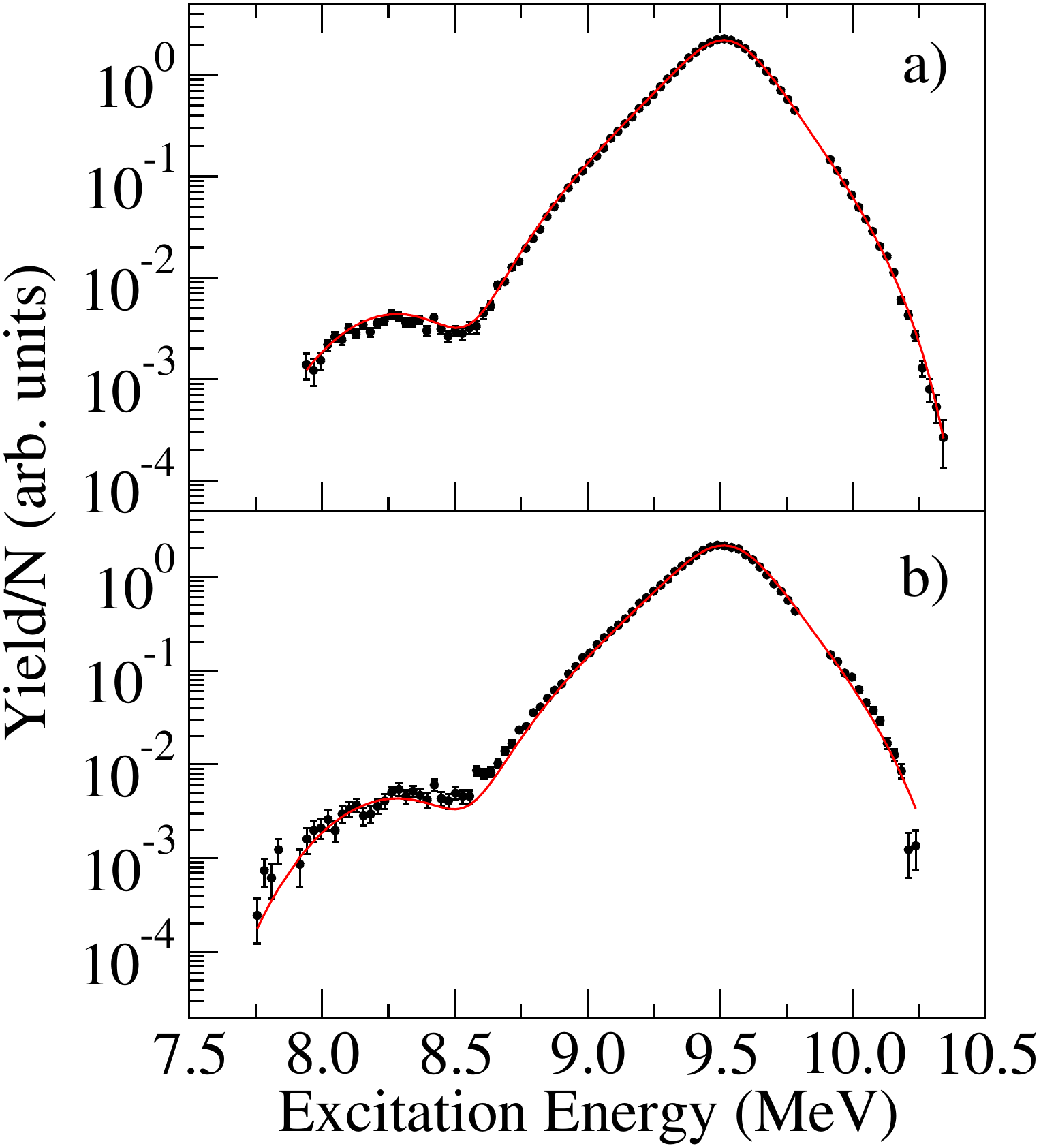}
\caption{(Color online) Simultaneous fit including the $^{16}$N$(\beta\alpha)^{12}$C data of both \textcite{PhysRevC.50.1194} a) and \textcite{PhysRevC.81.045809} b). In the fitting procedure the ANCs were fixed to the values of the transfer measurements. The solid red line represents the $R$-matrix cross section convoluted with the resolution function given by Eq.~\ref{eqn:target_effs} and the specific parameters given in Table~\ref{tab:beta_conv}. \label{fig:beta_delayed_alpha_with_bgp}}
\end{figure}

\begin{figure}
\includegraphics[width=1.0\columnwidth]{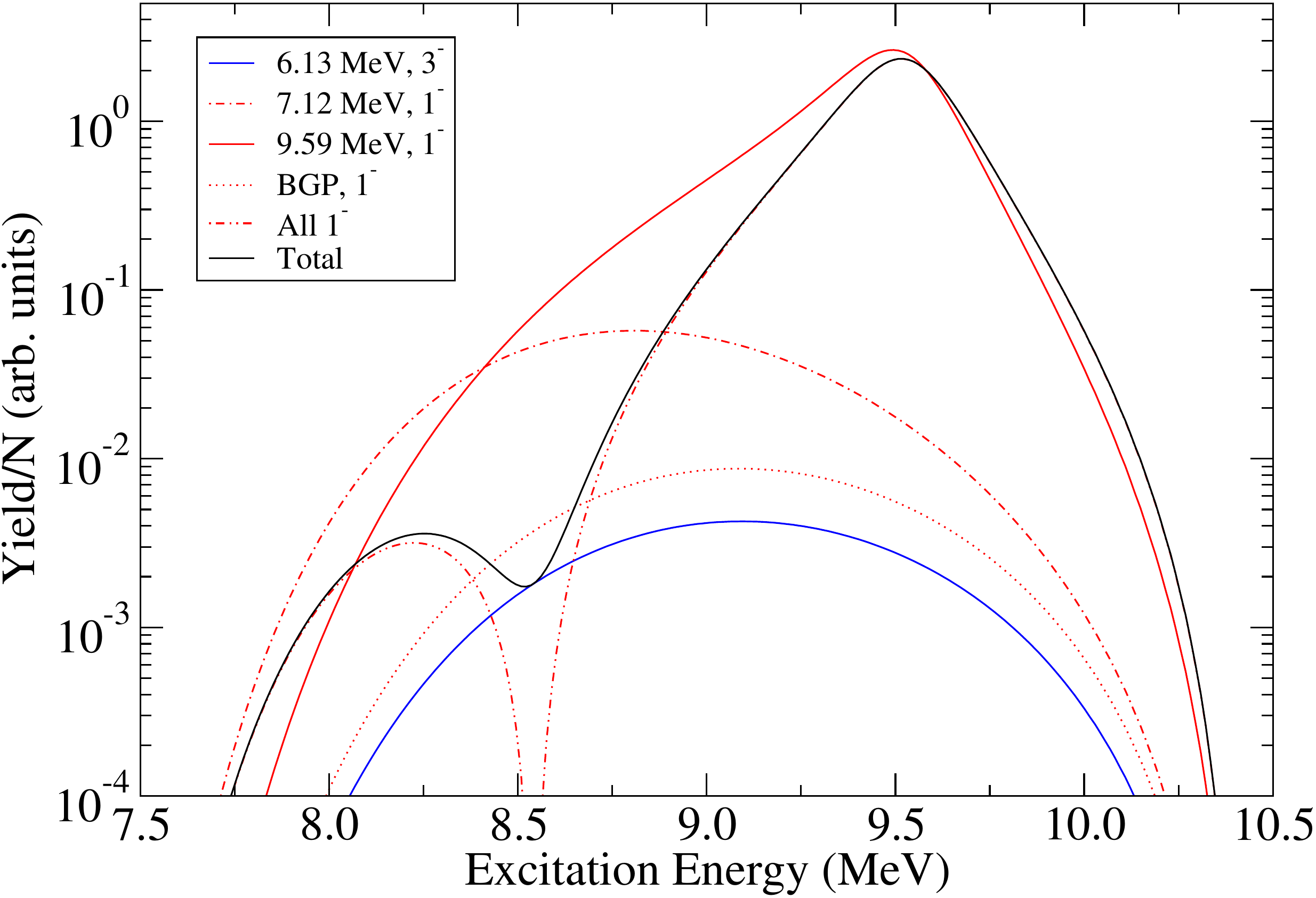}
\caption{(Color online) Components of the $R$-matrix fit to the $^{16}$N$(\beta\alpha)^{12}$C data of \textcite{PhysRevC.50.1194, PhysRevC.81.045809}. The different red curves (except the dashed-dot-dotted line) represent the 1$^-$ components of the fit, which add coherently to give the interference pattern shown by the red dashed-dot-dotted line. The individual 1$^-$ contributions come from the $E_x$ = 7.12 MeV, 9.59 MeV, and background pole (BGP) at 20 MeV. An additional 3$^-$ component adds incoherently to give the total yield, which is given by the solid black line. Note the similar energy dependence of the subthreshold state and BGP contributions.  \label{fig:components_beta}}
\end{figure}

        
A fit to both of these $^{16}$N$(\beta,\alpha)^{12}$C data sets simultaneously, along with the rest of the data from other channels, is shown in Fig.~\ref{fig:beta_delayed_alpha_with_bgp}. The fit includes contributions from the 1$^-$ levels at $E_x$~=~7.12 and 9.59~MeV, the 3$^-$ level at $E_x$~=~6.13 MeV. Additionally, as in both \textcite{PhysRevC.50.1194, PhysRevC.81.045809}, a 1$^-$ background pole (at $E_x$~=~20 MeV) is included. The different components for the fit are shown in Fig.~\ref{fig:components_beta}.

While the quality of the fit looks reasonable by eye, the small error bars reveal significant differences between the two data sets even after the convolution correction to the $R$-matrix curve. The resulting reduced $\chi^2$ of the fit is rather large as detailed in Table~\ref{tab:beta_simul}. However the resulting $\log(ft_{1/2})$ values are in reasonable agreement with those given in the compilation, with the exception of the 3$^-$ subthreshold state that is too large by about 3 standard deviations.

\begin{table*}
\caption{Summary of results for fits considering different $^{16}$N$(\beta\alpha)^{12}$C data. The half-life of $^{16}$N was taken as $t_{1/2}$ = 7.13(2)~s with a $\beta\alpha$ branching ratio of 1.20(5)$\times$10$^{-5}$ \cite{Tilley19931}. } \label{tab:beta_simul}
\begin{ruledtabular}
\begin{tabular}{ c c c c c } 
 & & & \multicolumn{2}{ c }{$\log(ft_{1/2})$} \\ \cline{4-5}
$\tilde{E}_x$ (MeV) & $J^\pi$ & ANC$_\alpha$(fm$^{-1/2}$) or $\tilde{\Gamma}_\alpha$ (keV) & \textcite{PhysRevC.50.1194, PhysRevC.81.045809} & \textcite{Tilley19931}\\
\hline
6.13 & 3$^-$ & 139(9) & 4.59 &  4.48(4) \\
7.12 & 1$^-$ & 2.08(17)$\times$10$^{14}$ & 5.08 & 5.11(4) \\
9.59 & 1$^-$ & 382 & 6.15 & 6.12(5) \\
20 (BGP) & 1$^-$ & 15600 & -5.70\footnotemark[1] & \\
\hline
$\chi^2$ (N) & &  & 496 (87), 793 (88) & \\
\end{tabular}
\end{ruledtabular}
\footnotetext[1]{The minus sign indicates the sign of the interference on the reduced width amplitude.}
\end{table*} 

To investigate the differences further, the spectra were fit independently (with the rest of the data). The $\chi^2$ value of the \textcite{PhysRevC.81.045809} data decreased substantially, demonstrating the tension between the two data sets. It is interesting to note that the $\chi^2$ of the fit to the data of \textcite{PhysRevC.50.1194} actually becomes larger. This is caused by the tension between the data in the other channels. The results are given in Table~\ref{tab:beta_compare2}.

\begin{table*}
\caption{Summary of fits to the $^{16}$N$(\beta\alpha)^{12}$C data but for each spectrum individually. \label{tab:beta_compare2}}
\begin{ruledtabular}
\begin{tabular}{ c c c c c c } 
 & & & \multicolumn{3}{ c }{$\log(ft_{1/2})$} \\ \cline{4-6}
$\tilde{E}_x$ (MeV) & $J^\pi$ & ANC$_\alpha$(fm$^{-1/2}$) or $\tilde{\Gamma}_\alpha$ (keV) & \textcite{PhysRevC.50.1194} & \textcite{PhysRevC.81.045809} & \textcite{Tilley19931}\\
\hline
6.13 & 3$^-$ & 139(9) & 4.44 & 4.80 & 4.48(4) \\
7.12 & 1$^-$ & 2.08(17)$\times$10$^{14}$ & 5.06 & 5.06 & 5.11(4) \\
9.59 & 1$^-$ & 382 & 6.13 & 6.18 & 6.12(5) \\
20 (BGP) & 1$^-$ & 15600 & -5.73 & -6.53 & \\
\hline
$\chi^2$ (N) & & & 519 (87) & 466 (88) & \\
\end{tabular}
\end{ruledtabular}
\end{table*} 

What is most interesting is the large difference in the $\log(ft_{1/2})$ values of the 3$^-$ state that result from the individual fits. This is caused by the difference in the data in the vicinity of the interference region around $E_x \approx$~8.5~MeV. With the $\alpha$ ANC fixed, the fit attempts to compensate this difference by changing the $\log(ft_{1/2})$ value. Given the deconvolution methods used, it is clear that the data of \textcite{PhysRevC.50.1194} are in the best agreement with the ANCs of \textcite{PhysRevLett.114.071101}.

\begin{figure}
\includegraphics[width=1.0\columnwidth]{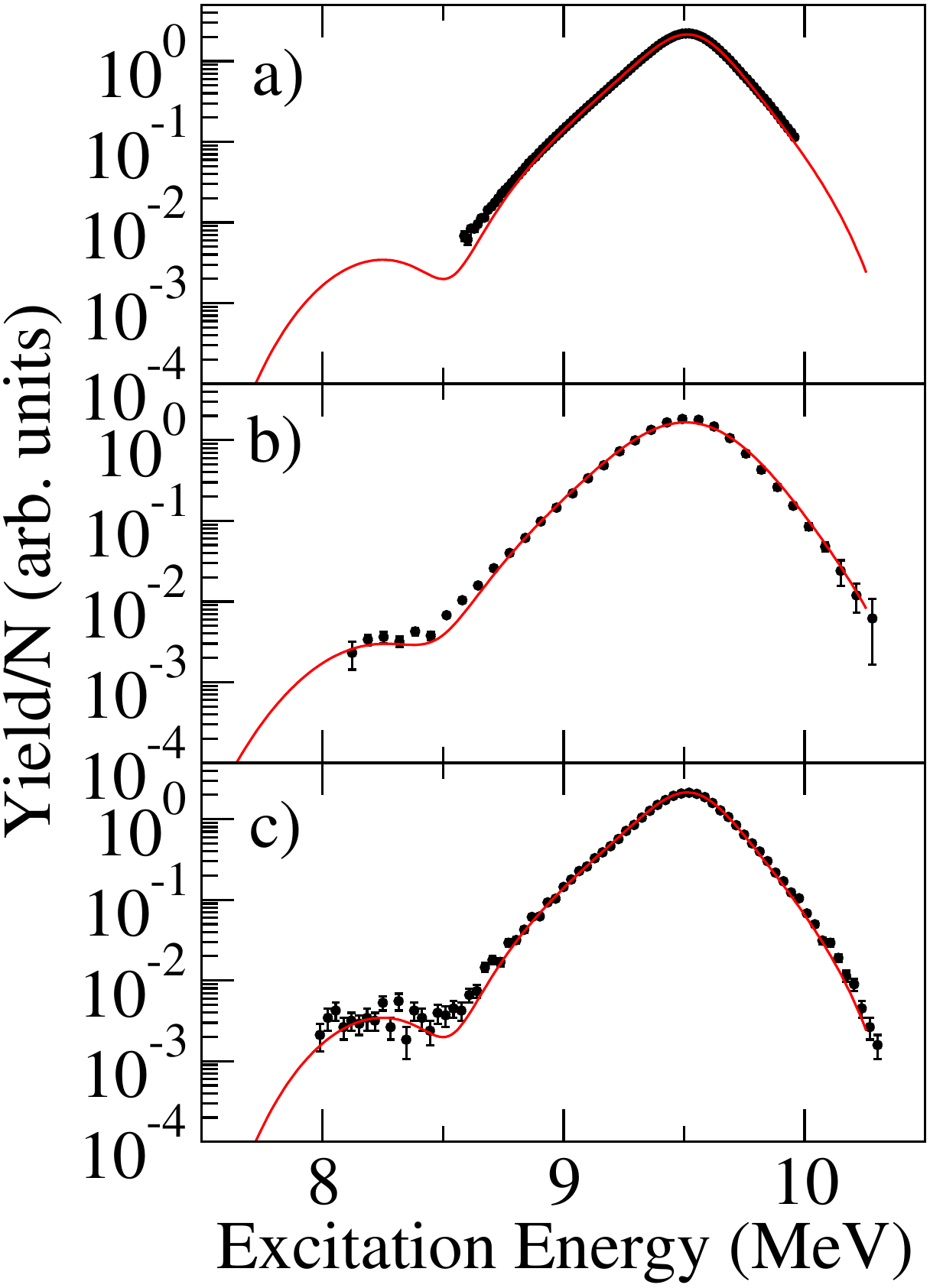}
\caption{(Color online) Comparison of the $^{16}$N$(\beta\alpha)^{12}$C spectra of Mainz a), Yale b), and Seattle c) to the best fit to the data from \textcite{PhysRevC.50.1194} (TRIUMF) and \textcite{PhysRevC.81.045809} (Argonne), but with the unknown convolution parameters varied to best match the data. \label{fig:beta_others}}
\end{figure}

A comparison is made using the best fit parameters from the simultaneous fit to the data of \textcite{PhysRevC.50.1194, PhysRevC.81.045809} with the remaining data sets from the Mainz, Seattle, and Yale experiments, where the data have been taken from \textcite{PhysRevC.75.065802}. Since these three data sets retain significant experimental effects, the convolution value of each data set was varied while the $R$-matrix parameters where kept constant. The resulting calculations are shown in Fig.~\ref{fig:beta_others} and the convolution parameters are given in the bottom half of Table~\ref{tab:beta_conv}. The data are reasonably reproduced by the convolution effects except near the interference dip around $E_x$~=~8.5~MeV. Here the fit under predicts the data for all three data sets. This seems to indicate a better agreement with the data of \textcite{PhysRevC.81.045809}. 

These seeming contradictions in the data, that seem to place the current data into two camps, are a critical puzzle that needs to be resolved. It seems that this can only be done through future measurements considering the number of re-analysis that have been unable to resolve the issues. The $^{16}$N$(\beta\alpha)^{12}$C spectrum could provide one of the most stringent constraints on several of the level parameters critical for the determination of the $E1$ ground state cross section. However, with these unresolved experimental effects and contradictions between different data sets, the uncertainties are substantially increased and are difficult to quantify.

As a final point, the analysis of the $^{16}$N$(\beta\alpha)^{12}$C spectrum utilizes $\beta$ decay branching ratios from the literature to constrain the values of the $R$-matrix $\beta$ decay fit parameters. In particular, these are the branching ratios to the $E_x$~=~6.13~(3$^-$), 7.12~(1$^-$), and 9.59~(1$^-$)~MeV levels in $^{16}$O. While the compilation \cite{Tilley19931} reports these values with small uncertainties, it is unclear where some of them actually originated. Most of the branching ratios date back to measurements at Brookhaven national laboratory from the 1950's (\textcite{PhysRev.116.939} and references therein). While the fact that these are older measurements does not mean they are incorrect, verification studies seem overdue since they play a rather important role in the analysis of the $^{16}$N$(\beta\alpha)^{12}$C spectrum. 

\subsection{$\alpha$ Scattering \label{sec:scat}}

In Sec.~\ref{sec:data} it is described how several scattering experiments \cite{Clark1968481, dagostino, PhysRevC.86.015805, PhysRevC.79.055803} have been performed in order to constrain the values of the ANCs of the subthreshold states as well as the particle widths of the unbound states. Here only the data of \textcite{PhysRevC.79.055803} (the higher energy analysis was worked out subsequently in \textcite{deboer2012}) are considered as they represent the most precise and accurate measurement, cover the broadest energy range, and are in good agreement with previous studies. In the present analysis, the original yield ratio data are fit directly instead of the phase shifts extracted in \textcite{PhysRevC.79.055803}. 

Some previous analysis attempted to limit the complexity of the fitting problem by only considering individual phase shifts. In particular, only the $s$ and $d$-waves since they are the only partial waves directly connected to the $E1$ and $E2$ capture cross sections. However, \citet{PhysRevC.54.393} realized that this was poor practice since it neglected the statistical correlations inherent in the full set of phase shifts. This also extends to the propagation of the uncertainties from the yield ratios to the phase shifts. 

However, fitting to the full set of the original data comes at a price. While the phase shift data only consist of 2814 (402 energies, $l$ = 0 to 6) data points there are 12864 yield ratio data points since the measurements were made at 32 different angles. In addition, the $R$-matrix cross section must be convoluted with the function given by Eq.~(\ref{eqn:target_effs}) to correct for beam energy resolution and energy loss through the target. This results in a significant increase in computation time. In order to avoid large convolution corrections, the data in the vicinity of narrow resonances, which are not fit in the analysis, have been neglected. This limits the data in the present analysis to 304 energies or 9728 data points. An example excitation yield ratio curve of this data is shown in Fig.~\ref{fig:level_diagram}.

In principle the scattering data can provide significant constraint on the value of the ANCs of the subthreshold states (see Sec.~\ref{sec:data}). At the current level of precision, the data only constrain the ANCs of the 2$^+$ and 1$^-$ states, which are closest to threshold. Since the ANC of the 1$^-$ state is already strongly constrained by the $^{16}$N$(\beta\alpha)^{12}$C data, the goal of recent scattering measurements has been to better constrain that of the 2$^+$ state (see Fig.~\ref{fig:S_0_summary_2}). However, as mentioned in Secs.~\ref{sec:modern_era} and \ref{sec:fitting}, the large background poles that are necessary to fit the data also tend to lessen the constraint of the fit on the ANCs. This is because both contributions to the cross section have a similar energy dependence, hence the ANCs and the parameters of the background poles are strongly correlated.

\begin{figure}
\includegraphics[width=1.0\columnwidth]{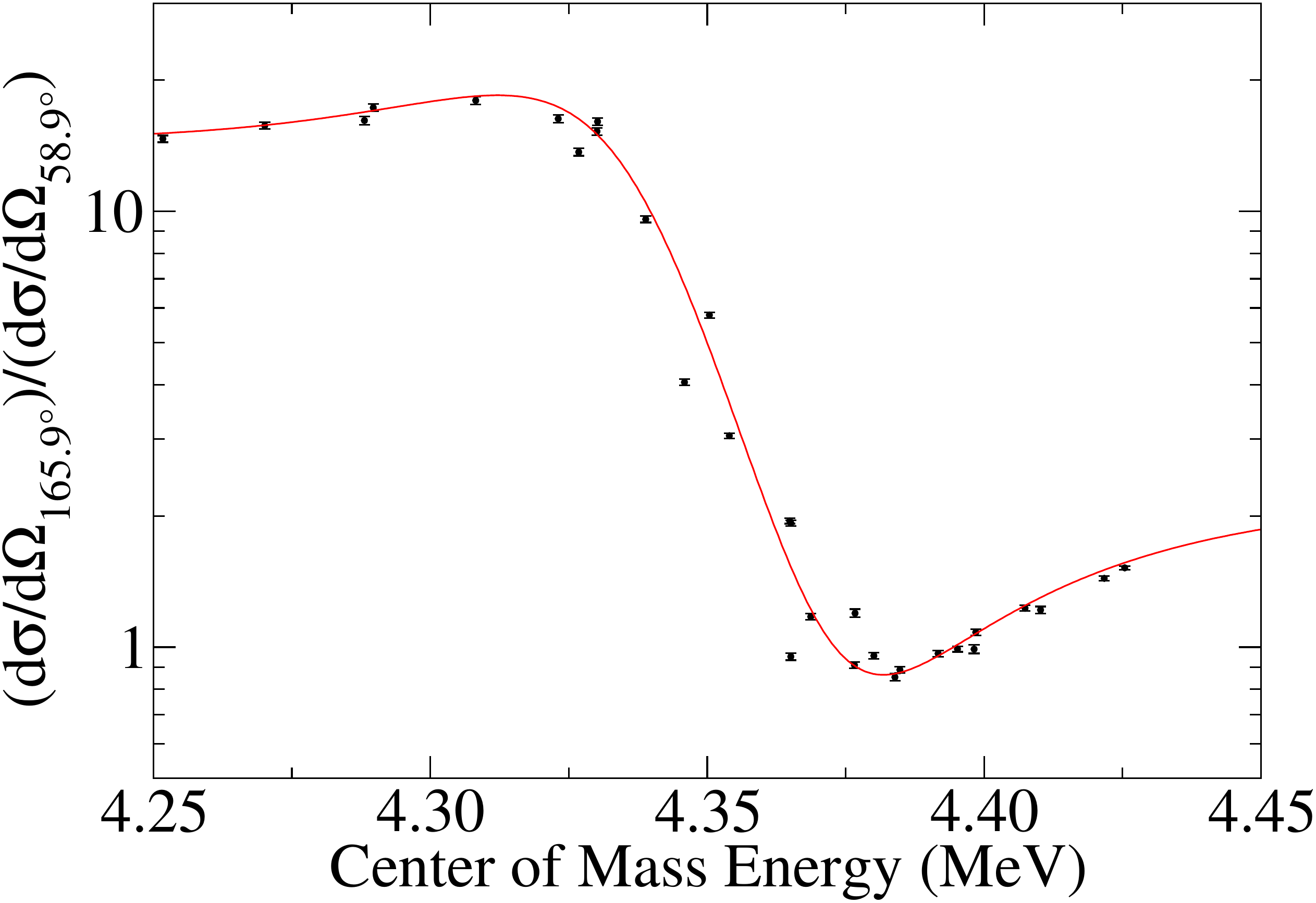}
\caption{(Color online) Sample region of the scattering data of \textcite{PhysRevC.79.055803} in the vicinity of the resonances that correspond to the levels at $E_x$~=~11.49~(3$^-$) and 11.51~(2$^+$)~MeV. The cross section ratio changes rapidly with energy, causing the effect of energy uncertainties on the $\chi^2$ to be amplified. \label{fig:scattering_narrow_res}}
\end{figure}

The present fit to the scattering data does not result in as small of a $\chi^2$ (see Table~\ref{tab:normalizations}) as that achieved in \textcite{PhysRevC.79.055803}. The main difference is that in the present global fit, there is tension between the scattering data and other data sets for the values of the energies, widths and ANCs. In particular, this tension has a very large effect on the $\chi^2$ of the fit in the vicinity of the $E_x$~=~10.36~MeV (4$^+$) and the $E_x$~=~11.49 (3$^-$) and 11.51~MeV (2$^+$) doublet. In these regions the scatting cross section changes rapidly in energy and the uncertainties on the yield ratio data are about 2\%. Further, there are some points that show significant deviations from the expected cross section ratios over these regions as shown in Fig.~\ref{fig:scattering_narrow_res}. Since this effect is less pronounced in other more slowly varying cross section regions this may be attributed to an unaccounted for energy uncertainty that is not reflected in the yield uncertainties.

\subsection{Subthreshold States \label{sec:subthreshold}}

In the previous sections the importance of the bound states of $^{16}$O have been stressed. This section is devoted to a discussion of the current understanding of these states. The parameters that are of interest are the energies, $\gamma$ widths, and ANCs of the five bound states of $^{16}$O.

The energies of all of the bound states are known with a precision of at least 1~keV \cite{Tilley19931}. These uncertainties propagate into small uncertainties in the cross sections. Additionally, the separation energies of $^{16}$O (listed in Table~\ref{tab:r_const}) all have uncertainties that are less than a keV.

All of the bound states of $^{16}$O decay with nearly 100\% probability to the ground state. Except for the first excited 0$^+$ state at $E_x$~=~6.05~MeV, the primary decay mode is single $\gamma$ ray emission. Because both the ground state and first excited state of $^{16}$O are 0$^+$, the state cannot decay by single $\gamma$ emission. Its primary decay mode is therefore through $e^+e^-$ emission (or $\pi$ decay) \cite{PhysRev.56.840}. The $\gamma$ ray widths of the bound state levels are necessary in order to calculate the subtheshold state capture contribution through different transitions. While the ground state $\gamma$ decays through the $E_x$~=~6.92 and 7.12~MeV subthreshold states have the most significant contribution to the total capture cross section, small branching to other excited states make contributions to some of the cascade cross sections.
  
Several measurements have been made to investigate the ground state $\gamma$ widths of the $E_x$~=~6.92 and 7.12~MeV subthreshold states by way of inelastic electron scattering \textcite{Miska1975155, Stroetzel1968376} and nuclear resonance florescence \textcite{PhysRevC.31.2314, Evers1968423, Swann1970300}. The values are in relatively good agreement and the compilation \cite{Tilley19931} adopts a weighted average of all the past measurements. This results in uncertainties of 3.1\% and 5.4\% for the $\gamma$ widths of the $E_x$~=~6.92 and 7.12~MeV states respectively. For the latest discussion see \textcite{PhysRevC.31.2314}. 

Past measurements of the different cascade $\gamma$ rays have been made by \textcite{Miska1975155, PhysRev.173.995, Fuchs1965638, PhysRev.163.1060}. A recent measurement by \textcite{PhysRevC.78.065801} investigated the branching ratios of the decay from $E_x$~=~7.12~MeV state, giving a more accurate value for the branching to the $E_x$~=~6.13~MeV state and an upper limit to the $E_x$~=~6.92~MeV state. The $\gamma$ ray decay widths that lead to significant subthreshold contributions to the cross section have been measured to a precision that is smaller than or competitive with other uncertainties as summarized in Table~\ref{tab:gamma_widths}.

\begin{table*}
\caption{Summary of the known $\gamma$ decay widths for the bound states of $^{16}$O. Where values are averaged, the compilation is quoted. \label{tab:gamma_widths}}
\begin{ruledtabular}
\begin{tabular}{ c c c c c } 
$E_i$ ($J^\pi$) (MeV) & $E_f$ (MeV) & $\Pi L$ & $\Gamma_\gamma$ (eV) & Ref.\\
\hline
6.05 (0$^+$) & G.S. & $E0$ & forbidden & \\
\hline
6.13 (3$^-$) & G.S. & $E3$ & 2.60(13)$\times$10$^{-5}$ & \textcite{Miska1975155} \\
             & 6.05 & $E3$ & unobserved & \\
\hline
6.92 (2$^+$) & G.S. & $E2$ & 0.097(3) & \textcite{Tilley19931} \\
             & 6.05 & $E2$ & 2.7(3)$\times$10$^{-5}$ & \textcite{Tilley19931} \\
             & 6.13 & $E1$ & \footnotemark[1]$<$7.8$\times$10$^{-6}$ & \textcite{PhysRev.173.995} \\
\hline
7.12 (1$^-$) & G.S. & $E1$ & 0.055(3) & \textcite{Tilley19931} \\
             & 6.05 & $E1$ & $<$3.3$\times$10$^{-7}$ & \textcite{PhysRev.163.1060} \\
             & 6.13 & $E2$ & \footnotemark[1]4.6(3)$\times$10$^{-5}$ & \textcite{PhysRevC.78.065801} \\
             & 6.92 & $E1$ & \footnotemark[1]$<$1.1$\times$10$^{-6}$ & \textcite{PhysRevC.78.065801} \\
\end{tabular}
\end{ruledtabular}
\footnotetext[1]{Calculated from branching ratio and $\Gamma_{\gamma 0}$.}
\end{table*} 

At present, what seems to be the most reliable method of determining the $\alpha$ ANCs of bound states is via Sub-Coulomb transfer reactions (see Sec.~\ref{sec:data}). Recent measurements of the $\alpha$ ANCs of the $E_x$~=~7.12~(1$^-$) and 6.92~(2$^+$)~MeV states have been made by \textcite{PhysRevLett.83.4025, Belhout2007178, PhysRevC.85.035804} using the $\alpha$ transfer reactions $^{12}$C$(^6$Li$,d)^{16}$O and $^{12}$C$(^7$Li$,t)^{16}$O and have been found to be in excellent agreement as summarized in Table~\ref{tab:ANCs}. 

\begin{table*}
\caption{Summary of $\alpha$ particle asymptotic normalization coefficients of the two subthreshold states at $E_x$ = 6.92 and 7.12~MeV. Because the CN reaction data does not place strong constraints on the ANCs (and are sometimes inconsistent), the values obtained from transfer studies are adopted as the best estimates for this analysis. For comparison, theoretical values, and those deduced from $R$-matrix analysis are also shown. \label{tab:ANCs}}
\begin{ruledtabular}
\begin{tabular}{ l c c } 
     &  \multicolumn{2}{c}{ANC$_\alpha$(fm$^{-1/2}$)} \\ 
\cline{2-3}
Ref. & 6.92~MeV, 2$^+$ & 7.12~MeV, 1$^-$ \\
\hline
\multicolumn{3}{l}{{\bf Theory}} \\
\textcite{Descouvemont1987309} & \footnotemark[1]1.34$\times$10$^{5}$ & \\
\textcite{PhysRevC.69.034601} & 1.445(85)$\times$10$^{5}$ & \\
\textcite{PhysRevC.78.015808} & 1.26(5)$\times$10$^{5}$ & \\
\hline
\multicolumn{3}{l}{{\bf $R$-matrix}} \\
\textcite{BarkerKajino1991} & \footnotemark[2]3.19$\times$10$^{5}$ & \\
\textcite{PhysRevC.50.1194} & & 1.90$\times$10$^{14}$ \\
\textcite{PhysRevC.61.064611} & \footnotemark[1]4.02$\times$10$^{5}$ & \\
\textcite{PhysRevC.64.022801} & \footnotemark[1]2.28$^{+33}_{-37}\times$10$^{5}$ & \\
\textcite{PhysRevC.78.065801} & 2.3(4)$\times$10$^{5}$ & \\
\textcite{PhysRevC.79.055803} & \footnotemark[1]1.54(18)$\times$10$^{5}$ & \\
\textcite{PhysRevC.81.045809} & 1.67$\times$10$^{5}$ & 1.96$\times$10$^{14}$ \\
\textcite{Schurmann201235} & 1.5$\times$10$^{5}$ & 1.94$\times$10$^{14}$ \\
\textcite{PhysRevLett.109.142501} & 1.59$\times$10$^{5}$ & \\
\hline
\multicolumn{3}{l}{{\bf Transfer}} \\
\textcite{PhysRevLett.83.4025} & 1.14(10)$\times$10$^5$ & 2.08(20)$\times$10$^{14}$ \\
\textcite{Belhout2007178} & \footnotemark[3]1.40(50)$\times$10$^5$ & 1.87(54)$\times$10$^{14}$ \\
\textcite{PhysRevC.85.035804} & 1.44(28)$\times$10$^5$ & 2.00(35)$\times$10$^{14}$ \\
\textcite{PhysRevLett.114.071101} & 1.22(7)$\times$10$^5$ & 2.09(14)$\times$10$^{14}$\\
\end{tabular}
\end{ruledtabular}
\footnotetext[1]{$\gamma_\alpha$ transformed to ANC$_\alpha$ by \textcite{PhysRevC.69.034601}.}
\footnotetext[2]{$\gamma_\alpha$ transformed to ANC$_\alpha$ by \textcite{PhysRevC.78.015808}.}
\footnotetext[3]{Renormalized by \textcite{PhysRevC.85.035804}.}

\end{table*}  

The ANCs of the $E_x$~=~6.05 and 6.13~MeV states have received less attention because the $E_x$~=~6.13~MeV state is too weak to have a significant ground state subthreshold state contribution and the $E_x$~=~6.05 MeV to ground state transition is forbidden. However, the $E_x$~=~6.13~MeV~(3$^-$) state makes a weak yet important contribution to the $^{16}$N$(\beta\alpha)^{12}$C spectrum just in the sensitive interference region between the $E_x$~=~7.12 and 9.59~MeV 1$^-$ states. These ANCs have been recently measured for the first time using Sub-Coulomb transfer in \textcite{PhysRevLett.114.071101}. Their effects on fits to the capture data and the $\beta$ delayed $\alpha$ spectrum are discussed in Secs.~\ref{sec:cascades_12Cag} and \ref{sec:beta_delayed_alpha} respectively. The past values are summarized in Tables~\ref{tab:ANC_605} and \ref{tab:ANC_613}.

\begin{table}
\caption{Asymptotic normalization coefficients for the $E_x$ = 6.05~MeV bound state in $^{16}$O. \label{tab:ANC_605}}
\begin{ruledtabular}
\begin{tabular}{ l c } 
Ref. & ANC$_\alpha$(fm$^{-1/2}$) \\
\hline
\textbf{$R$-matrix} & \\
\textcite{Schurmann2011557} & 44$^{+270}_{-44}$ \\
\textcite{PhysRevC.87.015802} & 1800 \\
\textbf{Transfer} & \\
\textcite{PhysRevLett.114.071101} & 1560(100) \\
\end{tabular}
\end{ruledtabular}
\end{table}

\begin{table}
\caption{Asymptotic normalization coefficients for the $E_x$ = 6.13~MeV bound state in $^{16}$O. \label{tab:ANC_613}}
\begin{ruledtabular}
\begin{tabular}{ l c }
Ref. & ANC$_\alpha$(fm$^{-1/2}$) \\
\hline
\textbf{$R$-matrix} & \\
\textcite{PhysRevC.50.1194} & 121-225 \\
\textcite{PhysRevC.81.045809} & 191-258 \\
\textcite{PhysRevC.87.015802} & 150 \\
\textbf{Transfer} & \\
\textcite{PhysRevLett.114.071101} & 139(9)\\
\end{tabular}
\end{ruledtabular}
\end{table} 

The ground-state $\alpha$ ANC is outside the kinematic window for sub-Coulomb transfer but it can be deduced by other means, such as sequential breakup reactions \cite{Adhikari2009216}. \textcite{Morais20111} has also investigated the use of the scattering reaction $^{12}$C$(^{16}$O$,^{12}$C$)^{16}$O. In addition, the $R$-matrix analysis of \textcite{PhysRevLett.109.142501} contends that the ground state $E2$ external capture makes a significant contribution to that cross section and have given a value based on an $R$-matrix fit to ground state transition $E2$ data. The measurements giving estimates of the ground state $\alpha$ ANC are summarized in Table~\ref{tab:ANC_gs}. In this work, a much smaller value has been obtained (see Table~\ref{tab:fit_params_part}). The data that constrain the value are primarily the $E2$ ground state data of \textcite{Schurmann2011557} in the off-resonance region near $E_\text{c.m.} \approx$~3.5~MeV (see Fig.~\ref{fig:angle_int_12Cag}).

\begin{table}
\caption{Summary of previous measurements of the $\alpha$ particle asymptotic normalization coefficient of the ground state of $^{16}$O. The scatter in the values combined with different favored solutions of the data represents a systematic uncertainty in the $E2$ cross section. See text for details. \label{tab:ANC_gs}}
\begin{ruledtabular}
\begin{tabular}{ l c } 
Ref. & ANC$_\alpha$(fm$^{-1/2}$) \\
\hline
\textbf{$R$-matrix} & \\
\textcite{PhysRevLett.109.142501} & 709\\
\textbf{Transfer} & \\
\textcite{Adhikari2009216} & 13.9(24) \\
\multirow{ 3}{*}{\textcite{Morais20111}} & 1200 (WS2) \\
 & 4000 (WS1) \\
 & 750 (FP) \\
\end{tabular}
\end{ruledtabular}
\end{table}

This section has summarized the ``best fit'' for the $R$-matrix analysis and has described the wide range of experimental measurements that have been used to constrain the phenomenological model parameters. The parameters for the best fit can be found in Tables~\ref{tab:fit_params_part} and \ref{tab:fit_params_gamma} of Appendix~\ref{app:fit_parameters}. For convenience, the main contributions to the low energy cross section are summarized in Table~\ref{tab:main_params} to aid in a quick reproduction of the low energy cross section. Yet for any analysis of this kind, the best fit is only the beginning, the real challenge is the estimation of the uncertainties. In particular, how are the systematic differences in the data and the ambiguities inherent in the model dealt with.

\begin{table} 
\caption{Summary of the critical parameters necessary to reproduce the total $^{12}$C$(\alpha,\gamma)^{16}$O low energy cross section at $E_{c.m.}$~=~300~keV with about 7\% deviation (lower) from the full parameter set. Signs on the partial widths indicate the sign of the corresponding reduced width amplitude. See Tables~\ref{tab:fit_params_part} and \ref{tab:fit_params_gamma} of Appendix~\ref{app:fit_parameters} for further details. \label{tab:main_params} }
\begin{ruledtabular}
\begin{tabular}{ c c c c } 
J$^\pi$ & $E_x$ (MeV) & $\Gamma_\alpha$ (keV) / ANC (fm$^{-1/2}$) & $\Gamma_{\gamma_0}$ (meV) \\
\hline
0$^+$ & 0 & 58 &  \\
0$^+$ & 6.05 & 1560 &  \\
2$^+$ & 6.92 & 1.14$\times$10$^5$ & 97 \\
1$^-$ & 7.12 & 2.08$\times$10$^{14}$ & 55 \\
\hline
1$^-$ & 9.586 & 382 & -15 \\
2$^+$ & 11.5055 & 83 & -490 \\

\end{tabular}
\end{ruledtabular}
\end{table}

\section{\label{sec:unc_analysis} Uncertainty Analysis}

The total uncertainty of the capture cross section, and subsequent reaction rate, resulting from the $R$-matrix analysis has significant contributions from both the experimental observables and the phenomenological model. In the following sections, investigations of these uncertainties are made by way of different sensitivity studies.

The studies begin with an examination of the sensitivity of the fit to different sets of data. Then the uncertainties stemming from the model are explored (i.e. background poles, channel radii, goodness of fit estimate). A frequentist Monte Carlo analysis (see, e.g., \textcite{Gialanella2001} and \textcite{PhysRevC.90.035804}) is then performed to estimate the contributions from the statistical and overall systematic uncertainties of the data. The uncertainties from quantities that were fixed in the fitting, primarily the ANCs and $\gamma$ widths of the subthreshold states, are now varied so that their uncertainties can be propagated through the Monte Carlo analysis.  

Throughout this section the results of different fits that test the uncertainty of the $R$-matrix analysis are compared to the $S$-factor of the best fit at $E_\text{c.m.}$~=~300~keV by giving the deviation $\Delta S_\text{300 keV}$ (see the bottom of Table~\ref{tab:S-extrap_history} for the best fit value of this work). This is a shorthand comparison since different solutions produce different deviations from the best fit as a function of energy, but it serves to give a good measure of the effect at the region of greatest interest (see Table~\ref{tab:systematic_unc} for a summary). The full excitation curves were recorded for each different test calculation and then used for the total cross section and reaction rate uncertainties.

\subsection{\label{sec:data_sets} Sensitivity to Different Data Sets} \label{sec:sens_diff_data}

In Sec.~\ref{sec:gs_12Cag} it was shown that the ground state $E2$ data are not always well reproduced by the $R$-matrix fit and that they show significant discrepancies between one another. Since the global fit includes data for the total cross section, the ground state $E1$, and all cascade transitions, the $E2$ cross section should be highly constrained even with no $E2$ cross section data (i.e. $\sigma_{\text{total}} - \sigma_{\text{G.S.} E1} - \sigma_{\text{cascades}} = \sigma_{\text{G.S.} E2}$). In practice, this is with the caveat that the $E2$ cross section is significant compared to the total cross section compared to the uncertainties of the data, which it is over several regions, especially at higher energies. When the $E2$ data were completely neglected in the fit, a very similar result was obtained ($\Delta S_\text{300 keV}$~=~-0.6 keV b). This is largely because the $E2$ cross section is dominated by the subthreshold state at $E_x$~=~6.92~MeV. Since the parameters that describe this contributions are highly constrained by other types of data, it does not change significantly.

The ground state angular distribution data also show discrepancies, both between one another and with the fit at certain energies and angles (see Fig.~\ref{fig:ang_dist_big_plot}). Another test was made by completely eliminating the ground state angular distribution data from the fit, leaving only the derived $E1$ and $E2$ excitation curve data. This had a somewhat more pronounced effect resulting in a lower value in the extrapolation ($\Delta S_\text{300 keV}$~=~-5.2~keV~b). Finally both the $E1$ and $E2$ excitation curve data were removed, fitting only to the angular distribution data for the ground state transition. Again, only a small deviation ($\Delta S_\text{300 keV}$~=~-0.9 keV b) was observed from the standard fit.

\subsection{Limiting Interference Solutions} \label{sec:interferences}

One of the largest sources of uncertainty in an $R$-matrix analysis can come from different possible interference solutions that can not be ruled out by the data. These different solutions are a result of the different possible signs for the reduced width amplitudes. When more than one level is present in the sum of Eq.~(\ref{eq:rmatrix}), the difference in the sign of each element can produce drastically different values for the cross section. This sign can not usually be predicted by theory, therefore experimental capture data is critical. However, limiting the solutions is often challenging because the cross section must usually be measured in off-resonance regions where the experimental yields are small. For the case of the $^{12}$C$(\alpha,\gamma)^{16}$O reaction, different interference solutions have been proposed for both the $E1$ and $E2$ ground state transition and can produce large differences in the extrapolated value of $S_\text{300 keV}$. Different interference solutions have also been proposed for the cascade transitions and these are not as well established as the ground state transition. At this stage, the data have limited the possible solutions, at least over the lower energy range, to only a few. 

\begin{figure}
\includegraphics[width=1.0\columnwidth]{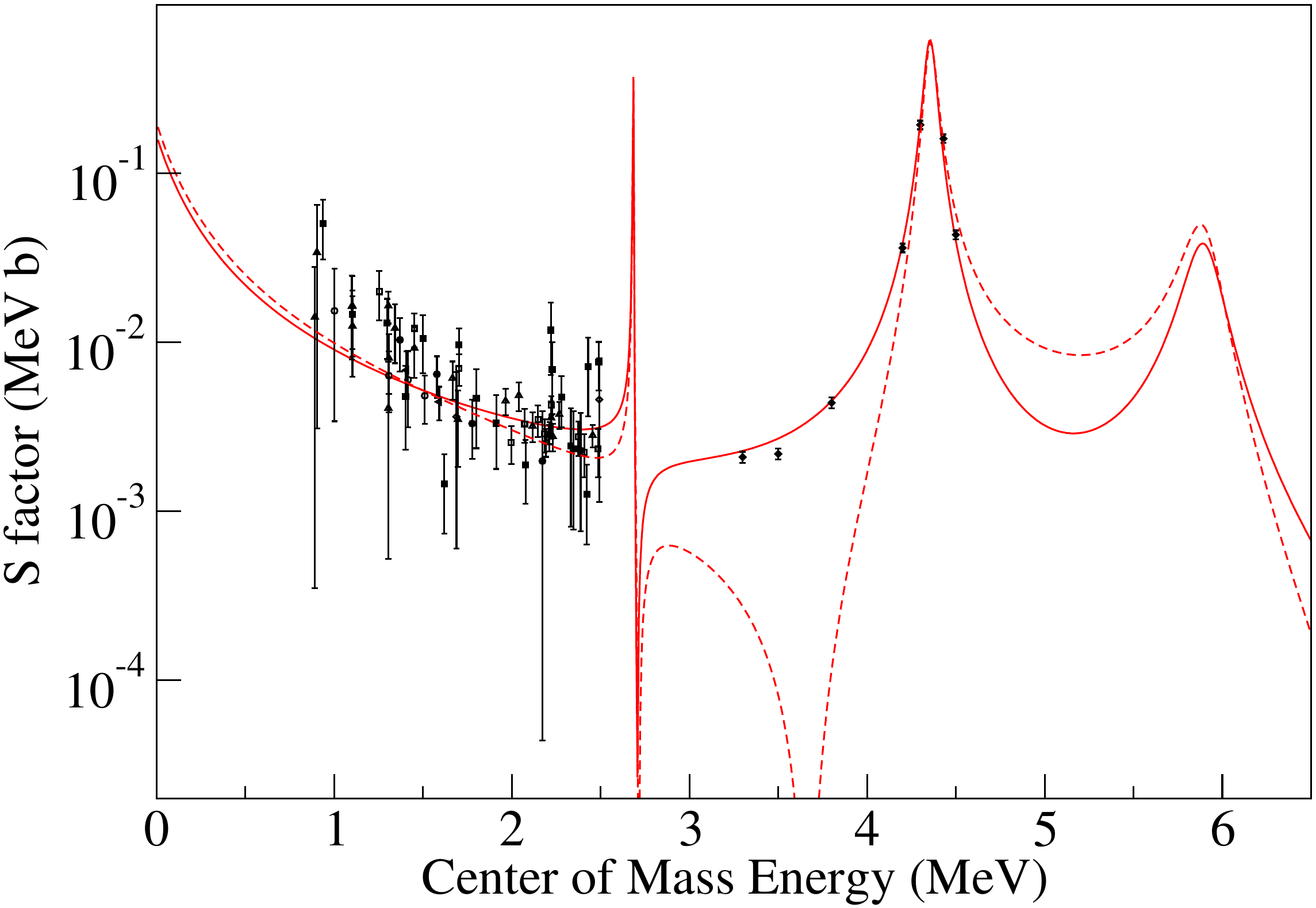}
\caption{(Color online) Comparison of the two allowed $E2$ interference solutions determined by \textcite{PhysRevLett.109.142501}. If the higher energy data of \textcite{Schurmann2011557} are also considered, only one solution remains viable. \label{fig:angle_int_12Cag_E2_only}}
\end{figure}

Starting with the ground state $E2$ cross section data, \textcite{PhysRevLett.109.142501} have made an extensive reanalysis of this low energy data where a statistical criterion was used to eliminate outlying data points. Further, \textcite{PhysRevLett.109.142501} analyzed the narrow resonance region in order to determine the relative interference signs of the $E2$ contributions to the cross section. The main result of \textcite{PhysRevLett.109.142501} was to limited the $E2$ interferences to two possible solutions. If the subsequent higher energy data from \textcite{Schurmann2011557} are now considered, only one interference solution remains viable as shown in Fig.~\ref{fig:angle_int_12Cag_E2_only}. This then gives a final interference pattern for the $E2$ cross section, at least between the $2^+$ subthreshold state, the narrow 2$^+$ at $E_x$ = 9.84~MeV, the first broad resonance at $E_x$ = 11.51~MeV, and the $E2$ external capture.

As an aside, in the analysis of \textcite{PhysRevLett.109.142501} the $R$-matrix fit was performed to only the ground state $E2$ data letting the ANCs of the ground state and $E_x$ = 6.92~MeV subthreshold state vary freely. The fit resulted in values of 709 and 1.59$\times$10$^5$~fm$^{-1/2}$ for the ANCs of the ground state and 2$^+$ subthreshold state respectively. This can be compared to the values of 14 \cite{Adhikari2009216} or 750-4000 \cite{Morais20111} for the ground state and 1.14(10)$\times$10$^5$ \cite{PhysRevLett.83.4025}-1.22(7)$\times$10$^5$ \cite{PhysRevLett.114.071101}~fm$^{-1/2}$ for the 2$^+$ subthreshold state. The two components interfere with each other destructively and can result in a range of values that produce a similar $E2$ cross section over the energy range of the data. The analysis of \textcite{PhysRevLett.109.142501} was made before the data of \textcite{Schurmann2011557} was available that significantly increases the sensitivity of the fit to the $E2$ data, especially the external capture that has a maximum effect in the off-resonance region at $E_{c.m.} \approx$~3~MeV. A measurement of the ground state ANC taken together with the capture data would then provide a consistency check for the value of the 2$^+$ subthreshold state ANC. Since large systematic differences occur for the $E2$ cross section, another method of verification is highly desirable. (see Secs.~\ref{sec:gs_12Cag} and \ref{sec:subthreshold}).

Now turning to the $E1$ cross section, a large source of uncertainty can come from the ambiguity in the interference sign between the 1$^-$ $E_x$ = 7.12~MeV subthreshold state and the unbound state at $E_x$ = 9.59~MeV. Most analyses have concluded (or assumed) that the constructive solution is favored, but most also do not provide detailed quantitative support for this decision. A few analyses, those of \textcite{PhysRevLett.69.1896}\footnote{However this seems to have been retracted in a subsequent publication \cite{PhysRevC.54.1982}.}, \textcite{Hale1997177}, \textcite{Gialanella2001}, have either ruled in favor or determined that the destructive solution can not be dismissed. Here the $E1$ destructive solution is investigated in detail in light of the present analysis and the most current data.

The situation has been investigated in detail by \textcite{Hammer2005514, Fey2004, Kunz_thesis}. However, approximations were made in these analyses that can now be improved upon and in fact are very significant to the analysis. The first is the neglect of the overall systematic uncertainties. This resulted in greatly inflated $\chi^2$ values for the fits of those works, regardless of the interference solution. This is because the systematic uncertainties are quite large compared to the statistical ones, at least near the peak of the low energy 1$^-$ resonance. The second issue is that the contributions from the two 1$^-$ levels at $E_x$~=~12.45 and 13.10~MeV are not explicitly included but are instead treated as a single background pole (the classic 3 level $E1$ fit). The single pole assumption leads to fits to the low energy data that are now found to be unphysical when the added constraints of the higher energy data are imposed. The last improvement is that the transfer reaction measurements have provided much tighter constraints on the values of the ANCs, further limiting the number of viable solutions.

\begin{figure*}
\includegraphics[width=2.0\columnwidth]{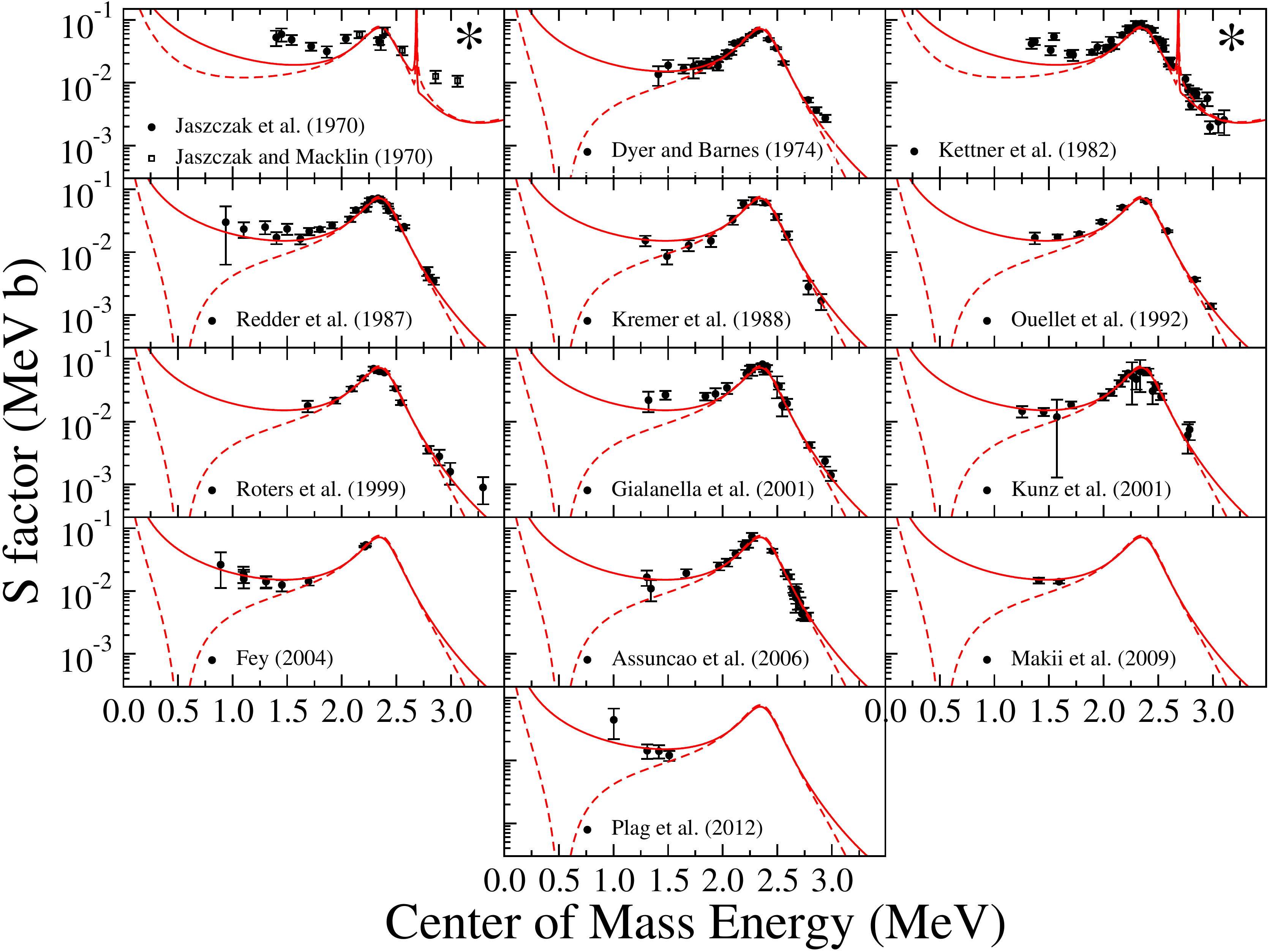}
\caption{(Color online) Comparison of destructive (red dashed lines) and constructive (red solid lines) $E1$ solutions. A 1$^-$ background pole is included in the ground state $\gamma$ ray channel in these cases. The energy of the pole is held constant in both cases and the fitted $\alpha$ widths are very similar. For the constructive case $\Gamma_{\gamma_0}$~=~1.8~eV, while the destructive case gives 21.7~eV. \label{fig:constructive_destructive_comparisons}}
\end{figure*}

These further constraints have a strong impact on the $E1$ destructive/constructive solutions. In fact, they limit the destructive solution to only one possibility, and it will be shown that this is also ruled out, in favor of the constructive one. Fig.~\ref{fig:constructive_destructive_comparisons} shows the result of an analysis of the destructive solution compared to the constructive one. In this fit, the ANC of the 1$^-$ subthreshold state has been fixed to a value of 2.08(20)$\times$10$^{14}$~fm$^{-1/2}$ \cite{PhysRevLett.83.4025}. This has been done because if it is allowed to vary freely, tension from other data sets cause the destructive solution result in unphysical values for the ANC. This on its own is one hint that this solution may not be viable. In particular, the fit would favor a very small ANC, many sigma removed from the value of \cite{PhysRevLett.83.4025}. Considering only the $\chi^2$ from the $E1$ data (165 data points), the constructive solution gives $\chi^2$ = 259.8 and the destructive solution gives $\chi^2$ = 583.8, a difference of $\Delta\chi^2$ = 324.0. The difference in $\chi^2$ for a 5$\sigma$ deviation for 78 fit parameters is $\Delta\chi^2_{5\sigma}$~=~169. Hence the destructive solution is ruled out under these conditions, and the ANC would have to be changed to a value far outside the acceptable range of the transfer measurements to recover a $\chi^2$ of less than 5$\sigma$. 

\begin{table}
\caption{Comparison of $\chi^2$ values for different $E1$ interference solutions. Only the data of \textcite{PhysRevLett.60.1475} favor a destructive solution. \label{tab:des_vs_cons_chi2}}
\begin{ruledtabular}
\begin{tabular}{ c c c c } 
 & & \multicolumn{2}{c}{$\chi^2$} \\ \cline{3-4}
Ref. & N & constructive & destructive \\
\hline
\textcite{Dyer1974495} & 24 & 69.6 & 135.9 \\
\textcite{Redder1987385} & 26 & 67.5 & 146.2 \\
\textcite{PhysRevLett.60.1475} & 12 & 18.2 & 16.9 \\
\textcite{PhysRevLett.69.1896} & 9 & 29.2 & 82.6 \\
\textcite{Roters1999} & 13 & 13.6 & 26.5 \\
\textcite{Gialanella2001} & 20 & 22.9 & 58.0 \\
\textcite{PhysRevLett.86.3244} & 19 & 12.1 & 36.8 \\
\textcite{Fey2004} & 11 & 4.5 & 39.0 \\
\textcite{PhysRevC.73.055801} & 24 & 19.4 & 26.3 \\
\textcite{PhysRevC.80.065802} & 2 & 0.2 & 5.5 \\
\textcite{PhysRevC.86.015805} & 4 & 1.9 & 5.2 \\
\hline
all & 164 & 259.8 & 583.8 \\
\end{tabular}
\end{ruledtabular}
\end{table}

\begin{figure}
\includegraphics[width=0.9\columnwidth]{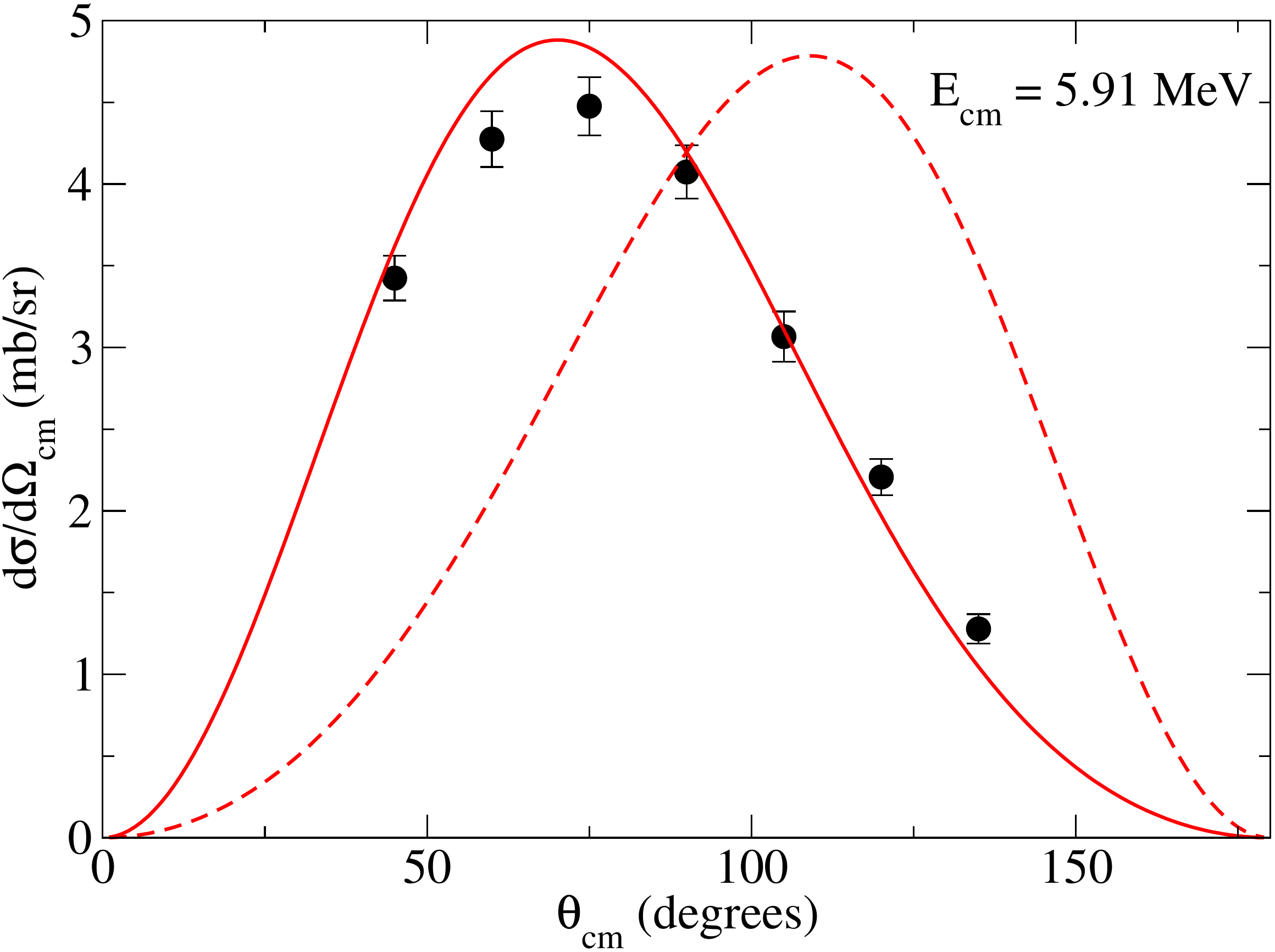}
\caption{(Color online) Example angular distribution for the $^{12}$C$(\alpha,\gamma_0)^{16}$O reaction at $E_{cm}$ = 5.91~MeV ($E_x$ = 13.07~MeV) from \textcite{Kernel1971352}. The red solid line shows the differential cross section with the preferred interference pattern while the red dashed line shows the differential cross section with the relative interference sign between the $E1$ and $E2$ contributions switched. \label{fig:E1_E2_int}}
\end{figure}

At higher energies, the possible interference combinations are further limited by the stronger $E1$/$E2$ interference in the angular distributions. This uniquely defines the interferences at high energy between the 1$^-$ levels at $E_x$~=~12.45 and 13.10~MeV and the 2$^+$ level at $E_x$~=~12.97~MeV. A similar situation exists for the $^{15}$N$(p,\alpha_0)^{12}$C data. Examples of these different interference solutions are shown in Fig.~\ref{fig:E1_E2_int} for the $^{12}$C$(\alpha,\gamma_0)^{16}$O reaction and Fig.~50 of \textcite{PhysRevC.87.015802} for the $^{15}$N$(p,\alpha_0)^{12}$C reaction.

\begin{figure}
\includegraphics[width=1.0\columnwidth]{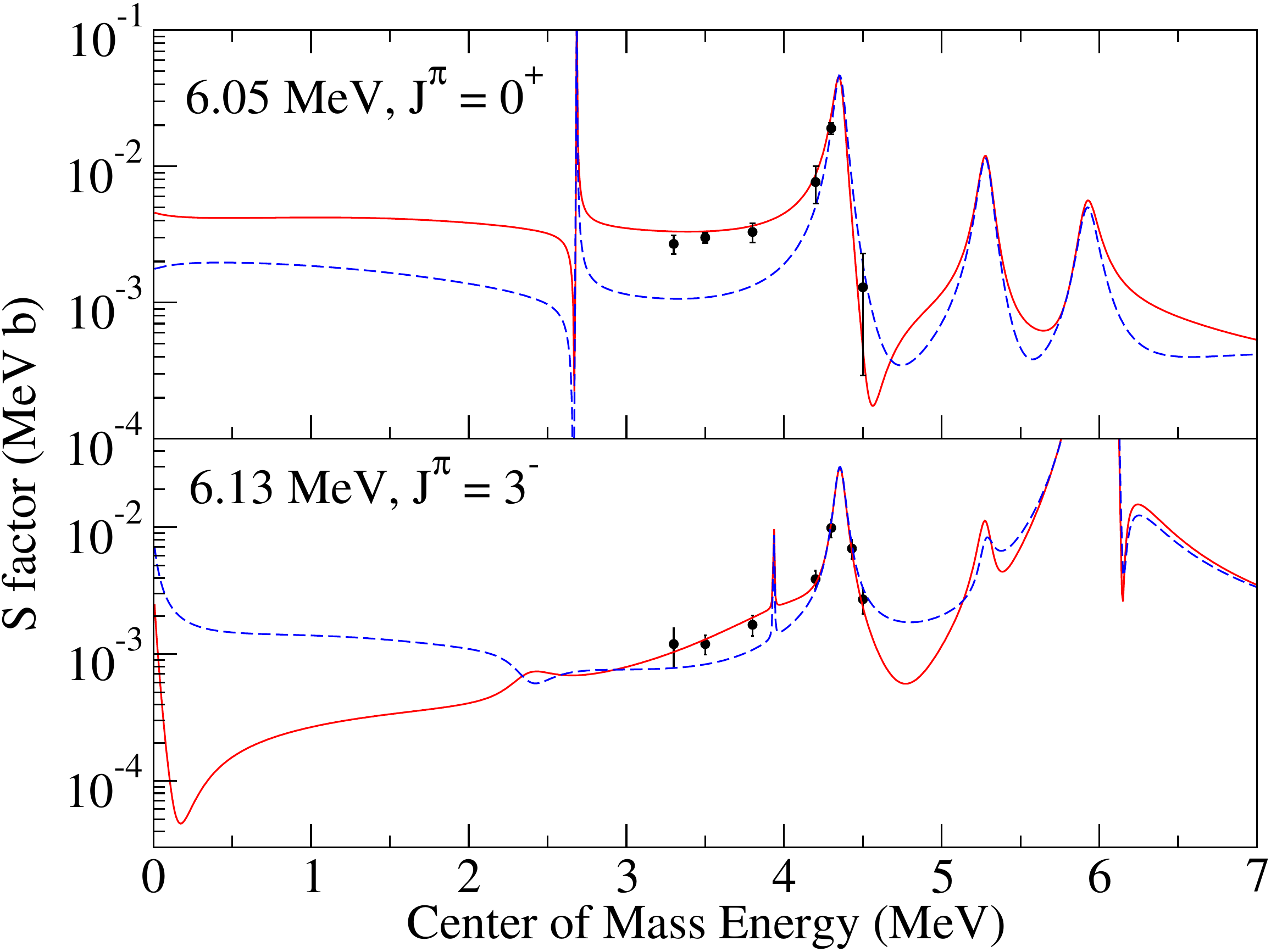}
\caption{(Color online) Comparison of the different possible interference solutions for the $E_x$~=~6.05 and 6.13~MeV transition capture cross sections as purposed by \textcite{PhysRevLett.114.071101}. The different solutions are compared with the higher energy data of \textcite{Schurmann2011557}, the only data available for these transitions. For the $E_x$~=~6.05 MeV transition, the constructive solution at low energy is highly favored. For the $E_x$~=~6.13 MeV transition the destructive solution is favored but their is more ambiguity, especially since the 3$^-$ state at $E_x$ = 11.49 MeV can contribute to the cross section over the energy range of the data. The calculation shows that additional measurements at just slightly higher energy than those \textcite{Schurmann2011557} could provide a more definitive solution. \label{fig:Avila_inter_solutions}}
\end{figure}

Turning to the cascade transitions, in \textcite{PhysRevLett.114.071101}, ANCs were measured for the $E_x$~=~6.05, 6.13, 6.92, and 7.12~MeV bound states, with the $E_x$~=~6.05 and 6.13~MeV ANCs being measured for the first time via Sub-Coulomb transfer. ANCs similar to those deduced in the global $R$-matrix fit of \textcite{PhysRevC.87.015802}, where the capture data of \textcite{Schurmann2011557} were fit to constrain the ANCs, were found. An interference ambiguity in the low energy $S$-factor for these two transitions was still left undefined in \textcite{PhysRevLett.114.071101}. If the external capture determined by these ANCs is combined with the higher energy capture data of \textcite{Schurmann2011557} the interference combination can be uniquely determined for the $E_x$~=~6.05~MeV transition, and is suggestive of a solution for the $E_x$~=~6.13~MeV transition as shown in Fig.~\ref{fig:Avila_inter_solutions}.

The most important result of this section is that only one viable interference solution is possible for both the $E1$ and $E2$ ground state cross sections. This drastically decreases the uncertainty, by ruling out the destructive $E1$ solution. While there is a limited amount of data for the cascade transitions, the current data do seem to constrain the major interference solutions. However, because of the limited data it is certainly possible that some of the solutions here are incorrect. The situation worsens at higher energy were the data become even more sparse. For the cascade transitions there is almost no data above $S_p$. While these different solutions do not have a direct effect on the low energy cross section determination, they may effect the extrapolation indirectly since the over all fit is quite sensitive to the total cross section data at higher energies. Hence further studies of the cascade data is highly desirable.

\subsection{\label{sec:systematic_more} Channel Radius and Background Poles}

A long standing complication with $R$-matrix theory is that it requires two sets of model parameters: channel radii and background poles. What complicates the situation is that these two sets of model parameters are correlated to one another, hence there is a range of viable solutions.

A phenomenological $R$-matrix fit must then be tested for its sensitivity to the choice of both the channel radii\footnote{Even the choice of how many different channel radii will be used is different for different $R$-matrix analyses. In principle a different channel radius can be chosen for each $s-l$ channel. While this is sometimes done, many analyses restrict themselves to different channel radii only for different particle partitions. That is the approach taken here.} and the background poles. The radius and the number of levels included in the analysis are closely linked as discussed in Sec.~\ref{subsection:rmatrixtheory}. In the strict $R$-matrix theory, the fit should be completely insensitive to the choice of channel radius, but this is in the limit of an infinite number of levels. It is also important to note that the channel radius does not correspond to a real nuclear radius, although the value used in phenomenological analyses is often rather close, which has caused much confusion over the issue. In practice the number of levels in an $R$-matrix calculations is truncated to only a few or even just one. Even if the tails of higher energy resonances do have an effect, their contributions can often be reproduce, up to some level of precision, with only a single background pole (for each $J^\pi$). However, as the data become more precisely measured, especially in the interference regions between resonances, it may be possible that more than one background pole is required.

While the sensitivity of the fit to the channel radius and background poles are closely linked, each is discussed separately in order to try to separately gauge their contributions. Further, while it is not often stated explicitly below, the many sensitivity tests that were made involved several different combinations of both radii and number/value of background poles.

A strong sensitivity in $\chi^2$ to the channel radius exists for the present fit. This is primarily the result of the fit to the scattering data as and has been detailed previously in \textcite{PhysRevLett.88.072501}. This was interpreted as a positive result in \textcite{PhysRevLett.88.072501}, giving a constraint on the radius that should be used. However, this is in direct conflict with the discussion above, where it has been argued that the value of the channel radius should be insensitive to the fit. This represents one of the remaining puzzles to be solved, not only for this case, but for the phenomenological $R$-matrix technique in general. 

It should be possible to decrease the sensitivity of the fit on the channel radius by adding more background poles. This procedure was performed, yet the sensitivity remained almost unchanged. Therefore, to gauge the sensitivity of the fit to the extrapolated value of the capture cross section, different fits were made at different channel radii varying between 4.5 and 6.5~fm and using various numbers and combinations of background poles. A radius of 5.43~fm was found to be the best fit value, in excellent agreement with the value of 5.42$^{+0.16}_{-0.27}$ of \textcite{PhysRevLett.88.072501}. Despite a rather large change in the overall $\chi^2$ of the fit, the extrapolated value of the $S$-factor only changed by $\pm$8~keV~b. This is because the sensitivity in $\chi^2$ comes mainly from the scattering data, while that of the capture data is much less so.  

Turning now to the background poles, one of the assertions of this analysis is that a reasonable fit can be obtained for the $^{12}$C$(\alpha,\gamma)^{16}$O data with no background pole contributions in the capture partition (they remain very necessary for the scattering partition). The reason for this assertion is that no strong higher energy resonances have been observed in the capture data up to $E_{c.m.} \approx$ 20~MeV \cite{PhysRevLett.32.1061}. This is the main reason for explicitly including the two 1$^-$ levels at $E_x$~=~12.45 and 13.09~MeV, which correspond to the final two strong $E1$ resonances observed in $^{12}$C$(\alpha,\gamma_0)^{16}$O, at least up to $E_x \approx$~20~MeV. It is therefore expected that higher energy background contributions will be weak for the ground state.

However, adding additional background poles is allowed by the data and does improve the quality of the fit. The question becomes whether the improvement is physically reasonable or is it simply the result of adding more free parameters to the fit. Additionally the improvement results largely from achieving a better fit to the very low energy data, but there are strong hints that much of this data may over estimate the cross section (see Sec.~\ref{sec:fitting}). 

Ground-state background contributions were considered for both the $E1$ and $E2$ cross sections using $J^\pi$~=~1$^-$ and $J^\pi$~=~2$^+$ poles respectively. The $J^\pi$~=~1$^-$ background pole contribution resulted in a significant improvement in the fit of the low energy $^{12}$C$(\alpha,\gamma)^{16}$O data. For example, it decreased the $\chi^2$ from 436 to 203 for the 164 data points considered in the low energy $E1$ capture data. This ambiguity in the strength of the 1$^-$ background pole is one of the most significant uncertainties in the fitting, producing a value of $S$(300 keV) 15.2~keV~b larger than the fit without. Inclusion of a $J^\pi$~=~2$^+$ background pole had only a small effect.

\begin{figure}
\includegraphics[width=1.0\columnwidth]{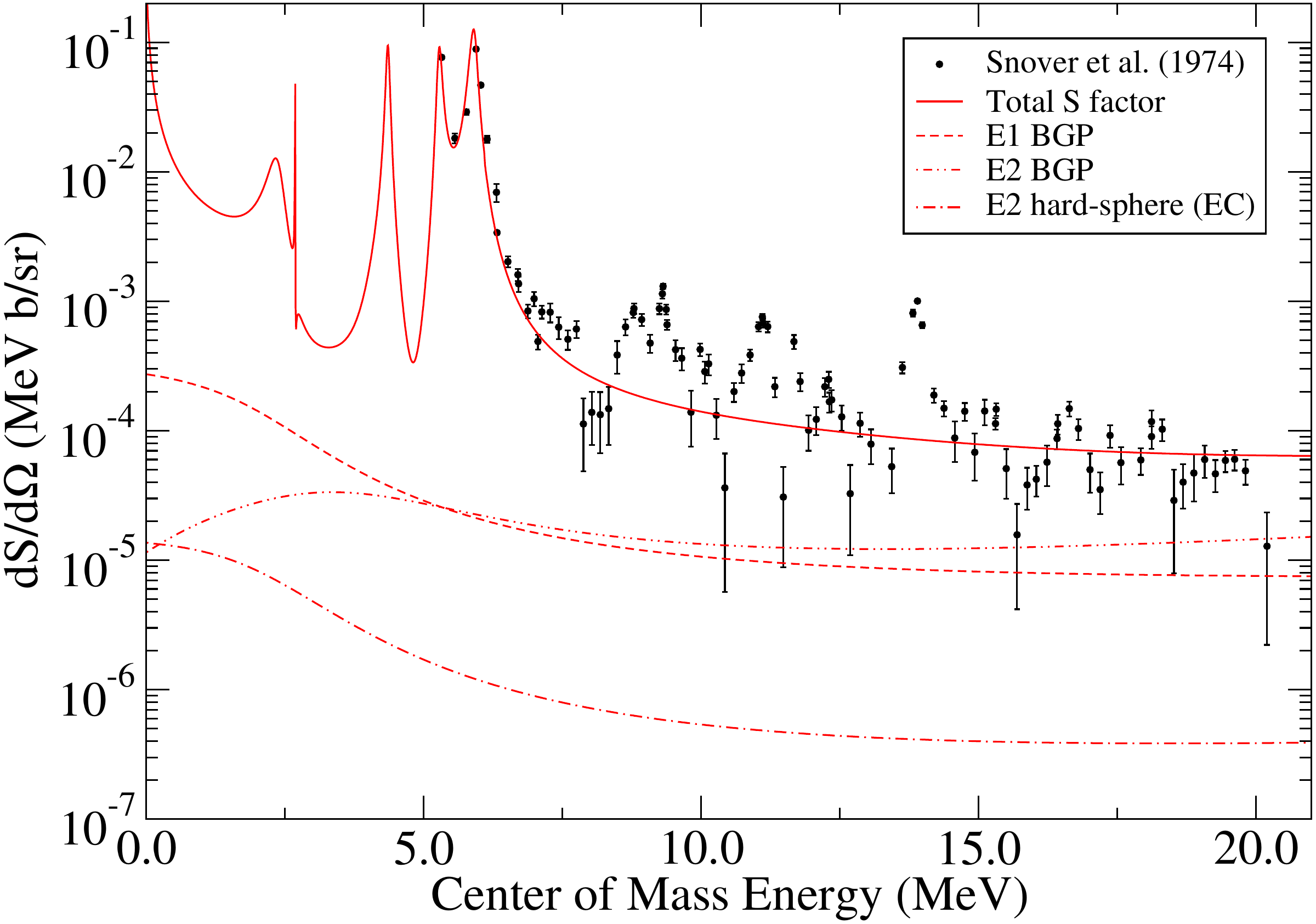}
\caption{(Color online) Differential $S$ factor of the data ($\theta_\text{lab}$~=~52$^\circ$) of \textcite{PhysRevLett.32.1061}. The $R$-matrix fit was made with the data fixed at the reported cross sections in order to test whether the amplitude of the free background poles were reasonable. The extrapolated values of the $S$-factor using the two different methods proved to be very similar. In the legend, $E2$ hard-sphere (EC) refers to the hard-sphere contribution to the external capture. \label{fig:Snover_bgp_compare}}
\end{figure}

One way to obtain further constraint on the background pole contributions is to continue to fit to higher energies. This becomes increasingly difficult in practice as the number of levels and channels increase quickly at higher energies. Even so, a test can be made to see if the magnitude of the background contributions is reasonable. There is one measurement by \textcite{PhysRevLett.32.1061}, which extends the ground state transitions cross section to much higher energies. The data show that the ground state cross section decreases substantially above the two strong 1$^-$ resonances at $E_x$~=~12.45 and 13.09~MeV. In Fig.~\ref{fig:Snover_bgp_compare} the data of \textcite{PhysRevLett.32.1061} are shown together with a fit that has all of the background poles placed now at higher energies ($E_x$~=~40~MeV). The fit gives a background contribution that is roughly consistent with this data in that it follows the off-resonance trend of the data. This produced a value for $S$(300 keV)~=~155.3~keV~b, very similar to the value of 153.7~keV~b found with out this higher energy data (with the lower energy background poles), demonstrating that this additional background contribution is at least physically reasonable.

It is somewhat surprising that such good fits can also be obtained for the cascade transitions without background poles. There is no higher energy data for these transitions so their higher energy behavior is unknown. In fact the higher energy (above $S_p$) constraint of this fit comes indirectly from $^{15}$N$(p,\gamma)^{16}$O cascade transition data \cite{imbriani2012}, and these have large uncertainties. One likely reason for such good fits is that external capture is the dominant contribution to the cascade transitions at low energy, lessening the effects of resonance interference.


\subsection{\label{sec:fit_methods} Different Fitting Methods}

In principle, if the data were expressed as true cross sections, were consistent with each another, and the uncertainties were completely characterized, performing a standard $\chi^2$ fit would be all that was required to achieve the best fit and accurate extrapolation of the low energy cross section. Unfortunately these conditions are hardly ever met in practice and therefore a blind $\chi^2$ fitting is likely to lead to an erroneous result. This is the reason why many different experimental approaches are critical; by examining the reaction mechanism through different methods, hidden systematic errors can be more readily revealed. 

While a standard $\chi^2$ fit analysis has been used as the standard fit procedure for this analysis, other methods have been investigated to check the robustness of the fitting and the uncertainty estimates. In particular, the tension between different data sets and the background pole contributions has the effect that the subthreshold parameters need to be fixed during the normal fitting.

An alternative fitting method investigated here is to base the goodness of fit on the reduced $\chi^2$ of each data set. The idea is based on that presented in \textcite{PhysRevC.15.518}, where a similar analysis involving many data sets was made. Eq.~\ref{eq:standard_chi_squared} is modified to
\begin{equation} \label{eq:reduced_chi_squared}
\chi^2 = \sum_i\left( \frac{\sum_j R_{ij}^2}{N_i-\nu}\right)+\frac{(n_i-1)^2}{\sigma^2_{\text{syst},i}}
\end{equation} 
where $N_i$ is the total number of data points in the $i^\text{th}$ data set and $\nu$ is the number of fit parameters. Fitting to the reduced $\chi^2$ puts each data set on a more equal footing regardless of the number of data points in that set. This is statistically incorrect, but has the result of putting each data set on a more equal footing, even if that data set has many experimental points. From a purely statistical view this doesn't make much sense, but from a practical standpoint this can be reasonable since it will lessen any systematic bias of a single data set over the rest, especially if that single data set has many points with small uncertainties.  

In the current analysis the scattering data of \textcite{PhysRevC.79.055803} dominate the normal $\chi^2$ function because they contain, by far, the largest number of data points and have small statistical uncertainties (see Sec.~\ref{sec:scat}). If the data were a perfect representation of the true cross section, this would be ideal because this would reflect the experimenter's ability to easily access this cross section. However, it is known that the data of \textcite{PhysRevC.79.055803} require experimental resolution corrections, which are quite significant at the statistical uncertainty level of the data points, particularly in regions where the cross section changes rapidly. If the method used for this correction is not precise enough or if there are any unaccounted for uncertainties in the data this will result in a bias in the fit from this data set. This issue undoubtedly exist in the data considered, and is not limited to the scattering data. The other main data sets where this effect is most likely are in the $^{16}$N$(\beta,\alpha)^{12}$C spectrum measurements of \textcite{PhysRevC.50.1194, PhysRevC.81.045809}.

Another different approach would be to also include the uncertainties on the subthreshold parameters in the $\chi^2$ fitting. This can be done by adding additional terms to the $\chi^2$ function
\begin{multline} \label{eq:reduced_chi_squared_and_sub_unc}
\chi^2 = \sum_i\left( \frac{\sum_j R_{ij}^2}{N_i}\right)+\frac{(n_i-1)^2}{\sigma^2_{\text{syst},i}} \\
 + \sum_k\frac{(P_{\text{fit},k} - P_{\text{exp},k})^2}{\sigma_{\text{exp},k}^2}
\end{multline}
where $P_{\text{fit},k}$ is the parameter value varied in the fit, $P_{\text{exp},k}$ is the experimentally determined value, and $\sigma_{\text{exp},k}$ is the experimental uncertainty.

This method, combined with using the reduced $\chi^2$ method, results in much more reasonable fits than the standard $\chi^2$ fitting when the subthreshold state parameters are allowed to vary freely. Using this approach the fit favors a larger ANC for the 1$^-$ subthreshold state (ANC~=~2.6$\times$10$^{14}$ fm$^{-1/2}$, $\Gamma_{\gamma_0}$~=~48 meV) and a smaller one for the 2$^+$ ANC (ANC~=~0.84$\times$10$^5$ fm$^{-1/2}$, $\Gamma_{\gamma_0}$~=~98.0 meV). This is a reflection of the tension between the $^{16}$N$(\beta,\alpha)^{12}$C, transfer, and elastic scattering measurements. The increasing 1$^-$ ANC and decreasing 2$^+$ ANC have canceling effects in the ground state cross section and in the cascade cross sections, resulting in a somewhat larger value for the total capture cross section of $S$(300 keV)~=~152.1~keV~b.  

In any analysis that contains a large amount of data there tend to be outlier data points. It has been shown that certain data sets are plagued by this problem in the current analysis. There are various methods of testing the sensitivity of fits to these points. For example, in \textcite{PhysRevLett.109.142501} Chauvenet's criterion was used to reject outliers in the $E2$ ground state data. Instead of outright rejection of data points, there are different methods of modifying the $\chi^2$ function to give less weight to outlier data. These methods are similar to increasing the uncertainties on the data points. One example is the method given by \textcite{Sivia2006} where instead of minimizing $\chi^2$, the function
\begin{equation} \label{eq:sivia}
L = \sum_j \log\left[\frac{1-e^{-R_{ij}^2/2}}{R_{ij}^2}\right]
\end{equation} 
is maximized. This alternative function has a broader probability density function leading to more conservative uncertainty estimates than the standard $\chi^2$ function. Fitting with this alternative approach produced a very similar fit as the standard $\chi^2$ method ($S$(300 keV)~=~146.2~keV~b) demonstrating that outlier data points have a minimal effect on the central value obtained for fit. However, as shown in the next section, the effect on the uncertainty estimation is quite significant.

\subsection{\label{sec:MC_analysis} Monte Carlo Uncertainty Analysis}

The best fit resulting from the $R$-matrix analysis has been subjected to a Monte Carlo (MC) uncertainty analysis. From the MC analysis, uncertainty contributions from the statistical and the overall systematic uncertainties of the experimental data were obtained for the fit parameters and the $^{12}$C$(\alpha,\gamma)^{16}$O cross section and corresponding reaction rate. However, in order for the fit to yield accurate uncertainties, the reduced $\chi^2$ of the fit should be approximately one. Given that this is not the case, as detailed in Table~\ref{tab:normalizations}, the alternative goodness of fit method given by Eq.~(\ref{eq:sivia}) was employed. In this way, the uncertainty from outlyer data points and discrepant data sets can be better estimated. At $E_\text{c.m.}$~=~300~keV, the difference in the uncertainty calculated with the standard $\chi^2$ function of Eq.~(\ref{eq:standard_chi_squared}) versus the $L$ estimator of Eq.~(\ref{eq:sivia}) was quite significant, inflating the uncertainty from about 10\% to about 15\%. Indeed, over other energy regions, especially near the low cross section area in the vicinity of $E_\text{c.m.} \approx$~3~MeV, the increase in the uncertainty was even more dramatic.

The MC technique was adapted from those of \textcite{Gialanella2001, Schurmann201235, PhysRevC.90.035804}. The following assumptions and steps were taken for this analysis:

\begin{enumerate}
\item The best fit from the $R$-matrix analysis is taken as the most probable description of the data. The $L$ method of Eq.~(\ref{eq:sivia}) is used to define the goodness of the fit.
\item The data are then subjected to a random variation based on their uncertainties. The data are varied, assuming a Gaussian probability density function, around the best fit cross section value. The uncertainty on the data point is scaled by the square root of the ratio of the cross section of the fit divided by the cross section of the Gaussian randomized cross section.
\item The systematic overall uncertainty for each experimental data set is also varied assuming a Gaussian probability density function.
\item The ANCs and $\gamma$ widths of the subthreshold states are also allowed to vary. Their uncertainty contributions are included using Eq.~(\ref{eq:reduced_chi_squared_and_sub_unc}). 
\item Background poles for $E1$ and $E2$ multipolarity are introduced to the capture channels to give further freedom in the fit. 
\item The $R$-matrix fit (the $L$ maximization) is then performed again. The initial values of the parameters are those from the original best fit.
\item Steps 3-6 are then repeated many times (referred to as the ``MC iterations"). For each of the MC iterations, an extrapolation of the cross section can be made using the best fit parameters from that iteration. This cross section is then numerically integrated to calculate the reaction rate as a function of temperature.
\end{enumerate} 

Histograms were then created from the MC procedure for the cross sections and reaction rates at specific energies or temperatures respectively. The lower and upper uncertainties were then defined by the 16 and 84\% quantiles. The central value is defined as the 50\% quantile.

\begin{figure}
\includegraphics[width=1.0\columnwidth]{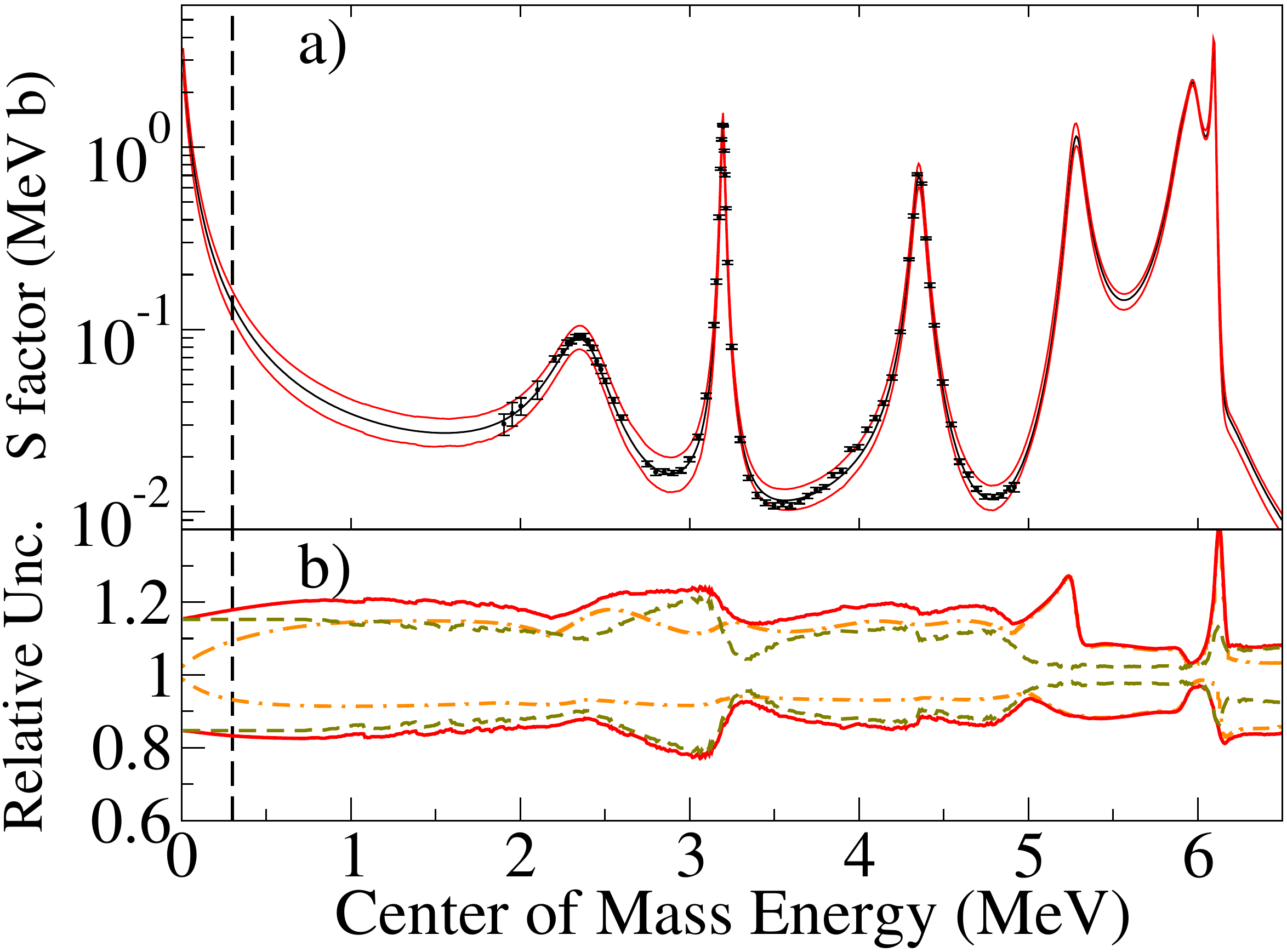}
\caption{(Color online) The uncertainty in the $S$ factor as derived by combining the MC analysis (which includes the subthreshold state uncertainties) and the model uncertainties is shown in Fig.~\ref{fig:XS_unc} a). The data from \textcite{schurmann2005} are shown for comparison. Fig.~\ref{fig:XS_unc} b) shows the uncertainties relative to the best fit value for the Monte Carlo analysis (olive colored dashed line) and the uncertainties derived from the model (dot-dashed orange line). The total uncertainty, taken as the MC and model uncertainties summed in quadrature, is shown by the solid red line. The black vertical dashed line marks the region of typical astrophysical interest at $E_\text{c.m.}$~=~300~keV. \label{fig:XS_unc}}
\end{figure}

The uncertainty in the cross section derived from the MC analysis is shown in Fig.~\ref{fig:XS_unc}. At low energies, the uncertainty budget is dominated by the uncertainties in the ANCs of the subthreshold states and is about 15\% at $E_\text{c.m.}$~=~300~keV. In the region above $E_{cm}$~=~5.0~MeV, the cross section is determined mainly indirectly through a combination of the $^{15}$N$(p,\gamma)^{16}$O, $^{12}$C$(\alpha,\alpha_0)^{12}$C, and $^{15}$N$(p,\alpha_0)^{12}$C data. In this region the uncertainty becomes much smaller, and was found to be in good agreement with that obtained from a standard $\chi^2$ analysis. At the highest energies, where the experimental data taper off, the uncertainty begins to increase again.  

Therefore, using Eq.~(\ref{eq:sivia}), it is believed that a conservative estimate of the uncertainty stemming from the experimental uncertainties has been obtained. Taking the approach considered here, where a large portion of the experimental data is considered, it is useful to compare with the other extreme where only a small subset of well defined data is considered as in \textcite{Schurmann201235}. Each approach has its advantages and drawbacks. Foremost among them, the method considered here likely errs on the side of including a significant amount of data that is incompatible, yet it is not subject to the bias of choosing the best data. On the other had, choosing a small subset of data can yield a smaller uncertainty that may be accurate, however the entire analysis hinges on choosing the ``correct'' data.

Both the uncertainties from the experimental data and the uncertainties from the model have been estimated. These results are now combined to give a best estimate of the total uncertainty on the $^{12}$C$(\alpha,\gamma)^{16}$O reaction rate.

\subsection{Summary and Total Uncertainty Estimate} \label{sec:total_unc}

The previous sections have investigated different sources of uncertainty in the extrapolation of the $^{12}$C$(\alpha,\gamma)^{16}$O cross section to the stellar energy range. The uncertainty analysis is complicated by data sets that lack good statistical agreement and by ambiguities inherent in the phenomenological model. For added clarification, the key results of the above sections are summarized here. 

\begin{itemize}
\item The reaction data now provide definitive solutions for the ground state interference patterns of both the $E1$ and $E2$ cross sections, eliminating a large source of uncertainty. However, many of the $E2$ measurements show large scatter with respect to one another as well as the $R$-matrix prediction for the cross section, that far exceeds their quoted uncertainties. The $E1$ data are in better agreement but still produce a reduced $\chi^2$ significantly greater than one.
\item The reduced $\chi^2$ values for the $\beta$-delayed $\alpha$ emission spectra and the scattering data are also significantly greater than one, likely a result of only approximate modeling of the remaining experimental effects reported in the data. This may even suggest that there are additional unaccounted for uncertainties in the data or, very likely, that the models used to correct for remaining experimental effects in the data are insufficiently accurate.
\item If experimental measurements and uncertainties are taken at face value and model uncertainties are ignored, and the uncertainty in the extrapolation of the $R$-matrix to low energy is calculated, an uncertainty of only a few percent is obtained. However, the large reduced $\chi^2$ values found for such a fit indicates that this method would significantly underestimate the uncertainty. For this reason an more conservative uncertainty estimator, that of Eq.~(\ref{eq:sivia}), was used for the MC analysis. Additionally, several sources of known model uncertainty were explored and found to make a significant contribution to the overall uncertainty budget.
\item The properties of the subthreshold states seem to be well known at present, both $\gamma$ widths and ANCs. Since the development of sub-Coulomb transfer measurements, different experimental measurements have yielded consistent results for the ANCs. In view of the previous points, the uncertainties in the ANCs and $\gamma$ widths of the subthreshold states have been included in the fitting and uncertainty analysis using Eq~\ref{eq:reduced_chi_squared_and_sub_unc}.
\end{itemize}

\begin{figure}
\includegraphics[width=1.0\columnwidth]{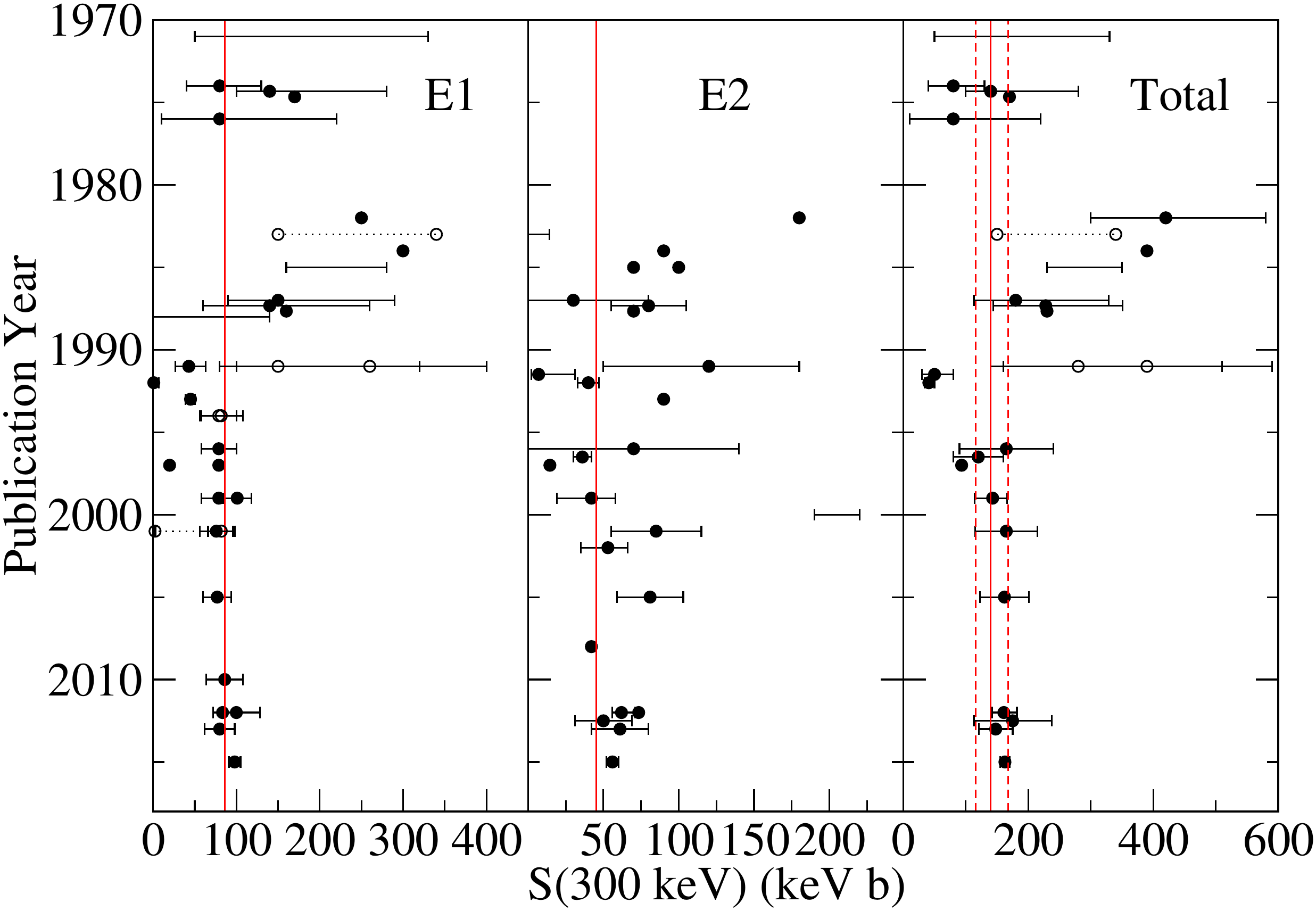}
\caption{(Color online) Plot of $S$(300 keV) values for all past estimates in the literature back to 1970. For the $E1$ extrapolation, the uncertainty is quite large before the early '90s. The dramatic decrease is due to the greatly improved constraint on the 1$^-$ subthreshold contribution provided by $beta$ delayed $\alpha$ emission data and later from sub-Coulomb transfer experiments. The extrapolation of the $E2$ data still has a somewhat larger spread but it too has seen significant reduction in its uncertainty thanks to high precision scattering and sub-Coulomb transfer experiments. \label{fig:S_0_summary_2}}
\end{figure}

The total uncertainty has thus been estimated by combining the uncertainties from the experimental data via the MC analysis and the model uncertainties summarized in Table~\ref{tab:systematic_unc}. This produces a total uncertainty of 15 to 20\% when both contributions are summed in quadrature over most of the energy region. The results of this analysis are compared to previous results of $S$(300 keV) in Table~\ref{tab:S-extrap_history} and Fig.~\ref{fig:S_0_summary_2}. With the fitting and uncertainties discussed in detail, the discussion can turn back to several important recent works that were neglected in Sec.~\ref{sec:modern_era}. 

\begin{table*}
\caption{Summary of the systematic uncertainties that were considered and their affects on $S$(300 keV). \label{tab:systematic_unc} }
\begin{ruledtabular}
\begin{tabular}{ l c } 
Source & Syst. Unc. Contribution (keV b) \\
\hline
Inclusion of $Q$-coefficients & +1.2 \\
Relativistic $\gamma$-ray angular distribution correction & -0.3\\
Fixed energy of $E_x$ = 9.5779 MeV & -1.2 \\
Different fitting functions & +7.7 \\
Fixed \textcite{Kunz_thesis} Cascade data normalizations & -0.5 \\
Fixed \textcite{Schurmann2011557, schurmann2005} data normalizations & +1.1 \\
No ground state $^{12}$C$(\alpha,\gamma)^{16}$O $E2$ excitation curve data & -0.6 \\
No ground state $^{12}$C$(\alpha,\gamma)^{16}$O $E1$ or $E2$ excitation curve data & -0.9 \\
No ground state $^{12}$C$(\alpha,\gamma)^{16}$O angular distribution data & -5.2\\
1$^-$ ground state $^{12}$C$(\alpha,\gamma)^{16}$O $E1$ BGP & +15.2 \\
2$^+$ ground state $^{12}$C$(\alpha,\gamma)^{16}$O $E2$ BGP & -3.0 \\
$\alpha_0$ channel radius variation & $\pm$8 \\
Alternative fitting approaches & +13.1 \\
\end{tabular}
\end{ruledtabular}
\end{table*}

\section{\label{sec:discussion} Discussion of recent works}

Fig.~\ref{fig:S_0_summary_2} gives an idea of the difficulty encountered in analyzing and then extrapolating the cross section of the $^{12}$C$(\alpha,\gamma)^{16}$O reaction to low energy by the wide range of values for $S$(300 keV) that have been estimated over the course of many works. As discussed in Sec.~\ref{sec:classical_era}, from the outset, Fred Barker realized the importance of a comprehensive analysis \cite{Barker1971} and it is an impressive feat that even his first work on the subject contains the key elements of the most sophisticated analyses published today: fit to capture, scattering, $\beta$ delayed $\alpha$ spectrum, and consideration of the reduced $\alpha$ widths from transfer reactions. However, the complexity involved in having to include data from many different reaction types is also why many analyses have only considered a subset of the data (or reactions). Even today, making a general analysis code that can simultaneously fit all of the data is quite challenging and simply compiling all of the experimental data is no small task. 

It is important to note that even the implementation of the $R$-matrix methods used over the years has undergone significant development. This is for the most part limited to the $\gamma$ ray channels, but it is important to realize that extrapolations using ``$R$-matrix" have not always been the same. In Sec.~\ref{sec:phenom_models} it was described how the hybrid $R$-matrix model was used for sometime but fell out of favor because it was unable to fit the scattering data as well as the standard theory. A significant improvement to the modeling of the external component of the capture cross section was provided by \textcite{BarkerKajino1991}, as discussed in detail in Sec.~\ref{sec:radiative_capture}. Yet even after this work, several analyses continued to neglected the external contributions.

\begin{figure}
\includegraphics[width=1.0\columnwidth]{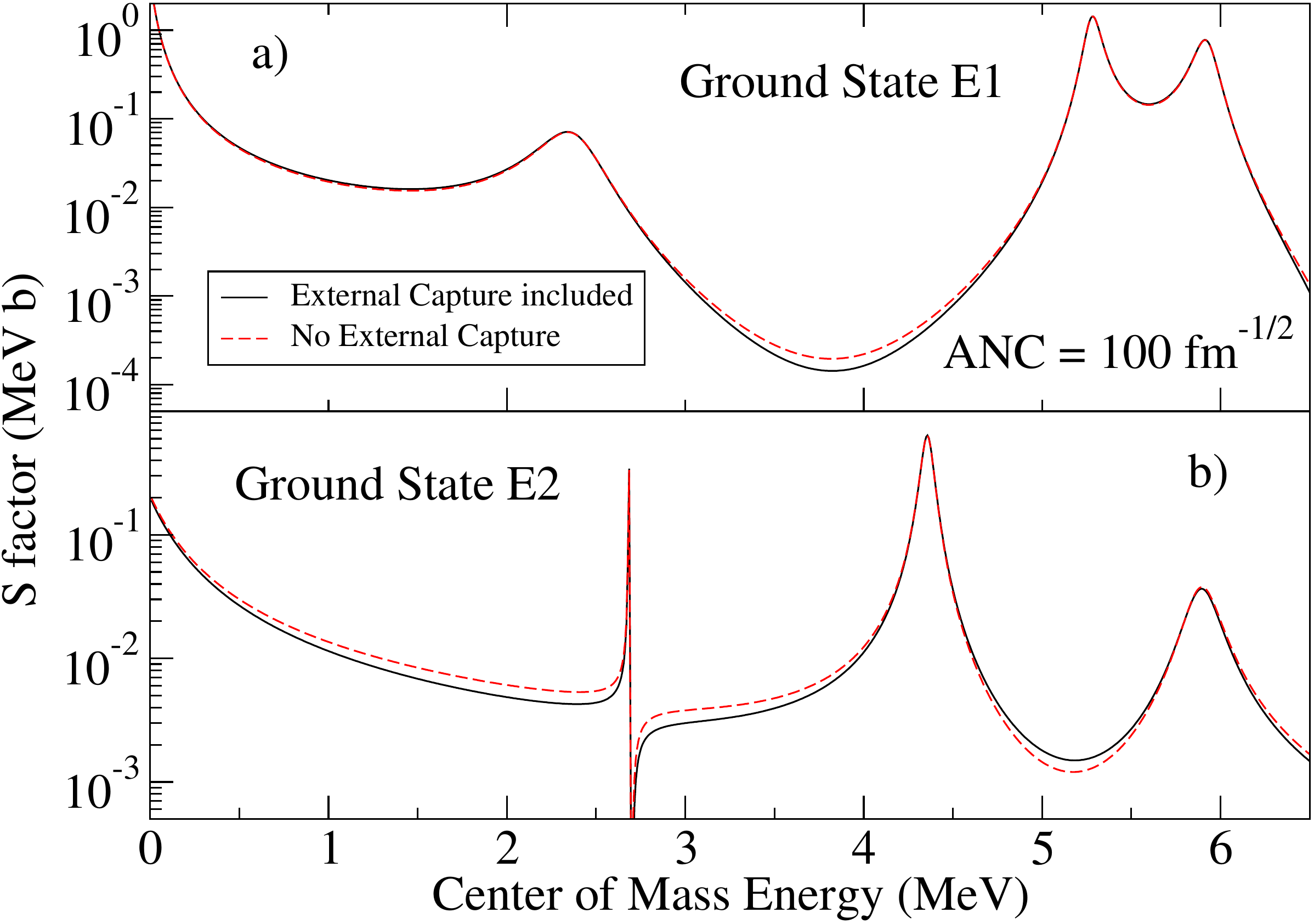}
\caption{(Color online) Comparison of calculations of the ground state transition capture $S$-factor made with (solid black line) and without (dashed red line) an external capture contribution to the $R$-matrix. Here no external capture signifies that only the internal part of Eq.~(\ref{eq:t-matrix_gammas}) is considered. A moderate value of 100~fm$^{-1}$ has been used as the value of the ground state ANC (see Table~\ref{tab:ANC_gs}). As expected, the $E1$ external capture is negligible over much of the energy range. Only in the very low $S$-factor region around $E_\text{c.m.} \approx$~3.75~MeV is there a significant effect. However, this energy range has proved largely experimentally inaccessible. For the $E2$ external capture, the lack of low energy resonances makes its contribution more important. Its interference with the subthreshold 2$^+$ resonance can produce a change in $S$-factor significant compared to the uncertainties of the data. As shown in Fig.~\ref{fig:components}, all of the cascade transitions have very significant hard-sphere contributions, making the inclusion of external capture critical for their modeling. The significance of the external capture for the different transitions is also reflected in the relative contributions of the internal and channel portions of the total reduced width amplitude as given in Table~\ref{tab:fit_params_grwa}. \label{fig:external_contributions}}
\end{figure}

Fig.~\ref{fig:external_contributions} demonstrates the effect of including external capture in the ground state transition of the $^{12}$C$(\alpha,\gamma_0)^{16}$O reaction in the $R$-matrix analysis. Since the $\alpha$ particle ANC of the ground state of $^{16}$O remains quite uncertain, a moderate value of 100~fm$^{-1/2}$ has been chosen for an example calculation. Fig.~\ref{fig:external_contributions} a) demonstrates that the $E1$ external capture can be neglected to a good approximation given the the data presently available. At $E_{c.m.}$~=~300~keV the effect is only about 2\% for the ANC used. The effect is maximum, $\approx$35\% difference, in the off-resonance region at $E_{c.m.} \approx$~4~MeV. Therefore if experimental measurements do access this region $E1$ external capture does become a necessary part of the calculation. 

In Fig.~\ref{fig:external_contributions} b) it is shown the the $E2$ external capture is much more significant. Here the effect is maximum, $\approx$30\%, in the region from 2~$\lesssim E_{c.m.} \lesssim$~3.5~MeV. Here certainly the $E2$ external capture can not be neglected since data have been measured throughout this region with uncertainties much less than 30\% in many cases. The effect lessens at $E_{c.m.}$~=~300~keV, but is still $\approx$10\%. Therefore, if $E2$ external capture is neglected, a fit may try to compensate by increasing the 2$^+$ subthreshold $\alpha$ particle ANC, which produces a similar energy dependence in the cross section from the tail of the subthreshold state. This would result in what would seem to be a tension between the ANC measured via transfer reaction and that deduced from the fit to the $E2$ capture data.

Further, as the main subject of \textcite{BarkerKajino1991}, the cascade transition cross sections can not be analyzed without external capture since it dominates the cross section over a wide energy range (see Fig.~\ref{fig:components}). Therefore any analysis that is making a global analysis of the $^{12}$C$(\alpha,\gamma)^{16}$O reaction must include a model for the direct capture process. 

This section now returns to the aforementioned recent works that were not discussed in Sec.~\ref{sec:modern_era} above, those of \textcite{Schurmann201235}, \textcite{PhysRevC.85.035804}, \textcite{Xu201361} (NACRE2), and \textcite{PhysRevC.92.045802}. Each of these different works has been made by a different research group and has performed an independent comprehensive analysis of the $^{12}$C$(\alpha,\gamma)^{16}$O reaction. 

\subsection{Sch\"{u}rmann {\it et al.} (2011)} \label{sec:compare_schurmann}

The global analysis of \textcite{Schurmann201235} is based on a long history of measurements and analyses performed by Claus Rolf's research group in M\"{u}nster \cite{springerlink:10.1007/BF01415851, Redder1987385, Roters1999, Gialanella2001, schurmann2005, Schurmann2011557} and Wolfgang Hammer's group at Stuttgart \cite{PhysRevLett.86.3244, Hammer2005363, Hammer2005514, PhysRevC.73.055801}. Much of the basis for the global $R$-matrix analysis was developed by Ralf Kunz and was published in \textcite{0004-637X-567-1-643} with many further details of the analysis in \textcite{Kunz_thesis}. It should be noted that this work was done with a completely independent $R$-matrix code to the one used here. This extremely comprehensive work was the first to push the $R$-matrix calculations up above the proton separation energy. There were several approximations that were made at that time that were subsequently improved upon in \textcite{Schurmann201235}. Important improvements have been an external capture calculation of the $E2$ ground state transition and a systematic uncertainty term in the $\chi^2$ function. However, even in \textcite{Schurmann201235}, the analysis does not include the proton and $\alpha_1$ partitions at higher energies. 

The main difference from the present analysis is that a very stringent rejection criterion for the data was adopted. This highlights a common philosophical difference in data analysis. One the one hand, the experimenter knows the details of his own data on a first hand basis, but must often rely on only the details of a publication for the re-analysis of other work. In \textcite{Schurmann201235} the criteria for their data analysis was so stringent that in effect only data measured by that group could be retained. On the other hand, despite the experimenter's confidence in their own data, it is often hard to justify why one measurement is correct over another. 

Another major difference is that \textcite{Schurmann201235} did not directly consider the transfer reaction results for the ANCs in their analysis. They were instead treated as fit parameters. However, unlike in the current analysis, because only a very limited amount of data was considered, the tension between different data sets was greatly reduced and their fitted values did not vary as widely as observed here. This may also be because only the phase shifts from \textcite{PhysRevC.79.055803} were used, the fewer number of data points then lessened the tension between the scattering data and other data sets. Also, only the $\beta$ delayed $\alpha$ data of \textcite{PhysRevC.50.1194} were considered eliminating the tension between those different data sets.

For the ground state transition, a larger value for the ANC of the 1$^-$ level than those of the higher precision transfer reactions of \textcite{PhysRevLett.83.4025, PhysRevLett.114.071101} was found. This is in general agreement with what has been found here as well when the the ANCs are allowed to vary freely (see Sec.~\ref{sec:fit_methods}). For the 2$^+$ ANC, excellent agreement was achieved (see Table~\ref{tab:ANCs}). The agreement with transfer is quite poor for the $E_x$~=~6.05~MeV $\alpha$ ANC, as has been discussed in Sec.~\ref{sec:cascades_12Cag}. 

For the final uncertainty analysis a Monte Carlo analysis similar to the one performed here (and originally applied to the case of $^{12}$C$(\alpha,\gamma)^{16}$O by \textcite{Gialanella2001}) was performed. Systematic uncertainties were also explored but few details were given regarding the details. In the end, a value of $S$(300~keV)~=~161$\pm$19$_\text{stat}$$^{+8}_{-2 \text{sys}}$ was found. 

This central value is very close to the upper 68\% confidence level of the current analysis. However, it is difficult to understand the values of $S$(300~keV) for the $E1$ and $E2$ contributions compared to the present results. The $E1$ values are very close (\textcite{Schurmann201235} 83.4, current work 86.3~keV~b). While naively these values are in good agreement, they should not be, since the 1$^-$ ANC used by \textcite{Schurmann201235} is significantly larger than that used here. Further, and equally perplexing, the $E2$ value is much larger than that deduced here (\textcite{Schurmann201235} 73.4, current work 45.3~keV~b), yet the 2$^+$ ANCs are nearly identical for the two analyses! Unfortunately \textcite{Schurmann201235} give few details as to the resonance parameters that were used so it is impossible to make an exact comparison. One explanation could be that very different background pole contributions were used. Since \textcite{Schurmann201235} did not include data at higher energies, more freedom should have been possible for their background contributions.

The overall uncertainty quoted by \textcite{Schurmann201235} for $S$(300~keV) is $\sim$13\%. Given the comparison with this work, this value seems reasonable, if one accepts that all of the data used are correct. It also may be that that some of the systematic uncertainties discussed here were overlooked since there are not many details given regarding this in \textcite{Schurmann201235}. 

\subsection{Oulebsir {\it et al.} (2012)}

A global $R$-matrix analysis was preformed as part of the transfer reaction study of \textcite{PhysRevC.85.035804}. Besides the analysis being performed from a transfer reaction experiment point of view, this analysis has been chosen for comparison because it represents the most recent calculations with Pierre Descouvemont's $R$-matrix code \texttt{DREAM} (see, for example, \textcite{Mountford2014359}). Another $R$-matrix code developed completely independently from the \texttt{AZURE2} code used here. While this analysis is limited to the more typical lower energy range, it considers all of the ground state transition $E1$ and $E2$ data. None of the cascade data are considered however. The fits do include some higher energy resonances explicitly in the $R$-matrix calculation with their parameters fixed to values in the compilation. This analysis follows a similar analysis technique as the ``best fit" of the present work in that the ANCs were treated as fixed parameters. However, in addition, all the resonance parameters of the unbound states were fixed to previously determined values. Only the background pole contributions widths were allowed to vary. Given that one of the largest uncertainties in the $R$-matrix analysis stems from the background poles, this seems to be a reasonable approximation. 

The fitting technique was done iteratively, first fitting to the scattering data to constrain the energy and $\alpha$ widths, then fitting to the capture data to constrain the ground state $\gamma$ widths of the background poles. Only $l$~=~1 and 2 phase shifts were fit instead of the actual scattering yields, which may cause difficulties in the uncertainty propagation as described in \textcite{PhysRevC.54.393}. The $^{16}$N$(\beta\alpha)^{12}$C spectrum was not fit but a comparison of the $l$~=~1 contributions was performed showing reasonable agreement. 

The extrapolated $S$ factors are in good agreement with the current analysis. This should be expected because one of the primary methodologies of the analysis is the same, fixing of the subthreshold state ANCs to the value measured through transfer reactions. The values for the $E1$ and $E2$ $S$ factors (100(28) and 50(19)~keV~b respectively) are systematically larger than the ``best fit" values of this work (86.3 and 45.3~keV~b respectively), although in good agreement considering the uncertainties. 

This difference is likely because there are no background poles included in the current ``best fit" and inclusion of the poles has only been found to increase the low energy cross section. For tests with background poles in this work, the extrapolated value $S$(300 keV)$_{E1}$ is nearly identical with the one found here (101.5~keV~b). For $S$(300 keV)$_{E2}$ the value found here remains nearly the same, actually slightly decreasing to 42~keV~b. This may be a result of the greater constraint imposed on the background poles by the inclusion of the higher energy data. 

The uncertainty in the $S$-factor quoted by \textcite{PhysRevC.85.035804} is significantly larger than that of the current analysis. This results from the larger uncertainties obtained in the ANCs from their transfer experiment ($\sim$20\%) over those of either \textcite{PhysRevLett.83.4025} and \textcite{PhysRevLett.114.071101} that have been adopted in this work ($\sim$10\%) (see Table~\ref{tab:ANCs}). Further, the lack of higher energy data, in particular the total cross section data of \textcite{schurmann2005}, results in significantly more freedom in the background contributions. 

\subsection{Xu {\it et al.} (2013) (NACRE2)} 

The analysis presented as part of \textcite{Xu201361} (NACRE2) provides an interesting comparison because it uses a potential model (PM) calculation instead of a phenomenological $R$-matrix method to predict the low energy $S$-factor. While still a phenomenological procedure, the PM method tries to take a step in a more fundamental direction by defining real Woods-Saxon potentials for each $J^\pi$. The parameters that define the potential, its magnitude, radius, and diffuseness, are treated as free parameters in the fit to match the resonances and binding energies of the experimental data. This has the advantage that it may lead to less fit parameters than the standard phenomenological $R$-matrix approach but it comes at the cost of less flexibility. Further, some states can not be produced by the potential model, as discussed in Sec.~\ref{sec:cluster_models}, and have been parameterized separately using approximate Breit-Wigner formulas. Interference between resonances must also be introduced ad hoc. The direct capture contribution to the cross section is modeled in a similar manner as that proposed by \textcite{Rolfs197329}. Additionally, each transition is fit independently, this includes the different ground state $E1$ and $E2$ fits. The phase shifts and the $^{16}$N$(\beta\alpha)^{12}$C data are not considered in the model.

It is clear from Fig.~64 of NACRE2 that the PM model used is not able to reproduce the experimental data to the same degree as the phenomenological $R$-matrix fits. However, considering the added constraints and approximations that are imposed, the reproduction of the data is impressive. While a $\chi^2$ fit to the data was likely made, the value is not given so an exact comparison is not possible. The final value of the extrapolated $S$-factor ($S_\text{total}$ = 148(27), $S_{E1}$ = 80(18), $S_{E2}$ = 61(19)~keV~b) is similar to that deduced here, although with all the considerations pointed out above, this maybe somewhat coincidental. The total uncertainty is quoted as about $\sim$18\%, but the details of how this is calculated are not given.

\subsection{An {\it et al.} (2015)} 

Baring the current work, \textcite{PhysRevC.92.045802} considers the largest amount of data over the broadest energy range. While a phenomenological $R$-matrix analysis was performed, it seems that the capture formalism was limited to that of internal contributions. The analysis is performed using the $R$-matrix code RAC, which has been used previously for evaluations of neutron capture data \cite{Carlson20093215}. The neglect of external contributions for the ground state $E1$ cross section has been justified (see, for example, \textcite{BarkerKajino1991}). However, for the ground state $E2$ cross section, which has an appreciable external capture component, this approximation is not valid for a high precision analysis as shown earlier in this section. Further, the cascade transition cross sections all have significant external contributions. Additionally, no mention was made of the corrections for experimental effects, which are quite important for the $^{16}$N$(\beta\alpha)^{12}$C and scattering data sets.

\textcite{PhysRevC.92.045802} make the bold claim that the low energy $S$ factor has been constrained to 4.5\%. However, the lack of an investigation of systematic effects and approximations that seem to have been made in the theory make this difficult to defend. The uncertainty determination that was used is an iterative fitting procedure that involves inflating the experimental error bars in order to achieve a fit with a reduced $\chi^2$ that approaches one. However, this has the underlying assumption that the average value of all of the data gives the best estimate of the cross section and it is far from clear that is the case. In this work it has been found that capture data likely over estimate the low energy cross section and that there is tension between the $^{16}$N$(\beta\alpha)^{12}$C, scattering, and transfer reaction data that translates into different preferred values for the subthreshold ANCs and by extension the low energy cross section. The results of the transfer reaction measurements of the ANCs were also not considered. 

It is also unclear how or if the experimental systematic uncertainties are included in the fit, since no $\chi^2$ function is given. A related issue is that the normalizations of some data sets were fixed and it is unclear how or if these uncertainties were propagated through into the final uncertainty budget. It is also unclear if systematic effects of the $R$-matrix model were investigated. It is stated that a channel radius of 6.5~fm was used for the fitting but there is no discussion of how different channel radii effect the fit and there is no discussion of how the background poles effect the fitting and extrapolation. The quality of the fits to the scattering data are also rather poor compared to the quality of a similar $R$-matrix analysis made by \textcite{PhysRevC.87.015802}. The reason for this is not discussed. Further, the definition of the $\gamma$ ray fit parameters given in Table~IV are unclear making a recalculation of the fit impossible.

In the end, not enough details are given by \textcite{PhysRevC.92.045802} to understand the fitting or uncertainty estimate procedure. It appears as though the many systematic uncertainties identified in the current work were neglected. In fact, if the model uncertainties and the tension between different data sets are ignored in the Monte Carlo uncertainty analysis described here, a result similar to the 4.5\% uncertainty of \textcite{PhysRevC.92.045802} is obtained. 

\section{\label{sec:RR} Stellar Reaction Rate and Implications}

The stellar reaction rate for $^{12}$C($\alpha,\gamma$)$^{16}$O was calculated as a sum of non-resonant, or broad resonant, $S$-factor contributions that were determined through the $R$-matrix analysis by numerical integration of Eq.~(\ref{eq:rr}), and narrow resonance contributions that were calculated through a Breit-Wigner approximation using Eq.~(\ref{eq:rres}). This separation was made to avoid numerical integration problems for the narrow resonances and because their uncertainties were better quantified experimentally as uncertainties on their strengths $\omega\gamma$ (see Sec.~\ref{sec:RR_intro}). The uncertainties in the rate were calculated from the MC analysis (see Sec.~\ref{sec:MC_analysis}) and from the model uncertainties (see Sec.~\ref{sec:unc_analysis}) discussed above. The experimental uncertainties in the energies and the strengths were likewise used to propagate the uncertainties stemming from the narrow resonances.

\begin{figure}
\includegraphics[width=0.9\columnwidth]{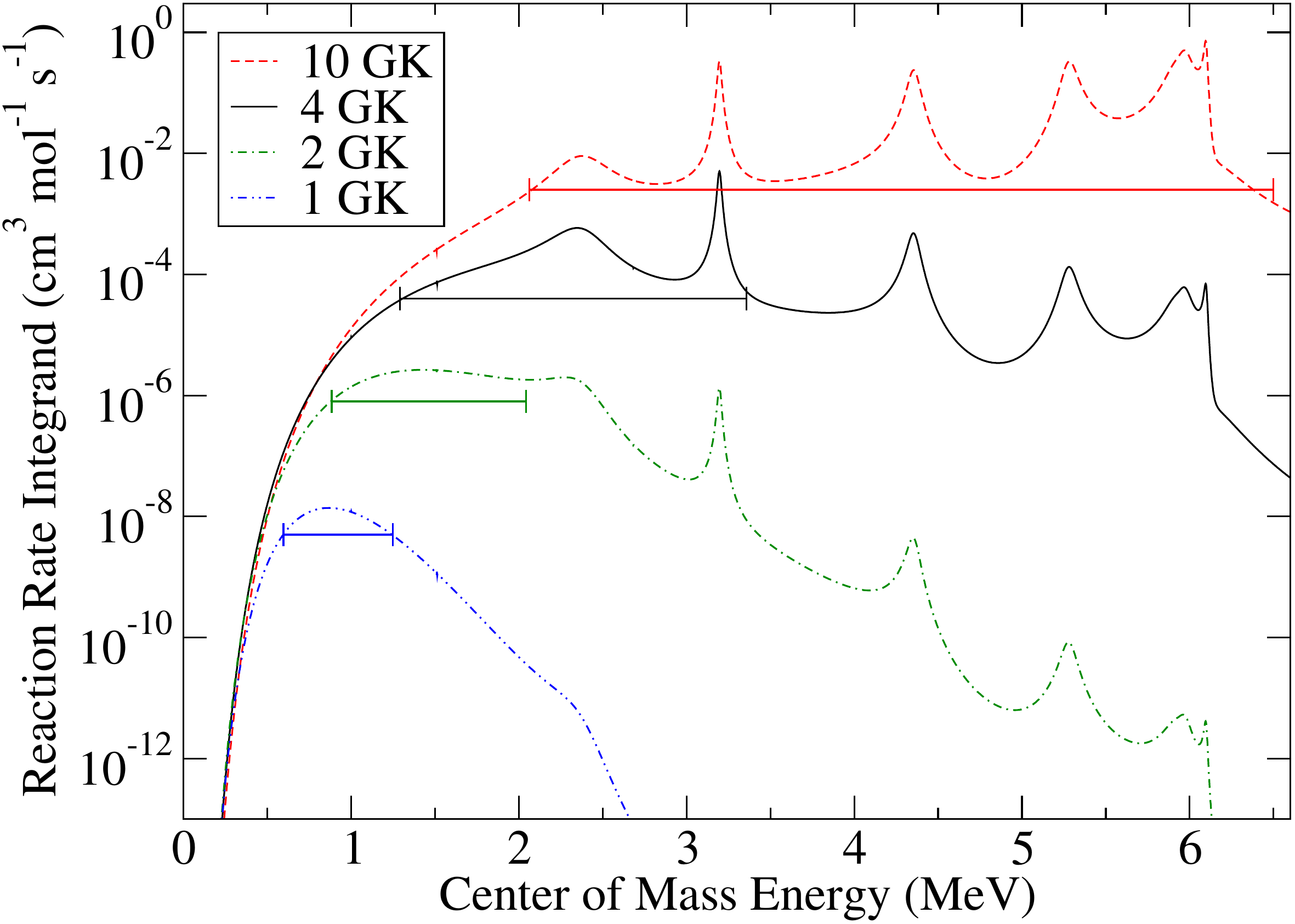}
\caption{(Color online) The reaction rate integrand as a function of CM energy for $T$~=~1, 2, 4, and 10~GK. At larger temperatures above $T$~=~1 GK several resonance contributions begin to dominate the rate. Above $T \approx$~4~GK it is estimated that higher lying resonance contributions (at $E_{\text{c.m.}} >$~6.5 MeV) not included in the present analysis could have a significant contribution compared to the quoted uncertainty. While the ground state transition has been shown to be fairly weak at these higher energies, limited information is available for the cascade transitions, and they may make significant contributions. For comparison, the Gaussian Gamow energy windows described by Eqs.~\ref{eq:Gamow} and~(\ref{eq:Gamow_delta}) are indicated by the horizontal error bars. \label{fig:RR_limit}}
\end{figure}

Fig.~\ref{fig:RR_limit} shows the Gamow window (see Eqs.~(\ref{eq:Gamow}) and (\ref{eq:Gamow_delta})) and the integrand of the $S$-factor with the Maxwell-Boltzmann energy distribution (the integrand of Eq.~(\ref{eq:rr})) for a range of temperatures. This depicts how different energy ranges of the cross section contribute to the reaction rate at different stellar temperatures.  

The narrow resonance contributions ($\Gamma_{\text{total}} \lesssim$~1~keV) are from the 2$^+$ resonance corresponding to the state at $E_{\text{x}}$~=~9.8445(5)~MeV ($\Gamma_\alpha$~=~0.62(10)~keV, $\Gamma_\gamma$~=~9.8(8) meV) and the 4$^+$ at $E_{\text{x}}$~=~11.0967(16)~MeV ($\Gamma_\alpha$~=~0.28(5)~keV, $\Gamma_\gamma$~=~5.6(14) meV) \cite{Tilley19931}. There is an additional narrow 0$^+$ state at $E_{\text{x}}$~=~12.049(2)~MeV ($\Gamma_{\text{total}}$~=~1.5(5)~keV). Its strength has been reported for the first time by \textcite{schurmann2005} ($\omega\gamma$~=~11.2(15) meV).

The narrow resonance contributions and their uncertainties have been calculated using \texttt{STARLIB} \cite{0067-0049-207-1-18}. As these resonances are narrow, their interferences with the broad states can be neglected to within the uncertainty of this analysis. The uncertainties are combined by summing (incoherently) the rate PDFs from the $R$-matrix Monte Carlo procedure with those from the Monte Carlo Breit-Wigner narrow resonance approximation calculations from \texttt{STARLIB}.

Of the narrow resonances, only the 2$^+$ has a significant affect on the rate. It makes a contribution of $>$2\% at $T >$~1.75~GK with a maximum contribution of 15\% at $T$~=~3.5 GK. It is estimated that this resonance can have a $>$2\% contribution up to $T \approx$~9~GK. It is estimated that the 4$^+$ resonance has a maximum contribution of $\approx$1\% at $T \approx$~6~GK. The 0$^+$ resonance is too weak to make a significant contribution even at $T$~=~10~GK.  

\begin{figure}
\includegraphics[width=1.0\columnwidth]{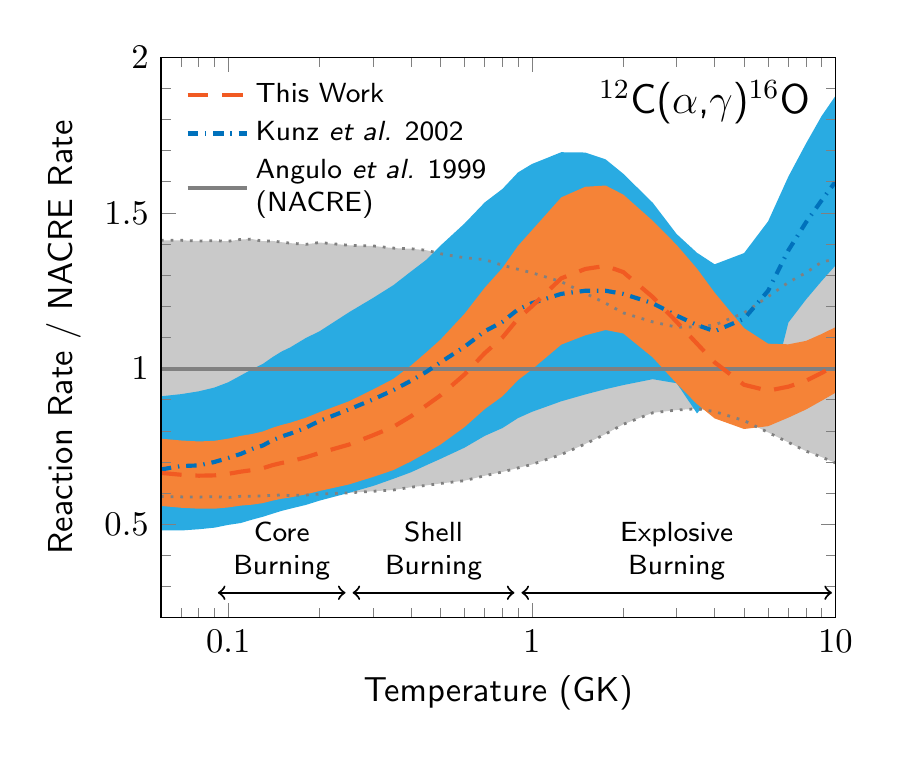}
\caption{(Color online) Comparison of the reaction rate and uncertainty calculated in this work (orange band, dash central line) and that from \textcite{0004-637X-567-1-643} (blue band, dash-dotted central line) normalized to the adopted value from \textcite{nacre} (NACRE compilation) (gray band, solid central line). The deviations at higher temperature are the result of the different narrow resonance and cascade transitions that were considered in the different works. Arrows at the bottom indicate temperature ranges for different Carbon burning scenarios. \label{fig:RR_compare}}
\end{figure}

The total reaction rate is compared to the NACRE rate \cite{nacre} in Fig.~\ref{fig:RR_compare}. The present rate is within the uncertainties of \cite{nacre} except at $T \approx$~2~GK where the present rate is larger because of the inclusion of the narrow 2$^+$ resonance and the cascade transitions. The uncertainty band is significantly smaller at low temperatures but is similar at higher temperatures reflecting the significantly increased constraint on the subthreshold parameters imposed by transfer reactions since the NACRE publication. The rate from \textcite{0004-637X-567-1-643} is also shown in Fig.~\ref{fig:RR_compare} for additional comparison. The reaction rate of this work is given in Table~\ref{tab:RR_absolute} of Appendix~\ref{app:RR}.

With the revised reaction rate now in hand, investigations can now be made to ascertain the effect of the smaller central value and smaller uncertainty band on stellar model calculations.

\section{Astrophysics implications}\label{sec:astro_2}

Returning at last to the discussions of Sec.~\ref{sec:astro}, the $^{12}$C($\alpha$,$\gamma$)$^{16}$O reaction is
responsible for the origin of oxygen in the universe and
for setting the profile of the carbon to oxygen ratio in stars. This, in turn,
affects subsequent stellar evolution and determines the nucleosynthesis phases of
quiescent and explosive burning events.

Attention is focused on models of a single star, with solar metallicity, and zero-age main-sequence
masses of 3, 15, and 25~M$_\odot$. These three masses are representative of different stellar evolution and nucleosynthesis paths.
The 3~M$_\odot$ models are of interest because they produce
CO white dwarf masses near the
M~$\simeq$~0.6~M$_\odot$ peak of the observed DA (hydrogen-dominated
atmosphere) and DB (helium dominated atmosphere) white dwarf mass
distributions \citep{eisenstein_2006_aa,kepler_2007_aa,kepler_2015_aa,kepler_2016_aa}.
The 15~M$_\odot$ models are chosen because, for a
classical Salpeter initial mass function with slope $\Gamma$~=~-1.35
\citep[e.g.,][]{salpeter_1955_aa,scalo_1986_aa,maschberger_2013_aa},
the average supernova mass by number is $\langle M_{SN}                                           
\rangle$~=~(1/2)$^{1/\Gamma}$~$M_{L}$ (here $M_L$ is the lower mass
limit for stars that become core-collapse supernovae). For
$M_L$~=~9~M$_{\odot}$, this yields $\langle M_{SN} \rangle$~=~15~M$_{\odot}$. Finally, the
25~M$_\odot$ models are motivated by their representation of the average nucleosynthesis supernova
for a Salpeter initial mass function; that is, the average
supernova mass by number weighted by the mass fraction ejected in heavy
elements.

The stellar models are evolved using the Modules for Experiments in Stellar
Astrophysics software instrument \citep[henceforth \texttt{MESA}, version 3372
for 3~M$_\odot$ models and version 7624 for 15 and 25~M$_\odot$
models,][]{paxton_2011_aa,paxton_2013_aa,paxton_2015_aa}.
The 3~M$_\odot$ models were computed from the pre-main sequence to the
AGB phase and through several thermal pulses (He-shell flashes). The
15 and 25~M$_\odot$ models were calculated from the pre-main sequence to
extinction of core He burning, defined as the time when the central mass
fraction of He has fallen below $1\times10^{-5}$. Other than the specified
$^{12}$C($\alpha$,$\gamma$)$^{16}$O reaction rate, models with the
same initial mass assume identical input physics assumptions
\citep[e.g., ][]{jones:15,fields_2016_aa,farmer_2016_aa}. An overview of the
model results using the $^{12}$C($\alpha,\gamma)^{16}$O reaction rate from this work is given in Table~\ref{tab:MESA-models}.

\begin{table}
\centering
\caption{Overview of the \texttt{MESA} stellar models.
$M_{\rm ini}$ is the initial mass in solar masses,
$\lambda_{12\alpha\gamma}$ is the lower (L), adopted (A), and upper (U)
$^{12}$C($\alpha,\gamma)^{16}$O reaction rate from this work.
$M_{\rm CO}$ is the carbon-oxygen core mass at the end of core helium burning.
X$_c$($^{12}$C) and X$_c$($^{16}$O) are the central mass fractions of $^{12}$C and $^{16}$O, respectively.
}
\label{tab:MESA-models}
\begin{ruledtabular}
\begin{tabular}{l c c c c c c}
        id & $M_{\rm ini}$  & $\lambda_{12\alpha,\gamma}$ &
        $M_\mathrm{CO}$ & X$_c$($^{12}$C) & X$_c$($^{16}$O) & X$_c$($^{12}$C)/X$_c$($^{16}$O) \\
\hline
m3l  & 3   & L   & 0.189  & 0.35 & 0.62 & 0.56 \\
m3a  & 3   & A   & 0.180  & 0.34 & 0.64 & 0.53 \\
m3u  & 3   & U   & 0.190  & 0.27 & 0.71 & 0.38 \\
m15l & 15  & L   & 3.07   & 0.30 & 0.67 & 0.45 \\
m15a & 15  & A   & 3.03   & 0.25 & 0.72 & 0.35 \\
m15u & 15  & U   & 3.08   & 0.22 & 0.75 & 0.29 \\
m25l & 25  & L   & 6.40   & 0.25 & 0.72 & 0.34 \\
m25a & 25  & A   & 6.45   & 0.20 & 0.76 & 0.27 \\
m25u & 25  & U   & 6.45   & 0.16 & 0.80 & 0.21
\end{tabular}
\end{ruledtabular}
\end{table}

Throughout this section a comparison to the rate from this work is made to that of \citet{0004-637X-567-1-643}, as they are propagated through different stellar models. The \citet{0004-637X-567-1-643} rate has been chosen because it has been widely excepted as one of the most accurate determinations of the $^{12}$C($\alpha$,$\gamma$)$^{16}$O reaction rate and has been used for several years in many rate libraries such as JINA Reaclib \cite{0067-0049-189-1-240}. As illustrated in Fig.~\ref{fig:RR_compare}, the rates of \citet{nacre} (NACRE), \citet{0004-637X-567-1-643}, and the present work are all in agreement, but their central values trend lower as do their uncertainties respectively.  

\subsection{\label{threemsun}Aspects of $^{12}$C($\alpha$,$\gamma$)$^{16}$O on 3~M$_\odot$ Stellar Models}

\begin{figure}
\includegraphics[width=1.0\linewidth]{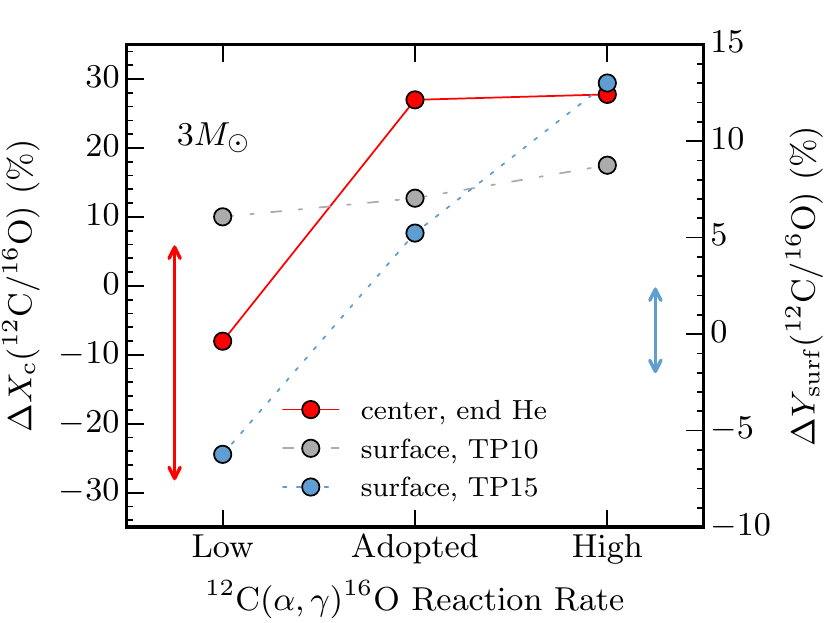}
\caption{(Color online) Ratio of $^{12}$C and $^{16}$O mass fractions at the
        stellar center at the end of core
        He burning (left axis; red circles) and number ratios at the surface
        (right axis; grey and blue circles) during the
        10$^{\rm th}$ and 15$^{\rm th}$ thermal pulses (He shell-flashes) in
        the 3~M$_\odot$ models. The points connected by lines are differences arising in the
        models when using the $^{12}$C$(\alpha,\gamma)^{16}$O rate of this work and that of \citet{0004-637X-567-1-643}.
        The red arrow along the left y-axis spans the range of
        $X_\mathrm{C}(^{12}\mathrm{C}/^{16}\mathrm{O})$ resulting from
        the uncertainty in the rate of this work. The blue arrow
        along the right y-axis is the equivalent range of
        $Y_\mathrm{surf}(^{12}\mathrm{C}/^{16}\mathrm{O})$ during the
        15$^\mathrm{th}$ thermal pulse.
}
\label{fig: 3_co}
\end{figure}

Fig.~\ref{fig: 3_co} summarizes some of the key differences between using
the $^{12}$C($\alpha$,$\gamma$)$^{16}$O reaction rate of this work and that of \citet{0004-637X-567-1-643}. The
x-axis is the lower, adopted, and upper reaction rate for each source
of the $^{12}$C($\alpha$,$\gamma$)$^{16}$O rate. The left y-axis is
the percentage difference in the central $^{12}$C/$^{16}$O mass fraction
ratio at the end of helium burning. The right y-axis is the percentage
difference in the surface $^{12}$C/$^{16}$O molar abundance ratio
for two thermal pulses. A molar abundance is related to the mass fraction
by the atomic weight (see Fig.~\ref{fig:heburn} caption). For both y-axes
the percentage difference is relative to \citet{0004-637X-567-1-643} (i.e.,
$\Delta A = 100 \cdot (A_\mathrm{this~work} - A_\mathrm{Kunz})/A_\mathrm{Kunz}$).

One of the key features shown in Fig.~\ref{fig: 3_co} is that the central $^{12}$C/$^{16}$O mass fraction
ratio is $\simeq$ 30\% larger when using the adopted or upper rates
from this work compared to those of
\citet{0004-637X-567-1-643}. This is because the latter reaction rates
are larger than those from this work at the relevant
temperatures (see Fig.~\ref{fig:RR_compare}). The trend is opposite,
a $\simeq$ -10\% decrease, for the lower limits because the rate from
this work is larger than that of \citet{0004-637X-567-1-643}. The red arrow along the left
y-axis in Fig.~\ref{fig: 3_co} gives the range of
$X_\mathrm{C}(^{12}\mathrm{C}/^{16}\mathrm{O})$, $\simeq$~35\%, resulting from the use of
the lower and upper $^{12}$C($\alpha$,$\gamma$)$^{16}$O rates of this work. 

Also highlighted in Fig.~\ref{fig: 3_co} is the impact of thermal pulses on
the surface abundances, which can be measured with stellar spectroscopy
\citep[e.g.,][]{smith_1990_aa, abia_2002_aa}. After a thermal pulse
the products of shell He-burning are brought to the surface regions by
convection. A larger $^{12}$C($\alpha$,$\gamma$)$^{16}$O rate results
in more $^{16}$O and less $^{12}$C during helium burning. Every time
the convective envelope dredges freshly burned helium up to
the surface, it mixes more $^{16}$O and less $^{12}$C with the
existing surface composition, decreasing the $^{12}$C/$^{16}$O ratio.

At the 10th thermal pulse (grey circles), the differences between the
$^{12}$C($\alpha$,$\gamma$)$^{16}$O rate of this work and that of \citet{0004-637X-567-1-643} ranges between $\simeq$5-9\%.
At the 15th thermal pulse (blue circles), the differences becomes
$\simeq$-5\% and +13\%. Both thermal pulses show
a linear trend toward larger differences as the two rates
are varied across their respective low, adopted, and high values.
Finally, the blue arrow along the right y-axis shows
the range, $\simeq$~5\% spread, of $Y_\mathrm{Surf}(^{12}\mathrm{C}/^{16}\mathrm{O})$, which represents the uncertainty in the \texttt{MESA} stellar models for the surface abundance ratio, given the uncertainty range of the rate from this work.

The treatment of convective boundary mixing in stellar models for the thermal
pulses has a major role on the surface abundances of $^{12}$C, $^{16}$O and
$s$-process elements, but remains uncertain
\citep[e.g.,][]{straniero_1995_aa,gallino:98,herwig:05,PAS:9305903,cristallo_2015_aa,
herwig_2007_aa,pignatari_2016_aa}. Several physics mechanisms have been
proposed that could dominate convective-boundary mixing including overshooting
\citep{herwig_1997_aa}, exponential decay of convective velocities
\citep{cristallo_2001_aa}, internal gravity waves
\citep{denissenkov_2003_ab,0004-637X-827-1-30}, and magnetic buoyancy
\citep{trippella_2016_aa}. A consensus on the solution to this puzzle has not
yet been reached.

\subsection{\label{onefivemsun}Aspects of $^{12}$C($\alpha$,$\gamma$)$^{16}$O on 15~M$_\odot$ and 25~M$_\odot$ Models}

\begin{figure}
        \includegraphics[width=1.0\columnwidth]{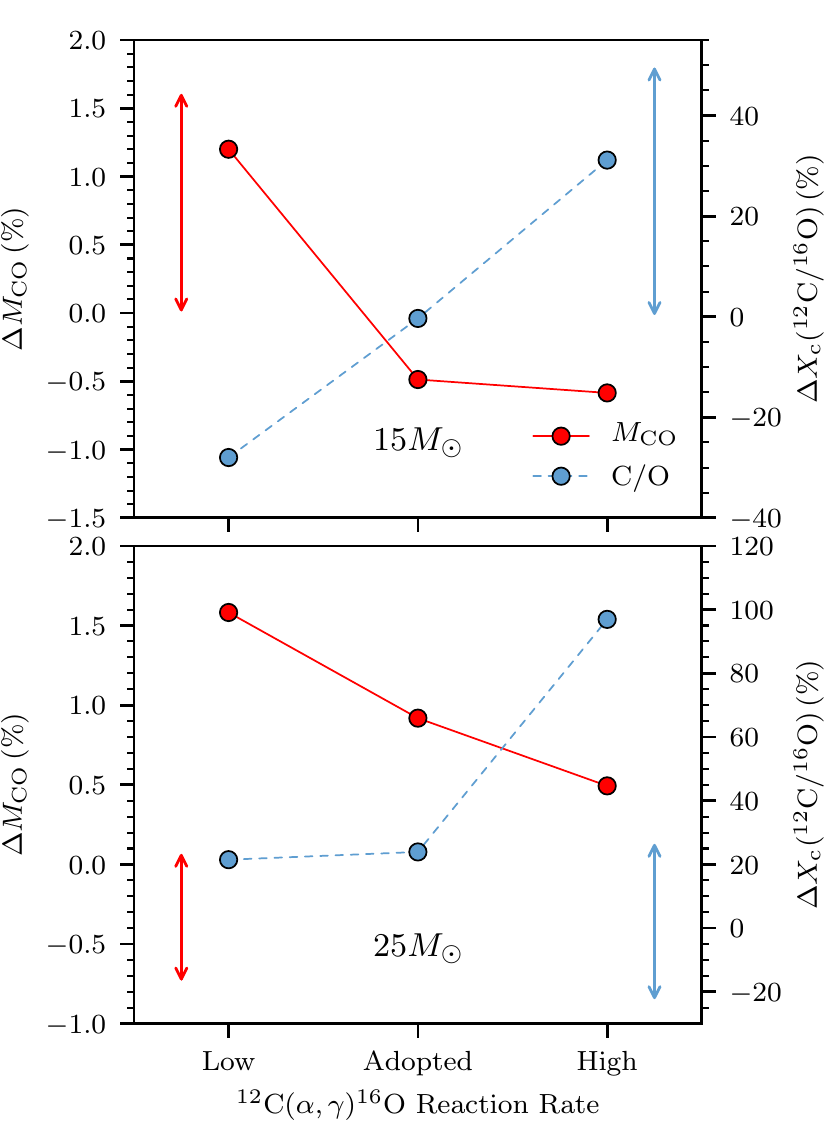}
        \caption{(Color online) Percent difference in the carbon-oxygen core
                mass, $\Delta M_{\rm{CO}}$ (left axis; red circles), and the central
                carbon/oxygen ratio, $\Delta X_{\rm{c}}$($^{12}$C/$^{16}$O)
                ratio (right axis; blue circles), between using the
                $^{12}$C($\alpha$,$\gamma$)$^{16}$O rate of this work and that of \citet{0004-637X-567-1-643} for 15~M$_{\odot}$ (top panel) and 25~M$_\odot$ (bottom panel) \texttt{MESA}
                models at core He depletion (defined as when the central He mass
                fraction falls below $10^{-5}$). The x-axis gives the low,
                adopted, and high $^{12}$C($\alpha$,$\gamma$)$^{16}$O rate for
                the respective rate selection. The red arrow along the left axis
                is the range of $M_\mathrm{CO}$ resulting from the uncertainty
                in the rate of this work. The blue arrow along
                the right y-axis is the equivalent range for the central carbon
                mass fraction.
}
\label{fig:15m25m}
\end{figure}

Fig.~\ref{fig:15m25m} shows the percent differences between using the
$^{12}$C($\alpha$,$\gamma$)$^{16}$O rate from this work and that of \citet{0004-637X-567-1-643} on the CO core
mass, $\Delta M_{\rm{CO}}$, and the central carbon/oxygen ratio,
$\Delta X_{\rm{C}}$($^{12}$C/$^{16}$O), for the 15
and 25~M$_{\odot}$ \texttt{MESA} models. The core masses are measured
as the mass coordinate where $X$($^{4}$He)~$<$~0.01
 and $X$($^{12}$C)~$>$~0.1 when $X_{\rm{c}}(^{4}$He) reaches
1$\times$10$^{-5}$. For the 15~M$_{\odot}$ models, $\Delta                  
M_{\rm{CO}}$ (red circles) ranges between $\simeq$1.2\% for the
respective low rates to $\simeq$-0.5\% for the respective adopted and
high rates. For the 25~M$_{\odot}$ models, $\Delta M_{\rm{CO}}$
ranges between $\simeq$1.7\% for the respective low rates to
$\simeq$1.0\% for the respective adopted rates and $\simeq$0.5\% for the
respective high rates. See Table~\ref{tab:MESA-models} for the
absolute values of the CO core masses resulting from the
$^{12}$C($\alpha$,$\gamma$)$^{16}$O rate from this work. The red arrow along the left y-axes in
Fig.~\ref{fig:15m25m} is the range of $M_\mathrm{CO}$ resulting from
the uncertainty in the rate from this work. The central CO
core masses have a spread of $\simeq$1.5\% for the 15~M$_\odot$ models
and $\simeq$0.7\% for the 25~M$_\odot$ models.

The 15~M$_\odot$ and 25~M$_\odot$ models show CO mass fraction
profiles that are nearly flat, due to convective mixing, from the
center to the inner edge of the shell He-burning region.
For the 15~M$_{\odot}$ models, $\Delta X_{\rm{C}}$($^{12}$C/$^{16}$O)
ranges from $\simeq$-30\% for the respective
low rates through $\simeq$~0\% for the respective adopted rates
to $\simeq$~30\% for the respective high rates.
For the 25~M$_{\odot}$ models, $\Delta X_{\rm{C}}$($^{12}$C/$^{16}$O)
ranges between $\simeq$~20\% for the respective
low and adopted rates to $\simeq$~100\% for the respective high rates.
The difference in the trends between these 15~M$_\odot$ and 25~M$_\odot$
\texttt{MESA} models is driven by the 15~M$_\odot$ models having
a larger electron degeneracy in the core than the 25~M$_\odot$ models,
and nonlinear couplings between the shell burning regions and the core
as the helium fuel depletes. The blue arrow along the right y-axes is the
spread of central carbon mass fraction achieved when using the lower, adopted
and upper rates from this work, and shows
larger spreads, $\simeq$~50\% for the 15~M$_\odot$ models and
$\simeq$~45\% for the 25~M$_\odot$ models.

The mass fractions of $^{12}$C and $^{16}$O in the core at the end of the
He-burning phase for the models using the rate from this work
are listed in Table~\ref{tab:MESA-models}. The models using the highest rate
have the lowest $^{12}$C abundance at the end of core He burning. This will
result in a smaller convective C-burning core in the subsequent evolution which
will in turn change slightly where the various shell-burning episodes will
ignite. Both the 15 and 25~M$_\odot$ models exhibit higher central C/O ratios
using the upper limit for the rate of this work compared with
the upper limit of \citet{0004-637X-567-1-643}, because the latter rate is
faster (see Fig.~\ref{fig:RR_compare}). The converse statement is true for the
lower limits.

\begin{figure}
\includegraphics[width=1.\columnwidth]{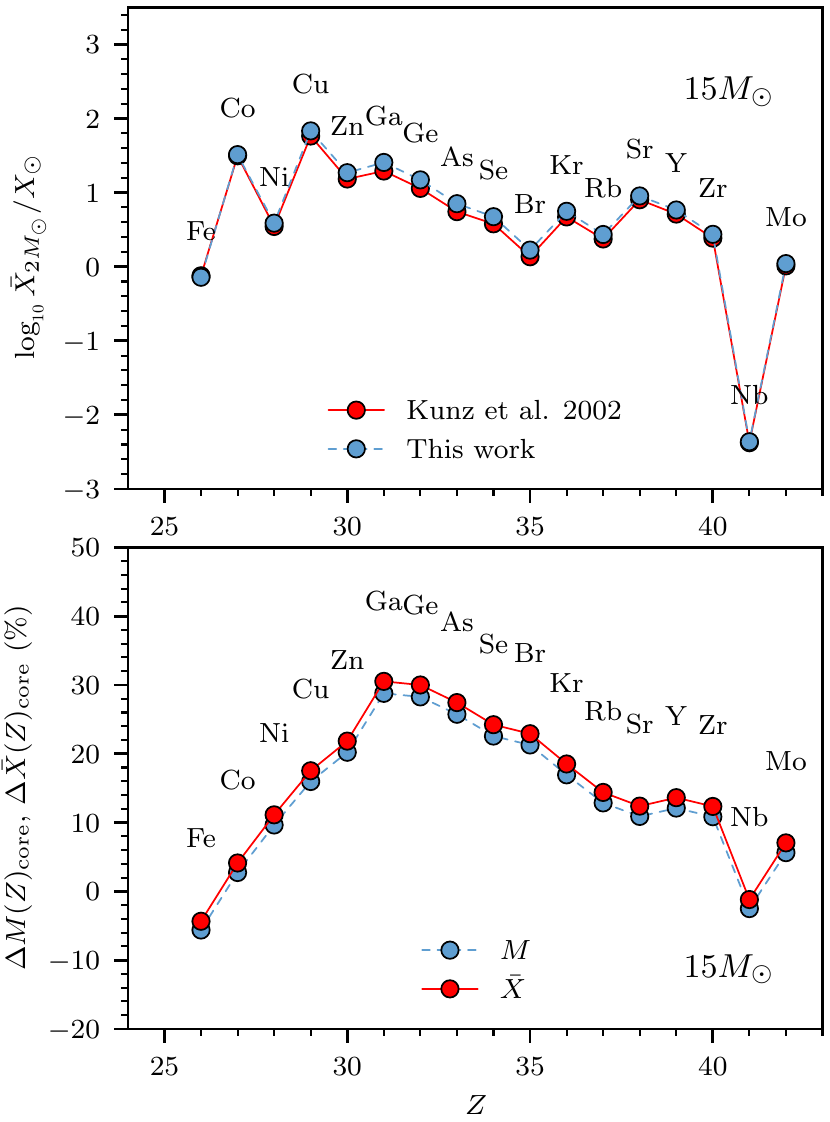}

\caption{(Color online) Top panel: mass-averaged mass fractions of the
        weak $s$-process elements in the central 2~M$_\odot$ of the 15~M$_\odot$
        models at the end of core He-burning using the
        $^{12}$C($\alpha$,$\gamma$)$^{16}$O rate of this work and that of \citet{0004-637X-567-1-643}. \\
        Bottom panel: percentage difference in weak $s$-process production at the
        end of core He-burning as both total elemental masses and mass-averaged mass
        fractions for the 15~M$_\odot$ models using the rates of this work and \citet{0004-637X-567-1-643}.}

\label{fig: he_core_sprocess}
\end{figure}

The nucleosynthesis from the 15~M$_\odot$ stellar models was computed by
post-processing the thermodynamic evolution of the models with a 1107-isotope
nuclear reaction network and mixing the species after every network time step
by solving the diffusion equation using the diffusion coefficient from the
\texttt{MESA} model. For this, the NuGrid software instruments were used
\citep{pignatari:12,jones:15}. In Fig.~\ref{fig: he_core_sprocess} the
abundances of elements with $26<Z<42$ (Fe--Mo) that are affected by the weak
$s$-process are shown at the end of the core He-burning phase. The top panel shows
the mass-weighted average of the elemental mass fractions in the inner-most
2~M$_\odot$ of the star as so-called overabundances (i.e. the abundance of each
element is normalized to the solar abundance, which in this case was the
initial composition of the models). Both models exhibit the familiar weak
$s$-process pattern that is made when $^{56}$Fe, the seed isotope, is depleted as
it captures neutrons being released by the $^{22}$Ne($\alpha$,n)$^{25}$Mg
reaction, producing the elements up to the neutron shell closure at $N=50$ (Sr,
Y, Zr).

The bottom panel in Fig.~\ref{fig: he_core_sprocess} shows the differences, as
percentages, in the abundances of the elements in the CO core at the end of
core He-burning when using the $^{12}$C($\alpha$,$\gamma$)$^{16}$O reaction
rate of this work compared with that of
\citet{0004-637X-567-1-643}. The differences are shown for both the total masses of
the elements in the CO core ($M$, blue circles) and the average mass fractions
across the CO core ($\bar{X}$, red circles). The two lines are similar, indicating that the difference in the actual CO
core masses resulting from the different $^{12}$C($\alpha$,$\gamma$)$^{16}$O
rates has no strong influence on the weak $s$-process element production, as for instance the $\alpha$ capture rates on the neutron source $^{22}$Ne (e.g. see for a recent impact study \citet{talwar:15}). The
weak $s$-process production is more efficient in the model that uses the adopted
$^{12}$C($\alpha$,$\gamma$)$^{16}$O rate from this work because the rate is
slower and therefore $^{12}$C is less of a competitor for $\alpha$ particles
than $^{22}$Ne, the neutron source. This can also be seen by simply noting that
$\Delta M(\mathrm{Fe}) < 0$ or $\Delta \bar{X}(\mathrm{Fe})_\mathrm{core} < 0$
in the bottom panel of Fig.~\ref{fig: he_core_sprocess}, i.e. the element Fe --
comprised mostly by the $s$-process seed isotope $^{56}$Fe -- is further
depleted in the model with the adopted rate of the present work than in the model
with the \citet{0004-637X-567-1-643} rate.

The difference in the weak $s$-process production arising from the different $^{12}$C($\alpha$,$\gamma$)$^{16}$O rates (this work vs. \citealp{0004-637X-567-1-643}) is significant and is as much as $\simeq 30\%$ in the total masses of Ga and Ge that are produced.
The variation in the weak $s$-process production that is due to the uncertainty of the $^{12}$C($\alpha$,$\gamma$)$^{16}$O rate from this work (not shown in Fig.~\ref{fig: he_core_sprocess}) is on the order of 20-30\%.

\section{Summary and Outlook}\label{sec:conc}

The $^{12}$C$(\alpha,\gamma)^{16}$O reaction has remained one of the most challenging problems in nuclear astrophysics. This is because it is extreme difficultly to access experimentally, it can not be accurately predict with theory, and even its impact on stellar environments is so convoluted with other uncertainties that its exact effect is difficult to quantify. Yet there is no doubt that it remains one of the most critical reactions for our understanding of nucleosynthesis and stellar evolution, and attempts to address these issues have remained at the forefront across the field. The major experimental improvements have been made through indirect techniques such as improved measurements of the $\beta$ delayed $\alpha$ emission spectra of $^{16}$N and sub-Coulomb transfer. This is apparent in Fig.~\ref{fig:S_0_summary_2} where $^{16}$N$(\beta\alpha)^{12}$C measurements drastically decreased the $E1$ uncertainties in the early 90's and the transfer measurements resulted in a large decrease in the $E2$ uncertainty around the early 2000's. The combination of these measurements together with a large amount of low energy data and the inclusion of higher energy data in the $R$-matrix analysis have thus reduced the uncertainty in the extrapolation for $S$(300~keV) to $\approx$20\% uncertainty level. However, improvements in the uncertainty are now hindered by the tension produced by inconsistencies between different measurements. This issue now bars the way to the smaller uncertainty level reflected by the statistical uncertainties of the individual measurements. To move forward, these inconsistencies must be resolved.

With the uncertainty in the rate suggested by this work and other recent global
analyses, is there still a real need for improvement? Some stellar modelers
have indicated that if the uncertainty in the $^{12}$C$(\alpha,\gamma)^{16}$O
rate reaches the 10\% level, it will be on par with other non-nuclear physics
uncertainties as well as that of the 3$\alpha$ reaction rate. However, it has
been pointed out that for AGB stars the impact of the present level of the
uncertainty in the $^{12}$C$(\alpha,\gamma)^{16}$O reaction rate is smaller
than other sources of uncertainties, e.g., the convective-boundary mechanisms
active at the He intershell, during the AGB phase. Yet for massive stars the
required level of precision must be at this level of about 10\% since post-helium burning
evolution is strongly affected by the $^{12}$C$(\alpha,\gamma)^{16}$O rate.
For example, in this work it has been suggested that relevant structure
differences are at the $\simeq$2\% uncertainty level in the carbon-oxygen core
mass, at the $\simeq$15\% level in the carbon and oxygen mass fractions, and at
the $\simeq$30\% level in the weak $s$-process yields. These estimates are only
for a limited number of stellar models, and we encourage the community to more
thoroughly investigate the impacts of this work's $^{12}$C$(\alpha,\gamma)^{16}$O reaction rate.

So how should we proceed? Of course any new method that allows for the more precise and accurate determination of the various different experimental data will improve the situation. For now we end finally with some suggestions for specific experimental measurements that could improve our knowledge of the $^{12}$C$(\alpha,\gamma)^{16}$O reaction that can likely be done with existing experimental techniques. 

\begin{itemize}
\item Measurement of ground state angular distributions to higher energies ($E_{c.m.} >$ 3~MeV), specifically in off-resonance regions, to place more stringent upper limits on $E1$ background terms and verify both $E1$ and $E2$ interference patterns.
\item Further verification of $\alpha$ ANCs from transfer reactions, perhaps by systematic studies of additional kinds of transfer reactions. 
\item Re-investigation of $^{16}$N$(\beta\alpha)^{12}$C spectrum measurements until constant measurements can be achieved.
\item Consistent measurements of ANC$^\alpha_{\text{G.S.}}$.
\item Recoil separator measurements to lower energy ($E_{\text{cm}}<$ 2 MeV) and to higher energies, continuing above $S_p$ ($E_{\text{cm}}>$ 5 MeV).
\item Cascade transition measurements over wide energy ranges.
\item Reporting of differential cross sections, not just $E1$ and $E2$ cross sections if the experiment permits.
\item Scattering measurements using thinner targets and covering a wide energy and angular range.
\end{itemize}

\begin{acknowledgments}
The authors would like to acknowledge many useful discussions with M. Avila (ANL), A. DiLeva (INFN), H.O.U. Fynbo (AU), G. Imbriani (INFN), L. Gialanella (CIRCE), G. Hale (LANL), A. Heger (Monash U.), C. Matei (NPL), M. Paris (LANL), A. Roberts (USD), G.V. Rogachev (TAMU), D. Sch\"{u}rmann (CIRCE), O. Straniero (INAF Teramo), F. Strieder (SDSMT), and A. Villano (UMN). RJD would like to recognize the support and hospitality of INFN during a portion this work. This research was supported in part by the Notre Dame Center for Research Computing and funded by the National Science Foundation through Grant No. Phys-0758100, and the Joint Institute for Nuclear Astrophysics through Grant No. Phys-0822648. MP and SJ acknowledge significant support from NuGrid via NSF grants PHY 02-16783 and PHY 09-22648 (Joint Institute for Nuclear Astrophysics, JINA), NSF grant PHY-1430152 (JINA Center for the Evolution of the Elements) and EU MIRG-CT-2006-046520. MP acknowledges support from the ``Lendulet-2014" Programme of the Hungarian Academy of Sciences and from SNF (Switzerland), and from the UK BRIDGCE network. NuGrid data is served by Canfar/CADC. SJ is a fellow of the Alexander von Humboldt Foundation and acknowledges support from the Klaus Tschira Stiftung. MP also acknowledges PRACE, through its Distributed Extreme Computing Initiative, for resource allocations on Sisu (CSC, Finland), Archer (EPCC, UK), and Beskow (KTH, Sweden); the support of STFC's DiRAC High Performance Computing Facilities. Ongoing resource allocations on the University of Hull's High Performance Computing Facility - viper - are gratefully acknowledged. The work of CRB was supported in part by the U.S. Department of Energy, Grant Nos. DE-FG02-88ER40387 and DE-NA0002905. CEF acknowledges partial support from a Predoctoral Fellowship administered by the National Academies of Sciences, Engineering, and Medicine on behalf of the Ford Foundation and an Edward J. Petry Graduate Fellowship from Michigan State University.
\end{acknowledgments}

\appendix

\section{\texorpdfstring{$R$}{R}-MATRIX FIT PARAMETERS} \label{app:fit_parameters}

The $R$-matrix parameters in this section represent the best fit of this work. It should be noted that this fit does not include any background poles for the capture channels, although it does for the particle channels. The value of $S_\text{total}$(300 keV) for the $^{12}$C$(\alpha,\gamma)^{16}$O reaction resulting from these parameters is 140~keV b. The fit used channel radii $a_{\alpha_0}$ = $a_{\alpha_1}$ = 5.43~fm and $a_{p_0}$ = 5.03~fm. No boundary conditions must be defined since the alternate parameterization of \textcite{PhysRevC.66.044611} is utilized.

The comprehensive fit used in this work is also quite complicated and includes many parameters. In order to aid in the reproduction of the results of this work, an \texttt{AZURE2} input file is provided in the Supplemental Material. The \texttt{AZURE2} code is open source and can be obtained at \url{azure.nd.edu}. 

\begin{table*}
\caption{Observable energies and particle widths (or ANCs) used for the $R$-matrix fit (see Eqs.~(\ref{eq:width_convert}) and (\ref{eq:ANC_to_rw})). Parameters marked in \textbf{bold} were treated as fit parameters. All others were held constant at their central or nominal values. For the uncertainty analysis, fixed parameters with uncertainties were varied assuming a Gaussian PDF. Minus signs on the partial widths correspond to the sign of the corresponding reduced width amplitude. Uncertainties are given in the form $\text{(central value)}^\text{+(stat, syst)}_\text{-(stat, syst)}$. \label{tab:fit_params_part}}
\begin{ruledtabular}
\begin{tabular}{ c c c c c c c c c }
& & & \multicolumn{6}{c}{(keV or fm$^{-1/2}$)} \\ \cline{4-9}
$J^\pi$ & \multicolumn{2}{c}{$\tilde{E}_x$ (MeV)} & \multicolumn{2}{c}{$\tilde{\Gamma}_{\alpha_0}$ or ANC$_\alpha$} & \multicolumn{2}{c}{$\tilde{\Gamma}_{p_0}$ or ANC$_p$} & \multicolumn{2}{c}{$\tilde{\Gamma}_{\alpha_1}$}  \\ \cline{2-3} \cline{4-5} \cline{6-7} \cline{8-9}
& this work & lit. & this work & lit. & this work & lit. & this work & lit. \\  
\hline
0$^-$ & \bf{12.7954}$^{(4,0)}_{(3,0)}$ & 12.7937 & & & \bf{40.9}$^{(15,1)}_{(15,2)}$ & 40 &  & \\
0$^+$ & & 0.0 & 58 &  & & 13.9(19) & & \\
0$^+$ & & 6.0494(10) & & 1560(100) & & & & \\
0$^+$ & & 12.049(2) & & 1.5(5) & & & & \\
0$^+$ & 15 (BGP) & & $\mathbf{-11.900}^{(220,50)}_{(165,110)}\mathbf{\times 10^3}$ & & \bf{420}$^{(37,70)}_{(47,20)}$ & & & \\
1$^-$ & & 7.1165(14) & & $2.08(20)\times 10^{14}$ & & 0.98(12) & & \\
1$^-$ & \bf{9.586}$^{(1,0)}_{(1,8)}$ & 9.585(11) & \bf{382}$^{(3,0)}_{(3,4)}$ & 420(20) & & & & \\
1$^-$ & \bf{12.4493}$^{(8,2)}_{(8,0)}$ & 12.440(2) & \bf{99.2}$^{(11,2)}_{(8,2)}$ & 91(6) & \bf{1.73}$^{(4,2)}_{(4,20)}$ & 1.1 & $\mathbf{-0.031}$$^{(2,4)}_{(2,2)}$ & 0.025 \\
1$^-$ & \bf{13.094}$^{(1,2)}_{(1,0)}$ & 13.090(8) & $\mathbf{-29.9}$$^{(1,0)}_{(1,6)}$ & 45 & \bf{110.4}$^{(2,5)}_{(2,0)}$ & 100 & \bf{0.636}$^{(27,5)}_{(43,0)}$ & 1 \\
1$^-$ & 17.09 (BGP) & & 500 & & & & & \\
1$^-$ & 20 (BGP) & & $\mathbf{15.6}^{(7,1)}_{(4,2)}\mathbf{\times 10^3}$ & & \bf{270}$^{(61,15)}_{(30,170)}$ & & & \\
1$^+$ & \bf{13.6646}$^{(7,0)}_{(7,2)}$ & 13.664(3) &  &  & $\mathbf{-10.3}$$^{(2,1)}_{(4,3)}$ & 8(3) & \bf{61.1}$^{(23,4)}_{(17,2)}$ & 55(3) \\
2$^+$ & & 6.9171(6) & & $1.14(20)\times 10^5$ & & 0.45(13) & & \\
2$^+$ & & 9.8445(5) & & 0.62(10) & & & & \\
2$^+$ & \bf{11.5055}$^{(3,0)}_{(5,1)}$ & 11.520(4) & \bf{83.0}$^{(6,1)}_{(3,0)}$ & 71(3) & & & & \\
2$^+$ & \bf{12.9656}$^{(14,5)}_{(28,2)}$ & 13.02(1) & $\mathbf{-349}$$^{(8,2)}_{(3,3)}$ & 150(10) & {\bf 1.82}$^{(8,4)}_{(8,18)}$ & & $\mathbf{-4.0}$$^{(7,0)}_{(9,0)}$ & \\
2$^+$ & 15 & & $\mathbf{5.5}^{(5,2)}_{(5,2)}\mathbf{\times 10^3}$ & & $\mathbf{-8.1}$$^{(20,0)}_{(15,5)}$ & & & \\
3$^-$ & & 6.12989(4) & & 139(9) & & 1.88(23) & & \\
3$^-$ & {\bf 11.5058}$^{(28,0)}_{(19,7)}$ & 11.60(2) & {\bf 902}$^{(5,1)}_{(6,1)}$ & 800(100) & & & & \\
3$^-$ & {\bf 13.1412}$^{(10,3)}_{(13,0)}$ & 13.129(10) & {\bf 72.9}$^{(42,9)}_{(24,8)}$ & 90(14) & $\mathbf{-1.11}$$^{(2,1)}_{(3,2)}$ & 1 & {\bf 20.5}$^{(13,0)}_{(16,1)}$ & 20 \\
3$^-$ & 13.2650 & 13.259(2) & {\bf 12.8}$^{(6,2)}_{(4,2)}$ & 9(4) & {\bf 3.66}$^{(10,1)}_{(8,2)}$ & 4.1 & {\bf 11.74}$^{(33,1)}_{(54,1)}$ & 8.2(11)\\
3$^-$ & 20 (BGP) & & $\mathbf{-22}^{(1,2)}_{(2,0)}\mathbf{\times 10^3}$ & & & & & \\
4$^+$ & {\bf 10.3581}$^{(1,0)}_{(1,1)}$ & 10.356(3) & {\bf 26.13}$^{(25,3)}_{(15,3)}$ & 26(3) & & & & \\
4$^+$ & & 11.0967(16) & & 0.28(5) & & & & \\
4$^+$ & 15 (BGP) & & $\mathbf{2.57\times 10^3}$$^{(22,2)}_{(14,4)}$ & & & & & \\
5$^-$ & 14.66 (BGP) & & $\mathbf{1.22\times 10^3}$$^{(10,12)}_{(6,9)}$ & & 8 & & & \\

\end{tabular}
\end{ruledtabular}
\end{table*}

\begin{table*}
\caption{Observable $\gamma$-ray widths of from the best fit of the $R$-matrix analysis as defined by Eq.~(\ref{eq:G_to_g_gammas}). Excitation energies are in MeV and correspond to those of Table~\ref{tab:fit_params_part}. \label{tab:fit_params_gamma}}
\begin{ruledtabular}
\begin{tabular}{ c c c c c c c c c c c c }
& & \multicolumn{10}{c}{(meV)} \\ \cline{3-12}
$J^\pi$ & $\tilde{E}_x$ & \multicolumn{2}{c}{$\tilde{\Gamma}_{\gamma_0}$} & \multicolumn{2}{c}{$\tilde{\Gamma}_{\gamma_{6.05}}$} & \multicolumn{2}{c}{$\tilde{\Gamma}_{\gamma_{6.13}}$} & \multicolumn{2}{c}{$\tilde{\Gamma}_{\gamma_{6.92}}$} & \multicolumn{2}{c}{$\tilde{\Gamma}_{\gamma_{7.12}}$}   \\ \cline{3-4} \cline{5-6} \cline{7-8} \cline{9-10} \cline{11-12}
 & & this work & lit. & this work & lit. & this work & lit. & this work & lit. & this work & lit. \\  
\hline
0$^-$ & 12.80 & \bf{3.28}$^{(24,2)}_{(19,2)}\mathbf{\times 10^3}$ & & & & & & & & & \\
1$^-$ & 7.12 & & 55(3) & & 3.3$\times$10$^{-4}$ & & 0.046 & & 1$\times$10$^{-3}$ & & \\
1$^-$ & 9.59 & $\mathbf{-15}$$^{(1,3)}_{(2,2)}$ & 15.6(12) & & & & & \bf{0.5}$^{(3,7)}_{(2,0)}$ & 1.4(14) & \bf{4.2}$^{(9,30)}_{(4,2)}$ & 7.8(16) \\
1$^-$ & 12.45 & $\mathbf{5.6}^{(2,0)}_{(2,9)}\mathbf{\times 10^3}$ & 9.5(17)$\times$10$^3$ & $\mathbf{-53}$$^{(18,8)}_{(16,4)}$ & 120(60) & \bf{30}$^{(15,2)}_{(18,9)}$ & & & & \bf{118}$^{(16,15)}_{(15,2)}$ & \\
1$^-$ & 13.09 & $\mathbf{42}^{(2,0)}_{(1,8)}\mathbf{\times 10^3}$ & 44(8)$\times$10$^3$ & 244 & & $-400$ & & & & & 1.35(4)$\times$10$^3$ \\
1$^-$ & 17.09 & & 500$\times 10^3$ & & & & & & & & \\
2$^+$ & 6.92 & & 97(3) & & $-0.027(3)$ & & $<$0.0078 & & & & \\
2$^+$ & 9.84 & & $-5.7(6)$ & & $-1.9(4)$ & & & & 2.2(4) & & 0.3 \\
2$^+$ & 11.51 & $\mathbf{-490}$$^{(40,100)}_{(70,30)}$ & 610(20) & \bf{49}$^{(27,19)}_{(23,0)}$ & 30(5) & $\mathbf{-33}$$^{(5,1)}_{(4,5)}$ & & \bf{24}$^{(7,5)}_{(5,0)}$ & 29(7) & \bf{14.0}$^{(4,1)}_{(4,6)}$ & $<$5 \\
2$^+$ & 12.97 & $\mathbf{-560}$$^{(120,0)}_{(60,190)}$ & 700 & & & & & & & & \\
3$^-$ & 6.13 & & 0.0260(13) & & & & & & & & \\
3$^-$ & 11.51 & {\bf 0.20 }$^{(20,64)}_{(18,20)}$ \footnotemark[1] & & & & {\bf 8}$^{(3)}_{(3)}$ & & {\bf 8}$^{(7,10)}_{(5,8)}$ \footnotemark[1] & & {\bf 21}$^{(4,7)}_{(4,7)}$ & \\
3$^-$ & 13.14 & 10 &  & & & 8000 & & & & & \\
3$^-$ & 13.26 & & & & & $\mathbf{-5.2}$$^{(6,3)}_{(6,0)}\mathbf{\times 10^3}$ & & & & & \\
4$^+$ & 10.36 & & 5.6(20)$\times 10^{-5}$ & & $<$1.0 & & & {\bf 42.7}$^{(2,7)}_{(6,2)}$ & 62(6) & & \\
4$^+$ & 11.10 & & & & & & 3.1(13) & & 2.5(6) & & \\   
\end{tabular}
\end{ruledtabular}
\footnotetext[1]{Consistent with zero, should be treated as an upper limit.}
\end{table*}

\begin{table*}
\caption{$\beta$-delayed $\alpha$-decay parameters resulting from the $R$-matrix fit. The half-life of $^{16}$N was taken as $t_{1/2}$ = 7.13(2)~s with a $\beta\alpha$ branching ratio of 1.20(5)$\times$10$^{-5}$ \cite{Tilley19931}. $\beta$-decay feeding factors $\tilde{B}_\lambda$ are those defined by \textcite{PhysRevC.66.044611} and can be compared directly with those found in Table~I of that work. $\log ft_{1/2}$ values were calculated using Eq.~(\ref{eq:logft}).} \label{tab:beta_params}
\begin{ruledtabular}
\begin{tabular}{ c c c c c } 
 & & & \multicolumn{2}{ c }{$\log ft_{1/2}$} \\ \cline{4-5}
 $J^\pi$ & $\tilde{E}_x$ (MeV) & $\tilde{B}_\lambda$ & this work & \textcite{Tilley19931}  \\
\hline
3$^-$ & 6.13 & 2.54 & {\bf 4.59}$^{(16,0)}_{(24,25)}$ &  4.48(4)  \\
1$^-$ & 7.12 & 1.27 & {\bf 5.08}$^{(0,1)}_{(2,2)}$ & 5.11(4) \\
1$^-$ & 9.59 & 0.451 & {\bf 6.15}$^{(2,1)}_{(2,0)}$ & 6.12(5)\\
1$^-$ & 20 (BGP) & $-$0.618 & $\mathbf{-5.70}$$^{(6,1)}_{(4,1)}$ & \\
\end{tabular}
\end{ruledtabular}
\end{table*} 

\begin{turnpage}
\begin{table*}
\caption{The internal reduced width amplitudes $\tilde{\gamma}_p$ and the external reduced width amplitudes $\tilde{\gamma}_p$(ch) calculated using Eq.~(\ref{eq:G_to_g_gammas}) from which the $\gamma$ widths in Table~\ref{tab:fit_params_gamma} were calculated. {\bf Bold} internal reduced width amplitudes correspond to the partial widths that were treated as fit parameters in Table~\ref{tab:fit_params_gamma} (i.e. they are not additional fit parameters). External reduced width amplitudes were always fixed as they correspond to fixed ANCs. Excitation energies are in MeV. \label{tab:fit_params_grwa}}
\begin{ruledtabular}
\begin{tabular}{ c c c c c c c c c c c c }
& & \multicolumn{10}{c}{(MeV$^{1/2}$)} \\ \cline{3-12}
$J^\pi$ & $\tilde{E}_x$ & \multicolumn{2}{c}{$E_f=0$} & \multicolumn{2}{c}{$E_f=6.05$} & \multicolumn{2}{c}{$E_f=6.13$} & \multicolumn{2}{c}{$E_f=6.92$} & \multicolumn{2}{c}{$E_f=7.12$}   \\ \cline{3-4} \cline{5-6} \cline{7-8} \cline{9-10} \cline{11-12}
 & & $\tilde{\gamma}_p$ & $\tilde{\gamma}_p$(ch) & $\tilde{\gamma}_p$ & $\tilde{\gamma}_p$(ch) & $\tilde{\gamma}_p$ & $\tilde{\gamma}_p$(ch) & $\tilde{\gamma}_p$ & $\tilde{\gamma}_p$(ch) & $\tilde{\gamma}_p$ & $\tilde{\gamma}_p$(ch) \\  
\hline
0$^-$ & 12.80 & \bf{0.295} & & & & & & & & & \\
1$^-$ & 7.12 & 0.0188 & 0.00567+$i$0.0 & & & & & & & & \\
1$^-$ & 9.59 & {\bf $-$0.00986} & $-$(3.8+$i$0.6)$\times 10^{-5}$ & & & & & {\bf 0.00939} & $-$(2.14+$i$1.03)$\times 10^{-4}$ & \bf{1.022} & 1.125+$i$0.472 \\
1$^-$ & 12.45 & {\bf 0.0588} & (639+$i$0.69)$\times 10^{-4}$ & {\bf $-$0.0321} & $-$(0.82+$i$1.35)$\times 10^{-4}$ & \bf{0.368} & 0.399+$i$0.109 & & & \bf{2.23} & 0.109+$i$0.158 \\
1$^-$ & 13.09 & 0.233 & 0.0610+$i$0.0047 & \bf{0.0563} & (4.0+$i$7.6)$\times 10^{-5}$ & \bf{$-$2.316} & 0.234$-i$0.0124 & & & {\bf 5.54} & $-$(0.0449+$i$0.0742) \\
1$^-$ & 17.09 & 0.587 & 0.0338+$i$0.0194 & & & & & & & & \\
2$^+$ & 6.92 & 0.974 & 0.0157+$i$0.0 & $-$3.31 & 0.401+$i$0.0 & 2.27$\times 10^{-3}$ & (5.81+$i$0.0)$\times 10^{-3}$ & & & & \\
2$^+$ & 9.84 & $-$0.0987 & (9.54+$i$1.06)$\times 10^{-4}$ & $-$0.650 & 0.0497+$i$0.0132 & & & 1.181 & 0.0560+$i$0.0163 & 7.54$\times 10^{-3}$ & $-$(4+$i$2)$\times 10^{-6}$\\
2$^+$ & 11.51 & $\mathbf{-0.610}$ & (5.42+$i$2.05)$\times 10^{-3}$ & \bf{0.991} & 0.230+$i$0.171 & $\mathbf{-0.0287}$ & $-$(5.0+$i$3.8)$\times 10^{-5}$ & \bf{1.057} & 0.252+$i$0.195 & \bf{0.0254} & $-$(3.4+$i$4.2)$\times 10^{-5}$ \\
2$^+$ & 12.97 & $\mathbf{-0.512}$ & 0.0243$-i$0.0038 & & & & & & & & \\
3$^-$ & 6.13 & 0.691 & & & & & & & & & \\
3$^-$ & 11.51 & $\mathbf{0.242}$ & 0.0+$i$0.0 ($E3$) & & & $\mathbf{-0.138}$ & 0.543+$i$0.221 & $\mathbf{0.0113}$ & $-$(4.22+$i$2.95)$\times 10^{-4}$ & $\mathbf{0.787}$ & 0.779+$i$0.481 \\
3$^-$ & 13.14 & {\bf 1.07} & 0.0+$i$0.0 ($E3$) & & & 9.70 & 0.0144+$i$0.0600 & & & & \\
3$^-$ & 13.26 & & & & & $\mathbf{-7.09}$ & 0.130+$i$0.0252 & & & & \\
4$^+$ & 10.36 & & & & & & & {\bf 3.18} & 0.729+$i$0.0819 & & \\
4$^+$ & 11.10 & & & & & 9.87$\times 10^{-3}$ & $-$(6+1)$\times 10^{-6}$ & 0.500 & 0.0416+$i$0.0094 & & \\   
\end{tabular}
\end{ruledtabular}
\end{table*}
\end{turnpage}

\section{TABULATED REACTION RATE} \label{app:RR}

The tabulated reaction rate is calculated at the same temperatures as in \textcite{nacre} (NACRE) for ease of comparison. In addition, the rate has been parameterized using the format recommended by JINA Reaclib \cite{0067-0049-189-1-240}. In order to fit the rate from 0.06~$< T_9< $~10~GK to a precision of better than 5\%, two instances of the rate parameterization formula,
\begin{equation} \label{eq:jina_reaclib_formula}
N_A\langle\sigma v\rangle = \exp\left[a_0+\sum_{i=1}^5a_iT_9^{\frac{2i-5}{3}}+a_6\ln T_9\right],
\end{equation}
were necessary: a non-resonant and resonance term (i.e. $N_A\langle\sigma v\rangle_\text{total} = N_A\langle\sigma v\rangle_\text{non-resonant}+ N_A\langle\sigma v\rangle_\text{resonance}$). The first follows the suggested format for a charged particle induced non-resonance contribution, while the second represents a single isolated narrow resonance term. The parameters are given in Table~\ref{tab:RR_jina_params}. Note that the parameter values do not necessarily correspond to physical quantities.

\begin{table*}
\caption{\label{tab:RR_absolute} The rate of the $^{12}$C$(\alpha,\gamma)^{16}$O reaction. Uncertainties are calculated using a combination of Monte Carlo analysis and investigation of systematic contributions from both data and model sources and are listed separately.}
\begin{ruledtabular}
\begin{tabular}{ c c c c } 
$T$ (GK) & Adopted Rate & Lower Rate & Upper Rate \\
\hline
0.06	&	6.78	$\times 10^{	-26	}$	&	5.69	$\times 10^{	-26	}$	&	7.90	$\times 10^{	-26	}$	\\
0.07	&	3.28	$\times 10^{	-24	}$	&	2.76	$\times 10^{	-24	}$	&	3.83	$\times 10^{	-24	}$	\\
0.08	&	8.00	$\times 10^{	-23	}$	&	6.71	$\times 10^{	-23	}$	&	9.35	$\times 10^{	-23	}$	\\
0.09	&	1.18	$\times 10^{	-21	}$	&	9.91	$\times 10^{	-22	}$	&	1.38	$\times 10^{	-21	}$	\\
0.1	&	1.20	$\times 10^{	-20	}$	&	1.00	$\times 10^{	-20	}$	&	1.40	$\times 10^{	-20	}$	\\
0.11	&	9.03	$\times 10^{	-20	}$	&	7.55	$\times 10^{	-20	}$	&	1.06	$\times 10^{	-19	}$	\\
0.12	&	5.38	$\times 10^{	-19	}$	&	4.50	$\times 10^{	-19	}$	&	6.31	$\times 10^{	-19	}$	\\
0.13	&	2.65	$\times 10^{	-18	}$	&	2.21	$\times 10^{	-18	}$	&	3.11	$\times 10^{	-18	}$	\\
0.14	&	1.11	$\times 10^{	-17	}$	&	9.28	$\times 10^{	-18	}$	&	1.30	$\times 10^{	-17	}$	\\
0.15	&	4.08	$\times 10^{	-17	}$	&	3.41	$\times 10^{	-17	}$	&	4.80	$\times 10^{	-17	}$	\\
0.16	&	1.34	$\times 10^{	-16	}$	&	1.12	$\times 10^{	-16	}$	&	1.58	$\times 10^{	-16	}$	\\
0.18	&	1.09	$\times 10^{	-15	}$	&	9.11	$\times 10^{	-16	}$	&	1.29	$\times 10^{	-15	}$	\\
0.2	&	6.64	$\times 10^{	-15	}$	&	5.53	$\times 10^{	-15	}$	&	7.83	$\times 10^{	-15	}$	\\
0.25	&	2.43	$\times 10^{	-13	}$	&	2.02	$\times 10^{	-13	}$	&	2.87	$\times 10^{	-13	}$	\\
0.3	&	3.73	$\times 10^{	-12	}$	&	3.10	$\times 10^{	-12	}$	&	4.43	$\times 10^{	-12	}$	\\
0.35	&	3.28	$\times 10^{	-11	}$	&	2.72	$\times 10^{	-11	}$	&	3.90	$\times 10^{	-11	}$	\\
0.4	&	1.96	$\times 10^{	-10	}$	&	1.62	$\times 10^{	-10	}$	&	2.33	$\times 10^{	-10	}$	\\
0.45	&	8.82	$\times 10^{	-10	}$	&	7.30	$\times 10^{	-10	}$	&	1.05	$\times 10^{	-9	}$	\\
0.5	&	3.22	$\times 10^{	-9	}$	&	2.66	$\times 10^{	-9	}$	&	3.85	$\times 10^{	-9	}$	\\
0.6	&	2.70	$\times 10^{	-8	}$	&	2.23	$\times 10^{	-8	}$	&	3.23	$\times 10^{	-8	}$	\\
0.7	&	1.47	$\times 10^{	-7	}$	&	1.21	$\times 10^{	-7	}$	&	1.76	$\times 10^{	-7	}$	\\
0.8	&	5.92	$\times 10^{	-7	}$	&	4.90	$\times 10^{	-7	}$	&	7.11	$\times 10^{	-7	}$	\\
0.9	&	1.92	$\times 10^{	-6	}$	&	1.59	$\times 10^{	-6	}$	&	2.31	$\times 10^{	-6	}$	\\
1	&	5.30	$\times 10^{	-6	}$	&	4.40	$\times 10^{	-6	}$	&	6.38	$\times 10^{	-6	}$	\\
1.25	&	4.10	$\times 10^{	-5	}$	&	3.42	$\times 10^{	-5	}$	&	4.93	$\times 10^{	-5	}$	\\
1.5	&	2.03	$\times 10^{	-4	}$	&	1.70	$\times 10^{	-4	}$	&	2.43	$\times 10^{	-4	}$	\\
1.75	&	7.65	$\times 10^{	-4	}$	&	6.46	$\times 10^{	-4	}$	&	9.14	$\times 10^{	-4	}$	\\
2	&	2.40	$\times 10^{	-3	}$	&	2.04	$\times 10^{	-3	}$	&	2.86	$\times 10^{	-3	}$	\\
2.5	&	1.57	$\times 10^{	-2	}$	&	1.32	$\times 10^{	-2	}$	&	1.88	$\times 10^{	-2	}$	\\
3	&	6.66	$\times 10^{	-2	}$	&	5.51	$\times 10^{	-2	}$	&	8.10	$\times 10^{	-2	}$	\\
3.5	&	2.09	$\times 10^{	-1	}$	&	1.71	$\times 10^{	-1	}$	&	2.55	$\times 10^{	-1	}$	\\
4	&	5.31	$\times 10^{	-1	}$	&	4.37	$\times 10^{	-1	}$	&	6.48	$\times 10^{	-1	}$	\\
5	&	2.38	$\times 10^{	0	}$	&	2.02	$\times 10^{	0	}$	&	2.84	$\times 10^{	0	}$	\\
6	&	7.93	$\times 10^{	0	}$	&	6.96	$\times 10^{	0	}$	&	9.22	$\times 10^{	0	}$	\\
7	&	2.11	$\times 10^{	1	}$	&	1.89	$\times 10^{	1	}$	&	2.41	$\times 10^{	1	}$	\\
8	&	4.64	$\times 10^{	1	}$	&	4.20	$\times 10^{	1	}$	&	5.26	$\times 10^{	1	}$	\\
9	&	8.75	$\times 10^{	1	}$	&	7.96	$\times 10^{	1	}$	&	9.86	$\times 10^{	1	}$	\\
10	&	1.46	$\times 10^{	2	}$	&	1.33	$\times 10^{	2	}$	&	1.64	$\times 10^{	2	}$	\\

\end{tabular}
\end{ruledtabular}
\end{table*} 

\begin{table*}
\caption{\label{tab:RR_jina_params} The sum of two instances of Eq.~\ref{eq:jina_reaclib_formula} are necessary to fit the reaction rate to better than 5\% accuracy over the range 0.06~$< T <$~10~GK and the parameters given here reproduce the recommend rate to better than 3.5\%. Not all parameters are necessary for the fitting and these have been set to zero. Parameters that were adjusted for the fit are marked in \textbf{bold}.}
\begin{ruledtabular}
\begin{tabular}{ l c c c c c c c }
term & $a_0$ & $a_1$ & $a_2$ & $a_3$ & $a_4$ & $a_5$ & $a_6$ \\
\hline
non-resonant & \textbf{24.1} & 0 & \textbf{-32} & \textbf{-5.9} & \textbf{1.8} & \textbf{-0.17} & -2/3  \\
resonance & \textbf{7.4} & \textbf{-30} & 0 & 0 & 0 & 0 & -3/2 \\ 
\end{tabular}
\end{ruledtabular}
\end{table*}

\newpage 
\bibliography{16O_paper}

\end{document}